\documentclass[twocolumn,showpacs,aps,floatfix,prd]{revtex4}

\usepackage{graphicx}
\usepackage{dcolumn}
\usepackage{amsmath}
\usepackage{epsfig}
\usepackage{hhline}

\newcommand{\BABARPubYear}    {01}
\newcommand{\BABARPubNumber}  {03}
\newcommand{\SLACPubNumber} {9060}

\newcommand{\deltam}{\ensuremath{\Delta m}}
\newcommand{\deltat}{\ensuremath{\Delta t}}
\newcommand{\deltae}{\ensuremath{\Delta E}}
\newcommand{\etaCP}{\ensuremath{\eta_{\CP}}}

\newcommand{\hbarps}{\,\ps^{-1}}

\def\D       {\ensuremath{D}}
\def\Dp      {\ensuremath{D^+}}
\def\Dm      {\ensuremath{D^-}}
\def\Dstar   {\ensuremath{D^{*}}}
\def\Dstarp  {\ensuremath{D^{*+}}}
\def\Dstarm  {\ensuremath{D^{*-}}}

\def\DDstarm {\ensuremath{D^{(*)-}}}
\def\rhop    {\ensuremath{\rho^+}}
\def\rhom    {\ensuremath{\rho^-}}
\def\aonep   {\ensuremath{a_1^+}}

\newcommand{\abslambda}{\ensuremath{|\lambda|}}
\newcommand{\etaimlambdaoverabslambda}{\ensuremath{Im\lambda
/\abslambda}}

\newcommand{\phanm}{\ensuremath{\phantom{-}}}

\newcommand {\ctby} {\ensuremath{\cos \theta_{B-\Dstar \ell}}}
\newcommand {\bzdstlnu} {\ensuremath{B^0 \ra D^{\ast-} \ell^+ \overline \nu}}
\newcommand {\bdstxlnu} {\ensuremath{B \ra D^{\ast} X \ell \overline \nu}}
\newcommand {\bzdstxlnu} {\ensuremath{B^0 \ra D^{\ast} X \ell^+ \overline \nu}}
\newcommand {\bchdstxlnu} {\ensuremath{B^+ \ra D^{\ast} X \ell \overline \nu}}
\newcommand {\bdstpilnu} {\ensuremath{B \ra D^{\ast} \pi \ell \overline \nu}}
\newcommand {\bbdstyl} {\ensuremath{B \overline B \ra D^{\ast} Y \ell}}
\newcommand{\ndstlnu}{ $29042\pm1500$}
\newcommand{\gssval}{\ensuremath{(4.5\pm0.3\pm2.2)\%}}
\newcommand{\gbbval}{\ensuremath{(4.8\pm0.4\pm2.2)\%}}

\def\ddstar  {\ensuremath{D^{\ast\ast}}}
\newcommand {\kin} {\ensuremath{{\cal{K}}}}

\def\result {\ensuremath{\stwob=0.59\pm 0.14\, {\rm (stat)} \pm 0.05\, {\rm (syst)}}} 
\def\dmresult {\ensuremath{\deltamd=0.516\pm 0.016\, {\rm (stat)} \pm 0.010\, {\rm (syst)} \hbarps}} 

\input pubboard/babarsym

\long\def\inst#1{\par\nobreak\kern 4pt\nobreak
    {\it #1}\par\vskip 10pt plus 3pt minus 3pt}

\begin{document}

\begin{flushleft}
\babar-PUB-\BABARPubYear/\BABARPubNumber \\
SLAC-PUB-\SLACPubNumber\\[5mm]
\end{flushleft}

\title{\large \bf
\Large \bf A Study of Time-Dependent {\boldmath \CP}-Violating Asymmetries\\
and Flavor Oscillations in Neutral {\boldmath \B} Decays at the {\boldmath \FourS}
\begin{center} 
\vskip 5mm
The \babar\ Collaboration
\end{center}
}

%
\author{B.~Aubert}
\author{D.~Boutigny}
\author{J.-M.~Gaillard}
\author{A.~Hicheur}
\author{Y.~Karyotakis}
\author{J.~P.~Lees}
\author{P.~Robbe}
\author{V.~Tisserand}
\affiliation{Laboratoire de Physique des Particules, F-74941 Annecy-le-Vieux, France }
\author{A.~Palano}
\author{A.~Pompili}
\affiliation{Universit\`a di Bari, Dipartimento di Fisica and INFN, I-70126 Bari, Italy }
% ...
\collaboration{The \babar\ Collaboration}
\noaffiliation

\date{\today}

\begin{abstract}
\noindent
We present a measurement of time-dependent 
\CP-violating asymmetries in neutral \B\ meson
decays collected with the \babar\ detector at the \pep2\
asymmetric-energy \BF\ at the Stanford Linear Accelerator Center.  
The data sample consists of 29.7\invfb\  recorded 
at the \FourS\ resonance and 3.9\invfb\ off-resonance.  
One of the neutral \B\ mesons, which are produced in pairs at the \FourS, is 
fully reconstructed in the \CP\ decay modes
$\jpsi \KS$, $\psitwos \KS$, $\chi_{c1}\KS$, 
$\jpsi\Kstarz$ ($\Kstarz\to\KS\piz$)
and $\jpsi\KL$, or in flavor-eigenstate
modes involving $D^{(*)}\pi/\rho/a_1$ and
$\jpsi\Kstarz$ ($\Kstarz\to\Kp\pim$).  
The flavor of the other neutral 
\B\ meson is tagged at the time of its decay, mainly with the charge of 
identified leptons and kaons.  
A neural network tagging algorithm is used to 
recover events without a clear lepton or kaon tag.
The proper time elapsed between the decays is determined 
by measuring the
distance between the decay vertices.
Wrong-tag probabilities, the time-difference resolution function, and the
\Bz-\Bzb\ oscillation frequency \deltamd\ are measured
with a sample of about $6350$ fully-reconstructed \Bz\ decays in
hadronic flavor-eigenstate modes. 
A maximum-likelihood fit to
this flavor eigenstate sample finds
$\dmresult$.
The value of the asymmetry amplitude \stwob\ is determined from
a simultaneous maximum-likelihood fit 
to the time-difference distribution of the flavor-eigenstate 
sample and about 642 tagged 
\Bz\ decays in the \CP-eigenstate modes. We find 
\result, demonstrating that \CP\ violation
exists in the neutral \B\ meson system.
We also determine the value of the \CP\ violation parameter
$\vert\lambda\vert = 0.93 \pm 0.09\ (\rm {stat}) \pm 0.03\ (\rm {syst})$, 
which is consistent with the expectation of
$\vert\lambda\vert= 1$ for no direct \CP\  violation.

\end{abstract}


\maketitle

\newpage
\newcommand{\secname}{}

\setcounter{footnote}{0}





\renewcommand{\secname}{Introduction}
\section{Introduction} 
\label{sec:\secname}

\CP\ violation has been a central concern of particle physics since its
discovery in 1964 \cite{EpsilonK}.  Interest was heightened
by Sakharov's observation \cite{Sakharov} 
in 1967 that without \CP\
violation, a universe that began as matter--anti-matter symmetric
could not have evolved into the asymmetric one we now see.  
An elegant explanation of the 
\CP-violating effects in \KL\ decays is provided by
the \CP-violating phase of the three-generation Cabibbo-Kobayashi-Maskawa 
(CKM) quark-mixing matrix~\cite{CKM}.
However, existing studies of \CP\ violation 
in neutral kaon decays and the resulting experimental constraints on the 
parameters of the CKM matrix~\cite{MSConstraints} do not provide a stringent 
test of whether the CKM phase describes \CP\ violation~\cite{Primer}.
Moreover, the Standard Model does not, through the CKM phase, incorporate
enough \CP\ violation to explain the current matter--anti-matter asymmetry
\cite{matter}.  Understanding
\CP\ violation thus remains a pressing challenge.

An excellent testing ground for \CP\ violation is provided by \B\ mesons through
particle--anti-particle mixing.  A particle that is purely $\Bz$ at time $t=0$
will oscillate between that state and $\Bzb$ with a frequency
$\deltamd$, the difference between the masses of the two neutral \B\ mass eigenstates. 
If decays to a \CP\ eigenstate $f$ 
are observed, any difference between the rates when starting
with a $\Bz$ or with a $\Bzb$ is a manifestation of \CP\ violation.  
In some circumstances, including those in the experiment
described here, the fundamental parameters of \CP\ violation in the
CKM model can be measured from such time-dependent rate asymmetries, unobscured by
strong interactions.
For example, 
a state initially produced as a \Bz\ (\Bzb) can decay to $\jpsi \KS$ directly 
or can oscillate into a \Bzb\ (\Bz) and then decay
to $\jpsi \KS$.  With little theoretical uncertainty in the Standard Model, the phase difference 
between these amplitudes is equal to twice the angle 
$\beta = \arg \left[\, -V_{cd}^{ }V_{cb}^* / V_{td}^{ }V_{tb}^*\, \right]$ of the
Unitarity Triangle.  The \CP-violating asymmetry can thus provide a crucial 
test of the Standard Model. 

The unitarity of the three-generation CKM matrix can be expressed in 
geometric form by six triangles of equal area in the complex plane. 
A nonzero area~\cite{Jarlskog} 
directly implies the existence of a \CP-violating CKM phase. The most 
experimentally accessible of the unitarity relations, involving the two 
smallest elements of the CKM matrix, $V_{ub}$ and $V_{td}$, has come to 
be known as the Unitarity Triangle. 
Because the lengths of the sides of 
the Unitarity Triangle are comparable, the angles  
can be large, leading to potentially large \CP-violating 
asymmetries 
from relative phases between CKM matrix elements.

In \epem\ storage rings operating at the \FourS\ resonance, a $\BzBzb$ pair
produced in an \FourS\ decay 
evolves in a coherent $P$-wave state. 
If one of the \B\ mesons, referred to as $B_{\rm tag}$, can be 
ascertained to decay to a state of known flavor, {\em i.e.} \Bz\ or
\Bzb, at a certain 
time $t_{\rm tag}$, 
the other \B, referred to as $B_{\rm rec}$, {\it at that time} must be of the 
opposite flavor as a consequence of Bose symmetry.
Consequently, the oscillatory probabilities for observing
$\BzBzb$, $\Bz\Bz$ and $\Bzb\Bzb$ pairs produced in
\FourS\ decays are a function of 
$\deltat = t_{\rm rec} - t_{\rm tag}$, allowing
mixing frequencies and \CP\ asymmetries to be determined
if \deltat\ is known.
The charges of identified leptons and kaons are the primary indicators of 
the flavor of
the tagging $B$,  
but other particles also carry flavor information that
can be exploited with a neural network algorithm.
The reconstructed neutral $B$ is found either
in a flavor eigenstate ($B_{\rm rec}=B_{\rm flav}$) 
or a \CP\ mode ($B_{\rm rec}=B_{CP}$) by full reconstruction of its
observed long-lived daughters.

At the PEP-II asymmetric $e^+e^-$ collider~\cite{pepii}, resonant production of
the \FourS\  provides a 
copious source of $\BzBzb$ pairs moving along the beam axis ($z$ direction)   
with an average Lorentz boost of $\left<\beta\gamma\right> = 0.55$.
Therefore, the proper decay-time difference $\deltat$ is, 
to an excellent approximation, proportional to 
the distance \deltaz\ between  the two \Bz-decay vertices 
along the axis of the boost, 
$\deltat \approx \deltaz / c\left< \beta \gamma \right>$.
The average separation between the two $B$ decay vertices is
$\deltaz = \left<\beta\gamma\right> c \tau_B = 260$\mum, while the RMS 
$\deltaz$ resolution of the detector is about 
180\mum.

\subsection{Measurement of \Bz\ flavor oscillations}

The phenomenon of particle--anti-particle mixing 
in the neutral $B$ meson system
was first observed almost fifteen years
ago~\cite{UA1,ARGUS}.  
The oscillation frequency in \Bz-\Bzb\ mixing
has been extensively studied with both time-integrated and 
time-dependent techniques~\cite{PDG2000}. 
By interchanging $b\overline d$ with $\overline bd$, \Bz-\Bzb\ mixing
changes the additive bottom quantum number by two units, {\it i.e.,}
$\vert \Delta B\vert=2$. In the Standard Model,
such a process is the result of ordinary $\vert\Delta B\vert=1$
weak interactions in second order
involving the exchange of charge-2/3 quarks, 
with the top quark contributing
the dominant amplitude.  A measurement of \deltamd\ is therefore
sensitive to the value of the CKM 
matrix element
$V_{td}$.  At present the sensitivity to $V_{td}$ is not limited by
experimental precision on \deltamd, but by theoretical
uncertainties in the calculation, in particular the quantity
$f_B^2 B_B$, where $f_B$ is the \Bz\ decay constant, and 
$B_B$ is the so-called bag factor, 
representing the $\Delta B=2$ 
strong-interaction matrix element. 
There may also be contributions from
interactions outside the Standard Model

Beyond these questions of fundamental interest, since the measurement
of \deltamd\ incorporates all elements of the analysis for time-dependent
\CP\ asymmetries, including \B\ reconstruction, tagging, and 
\deltat\ determination and resolution, it is a essential test of 
our understanding of these
aspects of the \stwob\ measurement.

\begin{figure}[!htb]
\begin{center}
 \includegraphics[width=\linewidth]{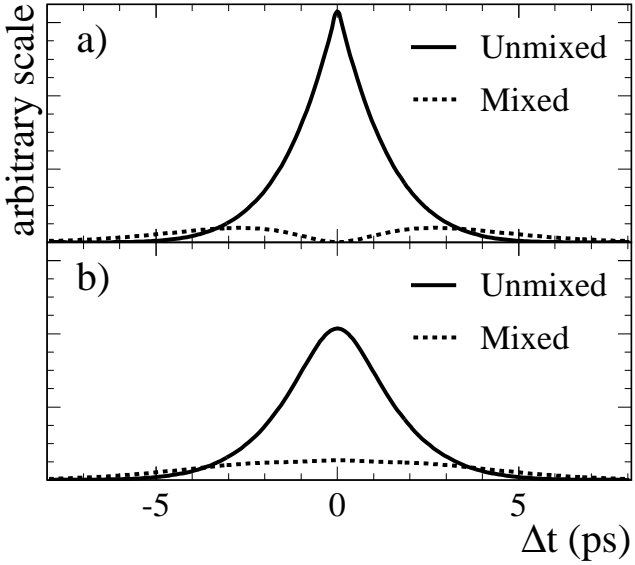}
\caption{Expected \deltat\ distribution for mixed and unmixed events
a) with perfect tagging and \deltat\ resolution, and b) with typical mistag
rates and \deltat\ resolution. }\label{fig:mixing_example}
\end{center}
\end{figure}

For the measurement of \deltamd, one neutral \B\ ($B_{\rm flav}$) is fully 
reconstructed in a 
flavor eigenstate~\cite{cc} as 
$D^{(*)-}\pi^+/\rho^+/a_1^+$ or
$\jpsi\Kstarz$ ($\Kstarz\to\Kp\pim$), while the second 
is tagged by its decay products.
The probability for \Bz-\Bzb\ mixing
is a function of \deltamd\ and
the proper time difference $\deltat$ between the two \B\ decays:
\begin{eqnarray}
\mbox{Prob}(\Bz\Bzb \rightarrow \Bz\Bz\ \mbox{or}\ \Bzb\Bzb, \Bz\Bzb) = && \nonumber \\
\frac{\Gamma}{4}\,{\rm{e}}^{-\Gamma|\deltat|} (1 \mp \cos \deltamd \deltat), & &
\label{eq:mix} 
\end{eqnarray}
where $\tau_{\Bz}=1/\Gamma$ is the \Bz\ lifetime.
The observed \BzBzb\ system produced in 
an \FourS\ decay can be classified 
as {\em mixed} or {\em unmixed} depending on whether the reconstructed 
flavor-eigenstate $B_{\rm rec}=B_{\rm flav}$ has the same or the opposite flavor as the 
tagging $B=B_{\rm tag}$. 
If the \deltat\ resolution and flavor tagging were perfect, 
the asymmetry as a function of $\deltat$ 
\begin{equation}
A_{\rm mixing}(\deltat) = 
\frac{N_{\rm unmix}(\deltat)-N_{\rm mix}(\deltat)}
{N_{\rm unmix}(\deltat)+N_{\rm mix}(\deltat)} 
\label{eq:asym}
\end{equation}
would describe a cosine function with 
unit amplitude.
The asymmetry goes through zero near 2.1 \Bz\ proper lifetimes and
the sensitivity to \deltamd, which is proportional
to $\deltat^2{\rm e}^{-\Gamma |\deltat|}\sin^2\deltamd\deltat$, reaches a
maximum in this region.
If the tagging algorithm incorrectly identifies the tag with a 
probability $\mistag$, the
amplitude of the oscillation is reduced by a dilution
factor ${\cal D}=(1 -2\mistag)$.  When more than one type of flavor tag is
used, each has its own mistag
rate $\mistag_i$.

Neglecting any background contributions, 
the probability density functions (PDFs) 
for the mixed $(-)$ and unmixed $(+)$ events, ${\cal H}_\pm$,
can be expressed as
the convolution of the underlying oscillatory physics distribution 
\begin{equation}
\label{eq:mixing_rate}
h_\pm(\deltat; \Gamma, \deltamd, \mistag )  = \frac{\Gamma}{4}\, 
{\rm e}^{ - \Gamma \left| \deltat \right| }
\left[  1 \pm {\cal D}\cos{ \deltamd \deltat } \right]
\end{equation}
with a 
time-difference resolution
function ${\cal {R}}(\delta_{\rm t}=\deltat-\deltat_{\rm true}; \hat {a})$ 
to give
\begin{eqnarray}
\label{eq:pdf}
&&{\cal H}_\pm(\deltat; \Gamma, \deltamd, \mistag, \hat {a} )  = \nonumber \\
&& \quad\quad h_\pm( \deltat_{\rm true}; \Gamma, \deltamd, \mistag) 
\otimes 
{\cal {R}}( \delta_{\rm t} ; \hat {a} ) ,
\end{eqnarray}
where $\deltat$ and $\deltat_{\rm true}$ are
the measured and the true time differences, and
$\hat a$ are parameters of the resolution function.
Figure~\ref{fig:mixing_example} illustrates the impact of typical mistag
and \deltat\ resolution effects on the \deltat\ distributions
for mixed and unmixed events.

A full likelihood function is then constructed by summing ${\cal {H}}_{\pm}$ 
over all mixed and unmixed events in a given uniquely-assigned
tagging category $i$ and over all tagging categories
\begin{widetext}
\begin{equation}
\label{eq:mixing_likelihood}
{\ln {\cal {L}_{\rm mix} } } = \sum_{ i}^{\rm tagging} 
\left[ \, \sum_{{\rm unmixed}}{  \ln{ {\cal H}_+( 
\deltat; \Gamma, \deltamd, \mistag_i, \hat {a}_i) } } + \sum_{ {\rm mixed} }{ \ln{ {\cal H}_-( 
\deltat ; \Gamma, \deltamd, 
\mistag_i, \hat {a}_i) } } \right].
\end{equation}
\end{widetext}
This can be maximized to extract the mistag fractions
$\mistag_i$ and resolution parameters $\hat {a}_i$ and, simultaneously, 
the mixing rate \deltamd.
The correlation between $\mistag_i$ and \deltamd\ is small, because
the rate of mixed events near $\deltat=0$, where
the \Bz-\Bzb\ mixing probability is small, is principally 
governed by the mistag rate. Conversely, the sensitivity 
to \deltamd\ increases at larger values of \deltat;
when \deltat\ is approximately twice the $B$ lifetime, half of
the neutral $B$ mesons will have oscillated.

\subsection{Measurement of \CP\ asymmetries}

For the measurement of \CP\ asymmetries, one \B\ ($B_{CP}$) is fully 
reconstructed in a 
\CP\ eigenstate with eigenvalue $\etaCP=-1$ ($\jpsi \KS$, $\psitwos \KS$, or 
$\chicone\KS$)
or $+1$ ($\jpsi\KL$), while the second 
is tagged with its decay products just as for the mixing measurement. 
The $B_{CP}$ sample is further
enlarged by including the mode $\jpsi\Kstarz$ ($\Kstarz\to\KS\piz$). However,
due to the presence of even ($L=0$, 2) and odd ($L=1$)
orbital angular momenta in the $\jpsi\Kstarz$ system, 
there are $\etaCP=+1$ and $-1$ contributions to its decay rate, respectively. 
When the angular information in the decay is ignored, 
the measured \CP\ asymmetry in $\jpsi\Kstarz$ is reduced 
by a dilution factor $D_{\perp} = 1-2R_{\perp}$, where
$R_{\perp}$ is the fraction of the $L=1$ component. We have measured 
$R_{\perp} = 0.160 \pm 0.032\pm 0.014$~\cite{babar0105} which, after 
acceptance corrections, leads to an effective 
$\eta_{\CP} = +0.65 \pm 0.07$ for the $\jpsi\Kstarz$ mode.

\def\hamiltonian{{\cal H}}
\def\bra#1{{\langle#1|}}
\def\ket#1{{|#1\rangle}}
\def\matrixelement#1#2#3{{\bra #1#2\ket#3}}
\def\Bzbar{\Bzb}
The expected time evolution for the tagged
$B_{CP}$ sample depends both on \Bz-\Bzb\ mixing and on the decay 
amplitudes of \Bz\ and \Bzb\ to the final state $f$ through 
a single complex parameter $\lambda$.
Mixing generates a lifetime difference as well as a mass difference
between the two neutral $B$ meson mass eigenstates, but the lifetime
difference is expected to be small since it is a consequence
of common final states in \Bz\ and \Bzb\ decays. Such common
states, which include the \CP\ eigenstates studies here, make up a very
small fraction of the decay width because they are quite suppressed by
CKM matrix elements. Dropping these, and thus ignoring
any lifetime difference, results in a simple expression
for $\lambda$ in terms of the $\vert\Delta B=1\vert$ and 
$\vert\Delta B=2\vert$ interactions,
\begin{equation}
\lambda=-{\frac{|\matrixelement\Bz{{\cal H}_{\Delta B=2}}\Bzbar|}
                  {\matrixelement\Bz{{\cal H}_{\Delta B=2}}\Bzbar}}
          {\frac{\matrixelement f{{\cal H}_{\Delta B=1}}\Bzbar}
                  {\matrixelement f{{\cal H}_{\Delta B=1}}\Bz}}.\label{eq:lambda}
\end{equation}
Redefining the states for \Bz\ and \Bzb\ by multiplying them by two different
phases has no effect on $\lambda$, which is thus phase-convention independent,
as every physical observable must be.
The decay distributions are
\begin{widetext}
\begin{equation}
f_\pm(\deltat) =  
\frac{\Gamma}{4}\,{\rm e}^{ - \Gamma | \deltat | }  
\left\{ 1
\pm {\cal D} \left[
\frac{2 Im\lambda}{1  + |\lambda|^2} \sin{ \deltamd  \deltat } 
-\frac{1-|\lambda|^2}{1  + |\lambda|^2} \cos{ \deltamd  \deltat }  
\right]  \right\}, \label{eq:direct}
\end{equation}
\end{widetext}
where the $+$ or $-$ sign indicates whether the $B_{\rm tag}$ is
tagged as a \Bz\ or a \Bzb, respectively. The dilution factor ${\cal D} = 1 - 2 \mistag$
accounts for the probability \mistag\ that the flavor of the tagging
\B\ is identified incorrectly.  

The distributions are much simpler when $|\lambda|=1$, which is the
expectation of the Standard Model for decays like $\Bz\to \jpsi\KS$.
If all the mechanisms that contribute to the decay have the same weak
phase then the ratio of the weak decay amplitudes in
Eq.~\ref{eq:lambda} is just $\etaCP {\rm e}^{2i\phi_{\rm dec}}$, where
$\phi_{\rm dec}$ is the weak phase for $\Bzb\to f$;
$\phi_{\rm dec}$ is convention dependent and unobservable.
The remaining factor introduces a phase due to
\Bz-\Bzb\ mixing. The combination of these phases
is convention independent and observable.

For decays such as $\Bz\to \jpsi\KS$, or more generally 
$(c\overline c)\KS$ and $(c\overline c)\KL$,
an explicit representation for $\lambda$ can
be found from the ratio of the amplitude for
$\Bzb\to(c\overline c)\Kzb$ to the interfering process
$\Bzb\to\Bz\to (c\overline c)\Kz\to(c\overline c)\Kzb$.
The decay $\Bz\to(c\overline c)\Kz$ involves a $\overline b\to
\overline cc\overline s$ transition with an amplitude
proportional to $[V_{cb}^*V_{cs}^{ }]$, while
$\Bzb\to(c\overline c)\Kzb$ provides analogously a factor
$\etaCP[V_{cb}^{ }V_{cs}^*]$.
Because $\Bzb\to\Bz$ mixing is dominated by the loop diagram with a 
$t$ quark, it introduces a factor
$[V_{td}^*V_{tb}^{ }/V_{td}^{ }V_{tb}^*]$,
while $\Kz\to\Kzb$ mixing, being dominated by the $c$-quark loop, 
adds a factor 
$[V_{cd}^{ }V_{cs}^*/V_{cd}^*V_{cs}^{ }]$.
Altogether, for
transitions of the type $b\to c{\overline c}s$,
\begin{eqnarray}
\lambda
&=&\etaCP
\left(\frac{V_{td}^{ }V_{tb}^*}{V_{td}^*V_{tb}^{ }}\right)
\left(\frac{V_{cb}^{ }V_{cs}^*}{V_{cb}^*V_{cs}^{ }}\right)
\left(\frac{V_{cd}^*V_{cs}^{ }}{V_{cd}^{ }V_{cs}^*}\right) \nonumber \\
&=&\etaCP
\left(\frac{V_{cb}^{ }V_{cd}^*}{V_{tb}^{ }V_{td}^*}\right)
\left(\frac{V_{tb}^*V_{td}^{ }}{V_{cb}^*V_{cd}^{ }}\right) \nonumber \\
&=&\etaCP {\rm e}^{-2i\beta}.
\end{eqnarray}
The time-dependent rate for decay of the $B_{CP}$ final state is then given by 
\begin{eqnarray}
\label{eq:TimeDep}
&&f_\pm(\deltat; \Gamma, \deltamd, \mistag, \sin{ 2 \beta } ) = \nonumber \\
&&\quad\quad\frac{\Gamma}{4} {\rm e}^{ - \Gamma \left| \deltat \right| }
\left[ 1 \mp
\etaCP {\cal D}\sin{ 2 \beta } \sin{ \deltamd \deltat } \right].
\end{eqnarray}
In the limit of perfect determination of
the flavor of the fully-reconstructed \B\ in the $B_{\rm flav}$
sample, which we assume throughout, the dilution here and in the
mixed and unmixed amplitudes of Eq.~\ref{eq:mixing_rate} 
arise solely from the $B_{\rm tag}$
side, allowing the values of the mistag fractions $\mistag_i$ to be
determined by studying the time-dependent rate of \Bz-\Bzb\ oscillations.

\begin{figure}[!htb]
\begin{center}
 \includegraphics[width=\linewidth]{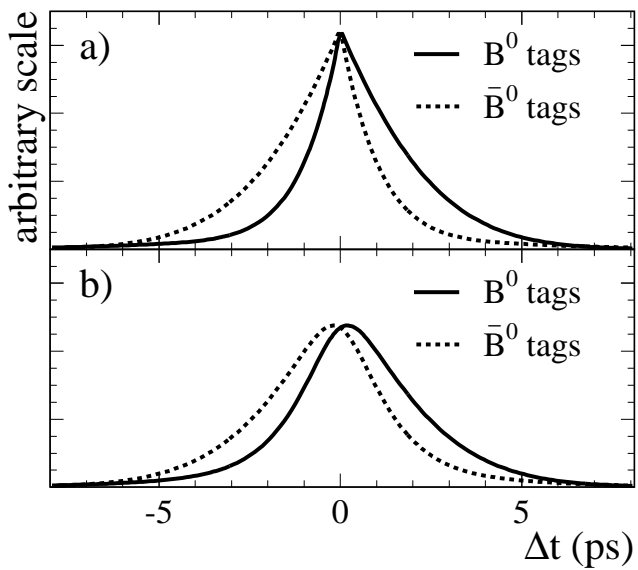}
\caption{Expected \deltat\ distribution for \Bz- and \Bzb-tagged \CP\ events
a) with perfect tagging and \deltat\ resolution, and b) with typical mistag
rates and \deltat\ resolution. }\label{fig:cp_example}
\end{center}
\end{figure}

To account for the finite resolution of the detector,
the time-dependent distributions $f_\pm$ for \Bz\ and \Bzb\ tagged events 
(Eq.~\ref{eq:TimeDep}) must be convolved with 
a time resolution function ${\cal {R}}( \delta_{\rm t}=\deltat - \deltat_{\rm true} ; \hat {a} )$ as described above for mixing,
\begin{eqnarray}
\label{eq:Convol}
&& {\cal F}_\pm(\deltat; \Gamma, \deltamd, \mistag, \sin{ 2 \beta }, \hat {a} ) = \nonumber \\
&& \quad\quad f_\pm( \deltat_{\rm true}; \Gamma, \deltamd, \mistag, \sin{ 2 \beta } ) \otimes
{\cal {R}}(\delta_{\rm t}; \hat {a}) ,
\end{eqnarray}
where $\hat {a}$ represents the set of parameters that describe the resolution function.  
In practice, events are separated into the same tagging categories 
as in mixing, 
each of which has a different mistag fraction $\mistag_i$, determined 
individually for each category.
Figure~\ref{fig:cp_example} illustrates the impact of typical mistag
and \deltat\ resolution effects on the \deltat\ distributions for
\Bz- and \Bzb-tagged \CP\ events.

It is possible to construct a \CP-violating observable 
\begin{equation}
  {\cal A}_{CP}(\deltat) = \frac{ {\cal F}_+(\deltat) - {\cal F}_-(\deltat) }
{ {\cal F}_+(\deltat) + {\cal F}_-(\deltat) } , 
\label{eq:asymmetry}
\end{equation}
which, neglecting resolution effects, is proportional to \stwob:
\begin{equation}
 {\cal A}_{CP}(\deltat) \propto -\etaCP {\cal D} \sin{ 2 \beta } \sin{ \deltamd \deltat }.
\label{eq:asymmetry2}
\end{equation}
Since no time-integrated \CP\ asymmetry effect is expected, an
analysis of the time-dependent asymmetry is necessary.  
The interference between the 
two amplitudes, and hence the \CP\ asymmetry, is maximal 
after approximately 2.1 \Bz\ proper lifetimes, when the mixing asymmetry
goes through zero. However, the maximum sensitivity to \stwob,
which is proportional to ${\rm e}^{-\Gamma |\deltat|}\sin^2\deltamd\deltat$, occurs
in the region of 1.4 lifetimes.

The value of the free parameter \stwob\ can be extracted with the 
tagged $B_{CP}$ sample
by maximizing the likelihood function
\begin{widetext}
\begin{equation}
\label{eq:Likelihood}
 \ln { {\cal {L} }_{CP} } = \sum_{i}^{\rm tagging}
 \left[ \sum_{\Bz\ {\rm tag} } {  \ln{ {\cal F}_+( \deltat; \Gamma, \deltamd, \hat {a}, 
\mistag_i, \sin{ 2 \beta } ) } } + 
 \sum_{\Bzb\ {\rm tag} }{ \ln{ {\cal F}_-( \deltat; \Gamma, \deltamd, \hat {a}, 
\mistag_i, \sin{ 2 \beta } ) } } \right] ,
\end{equation}
\end{widetext}
where the outer summation is over tagging categories $i$ and
the inner summations are over the \Bz\ and \Bzb\ tags within a given
uniquely-assigned tagging category. In practice, 
the fit for \stwob\ is performed
on the combined flavor-eigenstate and \CP\ samples with a likelihood constructed
from the sum of Eq.~\ref{eq:mixing_likelihood} and \ref{eq:Likelihood},
in order to determine \stwob, the mistag fraction 
$\mistag_i$ for each tagging category, and the vertex resolution
parameters $\hat {a}_i$. Additional terms are included in the likelihood
to account for backgrounds
and their time dependence.

The mistag rates can also be extracted with a 
time-integrated analysis as a cross check.  Neglecting 
possible background contributions and assuming the flavor of $B_{\rm flav}$ is 
correctly identified,
the observed time-integrated fraction of 
mixed events $\chi_{obs}$ can be expressed as a function of the \Bz-\Bzb\ mixing probability $\chi_d$:
\begin{equation}
\label{eq:TagMix:Integrated}
\chi_{obs} = \chi_d  + (1 - 2 \chi_d )\, \mistag,
\end{equation}
where 
$\chi_d = \frac{1}{2} x_d^2 / ( 1+ x_d^2 )=0.174 \pm 0.009$~\cite{PDG2000}
and $ x_d = {\rm \Delta} m_d /\Gamma $.
Taking advantage of the available decay time information, 
the statistical precision on $\mistag$ can be improved by selecting only events that 
fall into an optimized time interval $|\deltat| < t_0$, where $t_0$ is
chosen so that the integrated number of
mixed and unmixed events are equal outside this range.
With the use of such an optimized $\deltat$ interval
the time-integrated method achieves nearly the same statistical precision 
for the mistag rates as a full
time-dependent likelihood fit. 

\subsection{Overview of the analysis}
\label{sec:{\secname}Overview}

This article provides a detailed description of our published measurement of
flavor oscillations~\cite{babar0102} and
\CP-violating asymmetry~\cite{babar0118} in the neutral \B\ meson system. 
These measurements have six main components:
\begin{itemize}
\item
Selection of the $B_{\CP}$ sample
of signal events for neutral \B\ decays to \CP\ modes $\jpsi \KS$, 
$\psitwos \KS$, $\chicone\KS$, $\jpsi\Kstarz$ ($\Kstarz\to\KS\piz$), and $\jpsi\KL$; 
selection of the $B_{\rm flav}$ sample
of signal events for neutral flavor-eigenstate decays 
to $D^{(*)-}\pi^+/\rho^+/a_1^+$
and $\jpsi\Kstarz$ ($\Kstarz\to\Kp\pim$); 
selection of the \Bu\ control sample
in the modes $\Dbar^{(*)0}\pi^+$, $\jpsi K^{(*)+}$, $\psitwos\Kp$,
$\chicone\Kp$; and selection of a semileptonic neutral \B\ sample in the
mode $D^{*+}\ell^-\overline\nu$,
as described in Section~\ref{sec:Sample}; 
\item
Determination of the flavor of the $B_{\rm tag}$, as described 
in Section~\ref{sec:Tagging};
\item
Measurement of the distance \deltaz\ between the two \Bz\ decay 
vertices along the \FourS\ boost axis, and its conversion to \deltat, as described in Section~\ref{sec:vertexing};
\item
Construction of a log-likelihood function to describe the time evolution of signal and background 
events in the presence of mixing and \CP\ asymmetries, as described in Section~\ref{sec:Likelihood};
\item 
Measurement of the mixing rate \deltamd, mistag fractions $\mistag_i$, and vertex resolution
parameters $\hat {a}_i$ for the different tagging categories $i$,
with an unbinned maximum-likelihood fit to the
$B_{\rm flav}$ sample, as described
in Section~\ref{sec:Mixing};
\item
Extraction of a
value of \stwob, or more generally \etaimlambdaoverabslambda\ and \abslambda,
from the amplitude of the \CP\ asymmetry, the mistag fractions $\mistag_i$, 
and the
vertex resolution parameters $\hat {a}_i$ for the different tagging categories $i$, 
with an unbinned maximum-likelihood fit to the combined $B_{\rm flav}$ and 
$B_{\CP}$ samples,
as described in Section~\ref{sec:Sin2beta}.
\end{itemize}
Whenever possible, we determine time and mass resolutions, efficiencies and 
mistag fractions from the data. The measurement of \deltamd\ is
performed with a slightly reduced subset of the full $B_{\rm flav}$ sample, 
which is optimized for such a precision measurement. The $B_{\CP}$ sample
is not included, since this would add additional assumptions about the
resolution function without significantly improving the 
precision of \deltamd. The
measurement of \stwob\ is performed with
the full $B_{\rm flav}$ and $B_{\CP}$ samples, with a fixed value for
\deltamd\ and the \Bz\ lifetime. This strategy 
allows us to account correctly for
the small correlations among the mistag rates, \deltat\ resolutions
parameters, and \stwob. The same $B_{\rm flav}$ sample and vertex separation techniques
have been used to determine precision values for the charged and neutral 
\B\ lifetimes~\cite{babar0106}.

\renewcommand{\secname}{Detector}
\section{The \babar\ detector and data sets}
\label{sec:Detector}  

The data used in this analysis were recorded with the \babar\ 
detector~\cite{BabarNIM}
at the \pep2\ collider~\cite{pepii} in the period October 1999--June 2001.
The total integrated luminosity of the data set is equivalent to
29.7\invfb\ collected near the \FourS\ resonance and
3.9\invfb\ collected 40\mev\ below the \FourS\ resonance
(off-resonance data). The corresponding number
of produced \BB\ pairs is estimated to be about 32 million. The \FourS\
sample is sometimes divided into two subsamples for comparison
purposes: data recorded
in 1999-2000, about 20.7\invfb\ and referred to as ``Run 1'', and data
recorded in 2001, about 9.0\invfb\ and referred
to as ``Run 2''. These subsamples differ primarily in the quality of the
tracking system alignment and on the track-finding efficiency. 
The former requires a separate treatment of the 
\deltat\ resolution for the two periods, 
as discussed in Section~\ref{sec:run1vsrun2}, while
the latter results in substantially improved yields in Run 2 for 
reconstructed $B$ mesons.

\subsection{The \babar\ detector}

The \babar\ detector is a charged and neutral spectrometer with large
solid-angle coverage.
For this analysis, the most important detector capabilities include
charged-particle tracking, vertex reconstruction, and particle identification.
Charged particles
are detected and their momenta measured by a combination of 
a 40-layer, small-celled drift chamber (DCH)
filled with a 80:20 helium:isobutane gas mixture, and a
five-layer silicon vertex tracker (SVT), consisting of
340 AC-coupled double-sided silicon microstrip sensors. 
The cells of the DCH are organized into
10 superlayers within which the sense wires 
all have the same orientation, thereby allowing segment-based tracking. 
Both the DCH and the SVT lie
inside a 1.5-T solenoidal magnetic field. 
Beyond the outer radius of the DCH is a detector of internally reflected 
Cherenkov radiation (DIRC), which is used primarily for charged-hadron
identification. The device consists of 144 fused silica quartz bars in
which relativistic charged particles above the Cherenkov threshold
radiate photons while traversing
the material. The light is transported by total internal reflection down the length
of the bars to an array of 10752 photomultiplier
tubes mounted on the rear of the detector, where
the opening angle of the Cherenkov ring is
measured. A finely segmented
electromagnetic calorimeter (EMC), consisting of 6580 CsI(Tl) crystals, is used to 
detect photons and neutral hadrons, and also to identify electrons. 
The EMC is surrounded by a thin cylindrical superconducting 
coil and a segmented iron flux return, organized into a hexagonal barrel and
two endcaps. The instrumented flux return
(IFR) consists of multiple layers of resistive plate chambers (RPCs) 
interleaved with the flux-return iron and is
used in the identification of muons and neutral hadrons. 

\subsection{Charged particle reconstruction}

Charged track finding starts with pattern  
recognition
in the DCH, based on three different algorithms. 
The first uses the same fast algorithm employed by the Level-3 trigger
for finding and linking superlayer-based track segments from 
moderate-to-high \pt\ tracks originating from the interaction point. Two 
subsequent track finders then
work on superlayer segments not 
already attached to a reconstructed track. 
They are designed to find tracks with lower \pt, passing through fewer than the full ten  
superlayers of the chamber,
or originating away from the interaction point. 
At the end of this process, all tracks are refit with a  
Kalman-filter fitter~\cite{kalman}
that takes into account the detailed distribution of 
material in the detector and the non-uniformities  
in the detector magnetic field. These tracks are then projected into the SVT, and  
silicon-strip hits are
added if they are consistent within  
the extrapolation errors through the intervening material and field.
A search 
is performed for tracks that are reconstructed with the
remaining unused silicon clusters, again with two different algorithms.
At the end of the SVT-only track finding, an attempt is made to match SVT- and DCH-only
track segments, which may result when a hard 
scatter occurs in the support tube material between the two devices.

Charged-particle transverse momenta \pt\ are determined
with a resolution parameterized by 
$\sigma(\pt)/\pt = 0.0013 (p_T/\gevc) + 0.0045$. The SVT, 
with typical single-hit resolution of 10\mum, 
provides vertex information in both  
the transverse plane and in $z$, as well as the decay angles at the interaction
point. Decay vertices for \B\ meson candidates 
are typically reconstructed with a resolution of 50\mum\ in $z$ for 
fully reconstructed modes and about 
100 to 150\mum\ for the vertex of
the (unreconstructed) tagging \B\ meson in the event. 
The efficiency for finding tracks in hadronic
events that traverse the full DCH radius ($\pt>200\mevc$) is
about 90\% for Run 1 and 95\% for Run 2.

\subsection{Neutral reconstruction}

EMC clusters are formed around initial seed crystals containing at
least 10\mev\ of deposited energy.  Neighboring crystals are added 
to the cluster if their energy exceeds 1\mev. If the newly added
crystal has energy greater than 3\mev, its contiguous 
neighbors (including corners) are also considered for
inclusion in the cluster. In order to identify cases where 
several showers are in close proximity, such as unresolved 
photons from high-energy \piz\ decays,
local maxima within a cluster are identified. These local maxima are
defined as candidate crystals that have an energy
exceeding each of its neighbors by a fraction that 
depends on the number of crystals in the local neighborhood.
Clusters are then divided into as many ``bumps'' as there are local
maxima. The division is based on a two-dimensional
weighting scheme that assumes electromagnetic shower shapes 
to divide up the cluster energy. 
The position of each bump is calculated with a logarithmic weighting 
of crystal energies.

We determine whether a bump is associated with a charged or
neutral particle by projecting all tracks in the event to the 
inner face of the calorimeter. 
A bump is determined to be neutral, and therefore a photon candidate, if
no track intersects any of its crystals. A track intersection
is determined by computing the two-dimensional distance on the face 
of the calorimeter from the projected track impact point
to the bump centroid. A requirement is made on the difference between
the measured intersection distance and the Monte Carlo 
expectation for different particle species based on the measured 
track parameters.

The energy resolution in the EMC is measured directly with a radioactive source
at low energy under ideal low-background conditions and with electrons from
Bhabha scattering at high energy, from which we determine $\sigma(E)/E =
(5.0\pm 0.8)\%$ at 6.13\mev\ and $(1.9 \pm 0.07)\%$ at 7.5\gev.
The energy resolution can also be extracted from the observed mass
resolutions for $\piz$ and $\eta$ decays to two photons, which are measured 
to be around 7\mev\ and 16\mev, respectively.
A fit to the observed resolutions obtained from the \piz, $\eta$, and Bhabha samples gives 
a photon energy resolution parameterized by
$\sigma(E)/E = 0.023 {(E/\gev)}^{-1/4} \oplus 0.019$.

\subsection{Particle identification}
\label{sec:PID}  

Identification of electrons, muons and kaons is an essential
ingredient in both \B\ reconstruction and flavor tagging. 
Particle species can be distinguished by measurements of
the specific energy loss ($\dedx$) in the SVT layers and in the 
DCH gas along the particle trajectory, the number of Cherenkov 
photons and the Cherenkov angle in the DIRC, the electromagnetic 
shower energy in the EMC, and the particle penetration 
length in the IFR. Selection criteria are based on these quantities, 
on likelihood ratios derived from them, or on neural network 
algorithms combining different detector likelihoods.
Typically, looser selection criteria are applied for \B\ reconstruction
than for \B -flavor tagging.
Efficiencies and particle misidentification probabilities are
determined from data control samples with similar characteristics.

\begin{table*}[htb]
\caption{Criteria used for selecting the available categories of 
electron candidates. The difference between the measured mean 
\dedx\ and the expectation for an electron is required to lie 
within the interval specified in terms of the expected 
\dedx\ resolution $\sigma$.}
\begin{center}
\begin{tabular}{|l|c|c|l|}\hline
Category       & $\dedx$                 & $E/p$              & Cumulative additional requirements \\  
\hline\hline
{\tt VeryLoose} &  $[-3\sigma, 7\sigma]$   & $> 0.50$           & -- \\
{\tt Loose}     &  $[-3\sigma, 7\sigma]$   & $> 0.65$           & -- \\
{\tt Tight}     &  $[-3\sigma, 7\sigma]$   & $[0.75, 1.3]$ & Lateral shower shape \\ 
{\tt VeryTight} &  $[-2.2\sigma, 4\sigma]$ & $[0.89, 1.2]$ & Azimuthal shower shape; consistency \\
                &           &           & of DIRC Cherenkov angle ($3\sigma$) \\ \hline
\end{tabular}
\label{tab:epid}
\end{center}
\end{table*}

\begin{table*}[htb]
\caption{Criteria used for selecting available categories of muon candidates.}
\begin{center}
\begin{tabular}{|l|c|c|c|c|c|c|c|}\hline
Category & $n_\lambda$ & $\Delta n_\lambda$ & $\overline{n}_{hits}$ & $\sigma_{n_{hits}}$ & 
$\chi_{trk}^2 / n_{layers}$ & $\chi_{fit}^2 / n_{layers}$ &
$E_{EMC}$ [GeV] \\
\hline\hline
{\tt VeryLoose} & $>2.0$ & $<2.5$ & $<10$ & $<6$ & --   & --   & $<0.5$  \\
{\tt Loose}     & $>2.0$ & $<2.0$ & $<10$ & $<6$ & $<7$ & $<4$ & $<0.5$  \\
{\tt Tight}     & $>2.2$ & $<1.0$ & $< 8$ & $<4$ & $<5$ & $<3$ & [0.05, 0.4] \\
{\tt VeryTight} & $>2.2$ & $<0.8$ & $< 8$ & $<4$ & $<5$ & $<3$ & [0.05, 0.4] \\ \hline
\end{tabular}
\label{tab:mupid}
\end{center}
\end{table*}

\subsubsection{Electron identification}

Electron candidates are identified primarily by the ratio
of the bump energy in the electromagnetic calorimeter 
to the track momentum, $E/p$. They also must have a measured mean
\dedx\ in the DCH that is consistent 
with the electron hypothesis. 
In addition, for some applications, the lateral and azimuthal shape 
of the EMC shower~\cite{lat,a42}
and the consistency of the observed and expected Cherenkov angle 
in the DIRC are used for identification.
Four different categories of
electron candidates ({\tt VeryLoose}, {\tt Loose}, 
{\tt Tight}, and {\tt VeryTight}) are defined with the criteria listed in 
Table~\ref{tab:epid}. 
Candidates that are not matched to an EMC bump are retained 
as {\tt noCal} electron candidates if their
measured \dedx\ satisfies the same requirements as 
the {\tt VeryTight} selection.
Electron identification efficiencies in the momentum range 
$0.5 < p < 3.0\gevc$ vary between 88\% and 98\% for the 
criteria in Table~\ref{tab:epid}, while the pion misidentification 
rates are below 0.3\% for the {\tt VeryTight} selection.

\subsubsection{Muon identification}

Muon candidates are primarily identified by the measured number of 
hadronic interaction lengths $n_\lambda$ traversed from the 
outside radius of the DCH through the IFR iron, and
the difference $\Delta n_\lambda$ between $n_\lambda$ 
and the predicted penetration
depth for a muon of the same momentum and angle. 
Contamination from hadronic showers is rejected by a combination of the
average number $\overline{n}_{hits}$ and the variance $\sigma_{n_{hits}}$ 
of hits per RPC layer, the $\chi^2$ for the geometric 
match between the track extrapolation into the IFR and the RPC hits, 
$\chi_{trk}^2$, and the $\chi^2$ of a polynomial fit to the RPC hits,
$\chi_{fit}^2$. In addition, for those muons within the acceptance of the EMC, we require 
the calorimeter bump energy $E$ to be consistent with a minimum 
ionizing particle. Four different categories of
muon candidates 
({\tt VeryLoose}, {\tt Loose}, 
{\tt Tight}, and {\tt VeryTight}) are selected with the criteria listed in 
Table~\ref{tab:mupid}. 
In the forward region, which suffers from some machine 
background, additional requirements are made
on the fraction of RPC layers with hits.
Muon identification efficiencies in the momentum range
$1.1 < p < 3.0\gevc$ vary between 60\% to 92\% for the
criteria in Table~\ref{tab:mupid}, while pion misidentification
rates are about 3\% for the {\tt Tight} selection.

\subsubsection{Kaon identification}

Kaons are distinguished from pions and protons on the basis of specific
energy-loss measurements \dedx\ in SVT and DCH and the number of Cherenkov photons
and the Cherenkov angle in the DIRC.
The difference between the measured truncated-mean $\dedx$ in the DCH
and the expected mean for the pion, kaon and proton hypothesis, with typical resolution of 7.5\%, 
is used to compute likelihoods ${\cal L}_\pi$, ${\cal L}_K$ and ${\cal L}_p$
assuming Gaussian distributions.
Similarly, the difference between the measured 60\% 
truncated-mean $\dedx$ in the SVT and the expected \dedx\ is described by an asymmetric 
Gaussian distribution. For minimum-ionizing particles the resolution on the
SVT truncated mean is about 14\%.
In the DIRC, a likelihood is obtained for each particle hypothesis
from the product of two components: the expected number of Cherenkov photons, 
with a Poisson distribution, and the difference between the measured average 
Cherenkov angle to the expected angle for a given mass hypothesis, assuming a 
Gaussian distribution.

For \B-flavor tagging the likelihood variables from SVT, DCH and DIRC are
combined as inputs to a neural network whose output is a single discriminating variable
for kaon selection. The network is trained with Monte Carlo simulation of
generic \B\ decays.
The average efficiency of the selection is about 85\% for a
pion-misidentification probability of about 2.5\%.
Further details are described in Section~\ref{sec:NetTagger}.

The exclusive reconstruction of many \B\ meson final states does not generally
require explicit kaon identification.
For some channels a {\tt VeryLoose} kaon selection based on likelihood ratios
is imposed to reduce backgrounds to acceptable levels.
The combined likelihood uses the individual likelihoods from SVT and DCH
for momenta below 0.5\gevc, from DCH only for momenta between 0.5 and
0.6\gevc, and from DIRC only for momenta above 0.6\gevc.
Kaon candidates are rejected if the likelihood ratios satisfy
${\cal L}_K/{\cal L}_\pi < r$ and ${\cal L}_K/{\cal L}_p < r$, where
$r = 0.1$ for $p < 0.5\gevc$ and $r = 1$ for $p \ge 0.5\gevc$.
Tracks with no particle information are assumed to be kaons.
This {\tt VeryLoose} kaon requirement has a nearly constant 
kaon-identification efficiency of about 
96\% and a pion-misidentification probability of at most 15\%
for tracks in the transverse momentum range 1 to 2.5\gevc.
Tighter kaon selections require ${\cal L}_K/{\cal L}_\pi > r$, with
$r$ typically greater than one. For a loose pion selection, candidates
are rejected if they satisfy tighter kaon or lepton criteria.

\renewcommand{\secname}{Sample}
\section{Reconstruction of \boldmath \B\ mesons}  
\label{sec:\secname}  

Neutral \B\ mesons in flavor eigenstates are reconstructed in the hadronic final states 
$\Bz \to D^{(*)-} \pi^+$, $D^{(*)-} \rho^+$, $D^{(*)-} a_1^+$, and
$\jpsi \Kstarz(\Kp\pim)$, and the semileptonic decay mode $\Bz \to
D^{*-}\ell^+\nu$. The \CP\  sample is reconstructed in the channels 
$\Bz \to \jpsi \KS$, $\psitwos \KS$, $\chicone \KS$, $\jpsi\Kstarz$ 
($\Kstarz\to\KS\piz$) and $\jpsi\KL$.
In some cases, control samples of charged \B\ decays are studied, where the hadronic final states
$\Bu \to \Dbar^{(*)0} \pi^+$,
$\jpsi K^{(*)+}$, $\psitwos\Kp$ and  $\chicone\Kp$  are used.
All final-state particles, with the exception of the neutrino in
the semileptonic decay, are reconstructed.  
A number of \Dzb\ and \Dm\ decay modes are used to 
achieve reasonable reconstruction efficiency
despite the typically small branching fractions 
for any given \B\ or \D\ decay channel. A summary of
the various reconstructed \B\ samples and purities is provided in Table~\ref{tab:HadronicBYield}.

\begin{center}
\begin{table}[!htb]
\caption{
Event yields for the different samples used in this analysis, before any tagging or tagging vertex
requirements.
The yields, purity, and signal
size for \B\ decays to hadronic final states are obtained from a fit 
to the \mes\ distribution described
in Section~\ref{sec:Bselection}, after selection on $\Delta E$. 
Purities are quoted for $\mes >5.27$\mevcc. 
The results for $\jpsi\KL$ are obtained from a fit
to the \deltae\ distribution described in Section~\ref{subsec:sample_KL}.
The purity for $\jpsi\KL$ is quoted for events with $\deltae<10$\mev.
The results for $D^{*-}\ell^+\nu$ are obtained from a fit to the \ctby\ 
distribution described in Section~\ref{subsec:sample_semileptonic}.
Purity is quoted for $-1.1<\ctby<1.1$.
} 
\vspace{0.3cm}
\begin{tabular}{|l|l|c|c|} \hline
Sample &  Final state                            & Signal     & Purity \\ & & & (\%) \\ \hline \hline
$B_{\CP}$ & $\jpsi \KS$ ($\KS \to \pi^+\pi^-$)   & $461\pm 22$ & 99 \\
          & $\jpsi \KS$ ($\KS \to \pi^0 \pi^0$)  & $113\pm 12$ & 93 \\
          & $\psitwos \KS$                       &  $86\pm 17$ & 96 \\ 
          & $\chicone \KS$                       &  $44\pm 8$  & 98 \\
        & $\jpsi K^{*0}$ ($K^{*0} \to \KS \piz$) &  $64\pm 10$ & 74 \\
          & $\jpsi \KL$                          & $257\pm 24$ & 60  \\  \cline{2-4}
          & Total                                & $1025\pm 41$ & 83  \\ 
\hline\hline
$B_{\rm flav}$ & $D^{*-}\pi^+$                      & $2380\pm 57$ & 92 \\ 
 	  & $D^{*-}\rho^+$                          & $1438\pm 52$ & 84 \\ 
          & $D^{*-}a_1^+$                           & $1146\pm 45$ & 80 \\ 
          & $D^{-}\pi^+$                            & $2685\pm 65$ & 83 \\
          & $D^{-}\rho^+$                           & $1421\pm 57$ & 74 \\
          & $D^{-}a_1^+$                            &  $845\pm 44$ & 67 \\
          & $\jpsi K^{*0}$ ($K^{*0} \to K^+ \pi^-$) & $1013\pm 36$ & 95 \\ \cline{2-4}
          & Total                                   & $10941\pm 133$& 83 \\
\hline\hline
\Bu\      & $\Dzb \pi^+$                         & $6850\pm 102$  & 83 \\
          & $\Dstarzb\pi^+$                      & $1708\pm 51$   & 91 \\
          & $\jpsi K^+ $                         & $1921\pm 46$   & 97 \\
          & $\psitwos K^+ $                      & $292\pm 18$    & 98 \\
          & $\chicone K^+ $                      & $195\pm 29$    & 95 \\
          & $\jpsi K^{*+}$ ($K^{*+} \to K^+ \piz$) & $384\pm 25$  & 87 \\ \cline{2-4}
          & Total                                & $11343\pm 129$  & 86 \\
\hline\hline
Semi-     & $D^{*-}\ell^+\nu$                    & $29042\pm 1500$ & 78 \\
leptonic \Bz\ & & & \\
\hline
\end{tabular}
\label{tab:HadronicBYield}
\end{table}
\end{center}

\subsection{Event selection}

Multihadron events are selected by demanding a minimum of three reconstructed charged tracks 
in the polar angle range $0.41 < \theta_{lab} < 2.54$\rad.
Charged tracks must be reconstructed in the DCH and
are required to originate within 1.5\cm\ in $xy$ and 10\cm\ in $z$
of the nominal beamspot. 
A primary vertex
is formed on an event-by-event basis from a
vertex fit to all charged tracks in the fiducial volume. 
Tracks with a large
$\chi^2$ contribution to the vertex fit are removed until an overall 
$\chi^2$ probability greater than 1\% is obtained or only two tracks remain. 
The resolution achieved by this method is about 70\mum\ in $x$ 
and $y$ for hadronic events.
Events are required to have a primary vertex 
within 0.5\cm\ of the average position of the 
interaction point in the plane transverse to the beamline, 
and 6\cm\ longitudinally.
Electromagnetic
bumps in the calorimeter in the polar angle range
$0.410 < \theta_{lab} < 2.409\rad$ that are not associated with charged tracks,
have an energy greater than 30\mev, and
a shower shape consistent with a photon interaction are taken as neutrals.
A total energy greater than 4.5\gev\ in the fiducial regions for 
charged tracks and neutrals is required.
To reduce continuum background, 
we require the normalized second Fox-Wolfram moment~\cite{fox}  
$R_2$ of the event,
calculated with both charged tracks and neutrals, to be less 
than 0.5 (0.45) in hadronic (semileptonic) decay modes.  
The $\ell^{\rm th}$ Fox-Wolfram moment is the momentum-weighted sum 
of Legendre polynomial of the $\ell^{\rm th}$ order computed from the cosine 
of the angle between all pairs of tracks.
The ratio $R_2$ provides good separation 
between jet-like continuum events and more spherical \BB\ events. 

\subsection{Reconstruction of decay daughters}

The reconstruction of $B$ mesons typically involves the summation
of a set of related decay modes, with multiple decay chains for the charm
daughters or other short-lived decay products.
To simplify analysis of such complex decay chains, virtual {\em composite} 
particles and
their error matrices are constructed from the original daughter particles.
The composite particle then replaces
the daughters in subsequent fits and analysis. 
The three-momentum of the virtual
particle is fit directly, rather than computed from the updated
daughters, improving speed and numerical accuracy.

Vertex and kinematic fitting is used to improve four-momenta and position
measurements, as well as to measure the time difference between 
decaying \B\ hadrons in the $\FourS \to \B\Bbar$ decay. 
For example, in the case of
\bpsiks, the position measurement of the \Bz\ can be improved with the
constraint that the line-of-flight of the \KS\ intersects the \jpsi\ vertex.
The energy resolution of the \Bz\ can also be improved by applying 
a mass constraint to the \jpsi\ and \KS\ daughters. 
Generalized procedures have been developed and tested with constraints
implemented by the Lagrange-multiplier
technique. Possible constraints include
a common decay vertex, mass, energy, momentum,
beam energy (with and without smearing), beam-spot position and
line-of-flight. 

Non-linearities in the fits require the use of
an iterative procedure, where convergence is defined by
demanding that the change in $\chi^2$ between two successive
iterations is less than 0.01, within a maximum of six iterations.
Simple fits involving only vertex
constraints (except long-lived particles)
are, however, accurate enough with a single iteration.

\subsubsection{\piz\ selection}

Neutral pion candidates are formed from pairs of EMC
bumps with energy greater than 30\mev, assumed to be photons originating from 
the interaction point. The invariant mass of the photon pair is required to be 
within $\pm 20$\mevcc\ ($2.5\sigma$) of the nominal \piz\ mass, with a minimum 
summed energy of 200\mev. Selected candidates are subjected to a kinematic
fit with a \piz\ mass constraint. Within the acceptance of the EMC, 
efficiencies for this selection vary from about 55 to 65\% for 
\piz\ energies from 0.3 to 2.5\gev, typical of \B\ decays.

\subsubsection{\KS\ selection}

Candidates in the $\KS\to\pip\pim$ mode are selected by requiring an
invariant $\pip\pim$ mass, computed at the vertex of the two tracks, 
between 462 and 534\mevcc. The $\chi^2$ of the vertex fit must have a 
probability greater than 0.1\%. The angle between the
flight direction and the momentum vector 
for the \KS\ candidate is required to be smaller than 200\mrad. 
Finally, the transverse flight distance
from the primary vertex in the event, $r_{xy}$, must be greater than 2\mm.

Optimization for the reconstruction of
the \CP\ sample has produced slightly different
\KS\ selection criteria. The
$\pi^+ \pi^-$ invariant mass, determined at the vertex of the two tracks, 
is required to lie
between 489 and 507\mevcc\ and the three-dimensional flight
length with respect to the vertex of the charmonium candidate is
required to be greater than 1\mm.

Pairs of \piz\ candidates, each in the mass range 
100--155\mevcc\ ($-5\sigma$, $+3\sigma$)
and formed from non-overlapping EMC bumps, 
are combined to construct 
$\KS\rightarrow \piz\piz$ candidates. 
For each \KS\ candidate with an energy greater than 
800\mev\ and a mass between 300 and 700\mevcc\ at the interaction point, 
we determine the most probable \KS\ decay point along the path defined
by the initial \KS\ momentum vector and the \jpsi\ vertex by maximizing
the product of probabilities for the daughter \piz\ mass-constrained fits. Allowing for vertex resolution,
we require the distance from the decay point to the \jpsi\ vertex 
to be between $-10$ and $+40$\cm\ and 
the \KS\ mass, using the measured decay point, to be between 470 and 536\mevcc.

\subsubsection{\KL\ selection}

Candidates for \KL\ mesons are identified in the EMC and IFR detectors
as reconstructed clusters that cannot be associated with any
charged track in the event.  EMC candidates must have a cluster energy
between 200\mev\ and 2\gev\ and a polar angle $\theta$ that
satisfies $\cos \theta < 0.935$.  To suppress backgrounds from $\pi^0$
decay, \KL\ candidates consistent with a photon are paired with other
neutrals with $E_{\gamma}>30$\mev.  Any candidate with $100 <
m(\gamma \gamma) < 150$\mevcc\ is rejected.  Likewise, clusters with
more than 1\gev\ energy that contain two bumps are rejected if the bump
energies and shower shapes are consistent with two photons
from a $\piz$ decay.  Monte Carlo
simulation shows that clusters due to true \KL\ mesons are 
easily distinguished from $\piz$ candidates by these criteria.
The remaining background consists primarily of
photons and overlapping showers.  Isolated clusters produced by
charged hadrons are removed by the basic clustering algorithm, which
requires a minimum separation of about 20\cm\ between clusters.

IFR candidates are defined as clusters with hits in two or more RPC
layers that are not matched to any reconstructed charged track.  To
reduce beam-related backgrounds and to avoid regions where the charged
tracking efficiency is low, we require that the polar angle $\theta$ of 
the IFR cluster satisfy $-0.75 < \cos \theta < 0.93$, and eliminate
clusters that begin
in the outer 25\% of the forward IFR
endcap.  Due to the irregular structure of hadronic showers, some hits
from charged tracks are missed by the tracking association.  We
suppress these clusters by rejecting \KL\ candidates that lie within
$\pm 350$\mrad\ in polar angle and in the range $-750 (-300)$ to $+300
(+750)$\mrad\ in azimuth of the EMC intersection of any 
positively (negatively) charged track in the event.  The remaining
background is predominantly from charged particles and detector noise.

Some \KL\ candidates satisfy both the EMC and IFR selection requirements.
In the reconstruction of $\Bz\to\jpsi\KL$, additional criteria 
described in Section~\ref{subsec:sample_KL} 
are applied to resolve the classification of the corresponding \B\ candidates.
Extensive studies of \KL\ detection efficiencies have been conducted
with a control sample of radiatively produced $\phi$ mesons, decaying
to $\KL\KS$. 

\subsubsection{Selection of light resonances}

For $\rho^-$ candidates, the $\pim \piz$ mass is
required to lie within $\pm 150$\mevcc\ of the nominal $\rho^-$ mass.
The \piz\ from the \rhom\ decay is required to have an
energy greater than 300\mev. We reconstruct
\Kstarz\ candidates in the $K^+\pi^-$ and $\KS\piz$ modes, 
while \Kstarp\ candidates are reconstructed in 
the $K^+\piz$ and $\KS\pi^+$ modes. 
The invariant mass of the two daughters is required to be within 
$\pm 100$\mevcc\ of the nominal $K^*$ mass.
Candidates in the mode $\aonep\to\pip\pim\pip$
are reconstructed by combining three charged
pions, with invariant mass in the range of 1.0 to 1.6\gevcc. In
addition, the $\chi^2$ probability of a vertex fit of the \aonep\ candidate is required to be
greater than 0.1\%.

\subsubsection{Charmed meson and charmonium selection}

The decay channels $\Kp\pim$, $\Kp\pim\piz$, $\Kp\pip\pim\pim$ and $\KS
\pip\pim$ are used to reconstruct \Dzb\ candidates,
while \Dm\ candidates are selected in the $\Kp \pim \pim$
and $\KS\pim$ modes.  
Charged and neutral kaons are required to  have a momentum
greater than 200\mevc. The same criterion is applied to the pion in 
$\Bz\to\DDstarm\pip$, $\Bz\to\DDstarm\rhop$ decay.  
For the decay modes $\Bz\to\DDstarm\aonep$, the pions are required to have 
momentum larger than 150\mevc. 
We require \Dzb\ and \Dm\ candidates to lie within
$\pm 3\sigma$ of the nominal masses, where the error $\sigma$ is
calculated event-by-event.
The distributions of the difference between measured and nominal
\Dzb\ and \Dm\ meson masses, normalized by the measured error on the
candidate masses, are found to have an RMS in the range
1.1--1.2 when fit with a Gaussian distribution. 
For $\Dzb\to\Kp \pim \piz$, we only reconstruct the dominant resonant
mode $\Dzb\to\Kp \rhom$, followed by $\rhom\to\pim \piz$. 
The angle $\theta^*_{\Dz\pi}$ between the \pim\ and \Dzb\ in the $\rhom$ 
rest frame must satisfy $|\cos \theta^*_{\Dz \pi}| > 0.4$.
Finally, all \Dzb\ and \Dm\ candidates are required to  
have a momentum greater than 1.3\gevc\ in the \FourS\ frame 
and a $\chi^2$ probability for the topological vertex fit 
greater than 0.1\%.
A mass-constrained fit is applied to candidates satisfying these requirements.

We form \Dstarm\ candidates in the decay 
$\Dstarm\to\Dzb\pim$ by combining a \Dzb\ with a pion that has
momentum greater than 70\mevc. 
The soft pion is constrained to originate from the beamspot
when the \Dstarm\ vertex is computed.
To account for the small energy release in the 
decay $\FourS \to \B\Bbar$ (resulting in a small transverse flight of
the \B\ candidates), the effective vertical size of the beam 
spot is increased to  
40\mum. Monte Carlo simulation was used to verify that this 
does not introduce any significant bias in the selection or in the
$\Delta t$ measurement. After applying a mass-constrained fit to the \Dzb\ daughter,
\Dstarm\ candidates are required to have $m(\Dzb\pim)$
within $\pm 1.1$\mevcc\ of the nominal \Dstarm\ mass
for the $\Dzb\to\Kp \pim \piz$ mode and $\pm 0.8$\mevcc\ for all
other modes. This corresponds to about
$\pm 2.5$ times the RMS width of the signal distribution, which is 
estimated by taking a weighted
average of the core and broad Gaussian components of the observed $m(\Dzb\pim)$ distributions.

We form \Dstarzb\ candidates by combining a \Dzb\ with 
a \piz\ with momentum less than 
450\mevc\ in the \FourS\ frame. \Dstarzb\ candidates are 
required to have $m(\Dzb\piz)$
within $\pm 4$\mevcc\ of the nominal value, after applying 
a mass-constrained fit to the \Dzb\ daughter.

\begin{table*}[htb]
\caption{Particle identification and invariant mass requirements
for $\jpsi$ and $\psitwos \to \ell^+ \ell^-$ candidates.  The 
minimal
particle identification criteria are applied to both daughters,
while only one daughter must pass the restrictive requirement. 
Electron and muon selection requirements are defined in Section~\ref{sec:PID}.
Mass ranges are quoted in GeV/$c^2$ and MIP refers to a minimum-ionizing particle.}
\begin{center}
\begin{tabular}{|l|c|c|c|c|c|c|} \hline
 & \multicolumn{3}{c|}{$\epem$ candidates}

 & \multicolumn{3}{c|}{$\mu^+ \mu^-$ candidates} \\ \cline{2-7}
$B$ channel    &Minimal &Restrictive & $m(\epem)$ &Minimal &Restrictive & $m(\mu^+ \mu^-)$ \\ \hline \hline
$\jpsi \KS$    & None & {\tt Tight} or {\tt noCal}&2.95-3.14 & {\tt MIP} & {\tt Loose} & 3.06-3.14\\
$\psitwos \KS$ ($\ell^+ \ell^-$) & {\tt VeryLoose} & {\tt Tight} &3.436-3.736 & {\tt VeryLoose} & {\tt Loose} & 3.06-3.14\\
$\psitwos \KS$ ($\jpsi \pip \pim$) & {\tt VeryLoose} & {\tt Tight} &2.95-3.14 & {\tt VeryLoose} & {\tt Loose} & 3.06-3.14\\
$\chicone \KS$ ($\jpsi \gamma$) & {\tt Loose} & {\tt Tight} &2.95-3.14 & {\tt VeryLoose} & {\tt Loose} & 3.06-3.14\\
$\jpsi \Kstar$    & {\tt Tight} & {\tt Tight} &2.95-3.14 & {\tt Loose} & {\tt Loose} & 3.06-3.14\\
$\jpsi \KL$    & {\tt Loose} & {\tt VeryTight} &3.00-3.13 & {\tt Loose} & {\tt Tight} & 3.06-3.13\\
\hline
\end{tabular}
\end{center}
\label{tab:jpsireq}
\end{table*}

Candidates for $\jpsi$ and $\psitwos$ mesons are reconstructed in their 
$\epem$ and $\mu^+ \mu^-$ decay modes, while $\psitwos$ mesons are also
reconstructed in the $\jpsi \pi^+ \pi^-$ channel.
Table~\ref{tab:jpsireq} shows the particle identification 
and invariant mass requirements for the $\epem$ and $\mu^+ \mu^-$ daughters. 
These vary with reconstructed $B$ decay channel due to 
the differing levels of background encountered.  
For $\jpsi\to \epem$ and $\psitwos \to \epem$ decays, where the 
electron may have radiated
Bremsstrahlung photons, the missing energy is recovered by identifying
clusters with more than 30\mev\ lying within 35\mrad\ in polar angle 
and 50\mrad\ in azimuth of the electron direction
projected onto the EMC.

For the $\psitwos \to \jpsi \pi^+ \pi^-$ mode, \jpsi\ candidates are constrained to the nominal
mass and then
combined with pairs of oppositely-charged 
tracks considered as pions, with invariant mass between 400 and 600\mevcc. Candidates 
with $0.574 < m(\jpsi \pi^+ \pi^-)- m(\jpsi) < 0.604\gevcc$
are retained.  

Photon candidates used for the reconstruction 
of $\chicone\to\jpsi\gamma$ are required to lie within the 
calorimeter fiducial volume ($0.41<\theta_\gamma<2.41$\rad) and 
have an energy greater than 150\mev. 
In addition, the candidate should not 
form, in combination with any other photon in the event having 
at least 70\mev\ of energy, 
a \piz candidate with mass between 120 and 150\mevcc.
The invariant mass of the \chicone\ candidates is required to be greater than 3.476 and smaller than 3.546\gevcc. 

\subsection{{\boldmath \B} meson selection in fully-reconstructed modes}
\label{sec:Bselection}

We reconstruct \B\ candidates in all modes except $\Bz\to\jpsi\KL$ and 
\bzdstlnu\ using a pair of nearly uncorrelated 
kinematic variables, the 
difference \deltae\ between the energy of the \B\ candidate and the 
beam energy in the \FourS\ center-of-mass frame,
and the beam-energy substituted mass, \mes, defined as
\begin{equation}
\label{eq:mse}
      \mes =   \sqrt{  \left(  \frac{ \frac{1}{2} s + \mathbf{p} 
\cdot \mathbf{p}_{i}}{E_i} \right)^2  -  p^2 },
\end{equation}
where $s$ is the square of the center-of-mass energy,
$E_i$ and $\mathbf{p}_{i}$  are  the total energy
and the three momentum of the initial state in the laboratory frame, 
and $\mathbf{p}$ is the three momentum of the \B\ candidate 
in the same frame.
For the purpose of determining event counts and purities,
a signal region is defined in the $(\mes,\deltae)$ plane
as $5.27 < \mes <  5.29$\gevcc\ and $|\deltae|< 3\sigma(\deltae)$, 
where $\sigma(\deltae)$ is the resolution on \deltae.
Likewise, a sideband region is defined as $5.20 < \mes < 5.26$\gevcc\ and 
$|\Delta E|< 3\sigma(\deltae)$. 
The value of $\sigma(\deltae)$ is mode-dependent and 
varies between 7 to 40\mev\ as measured in the data.
When multiple \B\ candidates  (with $\mes > 5.20$\mevcc) are found in the same event, 
the candidate with the smallest value of $|\deltae|$ is selected.

Two types of background in the sample of selected \Bz\ candidates
are distinguished.
The first background, called combinatorial, arises from random 
combinations of charged tracks 
and neutral showers from both \B\ mesons in \BB\ events or from continuum events. 
This background is smoothly distributed in \mes\ and does not peak near the \B\ mass. 
The second, so-called ``peaking'' background, consists of events in which, for example, 
a slow pion from the reconstructed \B\ meson is replaced by a slow pion from the 
tagging \B, causing an enhancement near the nominal
\B\ mass. The peaking background from charged \B\ decays is considered as a specific background
source in the construction of the full likelihood function for \Bz-\Bzb\ mixing,
since these events have a particular time structure and set of effective 
dilutions. In this case, the peaking
background from other neutral \B\ decays has time-dependent properties and
dilutions that are essentially identical to the signal and is treated 
as such. For the likelihood describing the
\CP\ sample, the peaking background is simply
assumed to have zero effective \CP.

Suppression of continuum background, in addition to a general requirement
on $R_2$, is typically provided by 
restricting the 
thrust angle $\theta_{\rm th}$, 
defined as the angle between the thrust axis of the particles that form the
reconstructed $B_{\rm rec}$ candidate and the thrust axis of the remaining
tracks and unmatched clusters in the event, computed in the \FourS\ frame. 
The two thrust axes are almost uncorrelated in \BB\ events, 
because the \Bz\ mesons are nearly at rest in the \FourS\ rest frame.
In continuum events, which are more jet-like,
the two thrust axes tend to have small opening angles. Thus, a requirement
on the maximum value of $|\cos\theta_{\rm th}|$ is effective in continuum rejection.

Signal yields and sample purities are extracted from 
fits to the \mes\ distributions of \B\ candidates
with a Gaussian distribution for the
signal and an ARGUS background shape~\cite{ARGUS_bkgd} for the combinatorial background
with a functional form given by
\begin{equation}
 {\cal A}( \mes ; m_0,\xi  ) = A_B\, m_{\rm ES} \sqrt{ 1 - x_{\rm ES}^2 }\,
{\rm e}^{\displaystyle \xi \left( 1 - x_{\rm ES}^2 \right)},
\label{ARGUS_bkd}
\end{equation}
for $x_{\rm ES}=\mes/m_0<1$, 
where $m_0$ represents the kinematic upper limit and is
held fixed at the center-of-mass beam energy $E^{*}_b=5.291\gev$,
and $\xi$ and $A_B$ are free parameters.

We assign background and signal probabilities to each event
included in the likelihood fit based on the measured value
for \mes. However, it is the \mes\ sideband region, where the
background probabilities are essentially 100\%,
that dominates the determination
of the combinatorial background fraction and
\deltat\ structure for background events
under the \Bz\ signal peak. Monte Carlo simulation
shows a modest \mes\ dependence on the composition of 
the combinatorial background over the sideband
range $\mes>5.2\gevcc$ through the \B\ signal region, due to variation
of the fraction of continuum versus \BB\ contributions. Since these
two sources have different \deltat\ behaviors, the
changing composition leads to a small
correction and systematic error on the precision mixing measurement,
but is negligible for the \stwob\ extraction.
The fraction of peaking backgrounds from charged
\B\ decays are estimated with Monte Carlo simulation 
as described in the following sections.

\subsubsection{\Bz\ decays to flavor-eigenstates}
\label{subsec:sample_hadronicBz}
Candidates in the $B_{\rm flav}$ sample of neutral
flavor-eigenstate \B\ mesons
are formed by combining a \Dstarm\ or \Dm\ 
with a \pip, \rhop\ $(\rhop\to\pip\piz)$, \aonep\ $(\aonep\to\pip\pim\pip)$,
or by combining a \jpsi\ candidate
with a \Kstarz\ $(\Kstarz\to\Kp\pim)$. 
As described in Section \ref{sec:PID}, kaon identification is used to
reject background. For most \Bz\ modes, it is possible to achieve signal
purities of at least 90\% with the {\tt VeryLoose} selection, or
no particle identification at all. 
However, for the mode $\Bz\to\Dm\aonep$, the tighter kaon identification 
is required to reduce large combinatorial backgrounds.

For final states with a \Dstar\ and 2 (3) pions we require
$|\cos\theta_{\rm th}|<0.9$ (0.8) for the $\Dzb\to\Kp\pim$ and $\Kp\pim\piz$
modes and 0.8 (0.7) for  $\Dzb\to\Kp \pip \pim \pim$ and $\KS \pip
\pim$, while no requirement is made for 
the $\Bz \to \Dstarm \pip$ mode.
In modes which contain a \Dm\ and a \pip, $\rho^+$, or $a_1^+$ in the final
state, we require $|\cos\theta_{\rm th}|<0.9$, 0.8, or 0.7, respectively. 

\begin{figure}[!htbp]
\begin{center}
 \includegraphics[width=1\linewidth]{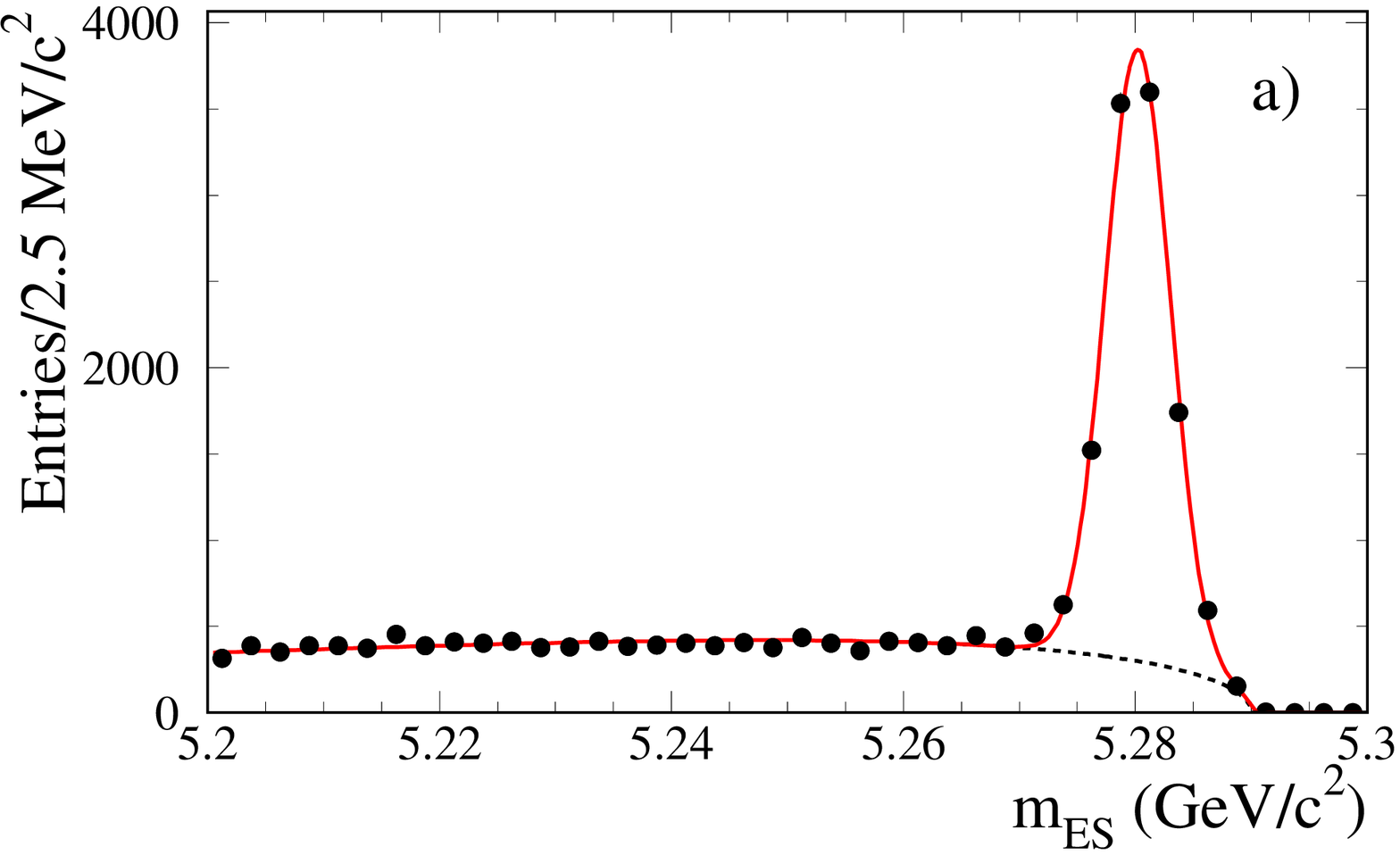}
 \includegraphics[width=1\linewidth]{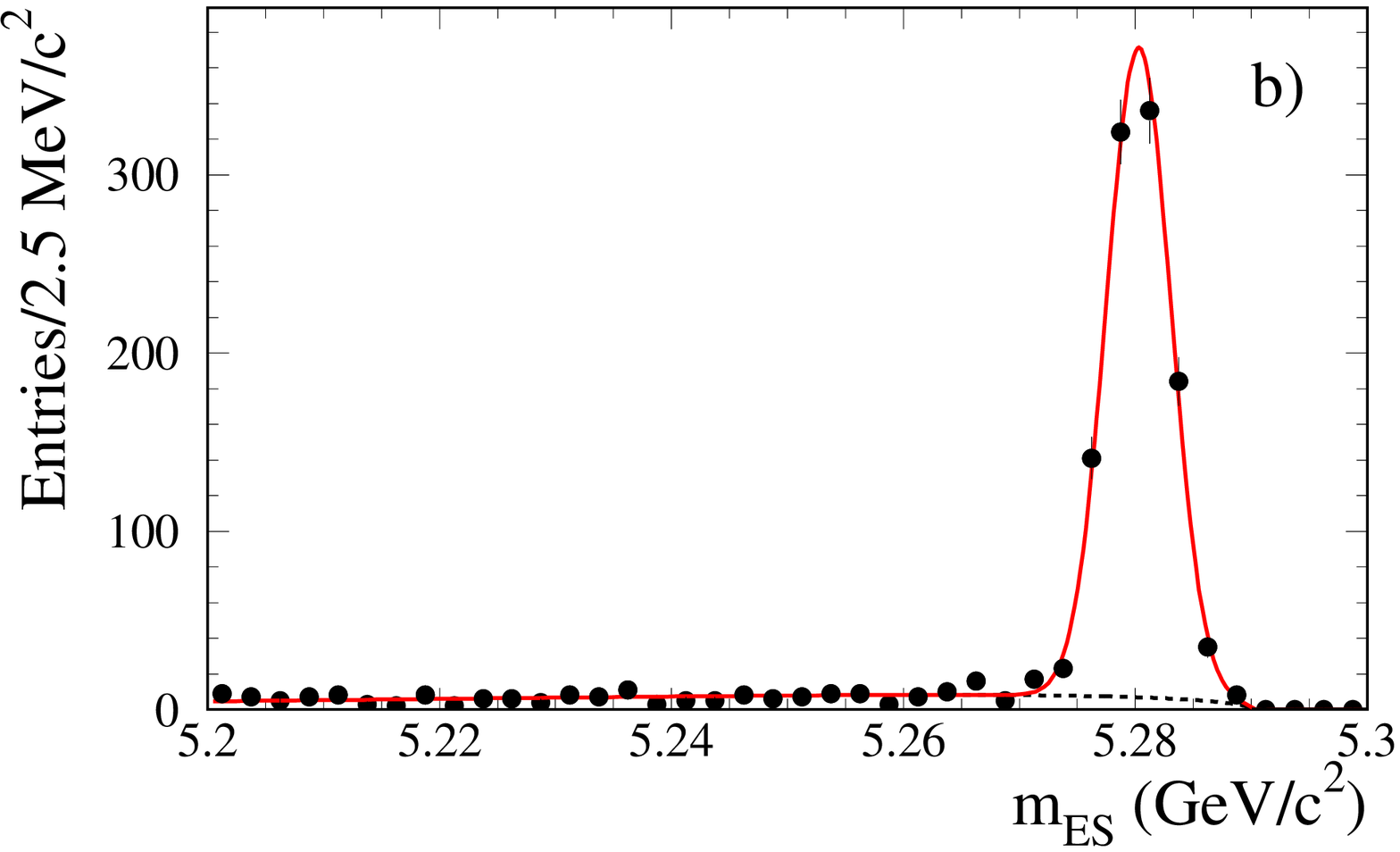}
\end{center}
\caption{
Distribution of \mes\ for all selected 
flavor-eigenstate \Bz\ candidates in hadronic decays 
to (a) open charm and (b) charmonium final states.
Overlaid in both cases is the result of a fit with a Gaussian distribution
for the signal and an ARGUS function for the background.
\label{fig:hadronicb0}}
\end{figure}

The \Bz\ signal yield and sample purity extracted from 
fits to the \mes\ distribution are summarized in 
Table~\ref{tab:HadronicBYield}. 
The net \Bz\ signal sample, before applying any decay vertex requirements, 
consists of $9922 \pm 129$
signal candidates in open charm decays with a purity of about 82\%, and 
$1013 \pm 36$ in the decay $\Bz\to\jpsi K^{*0}$ ($K^{*0} \to K^+ \pi^-$),
with a purity of about 95\%.
Figure~\ref{fig:hadronicb0}
shows the combined \mes\ distribution for all the hadronic
\Bz\ modes. Superimposed is the result of a fit with a Gaussian distribution for the
signal and an ARGUS background form~\cite{ARGUS_bkgd}.

The signal obtained by this method includes a small fraction of peaking background from
other charged and neutral \B\ decay modes. However, only the charged \B\ component 
needs to be determined, since it alone has a time structure that differs from the 
signal events. Therefore, the fraction of peaking background is estimated with a sample
 of $\FourS\to B^+B^-$ Monte Carlo events. The $\Bu$ mesons are forced to decay in the decay modes
\Dstarzb\ or \Dzb\ with a \pip, \rhop, or \aonep, since the main source is decay channels that 
have one more or one fewer pion in the final state than the signal modes of interest.
We then attempt to reconstruct neutral \B\ mesons in the channels used for the $B_{\rm flav}$
sample in data. A small peak at the \Bz\ mass, obtained with the charged \B\ Monte Carlo
sample, leads to an estimate of $(1.3\pm 0.3^{+0.2}_{-0.5})$\% as the peaking component in
the $B_{\rm flav}$ signal. This result is obtained from a fit with a Gaussian distribution, whose mean
and width are fixed by the \Bz\ signal parameters. The \deltat\ structure of the peaking background in Monte Carlo
is found to be consistent with the lifetime of the \Bu, as expected.

\subsubsection{\Bu\ control samples}
\label{subsec:sample_hadronicBch}

The \Bu\ control sample of charged \B\ candidates is 
formed by combining a \Dstarzb, \Dzb, \jpsi, or \psitwos\ candidate
with a \pip\ or \Kp.  
For the $\Dzb\pip$ final state, we require $|\cos\theta_{\rm th}|< 0.9$
for the $\Dzb\to\Kp\pim$ mode and 0.8 for all other \Dzb\ channels. In modes
that contain a $\Dstarzb\to\Dzb\piz$, the requirement is 
$|\cos\theta_{\rm th}|< 0.9$ for $\Dzb\to\Kp\pim$, 0.8 for the $\Kp\pim\piz$
and $\Kp\pim\pim\pip$, and 0.7 for $\KS\pip\pim$.

\begin{figure}[tbhp]
\begin{center}
 \includegraphics[width=1.0\linewidth]{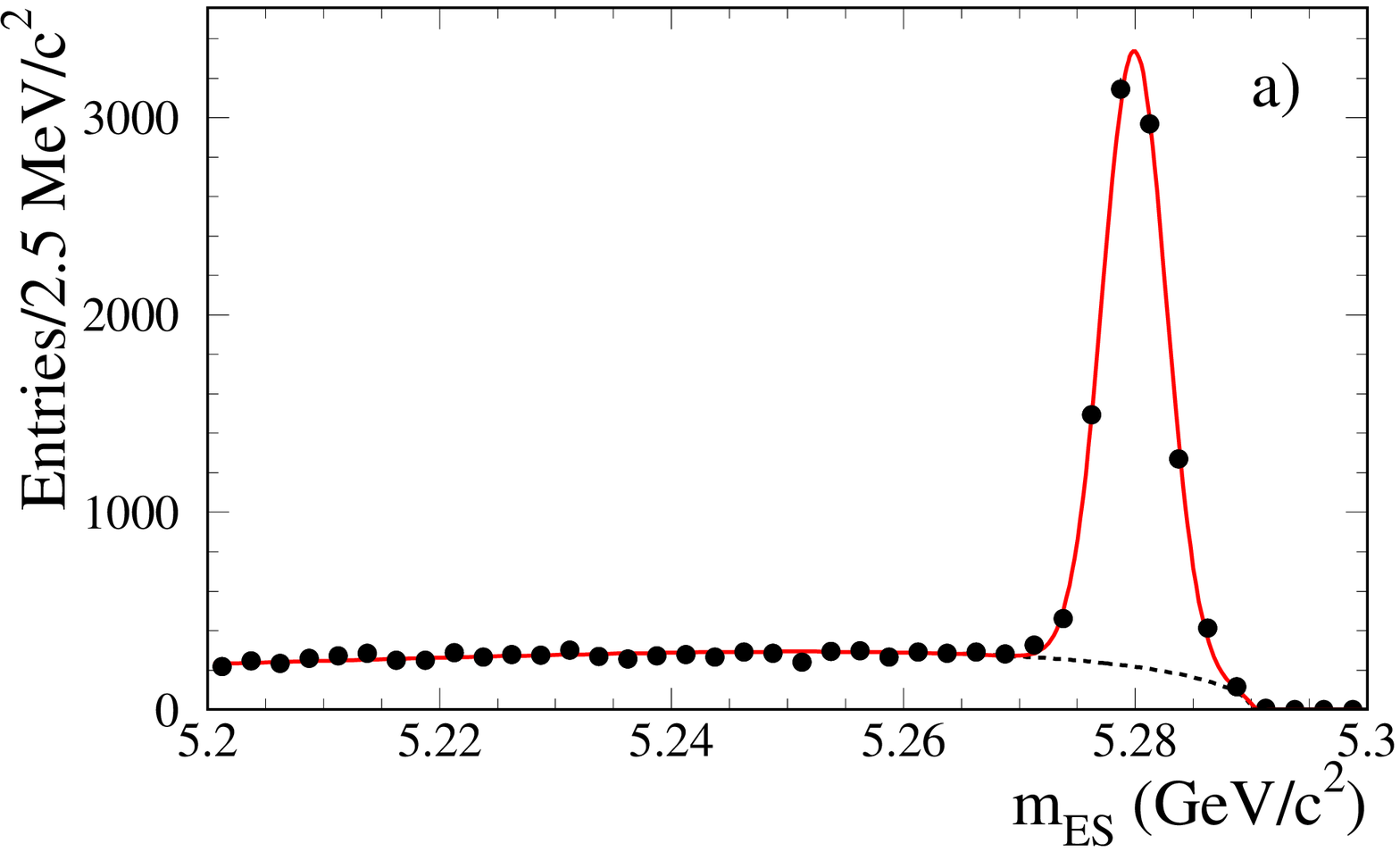}
 \includegraphics[width=1.0\linewidth]{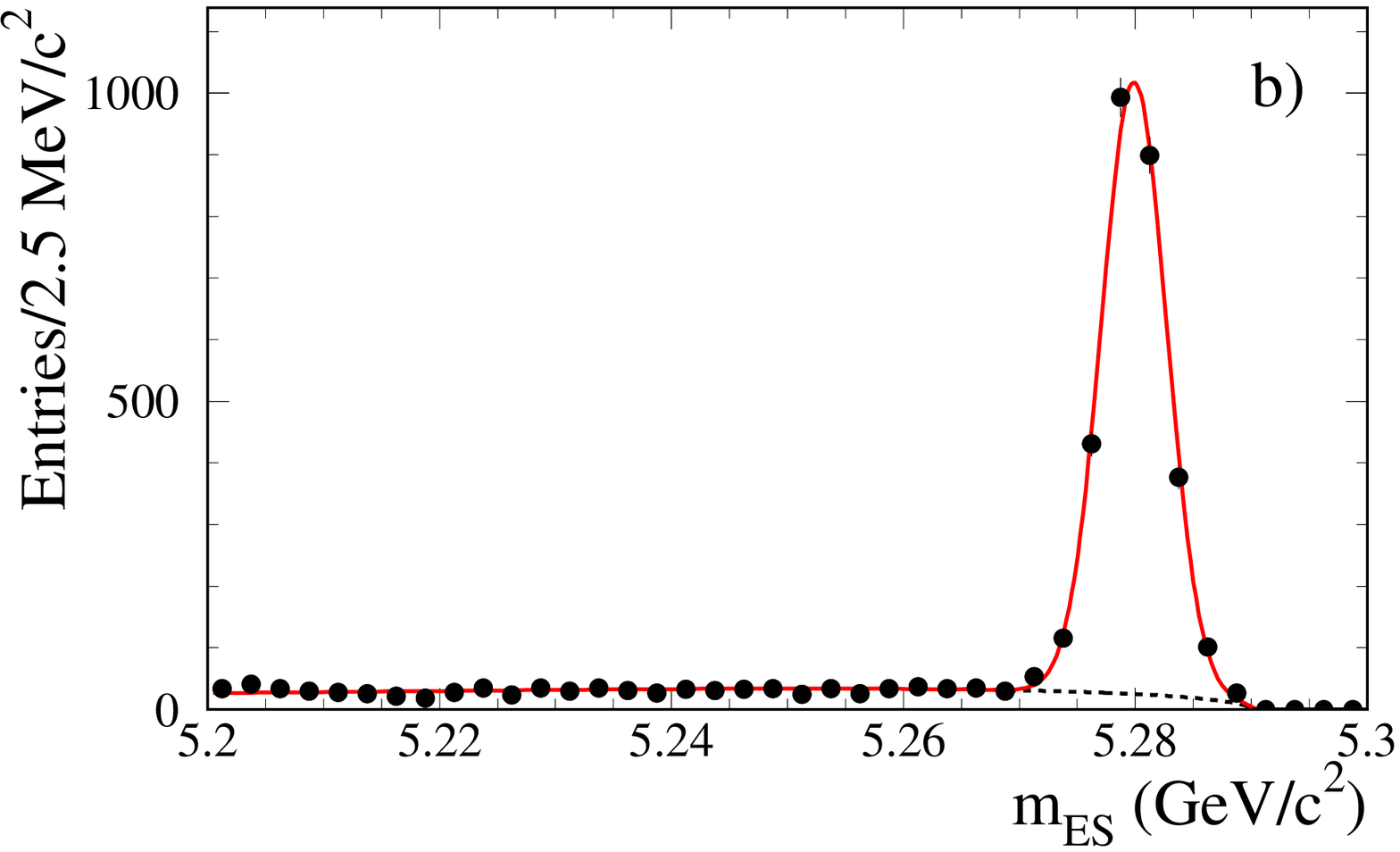}
\end{center}
\caption{
Distribution of \mes\ for all selected 
flavor-eigenstate \Bu\ candidates in hadronic decays 
to (a) open charm and (b) charmonium final states.
Overlaid in both cases is the result of a fit with a Gaussian distribution
for the signal and an ARGUS function for the background.
\label{fig:hadronicbch}}
\end{figure}

The \Bu\ signal yield and sample purity extracted from 
fits to the \mes\ distribution are summarized in 
Table~\ref{tab:HadronicBYield}. 
The net \Bu\ signal sample in open charm modes, before applying any decay-vertex requirements, 
consists of $2797\pm 62$ signal candidates in charmonium modes, with a purity of about 94\%,
and  $8547 \pm 115$ signal candidates in open charm modes, with a purity of about 84\%.
Figure~\ref{fig:hadronicbch}
shows the combined \mes\ distribution for all the hadronic
\Bu\ modes. Superimposed on the data is the result of a fit with a Gaussian distribution for the
signal and an ARGUS background form~\cite{ARGUS_bkgd}.

\subsubsection{\Bz\ decays to \CP\ modes involving \KS}
\label{subsec:sample_CP}
\begin{figure}[!htbp]
\begin{center}
 \includegraphics[width=\linewidth]{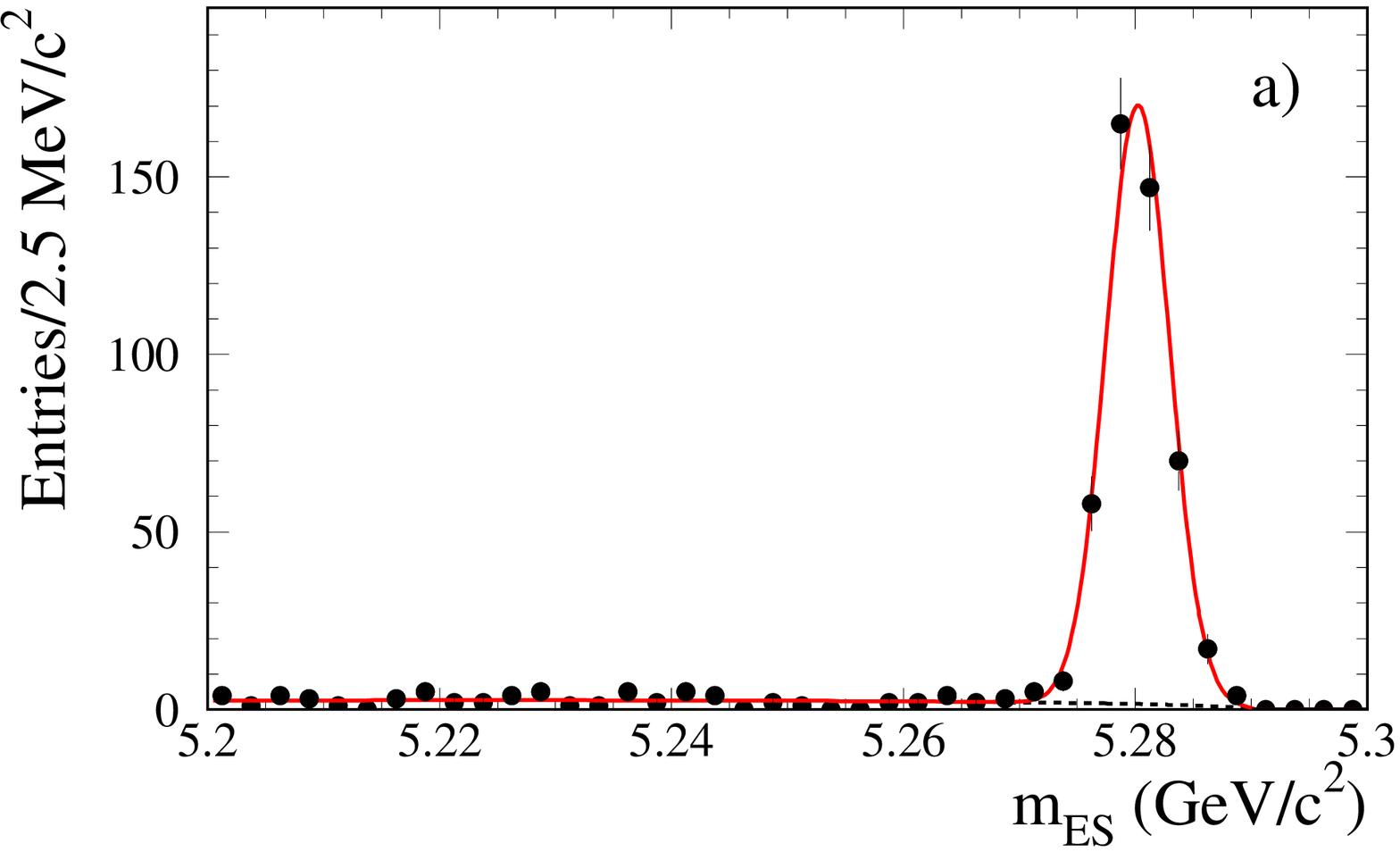}
 \includegraphics[width=\linewidth]{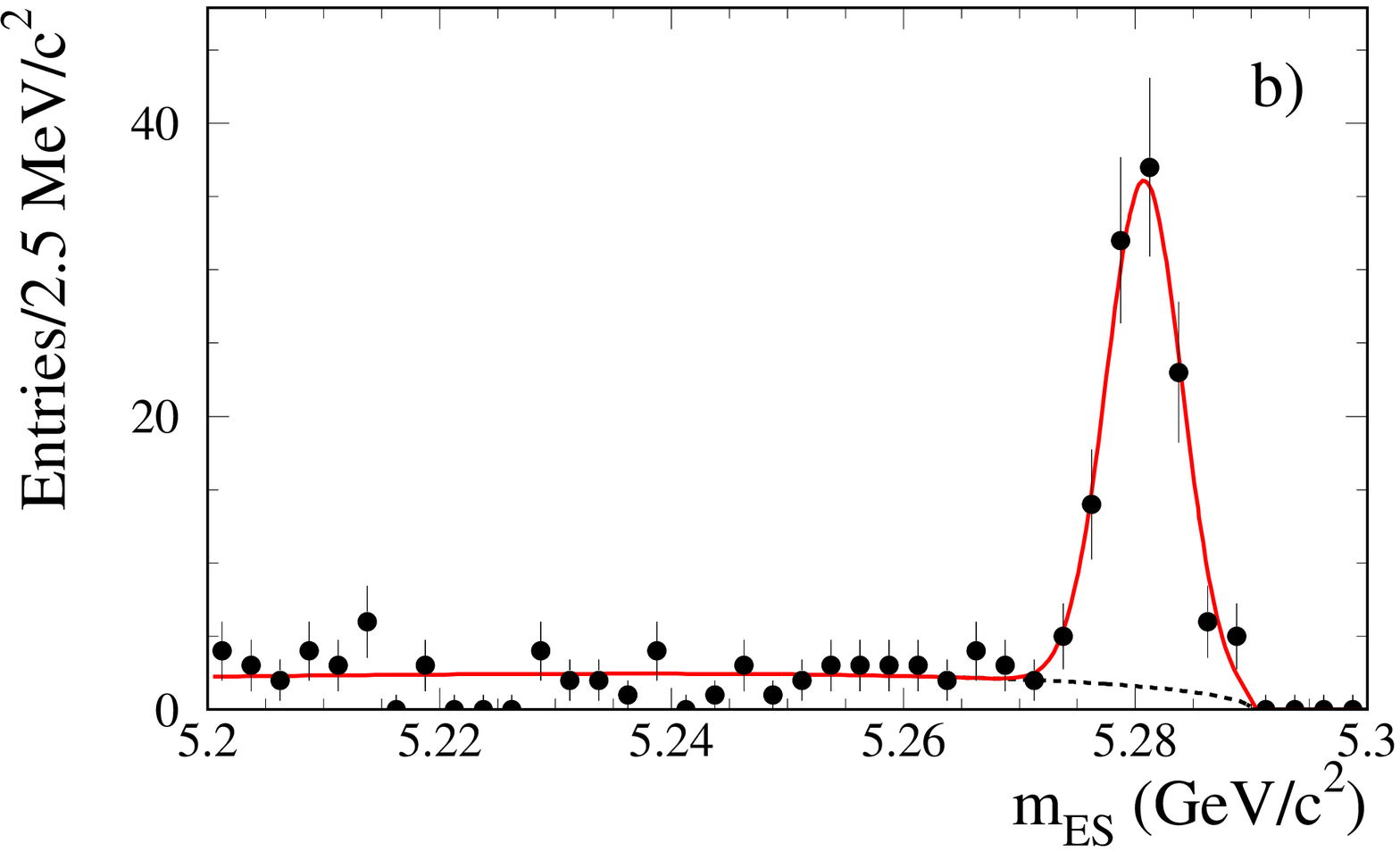}
 \includegraphics[width=\linewidth]{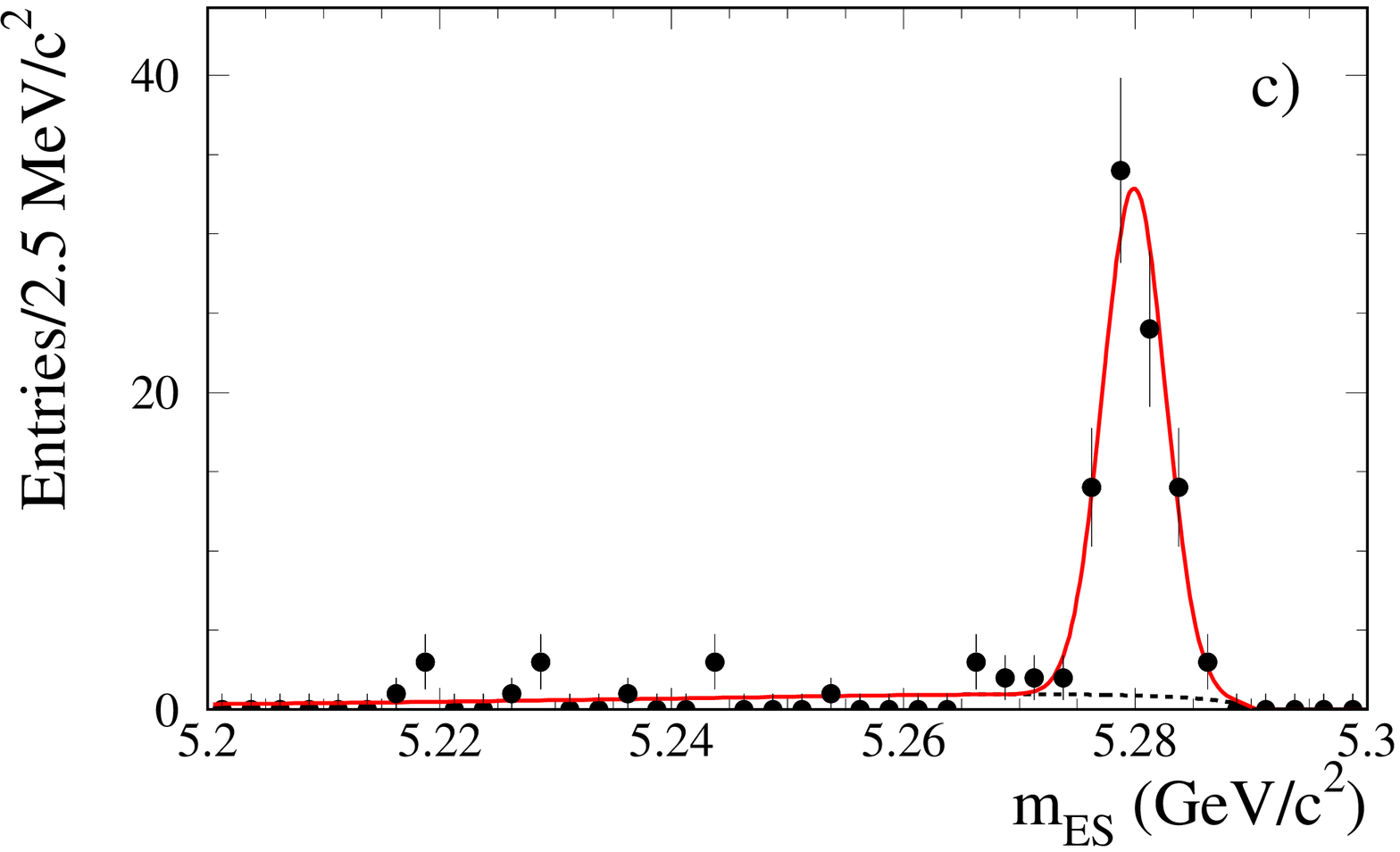}
 \includegraphics[width=\linewidth]{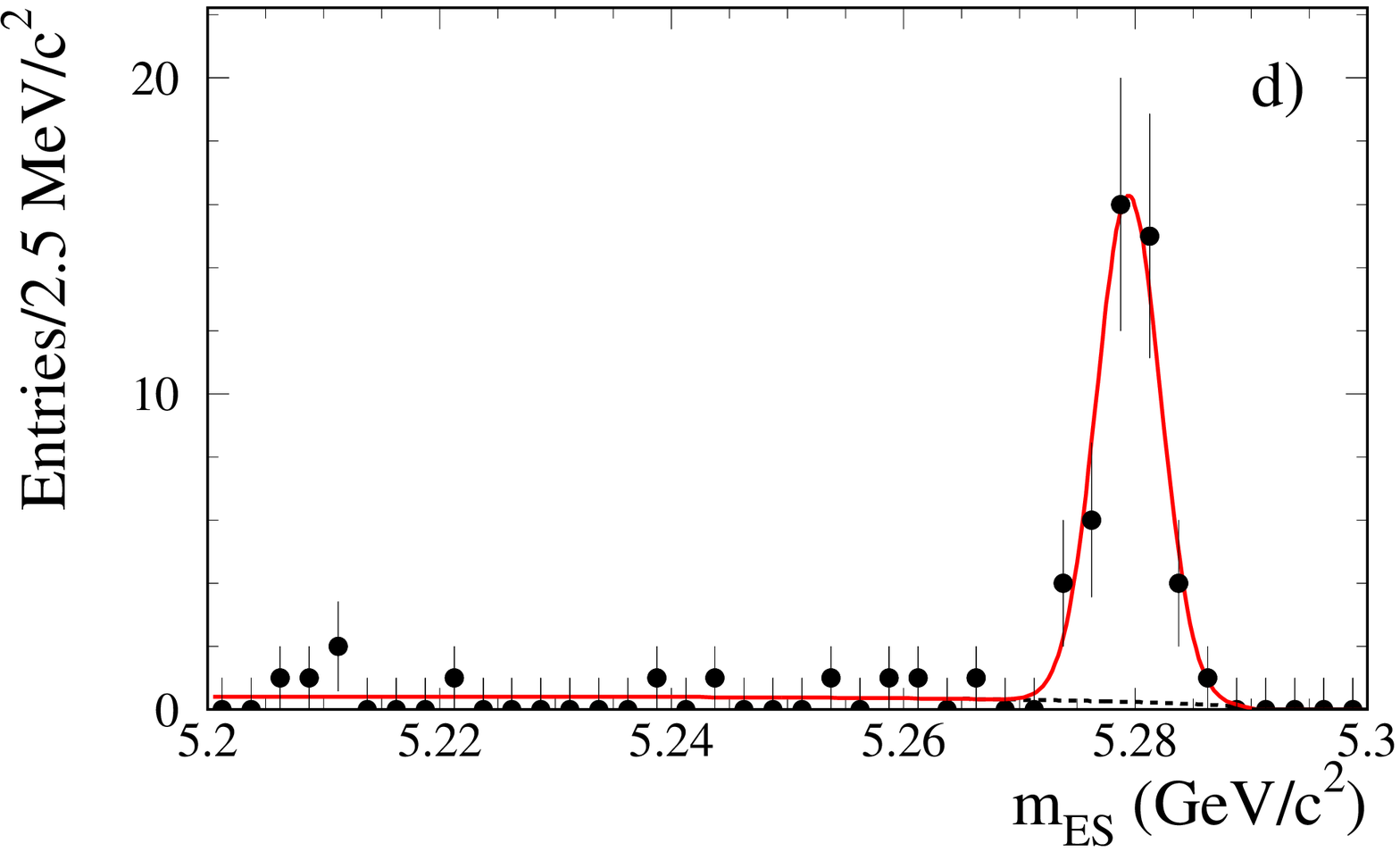}
\caption{
Candidates for $\Bz\to\jpsi \KS$ where \KS\ decays to a) $\pi^+ \pi^-$ or b) $\piz\piz$;
Candidates for c) $\Bz\to\psitwos \KS$ and d) $\Bz\to\chicone \KS$ ($\KS \to \pi^+ \pi^-$). 
Overlaid in each case is the result of a fit with a Gaussian distribution
for the signal and an ARGUS function for the background.
}
\label{fig:jkspm_00}
\end{center}
\end{figure}

We form the $B_{CP}$ sample of neutral \B\ candidates 
in charmonium modes with a \KS\ 
by combining mass-constrained \jpsi, 
\psitwos\ or \chicone\ candidates 
with mass-constrained \KS\ candidates, following the techniques of
our recent branching-fraction study~\cite{babar0107}.
The helicity angle $\theta_h$ of the \jpsi\ daughters with respect
to the \jpsi\ flight direction in the \B\ candidate rest frame should have
a $\sin^2\theta_h$ distribution. Therefore, we require that
$|\cos\theta_h|<0.8$ for the $\epem$ mode and 0.9 for the 
$\mu^+ \mu^-$ mode, as an efficient way of rejecting backgrounds.   
For the $\psitwos \KS$ candidates, $|\cos\theta_h|$
of the $\psitwos$ must be smaller than 0.9 for both leptonic modes.

Distributions of \mes\ are shown 
in Fig.~\ref{fig:jkspm_00} 
for the \CP\ samples.  
Signal event yields and purities, determined 
from a fit to the \mes\ distributions after selection 
on ${\rm \Delta} E$, are summarized in
Table~\ref{tab:HadronicBYield}. 

\begin{figure}[!htb]
\begin{center}
 \includegraphics[width=\linewidth]{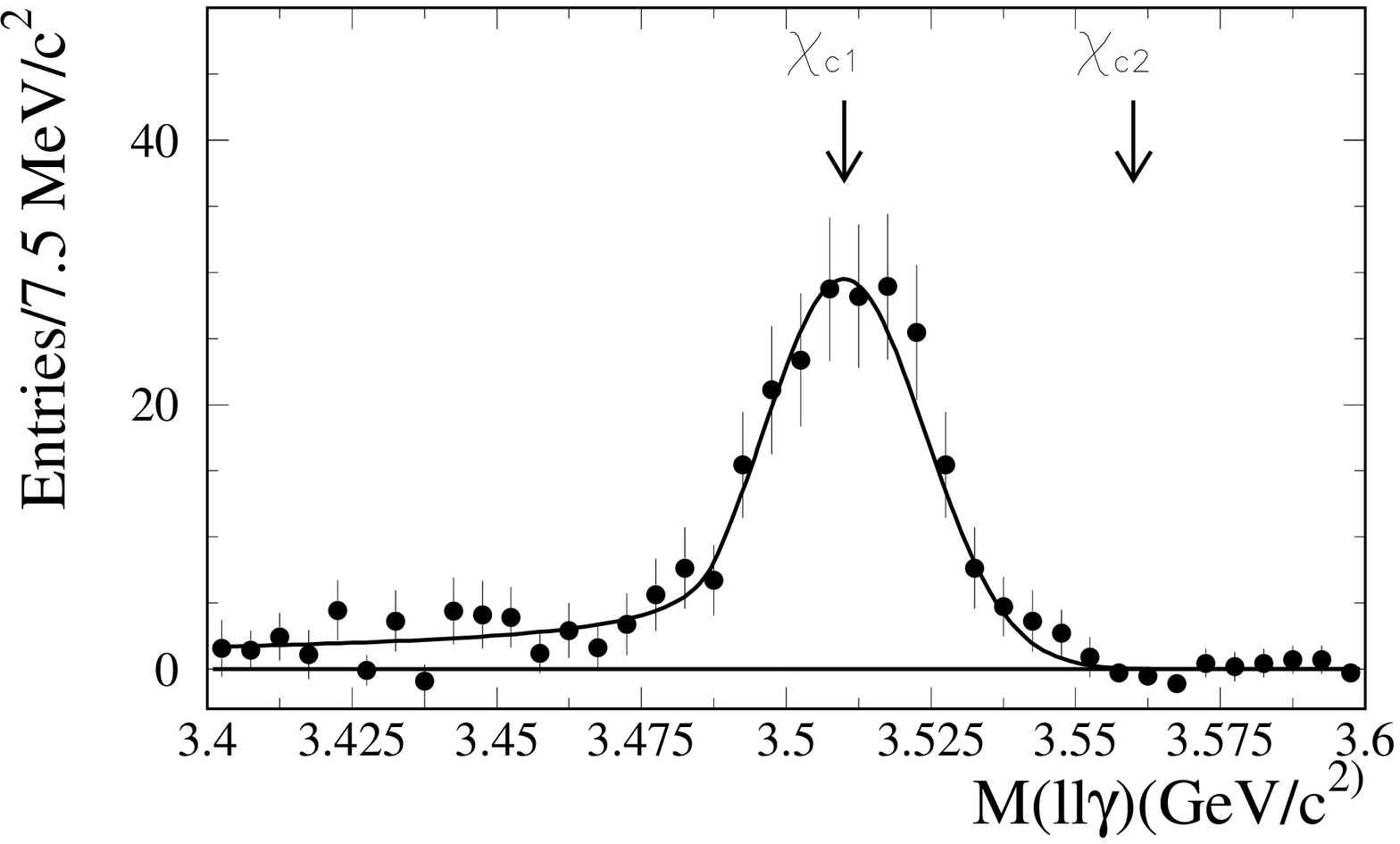}
\caption{
Distribution of $m(\ell\ell\gamma)$ for the \chicone\ 
daughters of $\Bp\to\chicone K^+$ and $\Bz\to\chicone \KS$ candidates. The
expected location of a \chictwo\ signal is indicated by the arrow.
}
\label{fig:chicExcl}
\end{center}
\end{figure}

The fraction of peaking background has been estimated with a sample of $B \to \jpsi X$ 
Monte Carlo events. The main source is decay channels that 
have one more or one less pion in the final state than the signal mode. 
The fractions are obtained by fitting the misreconstructed $B\to\jpsi X$
sample with a Gaussian distribution, whose mean and width are fixed by
the \Bz\ signal parameters.
The estimated contributions are $(0.41\pm 0.09)\%$,  $(1.2\pm 0.2)\%$,
$(2.9\pm 1.7)\%$, and $(1.1\pm 1.1)\%$ for the $\jpsi\KS$ ($\KS\to\pi^+\pi^-$),
$\jpsi\KS$ ($\KS\to\piz\piz$), $\psitwos\KS$ and 
$\chicone\KS$ channels respectively. 

In the case of the $\chicone\KS$
mode we have also explored the possibility of contamination from 
$\chictwo\KS$ events. These would have a very similar final-state signature, 
but opposite \CP. However, this decay mode has never been 
observed and the rate is expected to be highly suppressed 
due to angular momentum considerations. 
Figure~\ref{fig:chicExcl} shows the invariant mass difference,
$m(\ell\ell\gamma)-m(\ell\ell)$, for the \chicone\
daughters of the $\Bp\to\chicone K^+$ and  
$\Bz\to\chicone \KS$ candidates.
The distribution is background subtracted with the \mes\ sideband and a fit with two Crystal Ball distributions~\cite{CB}
is superimposed,
where the means have been fixed to the known $\chicone$ and $\chictwo$
masses and the widths are forced to be equal.
The fraction of $\chictwo K$ events in the selected sample is found
to be consistent with zero and, from the fit, an upper limit of 3.5\% at 95\% C.L. is set
on the fraction of  $B\to\chictwo K$ candidates in the selected sample.

\subsubsection{\Bz\ decays to the \CP\ mode $\jpsi\Kstarz$}
\label{subsec:sample_Kstar}
The $B_{\CP}$ sample is further enlarged by the addition of \Bz\ candidates
in the mode $\jpsi\Kstarz$ ($\Kstarz\to\KS\piz$). For this purpose,
mass-constrained \jpsi\ candidates are combined with $\Kstarz\to\KS\piz$ 
candidates to form a \Bz\ candidate.
To reduce the combinatorial background, the angle 
between the flight direction of the \KS\ and the 
vector connecting the reconstructed  vertices of the \jpsi\ and the \KS\ candidates is required to be 
less than 200\mrad. Cross-feed background from other $B\to\jpsi X$ modes 
involving a \piz\ (which includes cross-feed from the \CP\ mode itself)
is suppressed by requiring the cosine of the 
helicity angle of the \Kstarz\ in the \Bz\ meson rest frame 
to be smaller than 0.95. Further details of the selection and analysis
of this sample can be found in Ref.~\cite{babar0105}.

\begin{figure}[!htbp]
\begin{center}
 \includegraphics[width=\linewidth]{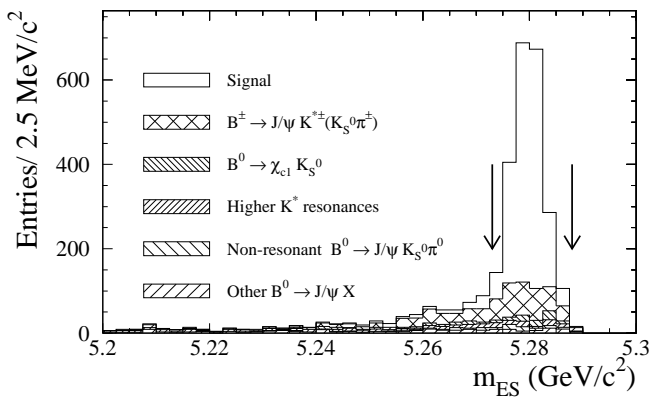}
\caption{
Distribution of \mes\ for selected $\jpsi \Kstarz$
combinations, where $\Kstarz\to\KS\piz$. The arrows indicate the
region between 5.273 and 5.288\gevcc\ that is used to define
the sample of \Bz\ candidates. Monte Carlo
estimates of the various background contributions
are also indicated.  
}

\label{fig:jkstar}
\end{center}
\end{figure}

The \mes\ distribution for $\jpsi\Kstarz$ ($\Kstarz\to\KS\piz$) combinations 
in data is shown in Fig.~\ref{fig:jkstar}.
Given the relatively tight criteria applied in the lepton identification 
of the daughters of the \jpsi\ candidates
(see Table~\ref{tab:jpsireq}), the background is dominated by true 
\jpsi\ mesons from \B\ decays. Its composition can therefore only 
be estimated with Monte Carlo simulation and the 
\mes\ distribution is not expected to follow the 
phase-space form of Eq.~\ref{ARGUS_bkd}.
Monte Carlo simulation of events with true $\jpsi$
candidates has been adjusted to match recent results for charmonium
branching fractions in \B\ decays and
takes into account the indication of 
$S$-wave $\Bz \to \jpsi \KS \piz$ decays and contributions due to
higher $\Kstar$ resonances reported in Ref.~\cite{babar0105}.

As a result, backgrounds are not estimated
with a fit to the observed \mes\ distribution (Fig.~\ref{fig:jkstar}),
but rather by extrapolation of Monte Carlo background distributions,
normalized to the number of produced \B\ mesons in the data.
All $\jpsi\KS\piz$ combinations
in the range $5.273<\mes<5.288$\gevcc\ are 
considered as candidates for this purpose.
Estimates of the signal and background contributions in the 
candidate sample, 
and the corresponding effective \CP, 
after acceptance correction for the signal selection,
is provided in Table~\ref{tab:kstarback},
while the signal yields and purities in data are listed in 
Table~\ref{tab:HadronicBYield}. The dominant source of cross-feed 
background, with zero effective \CP, is $\Bp\to\jpsi\Kstarp$ 
($\Kstarp\to\KS\pip$), where the daughter \pip\ is exchanged for a 
background \piz.

\begin{table}[htb]
\caption{Signal and background estimates for the selected 
$\Bz \to \jpsi \Kstarz$ ($\Kstarz\to\KS \piz$) sample. 
All the events within the range $5.273<\mes<5.288$\gevcc\ are 
considered as \Bz\ candidates and the background contributions are 
estimated with Monte Carlo simulation.
The quoted errors are derived from conservative
bound on the branching fractions and represent the size of the variation
used to estimate the systematic error on \stwob\ due to backgrounds.}
\begin{center}
\begin{tabular}{|l|c|c|c|}
\hline
Event type                            & Fraction (\%) & Effective \CP\ \\ \hline \hline
Signal                                &  $73.6 \pm 7.4$    & $+0.65 \pm 0.07$ \\
$\Bu \to \jpsi \Kstarp (\KS \pip)$  &  $17.4 \pm 1.7$    & $0$ \\
$\Bz \to \chicone \KS$                &  $2.4 \pm 0.7$     & $-1$ \\
Higher $\Kstar$ resonances            &  $2.6 \pm 1.3$     & $0 \pm 1$ \\
Non-resonant $\Bz \to \jpsi \KS \piz$ &  $1.8 \pm 0.9$     & $0 \pm 1$ \\
Other $\Bz \to \jpsi X$               &  $2.4 \pm 1.2$     & $0 \pm 1$ \\ \hline
Non $B \to \jpsi X$                   &  $0$ & $0$ \\ \hline
\end{tabular}
\end{center}
\label{tab:kstarback}
\end{table}

\subsection{\Bz\ decays to the \CP\ mode $\jpsi\KL$}
\label{subsec:sample_KL}
Candidates for the $B_{CP}$ sample in the mode $\Bz\to\jpsi\KL$ are obtained 
by combining mass-constrained
\jpsi\ and \KL\ candidates, following the methods in Ref.~\cite{babar0107}. 
The \jpsi\ candidates are 
required to have a momentum in the \FourS\ frame between 1.4 and 2.0\gevc. 
As the \KL\ energy is not well measured by the EMC or IFR detectors, 
the laboratory momentum of the \KL\ is determined by
its flight direction as measured from the EMC or IFR cluster and 
the constraint that the
invariant mass of the $\jpsi\KL$ system has the known \Bz\ mass.
The production angle $\theta_B$ of a \B\ meson
with respect to the $z$ axis in the \FourS\ frame 
follows a $\sin^2\theta_B$ distribution.
We require that $|\cos\theta_B|<0.9$. The \jpsi\ helicity angle 
is required to satisfy
$|\cos\theta_h|<0.9$ and the sum of $|\cos\theta_B|$ and 
$|\cos\theta_h|$ must be less than 1.3.
Events with a reconstructed charged or neutral \B\ decay to 
$\jpsi\KS$ ($\KS\to\pip\pim$ or $\piz\piz$), 
$\jpsi\Kstarz$ ($\Kstarz\to\Kp\pim$ or $\KS\piz$),
$\jpsi\Kp$, or 
$\jpsi\Kstarp$ ($\Kstarp\to\KS\pip$ or $\Kp\piz$)
are explicitly removed.
The total missing transverse momentum projected along the \KL\ direction,
where the total momentum is calculated with all charged tracks and
neutral clusters (without the \KL),
must be no more than $0.25$ ($0.40$)\gevc\ lower than the calculated
\KL\ transverse momentum for EMC (IFR) \KL\ candidates.

Events where multiple $\jpsi\KL$ combinations with 
$\Delta E<80$\mev\  satisfy these requirements
are treated as a special case.
A hierarchy is imposed where the highest energy EMC cluster
for multiple EMC combinations, or the IFR cluster with the largest number of 
layers for multiple IFR combinations is selected.
In cases where there are both an EMC and IFR combination, the EMC combination
is selected because it is expected to have better angular resolution.
Although the EMC information is used, such events are counted as IFR
events, since they have the same relatively high signal purity. 

\begin{table}[t]
\caption{Monte Carlo prediction for the composition of background 
channels containing a true $\jpsi$ that pass the 
$\Bz \to \jpsi \KL$ selection criteria. Events are required to 
have $|\deltae| < 10$\mev.  The quoted errors are derived from conservative
bound on the branching fractions and represent the size of the variation
used to estimate the systematic error on \stwob\ due to backgrounds.}
\begin{center}
\begin{tabular}{|l|c|c|c|}
\hline
Event type                           & EMC (\%) & IFR (\%) & Effective \CP\ \\ \hline \hline
$\Bz \to \jpsi \Kstarz (\KL \piz)$   & $23\pm 3$     & $26\pm 3$     & $-0.68\pm 0.07$ \\
$\Bu \to \jpsi \Kstarpm (\KL \pipm)$ & $28\pm 4$     & $45\pm 6$     & 0 \\
$\Bz \to \jpsi \KS$                  & $13\pm 2$     & $2\pm 1$      & $-1$ \\
$\Bz \to \chicone \KL$               & $3 \pm 1$     & $5 \pm 1$     & $+1$ \\
$\B \to \jpsi \KL \pi$               & $1^{+2}_{-1}$ & $1^{+2}_{-1}$ & $0$ \\
Other $\Bz \to \jpsi X$              & $32 \pm 16$   & $21 \pm 10$   & $0 \pm 0.25$ \\ \hline
\end{tabular}
\end{center}
\label{tab:psiklComposition}
\end{table}

\begin{table}[hb]
\caption{Results of the binned likelihood fit to the full 
\deltae\ distribution of the
$\Bz\to\jpsi\KL$ combinations.  All signal yields and background estimates are 
reported for the $B_{\CP}$ candidate range $|\deltae|<10$\mev.  }
\begin{center}
\begin{tabular}{|l|c|c|c|}
\hline
                    &  \multicolumn{3}{|c|}{ $\KL$ reconstruction type } \\ \cline{2-4}
                    &  EMC \& IFR     &      EMC      &  IFR  \\
\hline \hline
  Data events       &  427            &  228          &  199  \\
\hline
  Signal            &  257 $\pm$ 24   & 128 $\pm$ 17  &  129 $\pm$ 17  \\
  $\jpsi X$ bkgd    &  154 $\pm$ 15   &  89 $\pm$ 11  &   65 $\pm$ 10  \\
  non-$\jpsi$ bkgd  &   19 $\pm$  2   &  14 $\pm$  2  &    5 $\pm$  1  \\
\hline
 Signal fraction    & 0.60 $\pm$ 0.04 &   0.56 $\pm$ 0.05  &  0.65 $\pm$ 0.05 \\
\hline
\end{tabular}
\end{center}
\label{tab:psikldefit}
\end{table}

\begin{figure}[!htb]
\begin{center}
\includegraphics[width=\linewidth]{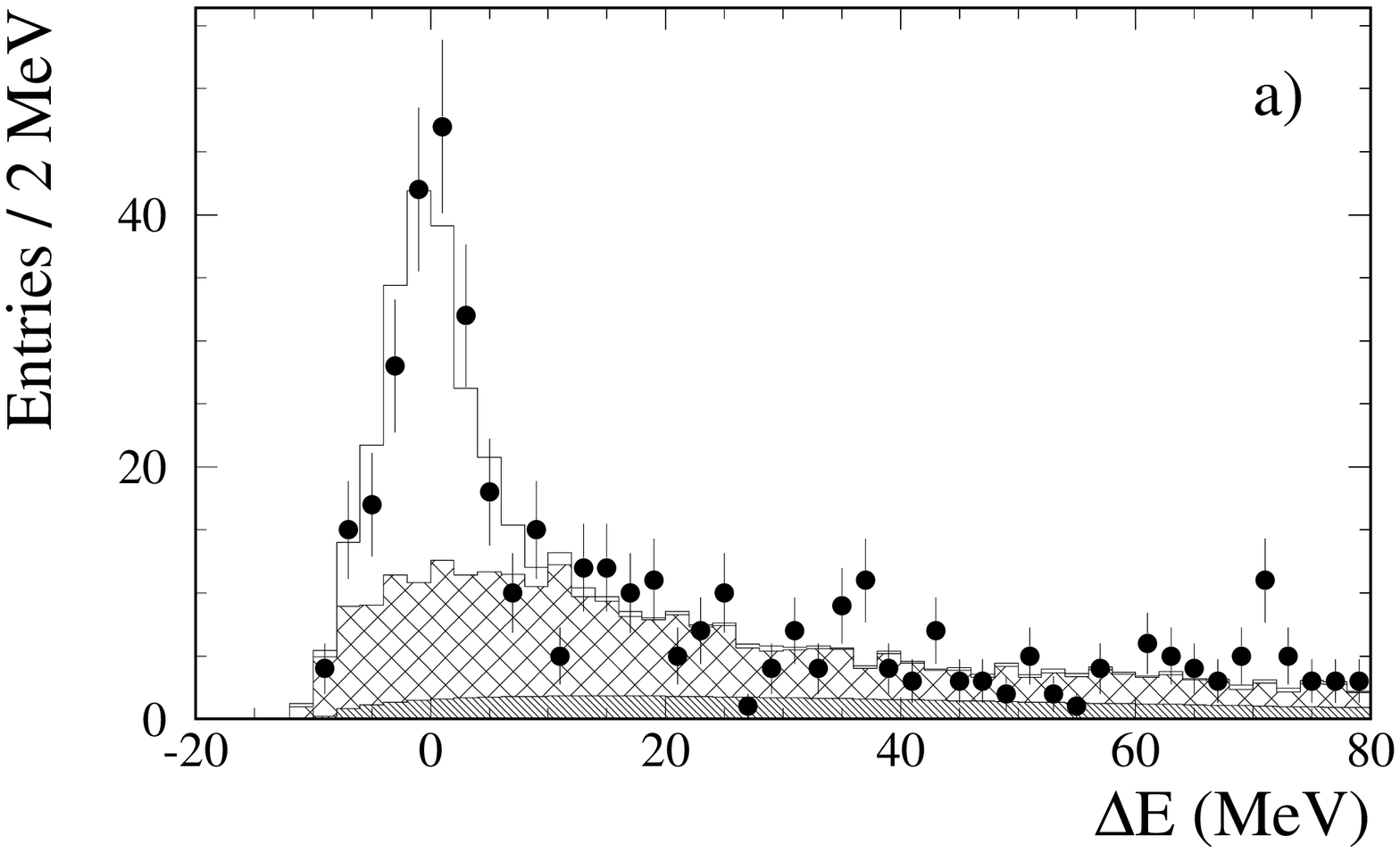}
\includegraphics[width=\linewidth]{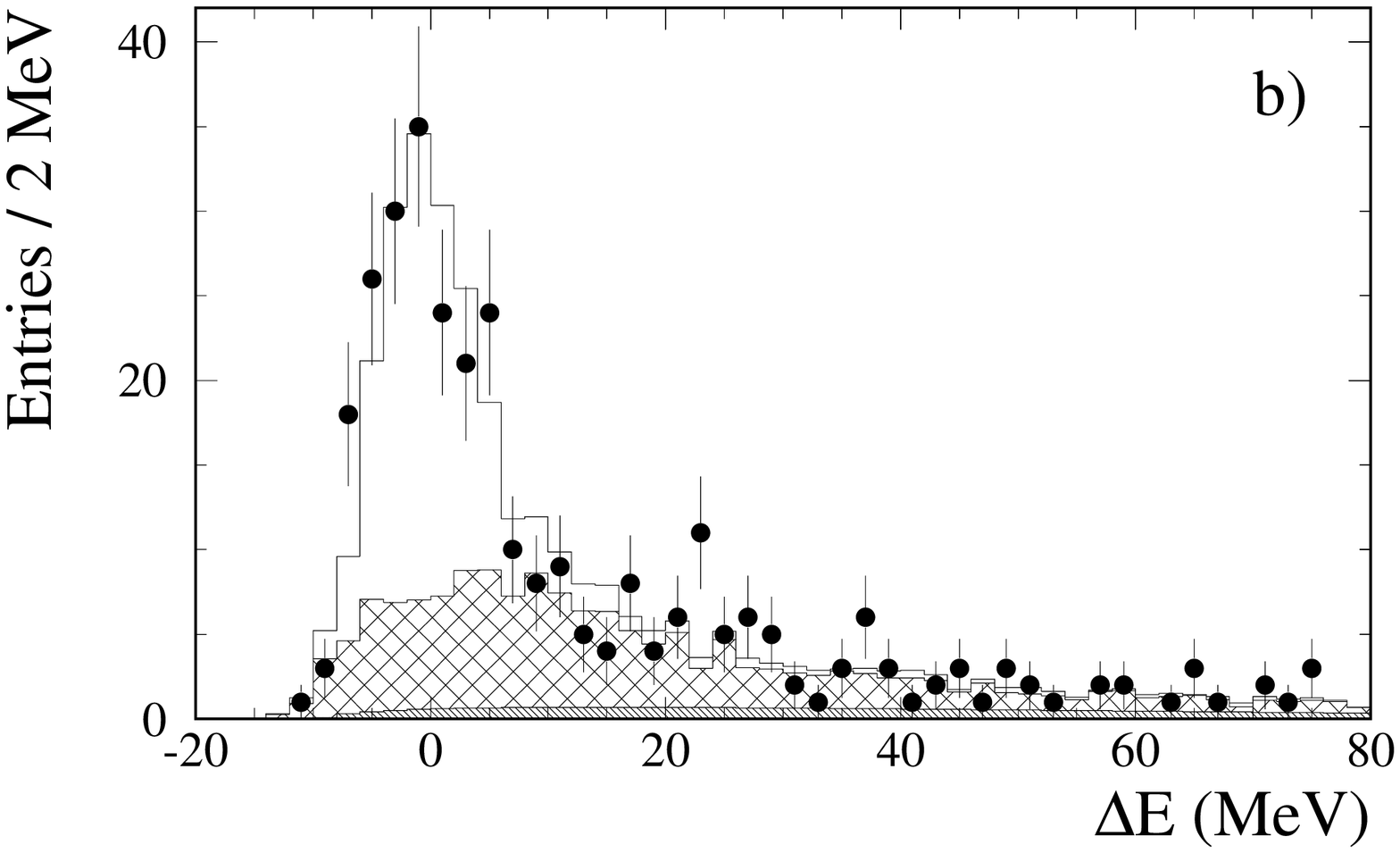}
\includegraphics[width=\linewidth]{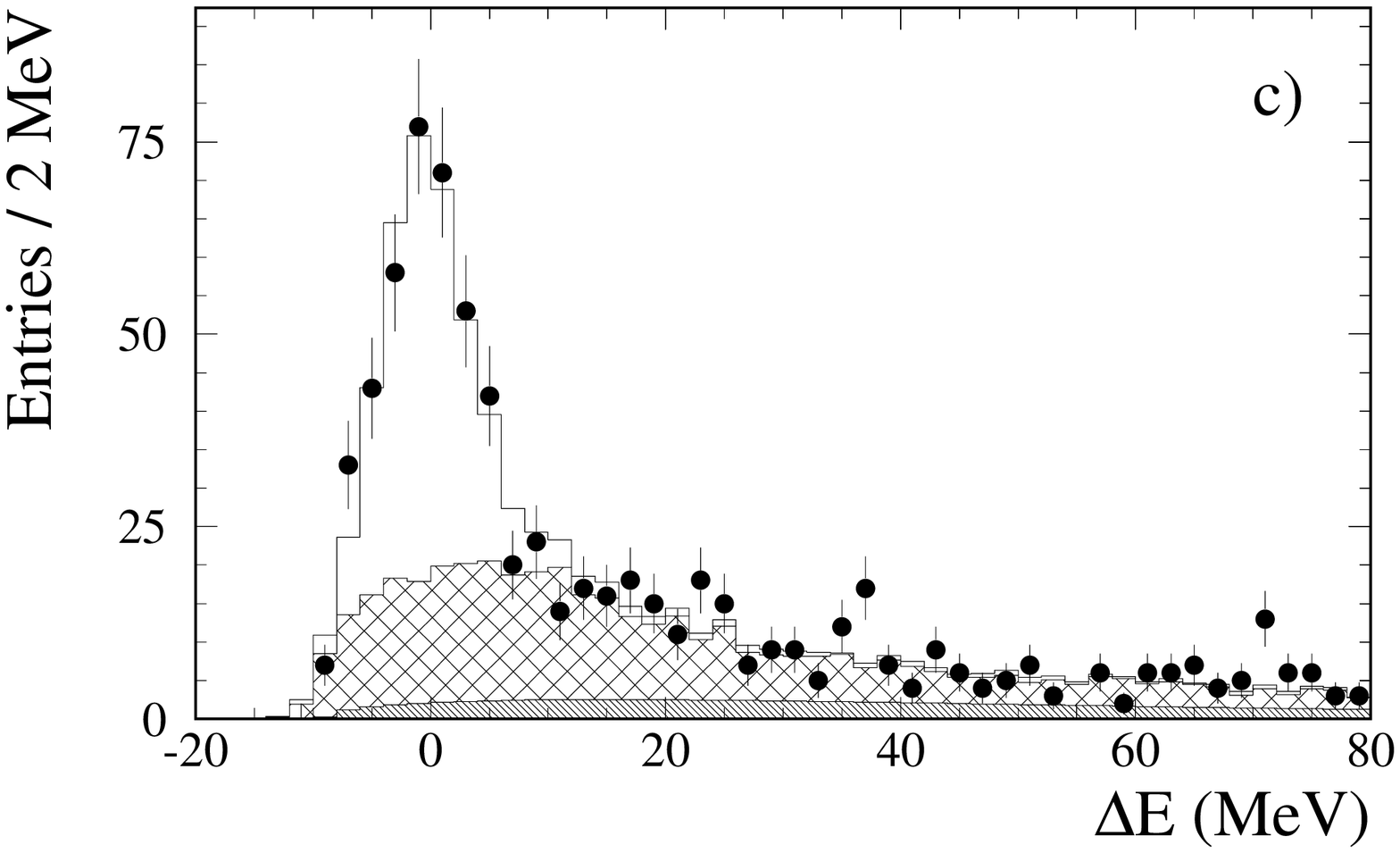}
\end{center}
\caption{Distribution of \deltae\ for selected $\Bz\to\jpsi\KL$ combinations
where the \KL\ is identified a) in the EMC, b) in the IFR,
or c) either subsample combined.
The points with error bars are the data.
The open histogram is the result
of a three-component binned likelihood fit, where the three components
are signal (open), inclusive \jpsi\ background (cross hatched),
and non-$J/\psi$ combinatorial background (dark shading).
The shapes of the signal and inclusive \jpsi\ background are taken
from Monte Carlo.  The shape of the non-$J/\psi$ combinatorial background
is taken from the $m(\ell\ell)$ sideband in data.
Candidates for the $B_{\CP}$ sample are selected in the
region $\vert\deltae\vert<10\mev$. }
\label{psikldata}
\end{figure}

The difference $\Delta E$ between
the energy of the $\jpsi\KL$ system and the 
beam energy in the \FourS\ frame is used
to discriminate between signal and backgrounds. 
The $\Delta E$ distribution 
of selected $\Bz\to\jpsi\KL$ combinations 
for the \FourS\ data is shown in Fig.~\ref{psikldata}.
Signal events are peaked within $\pm 10$\mev\ of $\Delta E=0$
while background events extend towards positive values of $\Delta E$. 
The small signal width
and the asymmetric distribution of the background in comparison 
with the \KS\ modes are both
consequences of the mass constraint
used to determine the \KL\ momentum.

The purity of the $\Bz\to\jpsi\KL$ sample is the lowest of the
\CP\ modes (60\%). Irreducible backgrounds are dominantly from 
$B \to\jpsi \KL X$ modes, which cannot be distinguished from signal 
due to imposition of the $m_B$ mass constraint
in determining the momentum of the $\KL$ candidate.
The largest single background contribution is from 
$\B \to \jpsi K^*$, where the
$K^*$ decays to $\KL\pi$. This mode and backgrounds from other 
$B \to \jpsi X$ decays are studied with Monte Carlo simulation.  The
composition of the events that are included in the $\jpsi \KL$ sample is
given in Table~\ref{tab:psiklComposition}.
The effective \CP, after acceptance correction for the signal selection,
is also provided.
The additional background 
from events with a misreconstructed $\jpsi \to \ell \ell$ candidate is studied
with the $m(\ell \ell)$ sidebands.

A binned likelihood fit to the $\deltae$ distribution
is performed separately for the EMC and IFR categories to 
determine the composition  of the $\Bz \to \jpsi \KL$ sample.  
There are three fit components:
the fraction of the data that is signal, 
the number of $B \to \jpsi X$ background events, and the 
number of non-$\jpsi$ background events.  
The $\deltae$ shapes for the signal and 
the $\jpsi X$ background are determined from Monte Carlo simulation.
The $\deltae$ shape of the non-$\jpsi$ component is taken from 
the $m(\ell\ell)$ sideband in the data.
A Poisson term, with mean given by
the expected number of non-$\jpsi$ events
in the $m(\ell \ell)$ signal region,
is included in the likelihood to constrain the normalization of
the non-$\jpsi$ component.  The result of the fit is shown in
Fig.~\ref{psikldata}, and the corresponding signal and background fractions
are reported in Table~\ref{tab:psikldefit} for the 
$\Bz\to\jpsi\KL$ combinations that are selected as $B_{\CP}$ candidates
in the interval $\vert\deltae\vert<10\mev$.
As expected from Monte Carlo studies,
the purity of the IFR sample is significantly better than the
EMC sample, mainly because the $\Bz\to\jpsi\KS$ ($\KS\to\pi^0\pi^0$)
background is significantly larger in the EMC sample.
Since the purities of the two subsamples are quite different, we obtain
better statistical sensitivity in the \stwob\ fit by treating
the EMC and IFR categories separately.

\subsection{Semileptonic \Bz\ decays}
\label{subsec:sample_semileptonic}
The semileptonic decay $\Bz\rightarrow D^{*-} \ell^+ \nu$,
with a measured branching fraction of $(4.60 \pm 0.27) \%$~\cite{PDG2000},  
is potentially a copious source of reconstructed \Bz\ mesons.  
However, since the neutrino cannot be detected, the background levels
in selected samples are generally larger and more difficult to characterize.
Likewise, the \deltaz\ determination cannot take advantage of the
beam spot and reconstructed \Bz\ direction.
As a consequence, we use a large sample of $\Dstarm\ell^+\overline\nu$
candidates only as a cross check on our determination of the 
mistag rates from the time structure of the $B_{\rm flav}$
and $B_{\CP}$ events. The selection criteria for this control sample
and the characterization of backgrounds is described here, while
the analysis of the mistag rates is reported in Section~\ref{sec:onebin}.
The semileptonic \Bz\ sample is obtained
by reconstructing the $D^{*-}$ through its decay to $\Dzb\pi^-$, 
where the three decay modes $K^+\pi^-$, $K^+\pi^+\pi^-\pi^-$ 
and $K^+\pi^-\pi^0$ are used to reconstruct the \Dzb.

\subsubsection{Event selection}

All reconstructed \Dzb\ candidates are required to  have an invariant
mass within
$\pm 2.5 \sigma$ of the nominal \Dz\ mass, based on the observed RMS width
of the signal.  
A vertex fit to the \Dzb\ candidate is required to have a $\chi^2$
probability greater than 0.001. 
There are no additional requirements for $\Dzb\rightarrow K^+\pi^-$.
For $\Dzb\rightarrow K^+\pi^+\pi^-\pi^-$ and $\Dzb\rightarrow K^+\pi^-\pi^0$
we require a {\tt VeryLoose} kaon and a {\tt Loose} pion 
particle identification as described
in Section~\ref{sec:PID}, and a minimum $\pi^0$ momentum of 200\mevc\ in
the laboratory frame.
In addition, the $K$ and $\pi$ candidates  are required to have
momenta greater than 200 and 150\mevc, respectively, for the mode
$\Dzb\rightarrow K^+\pi^+\pi^-\pi^-$.  The decay  
$\Dzb\rightarrow K^+\pi^-\pi^0$ occurs mostly through quasi-two-body channels.
The $\rho$ and $K^*$ resonances dominate and we use
weights calculated from the Dalitz-plot position
for each candidate~\cite{E687-Dalitz} to construct a probability per $\Dz$
and select events using
this quantity as a way of suppressing combinatorial background.

\Dzb\ candidates satisfying these requirements are combined with
all charged tracks with a minimum transverse momentum of 50\mevc\ and
charge opposite to that of the kaon from the \Dzb\ to form \Dstarm\ candidates.
The mass difference $m(\Dzb\pim) - m(\Dzb)$ is 
required to lie within $\pm 2.5 \sigma$ 
of the nominal value, based on the observed RMS width of the signal.

Finally, \Dstarm\ candidates are combined with electron or muon 
candidates 
satisfying the {\tt Tight} lepton-identification
requirements described in Section~\ref{sec:PID}. The lepton is required to 
have charge opposite to that of the $D^*$ and momentum greater than 1.2\gevc\
in the $\FourS$ frame.
A vertex fit to the $\Dstarm\ell^+$ candidate 
is required to converge and have a $\chi^2$ probability greater than 0.01.
The \Dstarm\ and lepton from a true \Bz\ decay tend to be back-to-back in the
\Bz\ rest frame, so we require 
$\cos\theta_{\Dstar - \ell} < 0$ where $\theta_{\Dstar -\ell}$ is the angle between
the \Dstarm\ and the lepton in the \FourS\ frame.
The cosine of the angle between the thrust axes of the
$(\Dstarm \ell^+)$ pair and the rest of the event is required to satisfy 
$\vert \cos \theta_{\rm th}\vert  \leq 0.85$, in order to reduce background
from $e^+ e^- \rightarrow c \overline c$ events. 

The neutrino cannot be reconstructed in the detector, but
we can determine whether the missing
four-momentum of the candidate is consistent with a particle of zero mass:
\begin{equation}
(p_B - p_{\Dstar} - p_{\ell})^2 = p_{\nu}^2 = 0.
\end{equation}
\noindent
Applying this relation in the \FourS\ frame, we obtain a constraint on
the angle between the
\Bz\ and the $\Dstarm\ell^+$ system:
\begin{equation}
\ctby={\frac{m_B^2 + m_{\Dstar \ell }^2 - 2E_BE_{\Dstar \ell}}
                          {2|{\mathbf p}_B| |{\mathbf p}_{\Dstar \ell}|}} .
\end{equation}
\noindent
The energy $E_B$ and magnitude of the momentum $|{\mathbf p}_B|$ of the initial-state
\Bz\ are known in the \FourS\ frame from the boosted beam energies. The energy
$E_{\Dstar \ell}$, the magnitude of the momentum $|{\mathbf p}_{\Dstar \ell}|$, and the invariant mass 
$m_{\Dstar \ell}$ of the $\Dstarm\ell^+$ system are obtained from the
four-momenta of the $\Dstarm$ and the lepton. 
The cosine of the 
angle \ctby\ should lie in the physical region 
$[-1,+1]$ for true
$\Dstarm\ell^+\overline\nu$ events.
Allowing for detector resolution effects in the reconstructed
momenta and angles,  we require
$|\ctby| < 1.1$.

\begin{figure}[!tbh]
\begin{center}
\includegraphics[width=\linewidth]{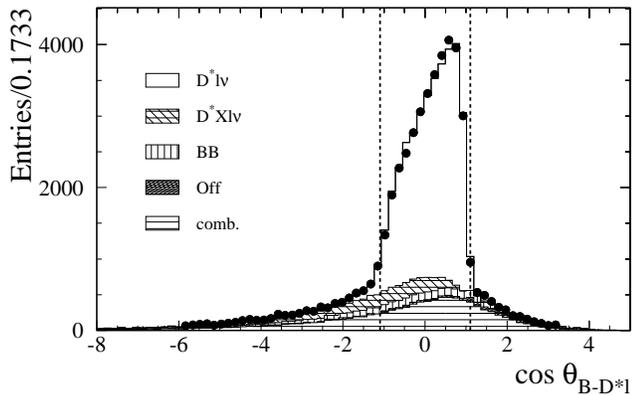}
\end{center}
\caption{Extraction of \BB\ and \ddstar\ backgrounds
  from the data. The
  points show the \ctby\ distribution for the reconstructed signal. 
  The histogram shows the result of the fit described in the text. The different 
  background contributions are indicated by the various hatchings.
\label{fig:dstlnu_data}}
\end{figure}

After applying these selection criteria, we obtain a sample of about 37500 
\bzdstlnu\ candidates. The $\ctby$ distribution of these candidates
is shown in Fig.~\ref{fig:dstlnu_data}.
 
\subsubsection{Sample composition}

The final sample contains a fraction $f_{\rm sig}$ of
\bzdstlnu\ signal events, as well as fractions $f_{\rm comb}$ of 
\Dstarm\ combinatorial background, $f_{\rm off}$ of
continuum background (a true \Dstarm\ and an identified 
lepton), $f_{\bchdstxlnu}$ of \bdstpilnu\ events, 
where the $\Dstar \pi$ can come from a radially or 
orbitally-excited $L=1$ state or non-resonant decay, 
and finally $f_{B\overline B}$ of background from
\BB\ events with a true \Dstarm\ and an identified lepton (\bbdstyl). 
Examples in this last category 
are cases where the \Dstarm\ and the lepton come from two different \B\ mesons or  
where the lepton and the \Dstarm\ are from the same \B\ but the lepton is produced in a 
charm decay.  

The combinatorial and continuum backgrounds can be extracted directly from the data. The 
former is determined with the $m(\Dzb\pi^-) - m(\Dzb)$ distribution itself. The latter is estimated with 
the off-resonance data sample, weighted by the ratio of the relative integrated 
luminosities for on- and off-resonance data.
The remaining three contributions
can be distinguished by their different distributions in
\ctby. The \bzdstlnu\ signal events are expected to lie in the region 
$-1<\ctby<1$, while contributions from \bdstpilnu\ semileptonic decay, due to 
missing particles, must extend below the
kinematic threshold $\ctby< -1$. Finally, \BB\ background events populate the
full \ctby\ distribution.
Thus, the region $\ctby>1$ contains mainly \BB\ background, while the region $\ctby<-1$ is mostly
populated by \bdstpilnu. The shape of the \ctby\ distributions for these three components is obtained 
from Monte Carlo
simulation and a fit to the full \ctby\ range is used to determine the two background fractions in the
signal region.

The orbitally-excited
resonances that can decay to $\Dstar \pi$ are the two narrow states
$D_1$, $D_2^*$ (observed with masses around 2420 and 2460\mevcc)
and the broad state $D_1^*$ (not yet seen, but with mass expected to be about
2420\mevcc\ and $\Gamma \ge 250\mev$).  Contributions from \bdstxlnu\ decays
with more than one pion are expected to be small and are more easily separated
from the signal with \ctby.
Isospin symmetry requires that the charged \B\
contribution be $2/3$ of the total $\Dstarm \pi$ pairs from \bdstpilnu\ decays, either from
orbitally-excited states or non-resonant 
decays.  The \ctby\ distribution obtained from Monte Carlo simulation for the different
channels is modeled with a general function that
is sufficiently flexible to describe both individual channels
as well as a superposition of
excited charm modes.

After subtraction of continuum and combinatorial backgrounds, a
fit is performed to the resulting \ctby\ distribution
over the full observed range $[-8,+5]$. However, it is only
the relative fraction of the various backgrounds in
the signal window $\ctby\in [-1.1,1.1]$ that we require. 
Furthermore, in the case of
\ddstar, only the charged \B\ contribution is a background for the
measurement of the mistag fraction and is assumed to be $2/3$ of the total
\ddstar\ contribution. The fitted fractions are defined by:
\begin{eqnarray}
  \label{eq:deff}
  g_{**}&=&N(\bchdstxlnu)/[N(\bzdstlnu)+ \nonumber \\
&& N(\bzdstxlnu)+N(\bchdstxlnu)] \nonumber \\
g_{BB}&=&N(\bbdstyl)/[N(\bzdstlnu) +\nonumber \\
&& N(\bzdstxlnu) + N(\bbdstyl)]
\end{eqnarray}
where the $N$ is the number of events from a given process
that survives the selection requirements.

The result of the fit to the full untagged sample is shown in Fig.~\ref{fig:dstlnu_data}, 
along with the Monte Carlo model for the
\ddstar\ component. 
The $\chi^2$ of the fit in the full \ctby\
range is 82 for 69 degrees of freedom.
The fitted contributions are
$g_{**}=\gssval$ and
$g_{BB}=\gbbval$,
where the first error is statistical and the second systematic.
To estimate the systematic error on these fractions, 
three extreme assumptions have been made concerning the
\bdstxlnu\ background:
all narrow \ddstar\ states, all broad \ddstar\ states, or all non-resonant decays. 
The largest deviation comes from the non-resonant
model. Another source of systematic uncertainty is the assumed form for the \ctby\
distributions in the Monte Carlo simulation.
The contribution from this effect has been estimated by
incorporating a 30\% fraction with a uniform 
distribution. 

The absolute background fractions in the untagged sample 
are given in Table~\ref{tab:D*lnu_bkgd}, where the 
uncertainties include both statistical and systematic contributions.
  
The sum of the fractions of signal
and background contributions is constrained to unity.
On this basis the signal component is found to be
$f_{\rm sig}=(78\pm4) \%$
leading to an estimated yield of \ndstlnu\ \bzdstlnu\ signal events.

\begin{table}[!tbh]
\caption{\label{tab:D*lnu_bkgd} Sample composition in data 
as determined from fits to the \ctby\ distributions. 
The fractions have been computed without a requiring tagging information.
The dominant errors are systematic except for
$f_{c \overline c}$, which is limited by the statistics.}
  \begin{center}
  \begin{tabular}{|c|c|c|c|}
      \hline
      $f_{\rm comb}$&$f_{off}$ & $f_{B \overline B}$ & $f_{\bchdstxlnu}$\\
      \hline\hline
      $0.139 \pm 0.028$  &$ 0.008 \pm 0.002$  &$0.039 \pm 0.018$  &$0.037 \pm 0.018$ \\
      \hline
    \end{tabular}
  \end{center}
\end{table}

\renewcommand{\secname}{Tagging}
\section{Flavor tagging}
\label{sec:Tagging}

After the daughter tracks of the reconstructed $B$ are removed, the
remaining tracks are analyzed to determine the flavor of the
$B_{\rm tag}$, and this ensemble is assigned a tag flavor, either \Bz\ or
\Bzb. 

We use four different types of flavor tag, or tagging categories, in
this analysis.  The first two tagging categories rely upon the
presence of a prompt lepton, or one or more charged kaons in the
event.  The next two categories exploit a variety of inputs with a
neural-network algorithm.  These tagging categories are hierarchical
and mutually exclusive.

To 
quantify the discriminating power of each tagging
category, we use as a figure of merit the effective tagging efficiency
$Q_i = \epsilon_i \left( 1 - 2\mistag_i \right)^2$, where
$\epsilon_i$ is the fraction of events associated to the tagging
category $i$ and $\mistag_i$ is the mistag fraction, the probability
of incorrectly assigning the 
tag to an event in this
category.  The statistical errors in the measurements of 
\stwob\ and \deltamd\ are
inversely proportional to $\sqrt{\sum_i Q_i}$.

The mistag fractions are measured with the $B_{{\rm flav}}$ data
sample. The results are shown in Section~\ref{sec:Mixing}.  
The performance of the tagging algorithm, $\sqrt{\sum_i Q_i}$, was
optimized and the neural networks were trained with Monte Carlo
simulations only. 
Differences between the tagging inputs in data and in simulation may
make the 
actual tagging algorithm somewhat non--optimal, 
but would not lead to a
bias because the wrong-tag fractions $\mistag_i$ are measured
directly with data, both for the mixing and \CP-violation
measurements.

\subsection{Lepton and kaon tags}
\label{sec:\secname:NOT}

The {\tt Lepton} and {\tt Kaon} tagging categories use the correlation
between the flavor of the decaying $b$ quark and the charge of a primary lepton from a semileptonic decay or
the charge of a kaon from the chain $b\to c\to s$. 

\begin{figure}[tbh]
\begin{center}
\includegraphics[width=\linewidth]{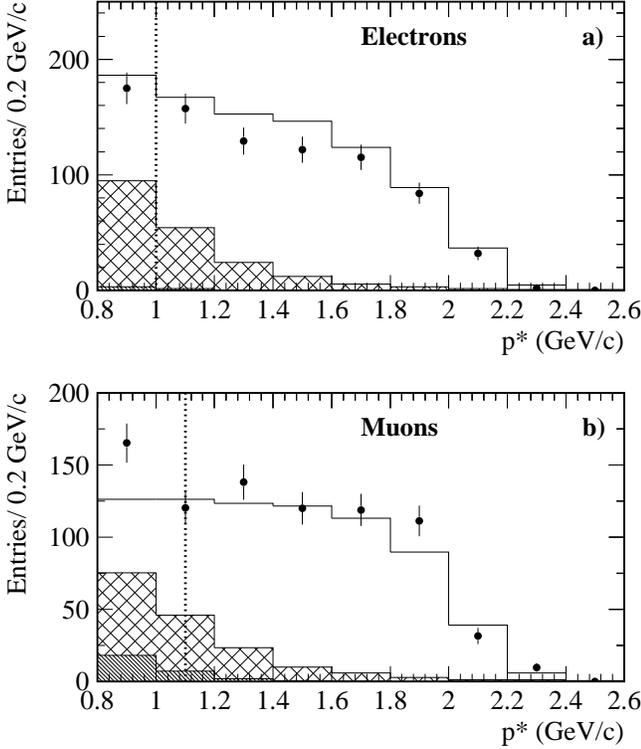}
\end{center}
\caption{Center-of-mass momentum distribution for a) electrons and b) muons. 
Data from the $B_{\rm flav}$ sample, after background
subtraction based on the \mes\ sideband, are
shown as points.  The open histogram shows primary leptons, the cross-hatched
histogram  cascade leptons, and the diagonally-hatched histogram fake leptons,
all from simulation. The simulation is
normalized, with a residual overall systematic error of 5\%,
to the total number of \Bz\ decays in data after
background subtraction, not to the number of observed leptons. 
The vertical lines at 1.0\gevc\ for electrons and 
1.1\gevc\ for muons indicate the requirement on center-of-mass 
momentum for the {\tt Lepton} category.
} \label{fig:tag.pstar}
\end{figure}

For the {\tt Lepton} category we use both electrons and muons, which
are required to pass the 
{\tt VeryTight} and {\tt Tight} selection, respectively (see 
Tables~\ref{tab:epid} and \ref{tab:mupid}).
A minimum requirement of 1.0 (1.1)\gevc\ on the electron (muon)
center-of-mass momentum is applied to reduce
the contamination from softer, opposite-sign leptons coming from cascade
semileptonic decays of charm mesons. The center-of-mass momentum 
spectra
for
electrons and muons are
compared to simulation in Fig.~\ref{fig:tag.pstar} for the $B_{\rm flav}$ sample, after
background subtraction based on the \mes\ sideband events.  
In each event, the electron or
muon with the greatest
center-of-mass momentum is used for flavor tagging;
for the tiny fraction of events with both an electron and muon, the
electron is used due to its smaller misidentification rate.

\begin{figure}[tbh]
\begin{center}
\includegraphics[width=\linewidth]{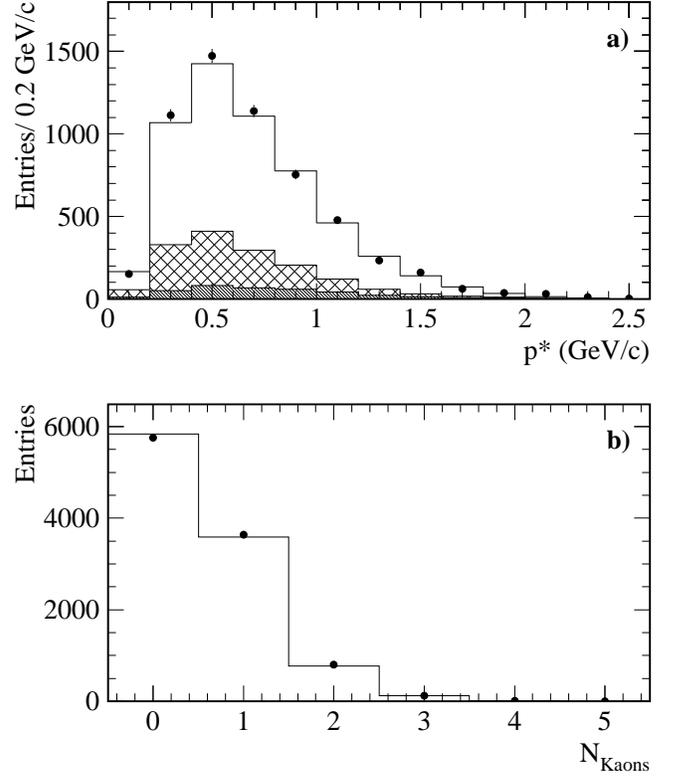}
\end{center}
\caption{a) Center-of-mass momentum distribution for kaons and b) kaon
multiplicity per event.  Data from the $B_{\rm flav}$ sample, after
background subtraction based on the \mes\ sideband, are shown as
points. The histograms are from simulation. In a), the diagonally-hatched
histogram is from fake kaons, 
the cross-hatched histogram is from 
kaons that have the wrong-sign charge, and the open histogram is from
kaons with the right-sign charge, all from simulation.
The simulation is
normalized to the total number of 
\Bz\ flavor candidates after
background subtraction, not to the number of observed kaons.}
\label{fig:tag.kaon}
\end{figure}

The kaon content of the event is evaluated by taking the sum of the
charges of all kaons identified with a neural network
algorithm ({\tt K} subnet described below in 
Section~\ref{sec:NetTagger}). The kaon
identification algorithm has been set to maximize the effective tagging
efficiency $Q$. 
There are 0.8 charged kaons per $B$ decay, and roughly
15\% of these have the wrong sign 
(e.g. $K^-$ from $\Bz$, rather than the expected $K^+$).
Wrong-sign kaons
occur primarily in $B$ decays to a charmed--anti-charmed pair of mesons.
The momentum
distributions are quite similar for right- and wrong-sign 
kaons; we find no kinematic quantity that distinguishes
between them. The center-of-mass momentum spectrum for charged kaons and
the distribution of
charged kaon multiplicity are shown in Fig.~\ref{fig:tag.kaon}
for the $B_{\rm flav}$ sample.

An event with an identified high-momentum lepton is assigned to the 
{\tt Lepton} category unless the sum of the charges of any kaons present has the opposite sign.
Next, events are assigned to the {\tt Kaon} category if the sum of the
kaon charges is non-zero.  
The charge of the lepton or
sum of kaon charges is used to assign the flavor of the $B_{\rm tag}$.
All remaining events, approximately 55\%  of the total including those with inconsistent lepton and kaon
charge (0.5\% of all events in simulation) 
and those with two oppositely-charged kaons (4.6\% of all events in
simulation), are passed to the
neural-network-based categories.

\subsection{Neural-network tags}
\label{sec:NetTagger}

Besides identified high-momentum leptons and charged kaons,
there are other features that can be used to determine the
flavor of the $B_{\rm tag}$, although not as easily or
cleanly distinguishing. 
These include soft pions from $D^*$
decays, high-momentum primary leptons that are not selected by the
electron or muon identification algorithms, lower-momentum primary
leptons, and charged kaons that are not selected by the kaon
identification algorithm.  These sources are combined with
a multivariate method; we use a sequence of neural networks to flavor-tag 
those events not assigned to the {\tt Lepton} or {\tt Kaon}
categories.

Three different track-based neural networks, called ``subnets'', are
trained, each with a specific goal.  The {\tt L}, {\tt K}, and {\tt
SoftPi} subnets are sensitive to the presence of primary leptons,
charged kaons and soft pions from $D^*$ decays, respectively. Each
subnet is applied to all tracks from the $B_{\rm tag}$.

\begin{figure}[tb]
\begin{center}
\includegraphics[width=0.9\linewidth]{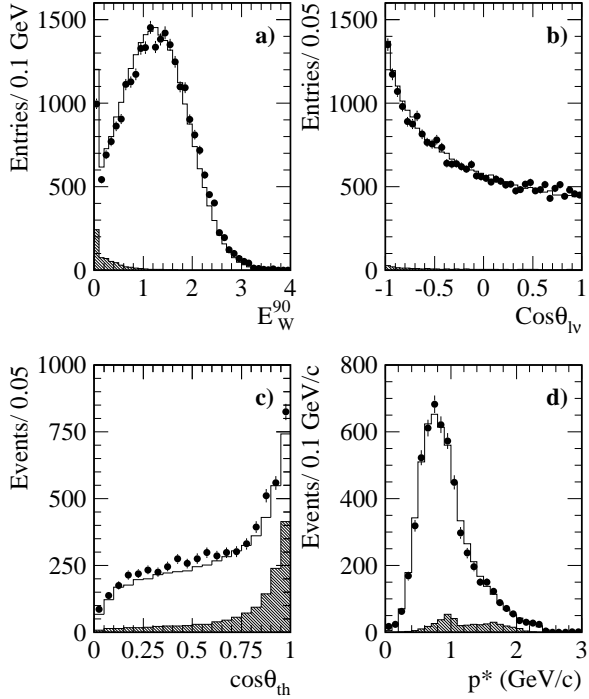}
\end{center}
\caption{
Inputs to the subnets: a) $E^{90}_{W}$, b)
  $\cos{\theta_{l\nu}}$, c) $\cos{\theta_{\rm th}}$ for low
  center-of-mass momentum tracks ($p^* < 0.18 $\gevc), and d) the
  center-of-mass momentum for all tracks.  The points are data from
  the $B_{\rm flav}$ sample after background subtraction based on the
  \mes\ sideband, and the histogram is simulation.  For
  $\cos{\theta_{\rm th}}$ the diagonally-hatched 
  histogram shows the contribution
  from soft-$\pi$ coming from $D^{*}$ decays, and for the other
  distributions shows the component from primary
  leptons.  The simulation is normalized to the total number of
  \Bz\ flavor candidates after the background subtraction. }
\label{fig:tag.netdv}
\end{figure}

The {\tt L} subnet uses the binary output of the electron and muon
identification algorithms on the input track, the center-of-mass
momentum of the input track, and a pair of kinematic variables, $E^{90}_{W}$
and $\cos{\theta_{l\nu}}$,  that
separate primary leptons from cascade leptons and other tracks.  

The isolation variable, $E^{90}_{W}$, is given by the sum of the
energies of all tracks within $90^\circ$ of the $W$ direction. 
The $W$ momentum is inferred as the sum of the input track momentum and the neutrino 
momentum, which we take to be the missing momentum in the
center-of-mass frame using all charged tracks in the $B_{\rm tag}$.
This variable is effective because in a semileptonic decay the
hadrons recoiling against the virtual $W$ would generally go off in
the opposite direction.

\begin{figure}[tb]
\begin{center}
\includegraphics[width=\linewidth]{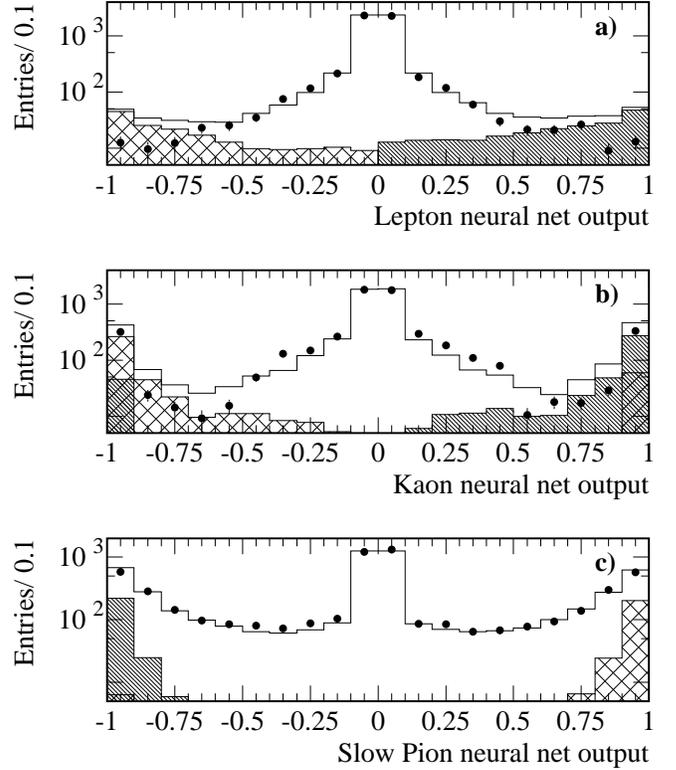}
\end{center}
\caption{
  Output of the subnets for events not assigned to the {\tt Lepton}
  or {\tt Kaon} categories: a) {\tt L} subnet, b) {\tt K} subnet, and
  c) {\tt SoftPi} subnet.  The points are data from the $B_{\rm flav}$
  sample after background subtraction based on the \mes\ sideband, and
  the histogram is simulation. For each distribution, the filled portion
  of the histogram shows the component with a \Bz (singly-hatched) or
  \Bzb\ (cross-hatched) tag from the full neural network algorithm that
  arises from true primary leptons, kaons, or soft-pions respectively.
  Note that the latter has the opposite charge correlation with the \Bz\ tag.
  The simulation is normalized to the total number of \Bz\ flavor candidates
  after background subtraction. }
\label{fig:tag.subnet}
\end{figure}

The other kinematic variable used, $\cos{\theta_{l\nu}}$, is the
cosine of the angle between the input track and the neutrino
direction.  The distributions in the $B_{\rm flav}$ sample and simulation
of $E^{90}_{W}$ and
$\cos{\theta_{l\nu}}$ are shown in Fig.~\ref{fig:tag.netdv}a and b,
for all events not in
the {\tt Lepton} or {\tt Kaon} category.
The corresponding distribution 
of the {\tt L} subnet output is shown in
Fig.~\ref{fig:tag.subnet}a.

The {\tt K} subnet uses the input track momentum in the laboratory
frame, together with the three relative likelihoods ${\cal L}_K/({\cal L}_{\pi} +
{\cal L}_K)$ for the SVT, the DCH and the DIRC. The SVT and DCH likelihoods
are derived from \dedx\ measurements, and the DIRC likelihood is
calculated from a global fit to the number of photons detected, their positions
and arrival times relative to the corresponding track. The distribution 
of the {\tt K} subnet output, again for
events not in the {\tt Lepton} or {\tt Kaon} category, is shown in
Fig.~\ref{fig:tag.subnet}b.

The {\tt SoftPi} subnet uses the center-of-mass momentum of the input
track, the cosine of the angle of the input track with the thrust axis
$\cos{\theta_{\rm th}}$, and the center-of-mass momentum of the
minimum momentum track.  The thrust axis is determined from all
charged tracks and neutral clusters in the $B_{\rm tag}$. The
direction of any $D^*$ in the decay of the $B_{\rm tag}$ is
approximated by the direction of the thrust axis.  Thus soft pions
from $D^*$ decays, which are aligned with the $D^{*}$ direction in the
center-of-mass frame, tend to be correlated with the thrust axis.  The
distribution of $\cos{\theta_{\rm th}}$ is shown for low
center-of-mass momentum tracks in Fig.~\ref{fig:tag.netdv}c,
comparing the $B_{\rm flav}$ sample with simulation
for all events not in
the {\tt Lepton} or {\tt Kaon} category.
The corresponding distribution 
of the {\tt SoftPi} subnet output is shown in
Fig.~\ref{fig:tag.subnet}c.

The outputs of the three subnets are among the inputs to a final neural
network, which is trained to distinguish between \Bz\ and \Bzb.  The
variables used as inputs to the final network include the maximal values of
the {\tt L} and {\tt SoftPi} subnet outputs, each multiplied by the
charge of the corresponding input track, and the highest and the
second-highest values of the {\tt K} subnet output again multiplied by
the charge of the corresponding input tracks. The two other inputs to the
final neural network are  the center-of-mass momentum of the maximum
momentum track multiplied by its charge, and the number of tracks with
significant impact parameters. 
The latter is an indicator of the presence of \KS\ mesons.
The distribution of the center-of-mass momentum for all
tracks is shown in Fig.~\ref{fig:tag.netdv}d.

\begin{figure}[tbhp]
\begin{center}
\includegraphics[width=\linewidth]{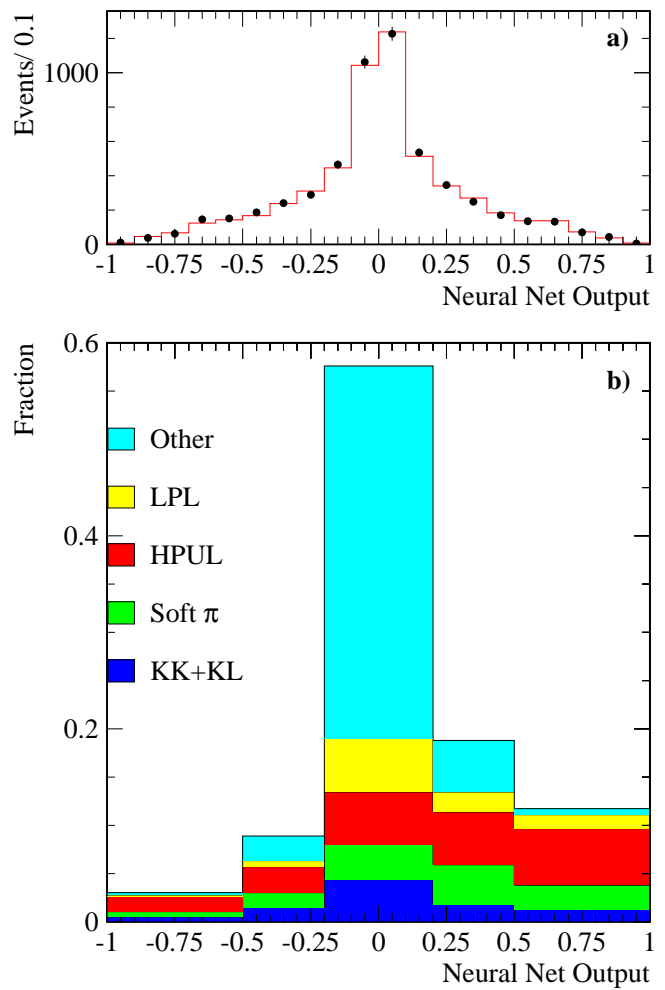}
\end{center}
\caption{a)  Output of the final neural network for $B_{\rm flav}$ events 
  that are not
  assigned to the {\tt Lepton} or {\tt Kaon} category, where the
  points are from the $B_{\rm flav}$ data after a background subtraction and
  the histogram is simulation; b)
  Neural network output from simulation of single \Bz\ decays
  with no time evolution, again for events not 
  in the {\tt Lepton} or
  {\tt Kaon} category. The breakdown
  from bottom to top is events with two kaons or a kaon and lepton (KK+KL),
  events with a soft pion (soft $\pi$),  events with a high momentum 
  unidentified
  lepton (HPUL), events with a lower momentum lepton (LPL), and all 
  remaining events.  
  The outermost bins correspond to the category {\tt NT1} and the next 
  to {\tt NT2}. Entries for $ x_{NT}>0.0$ represent correct tags, while
  those for $x_{NT}<0$ are mistags in each of the categories.
  The center bin contains events for which no tag is assigned.}
\label{fig:tag.net}
\end{figure}

The output from the final neural network, $x_{NT}$, is mapped onto the
interval $\left[ -1, 1 \right]$.  The assigned flavor tag is \Bz\ if
$x_{NT}$ is negative, and \Bzb\ otherwise.  Events with $\left| x_{NT}
\right| > 0.5$ are assigned to the {\tt NT1} tagging category and
events with $0.2 < \left| x_{NT} \right| < 0.5$ to the {\tt NT2}
tagging category.  Events with $\left| x_{NT} \right| < 0.2$, 
approximately 30\% of the total, have very
little tagging power and are rejected.  The distribution of $x_{NT}$
for all events not assigned to the {\tt Lepton} or {\tt Kaon} category
is shown in Fig.~\ref{fig:tag.net}a.

Most of the separation between \Bz\ and \Bzb\ in the {\tt NT1} and {\tt
NT2} tagging categories derives from primary leptons that are not
identified as electrons or muons and from soft pions from $D^*$ decays.
Simulation studies indicate that roughly 37\% of the effective tagging
efficiency $Q$ is due to events with unidentified primary leptons, 
28\% is due to events with a soft pion, a further 11\% from events
with a lower momentum primary lepton, and the remainder from a mixture
of the various inputs.  This classification is shown for the {\tt
  NT1} and {\tt NT2} categories in Fig.~\ref{fig:tag.net}b for a
simulation of \Bz\ decays.
The modest disagreements between Monte Carlo simulation and data
that are evident in the distributions of the subnet output variables shown in 
Fig.~\ref{fig:tag.subnet} lead to a difference
in the predicted value of $Q=(3.0\pm 0.1)\%$ and 
$(1.4\pm 0.1)\%$ for {\tt NT1} and
{\tt NT2} categories in simulation versus
$(2.5\pm 0.4)\%$ and $(1.2\pm 0.3)\%$ as measured in data.

\renewcommand{\secname}{Vertex}
\section{Time difference measurement}
\label{sec:vertexing}

The difference between $B$ decay times, 
$\Delta t = t_{\rm rec} - t_{\rm tag}$,
is determined from the measured separation $\Delta z$ 
between the vertex of the reconstructed $B$ meson ($B_{\rm rec}$) 
and the vertex of the flavor-tagging $B$ meson ($B_{\rm tag}$) 
along the $z$ axis. 
The $\Delta z$ resolution is
dominated by the $z$ position resolution for the
$B_{\rm tag}$ vertex.

\subsection{$\Delta z$  reconstruction}
\label{subsec:dz_reconstruction}

In the reconstruction of the $B_{\rm rec}$ vertex, 
we use all charged daughter tracks. Daughter tracks from
\KS\ and $D$ candidates are first fit to a separate vertex 
and the resulting parent momentum and position are used in the 
fit to the $B_{\rm rec}$ vertex. 
Mass constraints, which include neutral daughters,
are used for $D$ candidates but not for
$D^{*-}$, \jpsi\ and $\psi(2S)$ candidates.
The RMS resolution in $z$ for the $B_{\rm rec}$ vertex in 
Monte Carlo simulation is about 
$65\mum$ for more than 99\% of the $B$ candidates, 
and $40\mum$ for about 80\% of the candidates. 
As described in Section~\ref{sec:vtxchecks}, 
the resolution is about 
5\% worse in data than in Monte Carlo simulation.

\begin{figure*}[!htb]
\begin{center}   
\vskip 0.5cm
 \includegraphics[width=0.7\linewidth]{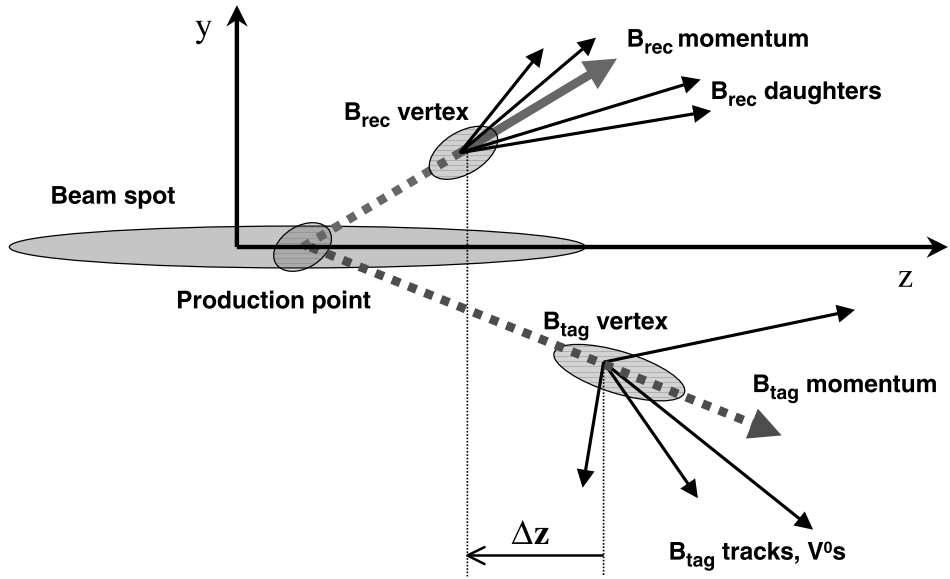} 
\end {center} 
   \caption{Schematic view of the geometry in the $yz$ plane for a $\FourS \to \B\Bbar$ decay.
   For fully reconstructed decay modes, the line of flight
   of the $B_{\rm tag}$ can be estimated from the
   (reverse) momentum vector and the vertex position of 
   $B_{\rm rec}$, and from the beam spot position
   in the $xy$ plane and the \FourS\ average boost. Note that the scale in the $y$
   direction is substantially magnified compared to that in the $z$ direction.
\label{fig:btagvtx}}
\end{figure*}

The vertex for the $B_{\rm tag}$ decay is constructed from all 
tracks in the event except the daughters of
$B_{\rm rec}$. 
For fully reconstructed modes,
an additional constraint is provided by the calculated
$B_{\rm tag}$ production point and three-momentum, with its associated 
error matrix. This is determined from the knowledge of the three momentum
of the fully reconstructed $B_{\rm rec}$ candidate, 
its decay vertex and error matrix, and from the knowledge of 
the average position of the interaction point and the \FourS\ average boost
(see Fig.~\ref{fig:btagvtx}).
These $B_{\rm tag}$ parameters are used as input to a
geometrical fit to a single vertex, 
including all other tracks in the event 
except those used to reconstruct $B_{\rm rec}$. 
In order to reduce bias and tails due to long-lived particles,
\KS\ and $\Lambda^0$ candidates are used as 
input to the fit in place
of their daughters. In addition, tracks consistent with photon conversions 
($\gamma \to e^+ e^-$) are excluded. 
To reduce contributions from charm decay products, 
which bias the determination of the vertex position,
the track with the largest vertex $\chi^2$ contribution greater than 
6 is removed and the fit is redone until no track fails the $\chi^2$ 
requirement.
In Monte Carlo simulation, the RMS of the core and tail Gaussian components of
the residual $\Delta z$ distribution
(measured $\Delta z$ minus true $\Delta z$) is $190\mum$.
We fit this residual distribution to the sum of three Gaussian
distributions and find that the RMS of the narrowest Gaussian,
which contains 70\% of the area, is about $100\mum$.
Only 1\% of the area is in the widest Gaussian.

The absolute scale of the measurement of $\Delta z$ depends on 
the assumed positions of the silicon wafers in the SVT.
These positions are determined from a combination of survey measurements
made before the SVT was installed and measurements of the positions of 
individual SVT modules, each containing several silicon wafers, made
with high-momentum charged particles that traverse the SVT.
We check the absolute scale comparing the known positions of 
distinct mechanical features at each end of the beampipe 
(about 18\cm\ apart)
with the apparent position measured
with charged tracks in the SVT.
The locations of these mechanical features are measured 
from the positions of track vertices at least 2\cm\ from the 
interaction point that contain a well-identified proton,
which are mainly due to $e^\pm$-nucleon interactions in material.
The measured distance in $z$ between these mechanical features is in
agreement with the known distance to a precision of 0.2\%. 
We conservatively enlarge this to 0.4\% to account
for any additional uncertainty in
extrapolating to the interaction point.

\subsection{\deltat\ measurement}
\label{subsec:deltat}

By far the dominant limitation on the accuracy with which 
\deltat\ is determined from the measured decay length
difference $\Delta z$ is
the experimental resolution on the  $\Delta z$ measurement.
The next most significant limitation is
the \B\ meson momentum of about 340\mevc\ in the \FourS\ rest frame.
We partially correct for this effect, as described below.
The impact on the \deltat\ measurement of
the spread in the two beam energies, which results in a distribution of 
\FourS\ momenta with a Gaussian width of about 6\mevc, is negligible.
Finally, we correct for the 20\mrad\ angle between the \FourS\ boost direction
(the $z$ axis in the following discussion)
and the axis of symmetry of the detector, along which we measure
the separation between vertices.

Neglecting the \B\ momentum in the \FourS\ frame,
we can write
\begin{eqnarray}
\Delta z & = & \beta\gamma c \Delta t,
\label{eq:dt_boost_approx}
\end{eqnarray}
where $\beta\gamma$ is the \FourS\ boost factor.
The average value for the boost factor is $\beta\gamma = 0.55$.
The boost factor is calculated directly from the beam energies,
which are monitored every 5 seconds, and has an uncertainty of 0.1\%.

In the case of a fully reconstructed $B_{\rm rec}$, we measure with good 
precision the momentum direction of the reconstructed candidate, which can
be used to correct for the \B\ momentum in the \FourS\ frame. 
However, the correction depends on the sum of the decay times,
$t_{\rm rec} + t_{\rm tag}$, which can only be determined with
very poor resolution.
We use the estimate  
$t_{\rm rec} + t_{\rm tag} = \tau_B + | \Delta t | $
to correct for the measured $B_{\rm rec}$ momentum direction
and extract $\Delta t$ from the following expression:
\begin{equation}
\Delta z = \beta \gamma \gamma^*_{\rm rec} c \Delta t +
               \gamma \beta^*_{\rm rec} \gamma^*_{\rm rec} \cos \theta^*_{\rm rec}
               c (\tau_B + | \Delta t |),
\label{eq:dt_taub_approx}
\end{equation}
where $\theta^*_{\rm rec}$, $\beta^*_{\rm rec}$, and
$\gamma^*_{\rm rec}$
are the polar angle with respect to the beam direction,
the  velocity, and the boost factor  of the $B_{\rm rec}$ 
in the \FourS\ frame.
The difference between $\Delta t$ calculated with Eq.~\ref{eq:dt_boost_approx}
 and Eq.~\ref{eq:dt_taub_approx} is very small because 
$\gamma^*_{\rm rec}=1.002$ and $\beta^*_{\rm rec}=0.064$.
The event-by-event difference in $\Delta t$ calculated with
the two methods has 
an RMS of $0.20$\ps.
Equation~\ref{eq:dt_taub_approx} improves the \deltat\ resolution by about 5\%.
In addition, it removes a correlation between the resolution on 
$\deltat$ and the true value of $\deltat$.
This correlation is due to the fact that the RMS of the second 
term in Eq.~\ref{eq:dt_taub_approx} depends on the expectation 
value of $(t_{\rm rec} + t_{\rm tag})^2$, which in turn depends on 
$|\deltat|$.
Equation~\ref{eq:dt_taub_approx} is used for all \B\ decays to
hadronic final states, while
Eq.~\ref{eq:dt_boost_approx} is used for semileptonic modes since
the $B$ direction cannot be measured for these decays.

\subsection{Vertex quality requirements}

A number of requirements are made in order to ensure a 
well-determined vertex separation.
The fit for both the $B_{\rm rec}$ and 
$B_{\rm tag}$ vertex must converge. 
Also, the error on $\Delta t$ determined from the vertex fit
must be less than 2.4\ps 
and $|\Delta t|$ must be less than 20\ps.
The efficiency for passing these requirements in data and 
Monte Carlo simulation is about 97\% for all $B_{\rm rec}$ modes. 
From  the Monte Carlo simulation, we find that the
reconstruction efficiency does not depend on the true value of 
$\Delta t$. 

The $B_{\rm rec}$ sample is used both to extract the 
\Bz-\Bzb mixing frequency and to measure the mistag probabilities
for the analysis of time-dependent \CP-violating asymmetries.
While the \CP measurement is statistically limited, 
the mixing measurement has a statistical precision of a few percent.
Therefore, in order to reduce systematic uncertainties in the 
mixing measurement,
more restrictive vertex criteria are imposed for 
the $B_{\rm rec}$ sample used for the mixing measurement 
than for the \CP and $B_{\rm rec}$ samples used for the \CP measurement.
However, as described in Section~\ref{sec:algchecks}, the more
restrictive criteria are also used as a cross-check in the
\CP measurement.
In order to reduce further the contributions 
from charm decay products in the mixing analysis, 
a track is not used in the reconstruction of the $B_{\rm tag}$ vertex 
if it is identified as a kaon according to the kaon identification
algorithm used for tagging (see Section~\ref{sec:Tagging:NOT}).
The maximum allowed error on \deltat\ determined from the 
vertex fit is decreased from 2.4\ps\ for the samples used for
the \CP measurement to 
1.4\ps\ for the sample used for the mixing measurement.
The efficiency to  pass these two additional criteria 
is about 87\% in data.
All figures in this section are obtained with the
vertex selection criteria applied in the \CP analysis.

\subsection{\deltat\ resolution function}
\label{sec:vtxresolfunct}

\begin{figure}[!htb]
\begin{center}     
    \includegraphics[width=\linewidth]{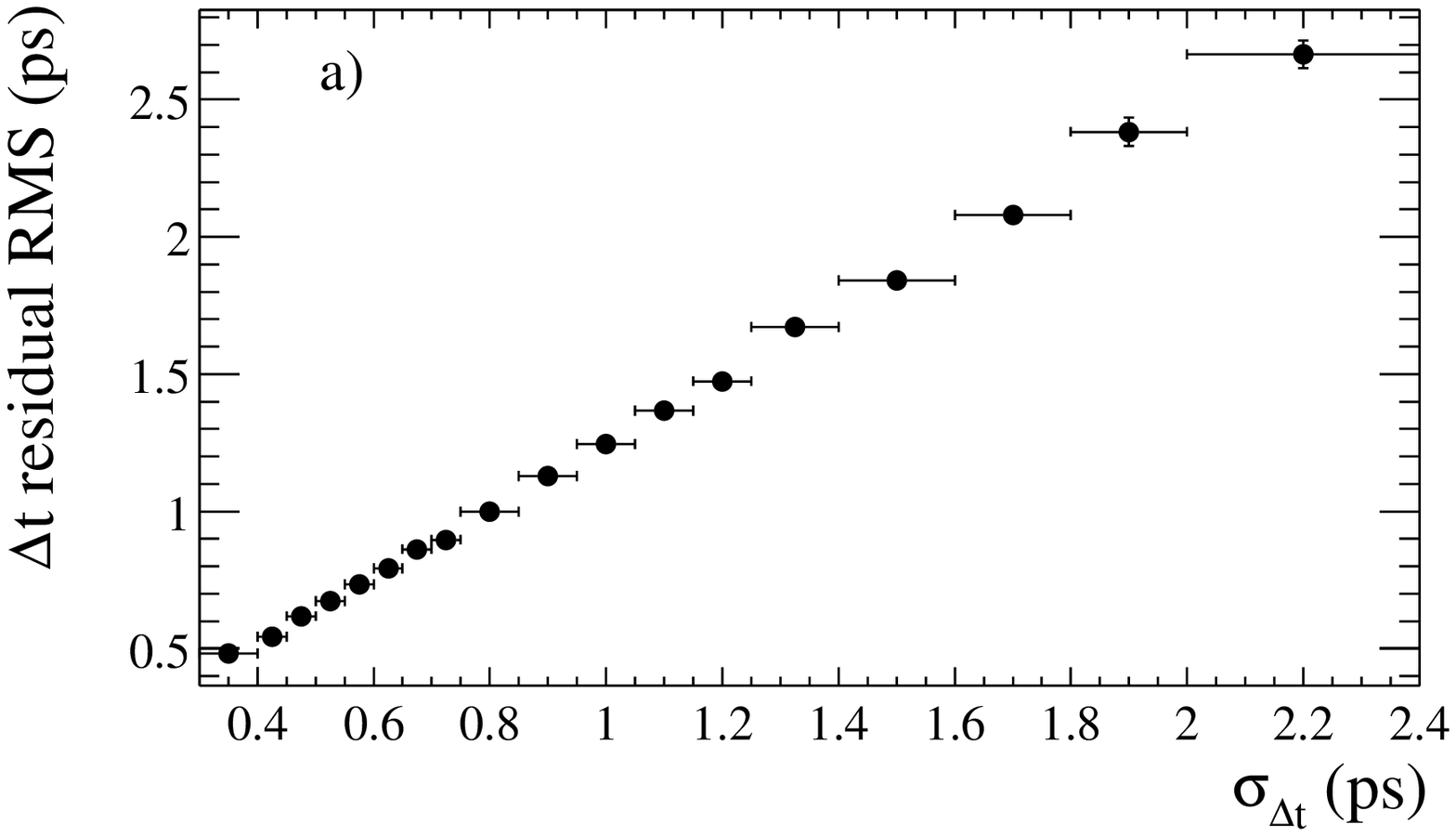}
    \includegraphics[width=\linewidth]{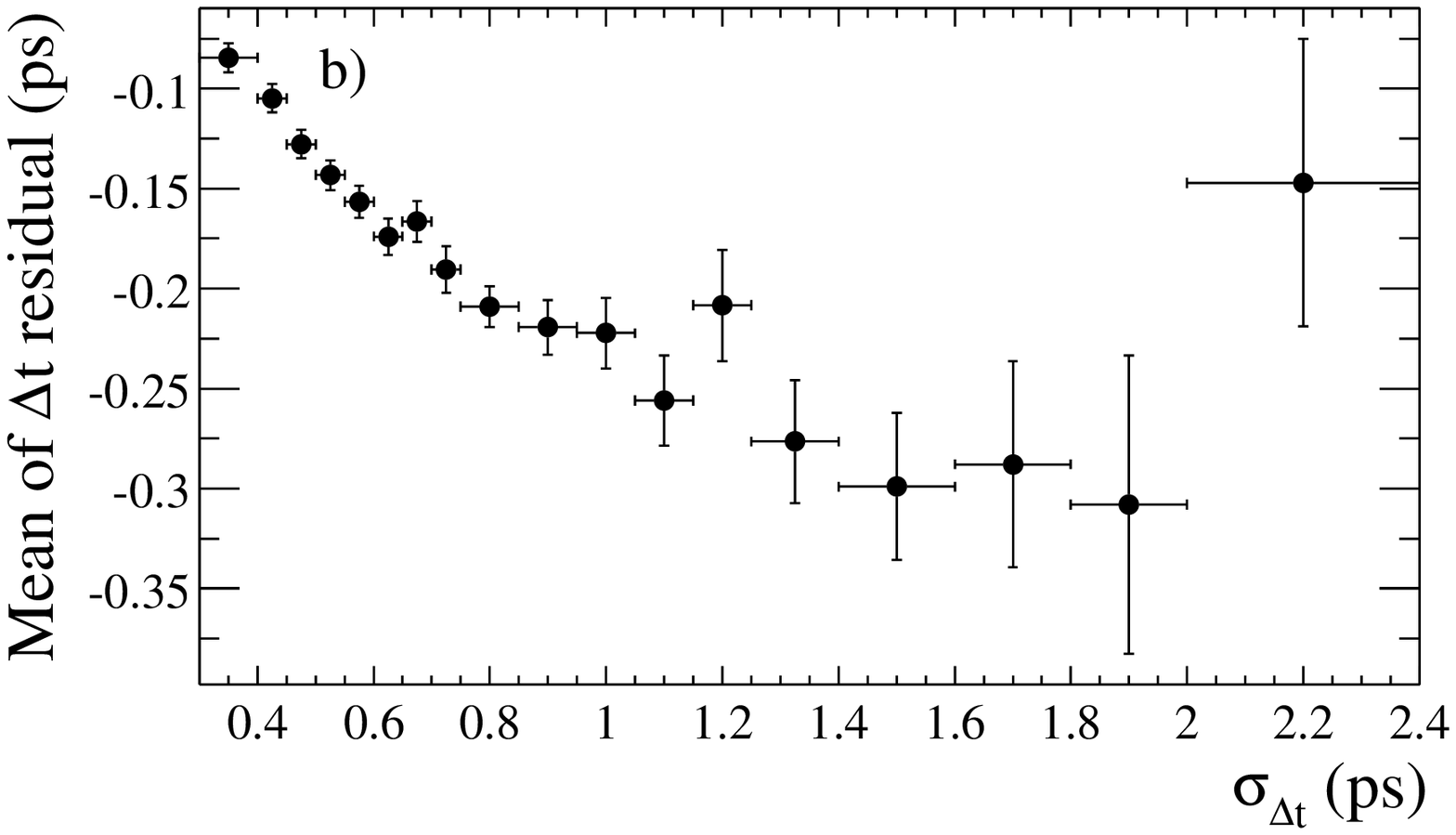}
\end{center} 
\caption{Correlation between the event-by-event error on 
\deltat ($\sigma_{\deltat}$) and a) the observed RMS and b) offset of 
the mean for $\delta_{\rm t}=\deltat - \deltat_{\rm true}$ from Monte Carlo 
simulation.
\label{fig:vtx_correlations}}
\end{figure}

The \deltat\ resolution function is represented in terms
of $\delta_{\rm t} = \deltat-\deltat_{\rm true}$ by a
sum of three Gaussian distributions 
(called the core, tail and outlier components)
with different means and widths:
\begin{widetext}
\begin{eqnarray}
{\cal {R}}( \delta_{\rm t} ; \hat {a} ) &=&  \sum_{k=1}^{2} 
{ \frac{f_k}{S_k\sigma_{\deltat}\sqrt{2\pi}} \, {\rm exp} 
\left(  - \frac{( \delta_{\rm t}-b_k\sigma_{\deltat})^2} 
 {2({S_k\sigma_{\deltat}})^2 }  \right) }  + 
{ \frac{f_3}{\sigma_3\sqrt{2\pi}} \, {\rm exp} 
\left(  - { \delta_{\rm t}^2 \over 
 2{\sigma_3}^2 }  \right) }.
\label{eq:vtxresolfunct}
\end{eqnarray}
\end{widetext}
For the core and tail Gaussians, 
we use the measurement error $\sigma_{\deltat}$ derived from the 
vertex fit for each event but allow two separate scale factors 
$S_1$ and $S_2$ to accommodate an overall 
underestimate ($S_k > 1$) or overestimate ($S_k < 1$)
of the errors for all events.
Figure~\ref{fig:vtx_correlations}a illustrates the correlation
between the RMS of $\delta_t$ and $\sigma_{\deltat}$ in Monte Carlo
simulation.
The core and tail Gaussian distributions are allowed to have a 
nonzero mean to account for residual charm decay products included in
the $B_{\rm tag}$ vertex. 
In the resolution function, 
these mean offsets are scaled by the event-by-event measurement error 
$\sigma_{\deltat}$
to account for an observed correlation shown in 
Fig.~\ref{fig:vtx_correlations}b between the mean of
the $\delta_{\rm t}$ distribution and the measurement error
$\sigma_{\deltat}$ in Monte Carlo simulation.
This correlation is due to the fact that, in $B$ decays, 
the vertex error ellipse for the $D$ decay products is oriented
with its major axis along the $D$ flight direction,
leading to a correlation between the $D$ flight direction and the 
calculated uncertainty on the vertex position in $z$ for the
$B_{\rm tag}$ candidate.
In addition, the flight length of the $D$ in the $z$ direction 
is correlated with its flight direction.
Therefore, the bias in the measured $B_{\rm tag}$ position due to 
inclusion of $D$ decay products is correlated with the $D$ flight 
direction.
Taking into account these two correlations, we conclude that
$D$ mesons that have a flight direction perpendicular to the 
$z$ axis in the laboratory frame will have the best $z$ resolution and 
will introduce the least bias in a measurement of the $z$ position of
the $B_{\rm tag}$ vertex, while
$D$ mesons that travel forward in the laboratory will have poorer 
$z$ resolution
and will introduce a larger bias in the measurement of 
the $B_{\rm tag}$ vertex.

Monte Carlo simulations confirm the expectation that the 
resolution function
is less biased for events with a primary lepton tag than those 
with a kaon tag. 
Therefore, the mean of the core Gaussian is allowed to be different 
for each tagging category.
One common mean is used for the tail component.
The third Gaussian has a fixed width of 
8\ps\ and no offset; it  accounts
for the fewer than 1\% of events 
with incorrectly reconstructed vertices.
The resolution parameters extracted from the full likelihood fits to the
\deltat\ distributions are shown in Table~\ref{tab:mixing-likelihood}
for the mixing analysis and in Table~\ref{tab:summary-sin2beta} for the
\CP analysis.

\begin{figure}[!htb]
\begin{center}     
    \includegraphics[width=\linewidth]{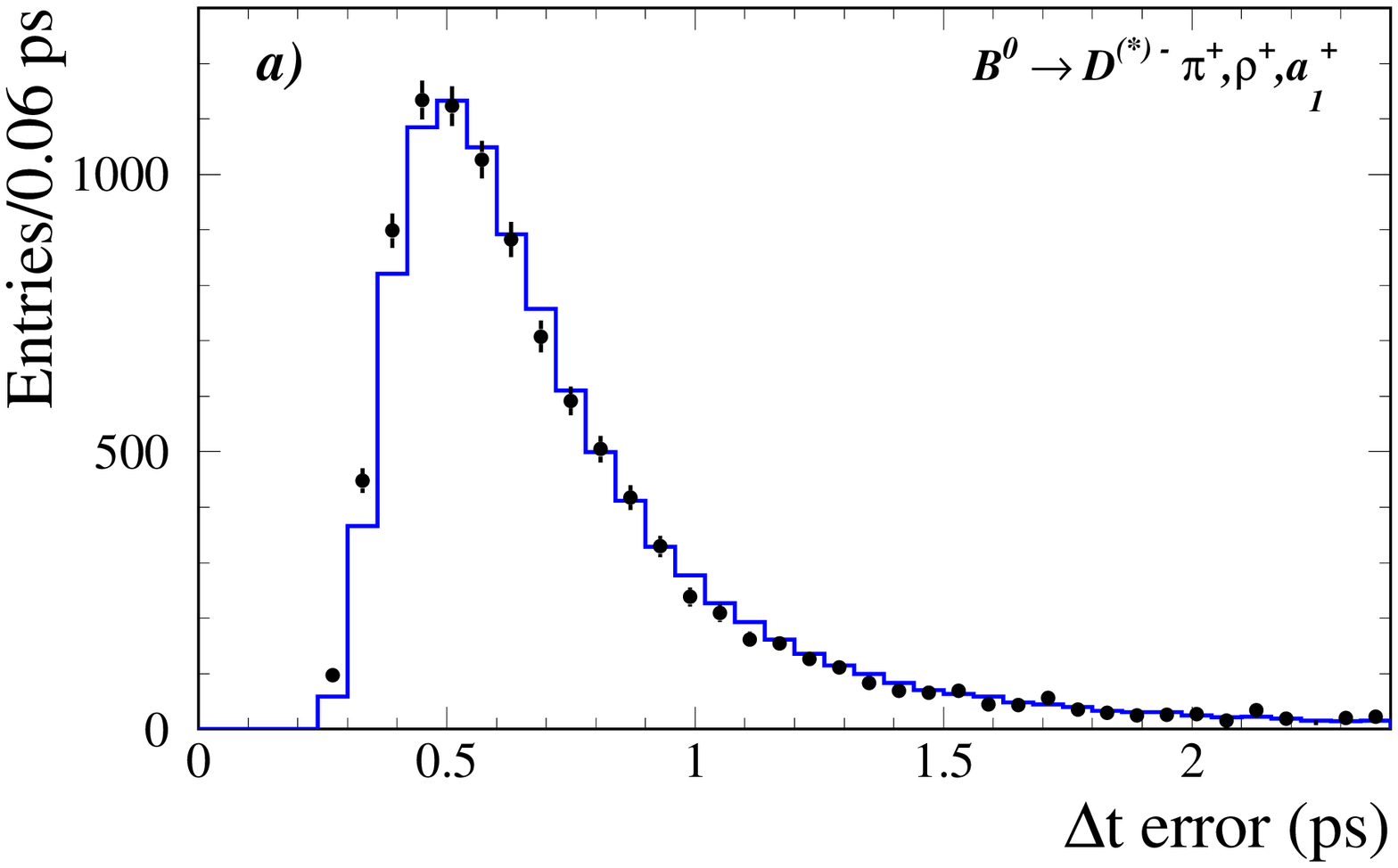}
    \includegraphics[width=\linewidth]{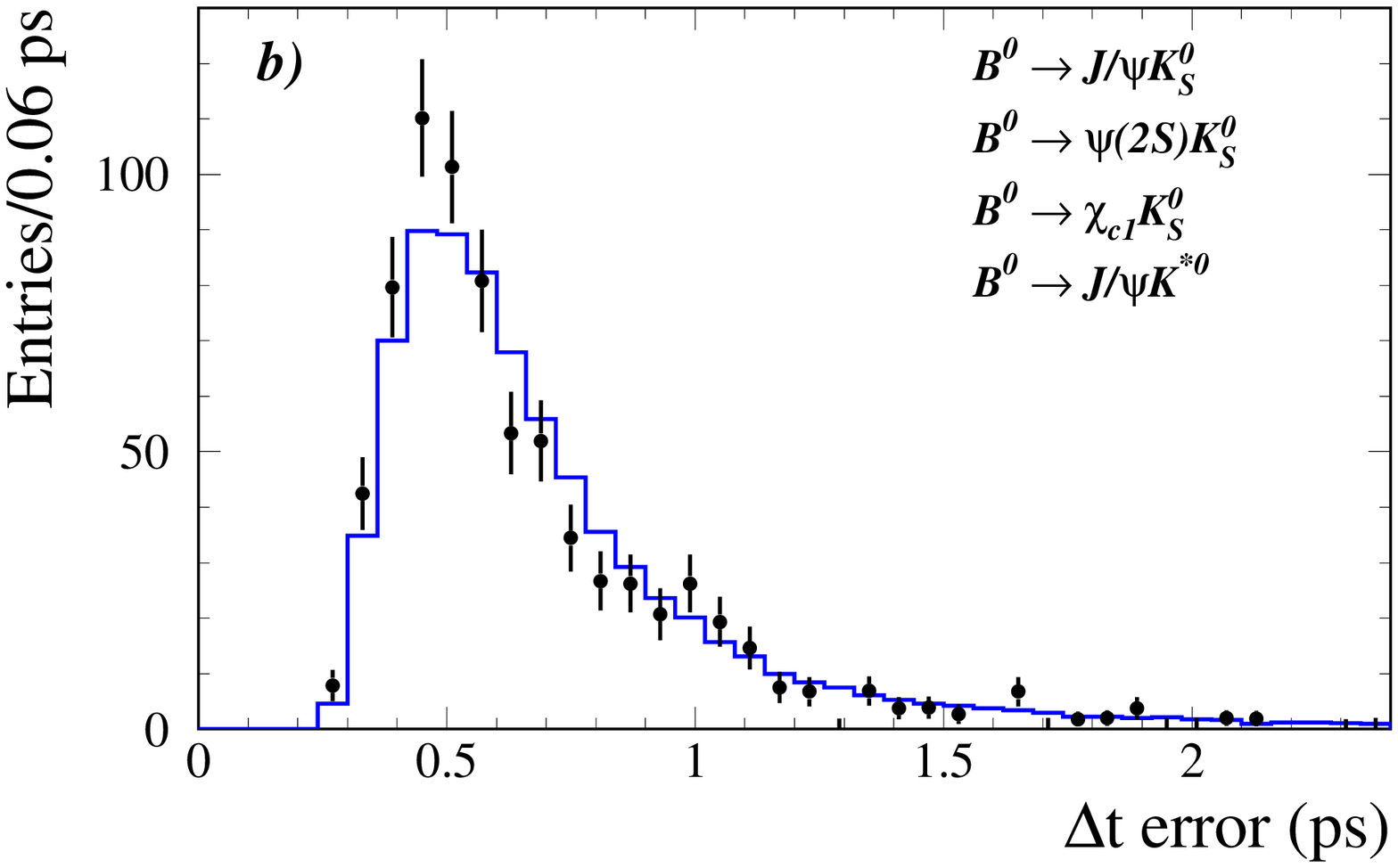}
\end{center} 
   \caption{Distribution of event-by-event uncertainties 
   on $\Delta t$ ($\sigma_{\Delta t}$) for
   a) the sample of neutral $B$ decays to flavor eigenstates 
   other than $\jpsi \Kstarz(\Kp\pim)$
   and b) the combined $\etaCP=-1$ ($\jpsi \KS$, $\psitwos \KS$, 
   $\chic1 \KS$) and $\jpsi \Kstarz$ samples. 
   The histogram corresponds to Monte Carlo  simulation and
   the  points with error bars to data. 
   All distributions have been background-subtracted with events from the
   \mes\ sideband.
   The Monte Carlo distribution has been normalized to the same 
   area as the data distribution. 
\label{fig:dt_error_data_mc}}
\end{figure}

Figure~\ref{fig:dt_error_data_mc} 
shows the distribution of the uncertainties
on \deltat\ calculated from the fit to $\Delta z$
for the flavor-eigenstate sample, and the combined $\etaCP=-1$ 
($\jpsi \KS$, $\psitwos \KS$, $\chic1 \KS$) and $\jpsi \Kstarz$ samples,
and compares data with Monte Carlo predictions.
Since the $B_{\rm tag}$ vertex precision dominates the
\deltat\ resolution, no significant differences between the 
$\Delta t$ resolution function
for the flavor-eigenstate sample and the \CP-eigenstate sample are 
expected. Hence, identical resolution functions are used for all modes.

\subsection{Checks and control samples}
\label{sec:vtxchecks}

Two of the fundamental assumptions in this analysis are 
that the \deltat\ resolution function for the sample of 
flavor-eigenstate modes is the same as that for
\CP\ events, and that the event-by-event vertex errors provide a good 
measure of the relative uncertainty on the $\Delta z$ measurement
from event to event.
In this section, we describe several studies that have been done to 
validate these assumptions.
We compare various distributions for
the \CP\ and flavor-eigenstate samples,
in both data and Monte Carlo simulation.
We take advantage of the small 
vertical size of the beam
to measure the resolution for the
$B_{\rm rec}$ and the $B_{\rm tag}$ vertices in the vertical direction.
We also study the vertex resolution
for a sample of $D^{*+}$ candidates from 
$c\overline c$ events and for continuum events in data and Monte Carlo
simulation.

\subsubsection{Comparison of flavor-eigenstate and \CP\ samples}

\begin{figure}[!htbp]  
\begin{center}
    \includegraphics[width=0.9\linewidth]{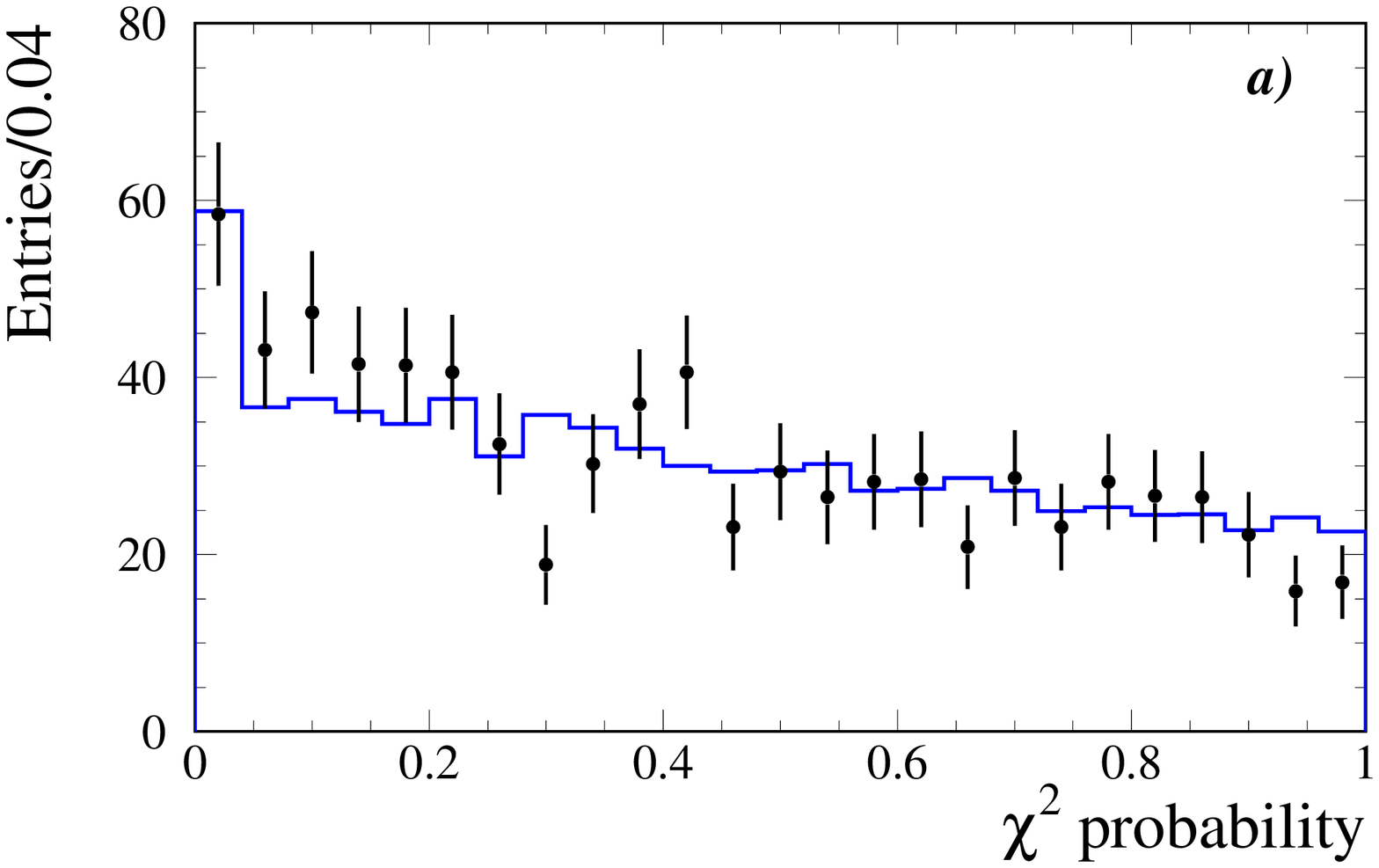}
    \includegraphics[width=0.9\linewidth]{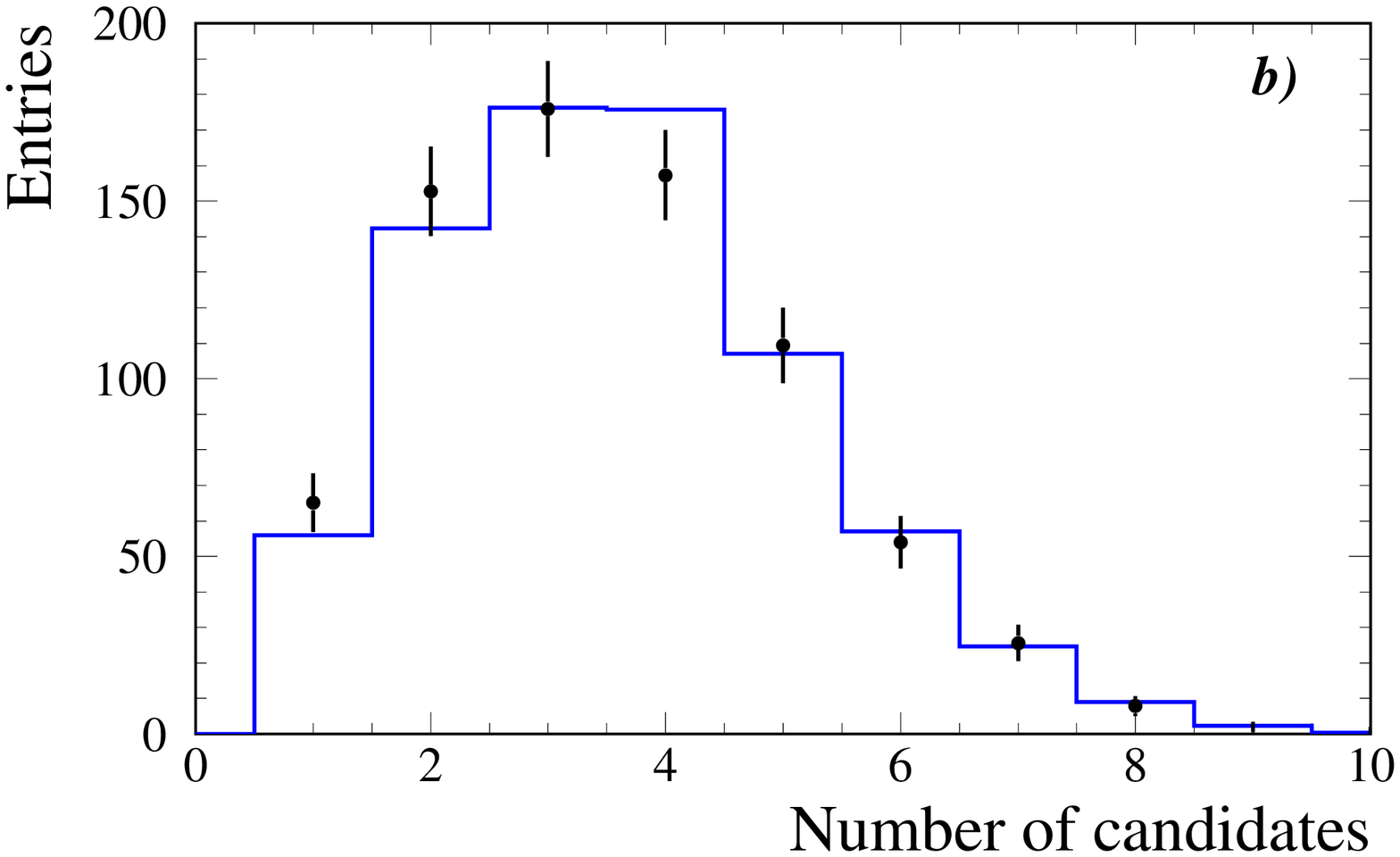}
    \includegraphics[width=0.9\linewidth]{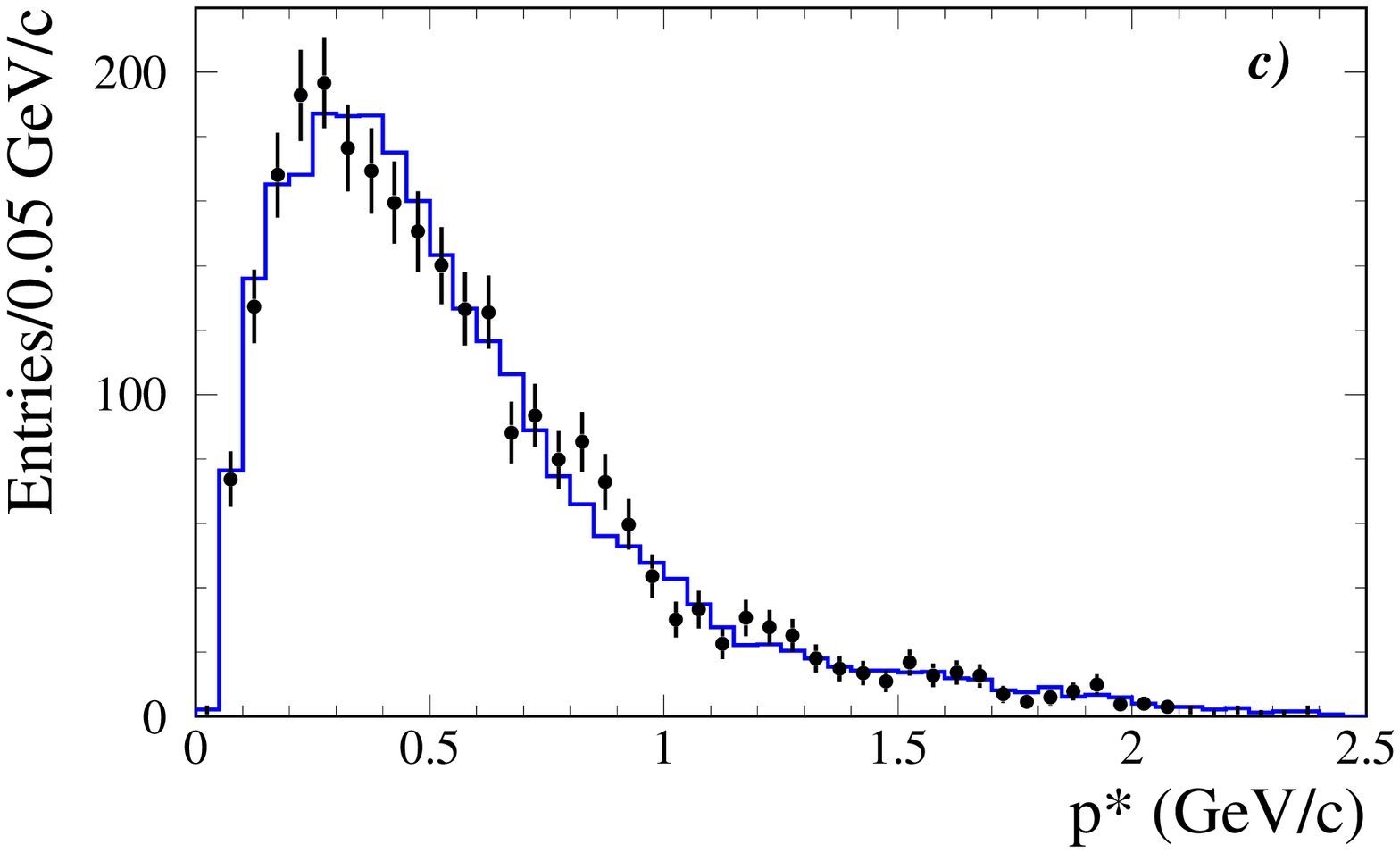}
    \includegraphics[width=0.9\linewidth]{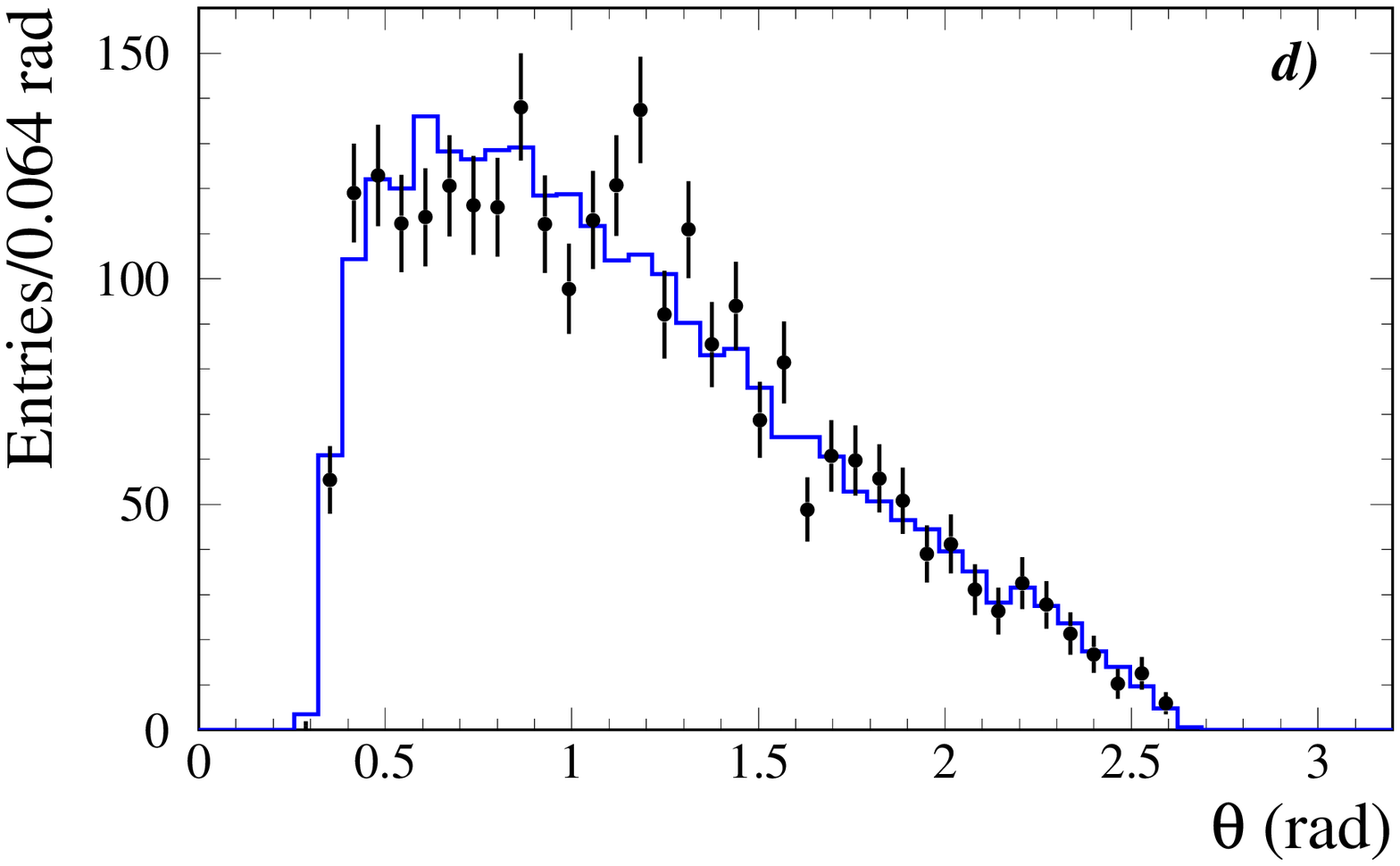}
\end{center}
\caption{ Distributions of
a) $\chi^2$ probability of the $B_{\rm tag}$ vertex fit, 
b) number of charged tracks and $V^0$ candidates used in the
$B_{\rm tag}$ vertex, c) momentum in the center-of-mass frame, and
d) polar angle in the laboratory frame
for tracks in the $B_{\rm tag}$ vertex, for the
flavor-eigenstate (histograms) and the combined $\etaCP=-1$ and $\jpsi \Kstarz$
(points with errors bars) data samples. 
All distributions have been background-subtracted with events from the
\mes\ sideband.
The flavor-eigenstate distributions have been normalized to
the same area as the distributions from
the combined $\etaCP=-1$ and $\jpsi \Kstarz$ samples.
\label{fig:bflav_cp_comparison}}
\end{figure}

In Fig.~\ref{fig:bflav_cp_comparison}, we compare various properties of
the flavor-eigenstate sample with
the combined $\etaCP=-1$ and $\jpsi \Kstarz$ samples.
These include the $\chi^2$ probability for the vertex fits,
the number of tracks used in the $B_{\rm tag}$ vertex,
and the momentum in the \FourS\ rest frame 
and polar angle in the laboratory frame 
of tracks used in the $B_{\rm tag}$ vertex. 
Good agreement in all variables
is observed between the two data samples.

A similar comparison of
the momentum and polar-angle distribution of tracks
in data and Monte Carlo simulation also shows good agreement.
However, there are modest discrepancies for the  
$\chi^2$ probability for the vertex fits
and the number of tracks used in the $B_{\rm tag}$ vertex.
The agreement improves when we include residual
misalignments between the SVT silicon modules 
in the Monte Carlo simulation.
Systematic uncertainties due to residual SVT misalignments in data
are discussed in Sections~\ref{sec:sigpropsyst} 
and \ref{subsec:stwob-systematics}.

As expected, there are no significant differences observed in
comparisons between the
\CP\ modes used in the \stwob\ analysis. However,
comparisons between the \CP\ and flavor-eigenstate samples, in data as
well as in the Monte Carlo simulation, show that the \CP\ events
have a slightly better $\Delta z$ resolution.
For example, in Monte Carlo simulation
the most probable value for $\sigma_{\deltat}$
is about 0.017\ps (3\%) worse for the $B_{\rm flav}$ sample,
as can be seen by comparing the distributions 
in Figures~\ref{fig:dt_error_data_mc}a and b. This is due to
the fact that the $B_{\CP}$ vertex is better determined 
because tracks in the lower-multiplicity \CP\ final states
generally have higher momentum. 
We account for this effect in the likelihood fit 
by using the calculated event-by-event errors, 
as described in Section~\ref{sec:vtxresolfunct}.
Indeed, for Monte Carlo simulation, the pull distributions
for $\sigma_{\deltat}$ (defined as the difference between 
the fitted and generated value divided by the calculated 
error) are nearly Gaussian with unit width
for both the $B_{\CP}$ and $B_{\rm flav}$ samples.
Any residual effect due to differences in the 
observed scale factors in data
is included as a systematic uncertainty 
(see Section~\ref{subsec:stwob-systematics})
and found to be negligible.

\subsubsection{Vertex resolution in vertical direction}

Since the size of the PEP-II beam is only about 10\mum\ in the
vertical ($y$) direction,
the measured distance $\Delta y$ between the $B_{\rm rec}$ or $B_{\rm tag}$
vertex and the nominal beam spot position in the $y$ direction 
can be used to compare the resolution for the \CP\ and flavor-eigenstate
samples, and to evaluate the accuracy of the event-by-event errors
$\sigma(\Delta y)$.
The average beam-spot $y$ position is determined with a precision of
better than a few microns
with two-track events
for each data run (approximately one hour of recorded data).
There is a non-negligible contribution to $\Delta y$ 
of $\approx 25\mum$ (RMS) due to
the \B\ lifetime and the transverse momentum of the \B.

\begin{figure}[!htb]
\begin{center}     
    \includegraphics[width=\linewidth]{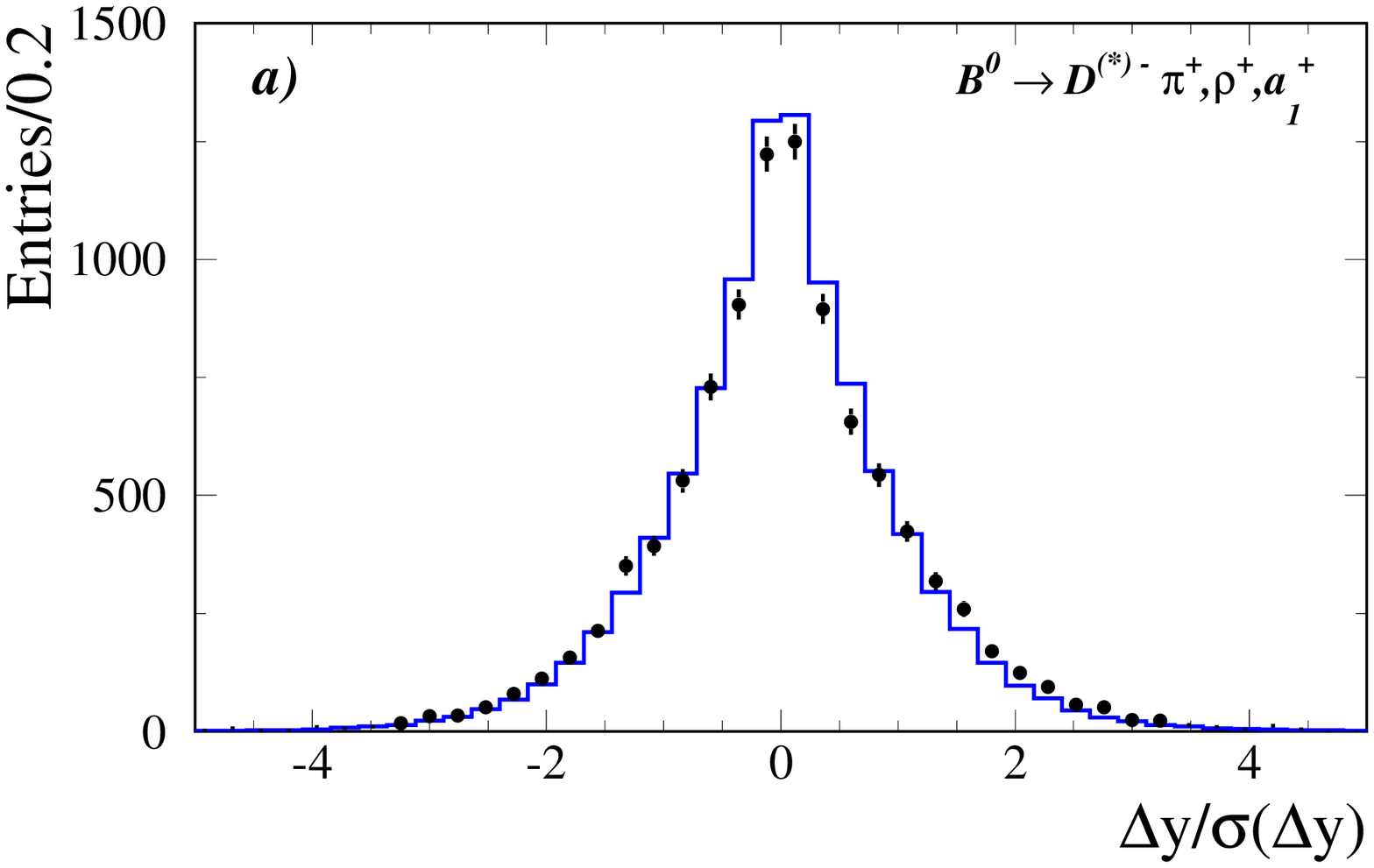}
    \includegraphics[width=\linewidth]{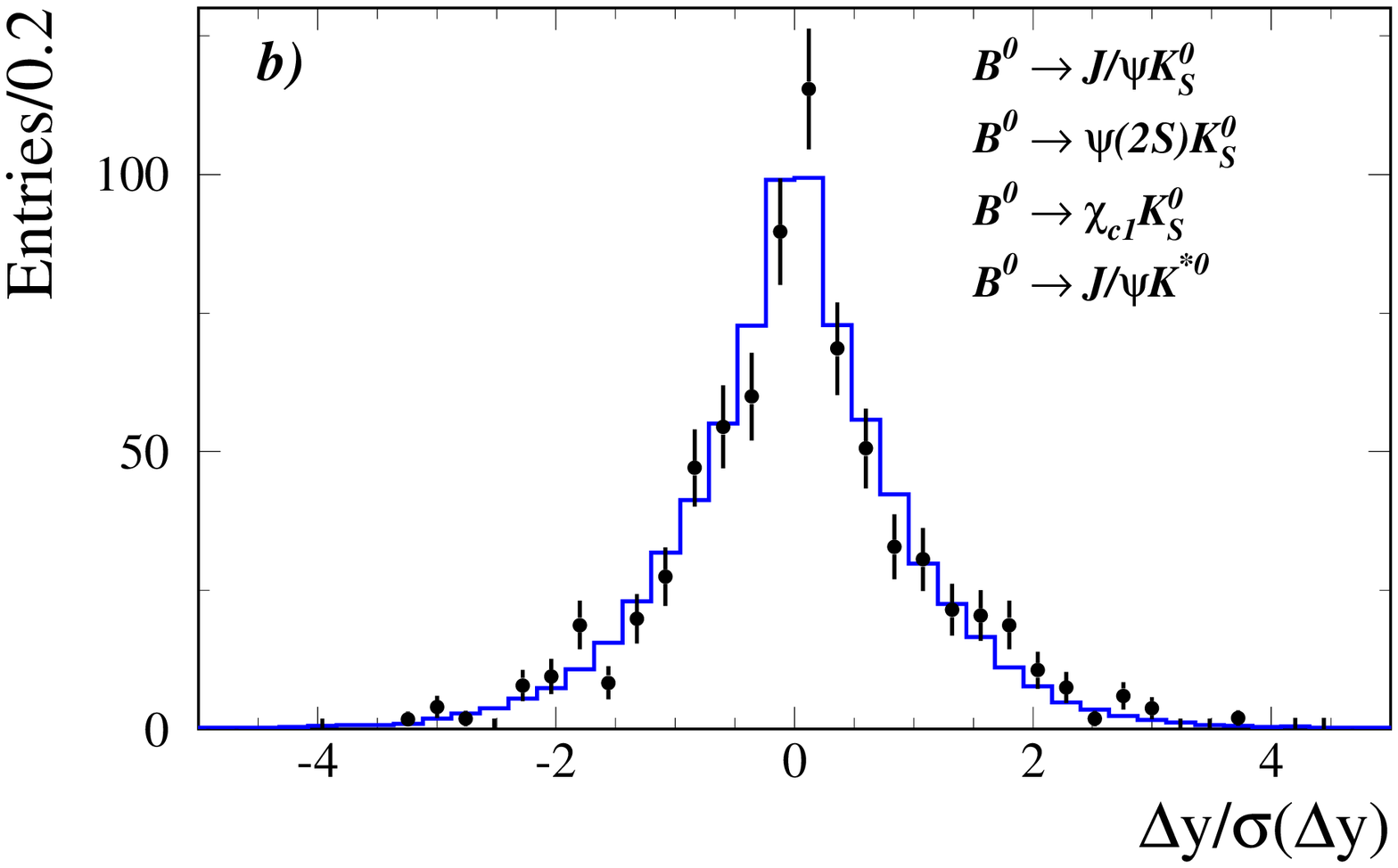}
\end{center} 
   \caption{ Distributions of the measured
distance in the vertical direction $\Delta y$ between the
$B_{\rm tag}$ vertex and the beam spot position, divided by the
event-by-event error on the measured distance $\sigma(\Delta y)$
for each event: 
a) Monte Carlo
simulation (histogram) and data (points with error bars) for the
flavor-eigenstate sample; b) flavor-eigenstate
(histogram), and $\etaCP=-1$ and $\jpsi \Kstarz$ (points with error bars) 
samples in data.
All distributions have been background-subtracted with events from the
\mes\ sideband.
In a), the data distribution has been normalized to the same area 
as the Monte Carlo simulation distribution; in b)
the combined
$\etaCP=-1$ and $\jpsi \Kstarz$ data distribution has been normalized to
the same area as the flavor-eigenstate distribution.
\label{fig:ytagbs_bflav_cp}}
\end{figure}

The distance in $y$ between the
$B_{\rm tag}$ vertex and the beam spot is used to measure the 
$B_{\rm tag}$ vertex resolution and bias in $y$. 
In Fig.~\ref{fig:ytagbs_bflav_cp}, we show the distribution of 
$\Delta y / \sigma(\Delta y)$ for the $B_{\rm tag}$ vertex 
for the flavor-eigenstate and \CP\ samples, 
in data and Monte Carlo simulation.
The RMS of the $\Delta y / \sigma(\Delta y)$ distribution is 
1.3 and 1.4
for Monte Carlo simulation and data, respectively.
No statistically significant biases are observed.

Similar results are obtained for the $B_{\rm rec}$ vertex resolution.
In addition, good agreement in the resolution on the $y$ position 
is observed between the flavor-eigenstate sample and the \CP\ sample.
The resolution is about 5\% worse in data than in Monte Carlo simulation.

\subsubsection{Vertex resolution in continuum events}
\label{sec:vtxrescont}

Two samples have been used to 
cross-check the reliability of the resolution function extracted 
from the likelihood fit as well as the discrepancies between data and 
Monte Carlo simulation:  
a sample of $109,000$ $D^{*+}$ candidates from $c\overline c$ events
and a sample of off-resonance data.

\begin{figure}[!htb]
\begin{center}     
    \includegraphics[width=\linewidth]{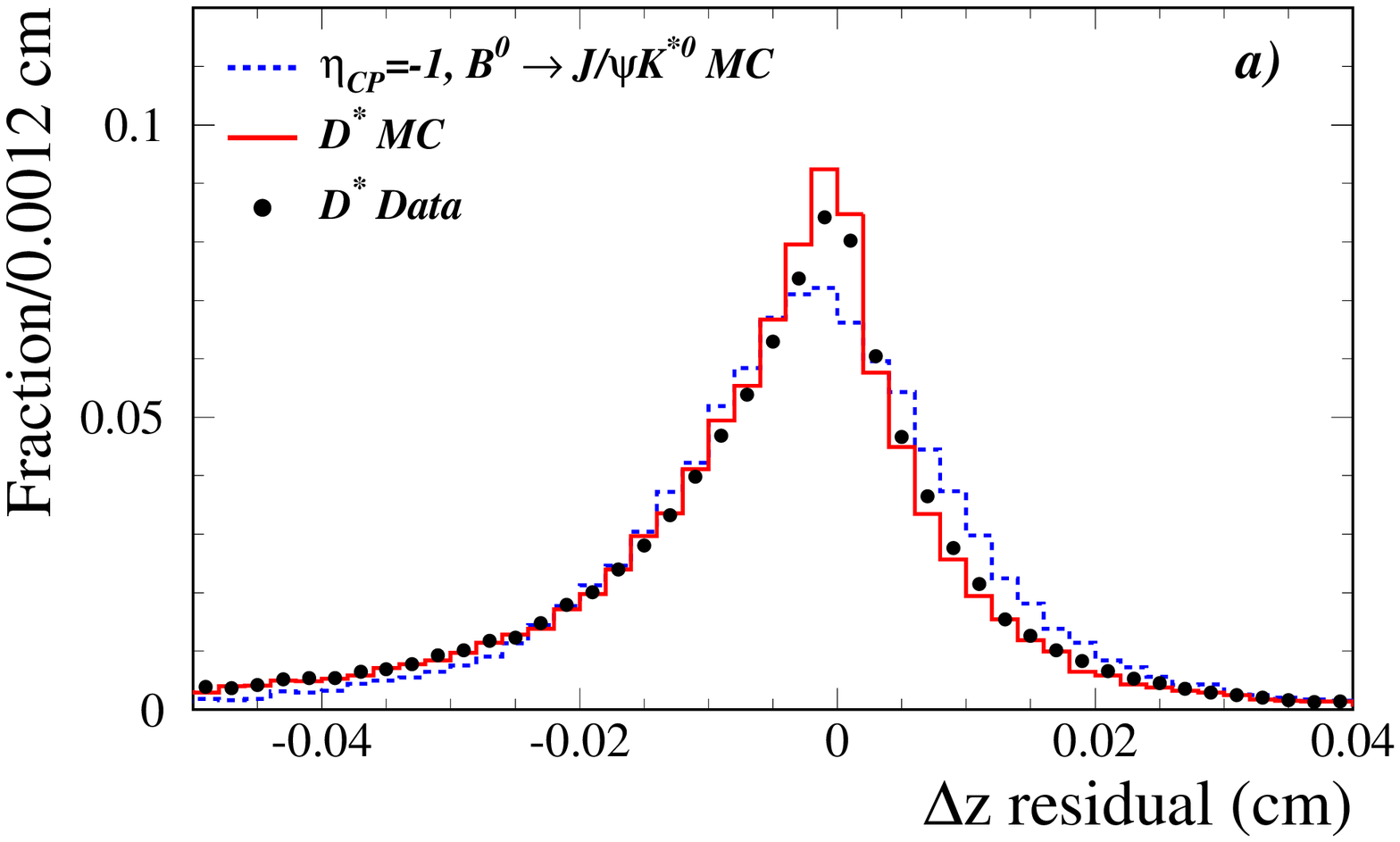}
    \includegraphics[width=\linewidth]{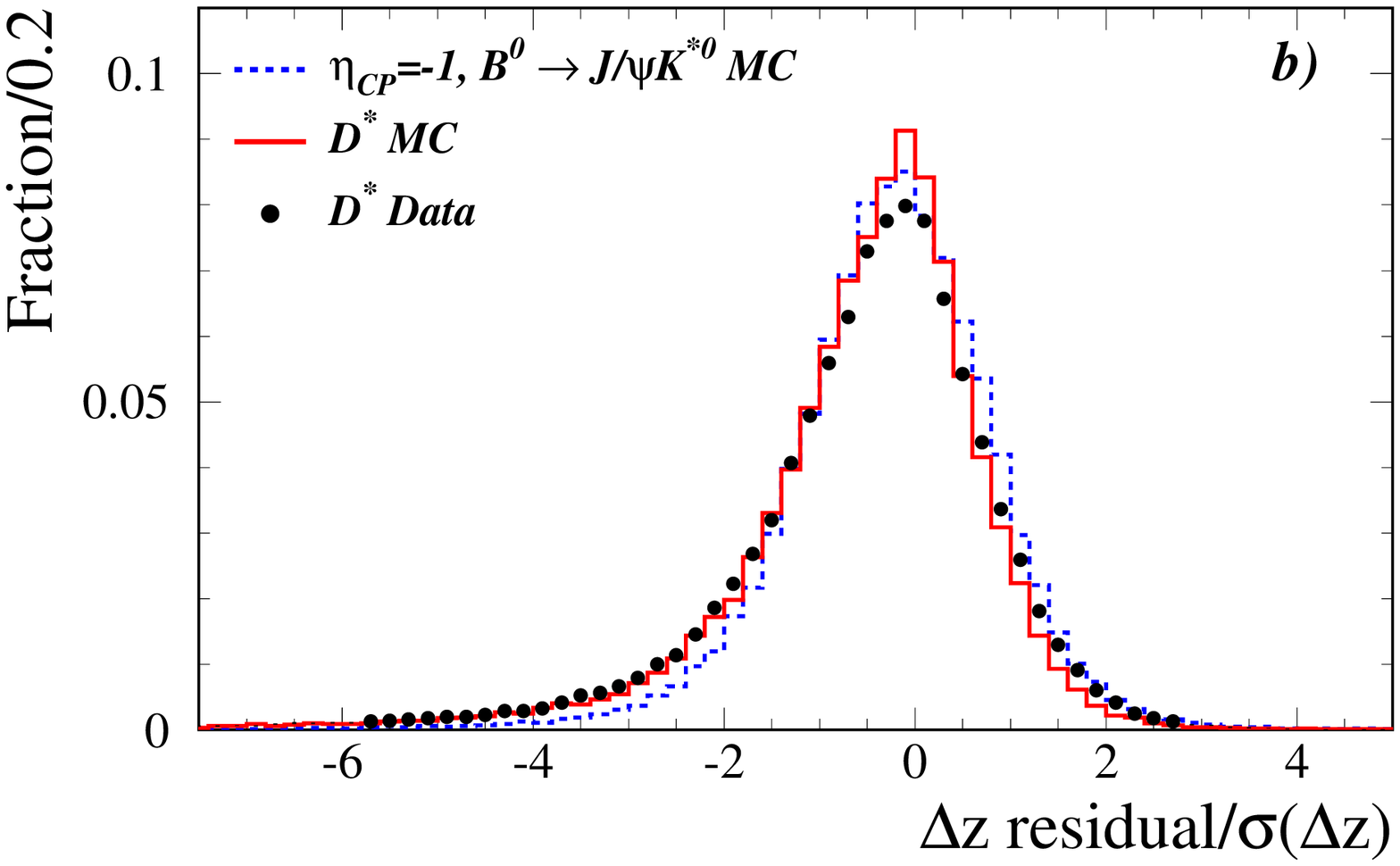}
\end{center} 
   \caption{a) $\Delta z$ and b) $\Delta z / \sigma(\Delta z)$
distributions for the $D^{*+}$ control sample in
data (points with error bars) and Monte Carlo simulation (solid histogram).
For comparison, the difference between the measured 
$\Delta z$ and true $\Delta z$ for $\etaCP=-1$ 
and $\jpsi \Kstarz$ events in Monte Carlo simulation
is also shown (dashed histogram). 
All distributions are normalized to unit area.
\label{fig:dzcontrolsample}}
\end{figure}

For the first sample,
we reconstruct  high-momentum $D^{*+}$ candidates in the mode $D^{*+}
\rightarrow D^0 \pi^+$, followed by $D^0 \rightarrow K^- \pi^+$,  $K^- \pi^+
\pi^0$, or $K^- \pi^+ \pi^- \pi^+$, and then  use the remainder of the charged
tracks in the event 
(fragmentation particles and recoil charm decay products)
to determine a vertex position 
with the standard $B_{\rm tag}$ vertex algorithm.
Since position information for the $D^{*+}$ vertex is
poor, due to scattering of the slow pion, and the $D^{*+}$ decay point 
coincides with the $e^+e^-$ interaction point,
a beam-spot constraint is used for the $D^{*+}$. 
In Fig.~\ref{fig:dzcontrolsample}, we show
the distribution of the distance along the $z$ axis 
between the $D^{*+}$ vertex and
the vertex formed from the rest of the tracks in the event $\Delta z$, as
well as $\Delta z$ divided by the event-by-event error on $\Delta z$, 
for both data and Monte Carlo simulation.

In Monte Carlo simulation, the resolution on $z$ for the $D^{*+}$ candidate
is $\approx 90\mum$, 
very similar to that for $B_{\rm rec}$ vertices.
However, the momentum spectrum of fragmentation 
tracks in $c\overline c$ events is softer than that for tracks from 
\B\ decays, while $D$ mesons are more energetic in the $D^{*+}$ control
sample than in $B$ decays.
Therefore, a slightly more
asymmetric resolution function is expected for the $D^{*+}$ control sample
compared to that for \B\ events, as shown in Fig. \ref{fig:dzcontrolsample}a.
Comparison of distributions of several sensitive
variables (such as the number of tracks 
used in the vertex, and the momentum and polar angle of the tracks) 
shows small differences between $D^{*+}$ and
\B\ events.

The RMS of the distance between the $D^{*+}$ vertex and 
the vertex formed from the rest of the tracks in the events is 
about 220\mum\ in the Monte Carlo simulation. Fitting this
distribution to the sum of three Gaussians, we find a
resolution of about 140\mum\ for 97\% of the events, compared to 
150\mum\ for 99\% of the $\Bz\Bzb$ events. Only small differences
are observed in the distribution of $\Delta z / \sigma(\Delta z)$,
as illustrated in Fig.~\ref{fig:dzcontrolsample}b.
Therefore, the sample can be used to confirm the resolution
and scale factors extracted from the likelihood fit, as well as
to compare how well the Monte Carlo simulation reproduces the data.

The distributions are fit to the sum of
three Gaussian distributions with different widths and means. The
width of the third Gaussian is fixed to 2.0\mm.
From the fit results, we come to the following conclusions:
\begin{itemize}
  \item The event-by-event errors on $\Delta z$
  are underestimated by 
  about 10\% in data
  (Fig.~\ref{fig:dzcontrolsample}b).
  \item The bias in the resolution function due to charm decay products that 
  is observed in data is well reproduced by the  Monte Carlo simulation
  as shown in Fig.~\ref{fig:dzcontrolsample}a and b.
  \item The resolution measured in the data is about 
  5\% worse than that predicted by the Monte Carlo simulation 
  (Fig.~\ref{fig:dzcontrolsample}a).
\end{itemize}
These results will be compared with those obtained from the
likelihood fit to the \B\ events, described in Section~\ref{sec:fitresults}.

The second control sample is obtained
from off-resonance data alone. 
Charged tracks from these continuum events are
randomly split into two sets, and the vertex of each set is 
found with the same algorithm used to determine the 
$B_{\rm tag}$ vertex. 
In this case the $B_{\rm tag}$ vertex reconstruction
strategy is applied to both vertices in the event, 
so that this sample provides
an unbiased estimation of the resolution, suitable for
comparisons between data and Monte Carlo simulation. 
Results from this study are compatible with those reported above.

\subsection{Comparison of 1999-2000 and 2001 performance}
\label{sec:run1vsrun2}

The internal alignment of the SVT has improved
significantly for the reconstruction of the 2001 data set (Run 2) compared
to 1999-2000 (Run 1).
Therefore, we expect better resolution and event-by-event errors on
\deltat\ for Run 2, which requires the use of separate resolution 
functions for the two data sets. 

\begin{figure}[!htb]
\begin{center}    
    \includegraphics[width=\linewidth]{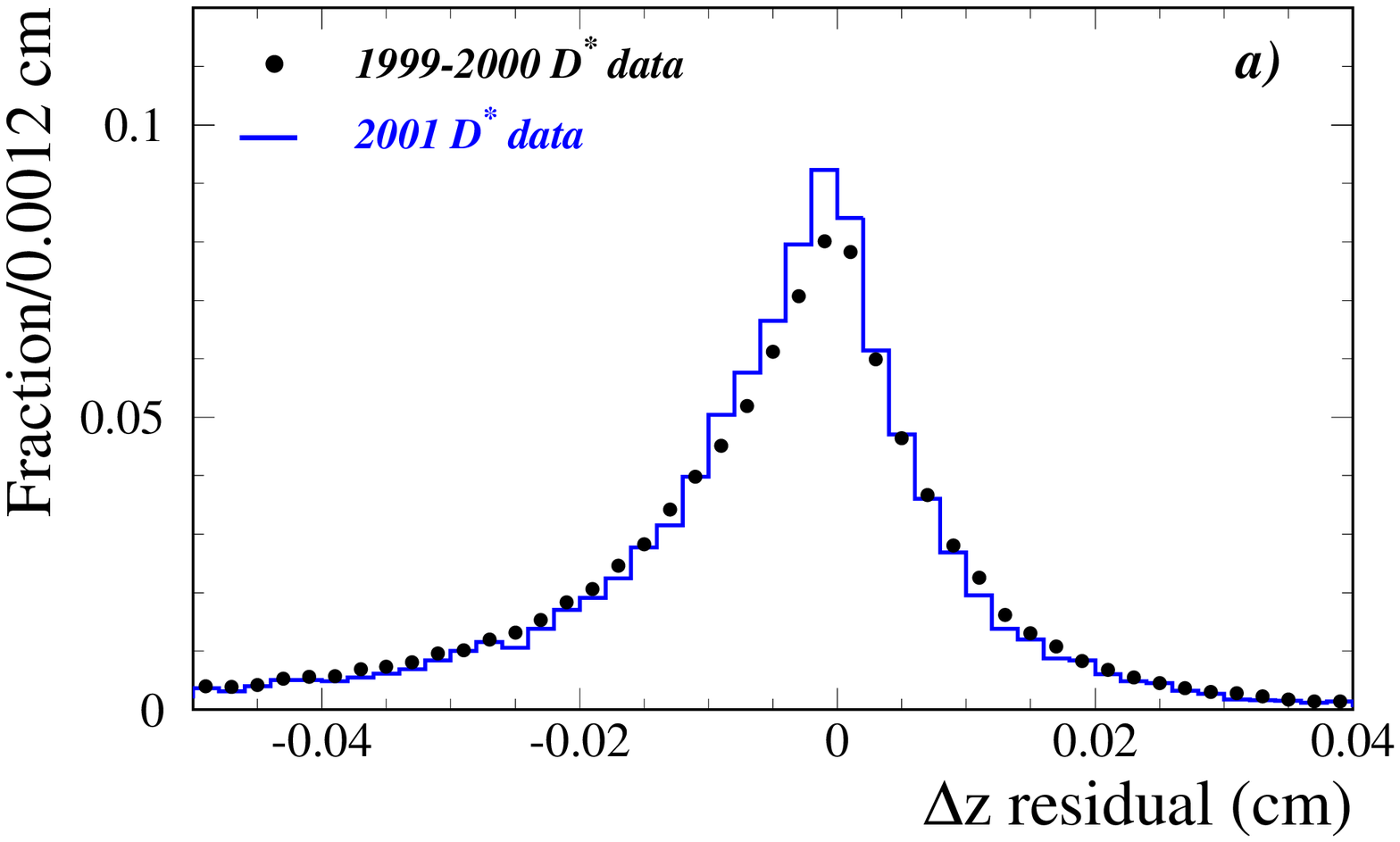}
    \includegraphics[width=\linewidth]{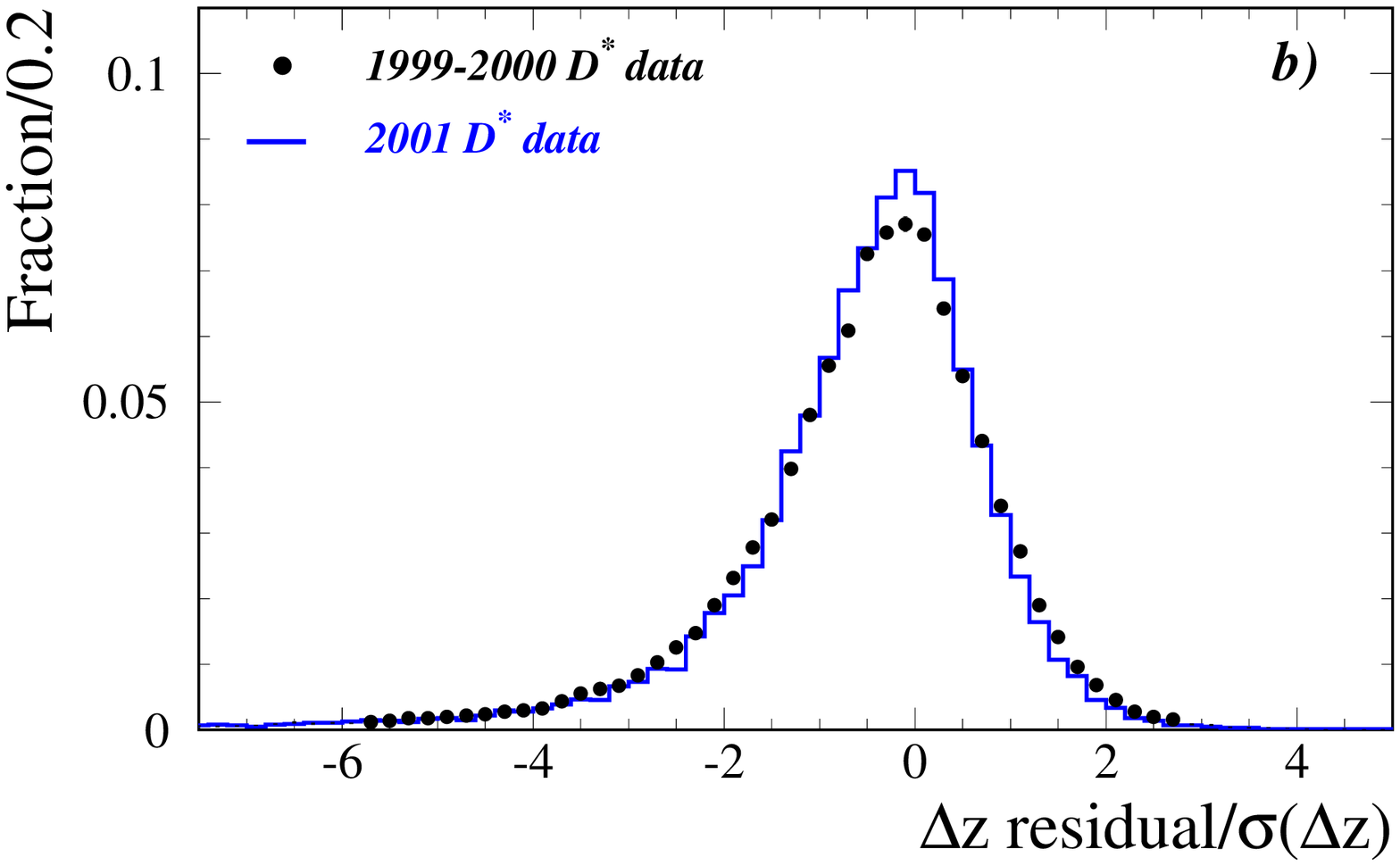}
\end{center} 
   \caption{
Comparison of the distributions of 
a) $\Delta z$ and b) $\Delta z / \sigma(\Delta z)$ for the $D^{*+}$ 
control sample described in Section~\ref{sec:vtxrescont}, for 
Run 1 (points) and Run 2 (histogram) data. All distributions are normalized to unit area.
\label{fig:Dstarrun1vsrun2}}
\end{figure}

The differences in resolution and event-by-event errors for Run 1 and Run 2 
are illustrated in Fig.~\ref{fig:Dstarrun1vsrun2}, where a comparison 
of the distributions for $\Delta z$ and $\Delta z / \sigma(\Delta z)$
in the $D^{*+}$ control sample described in Section~\ref{sec:vtxrescont}
is shown. From the separate analysis of the two data sets, 
we conclude the following:

\begin{itemize}
  \item The event-by-event errors on $\Delta z$
  are underestimated by 15\% (5\%) for the Run 1 (Run 2) data set
  (Fig.~\ref{fig:Dstarrun1vsrun2}b).
  \item There is no statistically significant difference in the bias 
  between the two data sets.
  \item The resolution for the Run 1 data set is about 15\% worse
  than for that for Run 2 (Fig.~\ref{fig:Dstarrun1vsrun2}a).
\end{itemize}

\begin{figure}[!htb]
\begin{center}   
    \includegraphics[width=\linewidth]{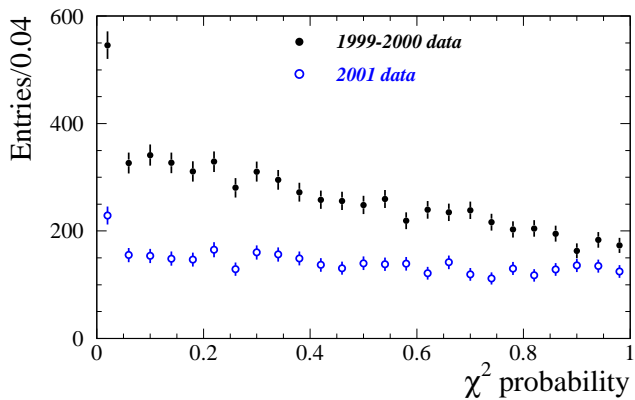}
\end{center} 
   \caption{Comparison of the $\chi^2$ probability distributions of
the $B_{\rm tag}$ vertex fit for the flavor-eigenstate data samples in
Run 1 and Run 2.
The distributions have been background-subtracted with events from the
\mes\ sideband.
The area of each distribution equals the total number of events in
the corresponding sample.
\label{fig:pchi2run1vsrun2}}
\end{figure}

The improved quality of the event-by-event errors in Run 2 is also 
illustrated in Fig.~\ref{fig:pchi2run1vsrun2},
where we compare the distributions of $\chi^2$ probability 
for the $B_{\rm tag}$ vertex fit with the flavor-eigenstate data sample
selected from the two different data periods.

\renewcommand{\secname}{Likelihood}
\section{Likelihood fit method}
\label{sec:Likelihood}

The value of \stwob\ is extracted from the tagged
$B_{\CP}$ sample with an unbinned
maximum-likelihood technique based on  $\ln { {\cal {L} }_{\CP} }$ 
and the probability density functions
${\cal F}_\pm$ of Eq.~\ref{eq:Convol}. However, the
dilutions ${\cal D}_i$ and 
\deltaz\ resolution parameters 
$\hat {a}_i$ are also needed for the measurement. Assuming that mistag 
rates and vertex resolutions do not depend on the particular
channel used to reconstruct the \B\ meson,
these parameters are best determined with the much larger 
mixing sample, since they also appear in  ${\cal {L}}_{mix}$. In order
to properly incorporate the correlations between these parameters and \stwob, 
the fit is performed by simultaneously maximizing the sum
\begin{equation}
 \ln { {\cal {L} }_{\CP} }+ \ln { {\cal {L} }_{\rm mix} }
\end{equation}
for the combined tagged 
$B_{\rm flav}$ and $B_{\CP}$ samples. The values of 
\Bz\ lifetime and \deltamd\ are kept fixed
in extracting \stwob.

The value of \deltamd\ is obtained with an unbinned
maximum-likelihood fit to the tagged 
$B_{\rm flav}$ sample alone, where the log-likelihood $\ln {\cal L}_{\rm mix}$
is maximized while keeping the \Bz\ lifetime fixed.

\subsection{Mistag asymmetries}

The probabilities of mistagging a \Bz or \Bzb meson are expected to be
very nearly, but not exactly, equal. For example, the response of the detector to
positive pions and kaons differs from its response to negative pions and 
kaons due to differences in total and charge-exchange cross sections.
To account for any possible mistag differences, we introduce separate 
mistag probabilities \mistag\ for \Bz\ and $\overline\mistag$ for \Bzb\ with the 
conventions
\begin{equation}
\label{Eq:deltaOmega}
\begin{array}{rcrrcr}
\langle \mistag\rangle &=&{\displaystyle{\frac 12}}(\mistag + {\overline \mistag});&\qquad
\Delta \mistag &=&(\mistag-{\overline \mistag})\nonumber \\
{\cal D}&=&1-2\mistag;&\qquad
{\overline {\cal D}}&=&1-2{\overline \mistag} \\
\langle {\cal D}\rangle &=&{\displaystyle{\frac 12}}({\cal D}+{\overline {\cal D}});&\qquad
\Delta {\cal D} &=&({\cal D}-{\overline {\cal D}})\nonumber \\
\end{array}  
\end{equation}
The time distributions for the mixing and \CP samples will thus depend on 
whether the tag was identified as a \Bz\ or a \Bzb, resulting in modifications to the
expressions for mixing time development (Eq.~\ref{eq:mixing_rate})
\begin{eqnarray}
h_{\pm,tag=\Bz} &\propto&[(1+{\displaystyle{\frac 12}}\Delta {\cal D})
                 \pm \langle {\cal D}\rangle \cos\deltamd\deltat]\nonumber\\
h_{\pm,tag=\Bzb}&\propto&[(1-{\displaystyle{\frac 12}}\Delta {\cal D})
                 \pm \langle {\cal D}\rangle \cos\deltamd\deltat],
\end{eqnarray}
where the $\pm$ in the index refers to mixed $(-)$ and unmixed $(+)$ 
events as before, 
and for \CP\ violation time development (Eq.~\ref{eq:direct})
\begin{equation}
f_{\pm}          \propto [(1\pm{\displaystyle{\frac 12}}\Delta {\cal D})
                 \mp\langle {\cal D}\rangle \eta_{\CP}\stwob \sin\deltamd\deltat],
\end{equation}
where the $\pm$ in the index refers to events where $B_{\rm tag}$ is
a \Bz\ $(+)$ and \Bzb\ $(-)$ and we have taken $|\lambda|=1$.

\subsection{Background modeling}

In the presence of backgrounds, the probability distribution functions
$\cal {H}_{\pm}$ of Eq.~\ref{eq:pdf} and ${\cal F}_\pm$ of Eq.~\ref{eq:Convol}
must be extended to include a
term for each significant background source.
The backgrounds for the flavor eigenstates and $\etaCP=-1$ 
modes are quite small and are mostly
combinatoric in nature. However, for the
$\Bz\to\jpsi\KL$ and $\Bz\to\jpsi\Kstarz$ channels the backgrounds
are substantial and originate mainly from other $B\to\jpsi X$ modes
that have, to a very good approximation, the same flavor tagging 
and $\Delta t$ resolution properties as
the signal.
The background properties of the flavor eigenstates, $\etaCP=-1$ modes, and the non-$J/\psi$ background in
the $\Bz\to\jpsi\KL$ channel are determined empirically from sideband events in the data.

\subsubsection{Background formulation for flavor eigenstates and {\boldmath $\etaCP=-1$} modes }
\label{cpm1background}
The background parameterizations are allowed to differ for each 
tagging category. Each event belongs to a particular tagging category $i$.
In addition, the event is classified as
either mixed ($-$) or unmixed ($+$)
for a flavor-eigenstate
or by whether $B_{\rm tag}$ was a $\Bz$ ($+$) or a $\Bzb$ ($-$)
for a \CP-eigenstate.
Thus background distributions $j$ must be specified for 
each possibility $(+/-,i)$, so that the full likelihood function becomes
\begin{widetext}
\begin{equation}
{{\cal H}_{\pm,i}} =
 f^{\rm flav}_{i, {\rm sig}}{\cal H}_\pm(\deltat;\Gamma,\deltamd,\mistag_i,\hat a_i) +
f^{\rm flav}_{i,{\rm peak}} {\cal{B}}^{\rm flav}_{\pm,i,{\rm peak}}(\deltat;\hat a_i) +
\sum_{j={\rm bkgd}} f^{\rm flav}_{i,j} {\cal{B}}^{\rm flav}_{\pm,i,j}(\deltat;\hat b_i)
\label{eq:H+-Bkg}
\end{equation}
for flavor-eigenstates, and
\begin{equation}
{{\cal F}_{\pm,i}} =
 f^{\CP}_{i, {\rm sig}}{\cal F}_\pm(\deltat;\Gamma,\deltamd,\mistag_i,\stwob, \hat a_i) + 
f^{\CP}_{i, {\rm peak}} {\cal{B}}^{\CP}_{\pm,i,{\rm peak}}(\deltat;\hat a_i) +
\sum_{j={\rm bkgd}} f^{\CP}_{i,j} {\cal{B}}^{\CP}_{\pm,i,j}(\deltat;\hat b_i)
\label{eq:F+-Bkg}
\end{equation}
for \CP-eigenstates.
\end{widetext}
The fraction of background events for each source and tagging category is a function
of $\mes$ and is given by  
$f_{i,j}$.  The peaking and combinatorial background PDFs, 
${\cal{B}}_{\pm,i,{\rm peak}}$ and ${\cal{B}}_{\pm,i,j}$, provide an empirical 
description of the $\Delta t$ distribution of the background events in the sample, including
a resolution function parameterized by $\hat a_i$ and $\hat b_i$, respectively. 
These distributions are normalized such that, for each $i$ and $j$,
\begin{equation}
\int_{-\infty}^\infty ({\cal {B}}_{+,i,j} + {\cal {B}}_{-,i,j})\,d\Delta t =1.
\end{equation}

The probability that a \Bz\ candidate is a signal or a background
event is determined from a separate fit to the observed \mes\ distributions of $B_{\rm flav}$ or $B_{\CP}$ candidates with $\etaCP=-1$.
We describe the \mes\ shape with a single Gaussian distribution ${\cal S}(\mes)$
for the signal and an ARGUS parameterization ${\cal A}(\mes)$ for
the background (Eq.~\ref{ARGUS_bkd}). Based on this fit, the event-by-event signal
and background probabilities that appear 
as the relative weights for the various signal and background terms in
Eq.~\ref{eq:H+-Bkg} and \ref{eq:F+-Bkg} are given by  
\begin{eqnarray}
f_{i, {\rm sig}}(\mes) &=&  \frac{(1-\delta_{\rm peak}){\cal S}(\mes)}{{\cal S}(\mes)+{\cal A}(\mes)}\nonumber \\
f_{i, {\rm peak}}(\mes) &=& \frac{\delta_{\rm peak} {\cal S}(\mes)}{{\cal S}(\mes)+{\cal A}(\mes)} \nonumber \\
\sum_{j={\rm bkgd}} f_{i,j}(\mes) &=& \frac{{\cal A}(\mes)}{{\cal S}(\mes)+{\cal A}(\mes)}
\label{eq:SigProb}
\end{eqnarray}
The fraction $\delta_{\rm peak}$ of the signal Gaussian distribution that is due to peaking
backgrounds is determined from
Monte Carlo simulation. 

Backgrounds arise from many different sources. Rather than attempting
to determine the various physics contributions we 
use an empirical description in the likelihood fit, allowing for background
components with various time dependencies. For the $B_{\rm flav}$ sample, 
the background time distributions considered, each with its own effective dilution 
factor ${{\cal D}_i}$ and either a common
resolution function ${\cal R}(\deltat; \hat b_i )$ or the
signal resolution function ${\cal R}(\delta_{\rm t}=\deltat-\deltat_{\rm true}; \hat a_i )$, are
\begin{eqnarray}
{\cal B}_{\pm ,i, 1}^{\rm flav}&=&\frac{1}{2}\,(1 \pm {\cal D}_{i,1}^{\rm flav} )\ \delta(\deltat_{\rm true})
\otimes {\cal {R}}( \delta_{\rm t} ; \hat b_i ), \nonumber \\
{\cal B}_{\pm ,i, 2}^{\rm flav}&=& \frac{1}{4}\,\Gamma_{i,2}^{\rm flav}(1 \pm {\cal D}_{i,2}^{\rm flav} )\times\nonumber \\
&& { \rm e}^{-\Gamma_{i,2}^{\rm flav} | \deltat_{\rm true}|} \otimes
 {\cal {R}}(\delta_{\rm t}; \hat b_i ), \nonumber \\
{\cal B}_{\pm ,i, 3}^{\rm flav}&=& \frac{1}{4}\,\Gamma_{i,3}^{\rm flav}(1 \pm {\cal D}_{i,3}^{\rm flav} \cos \deltam_{i,3} \deltat_{\rm true} )\times \nonumber \\
&& { \rm e}^{-\Gamma_{i,3}^{\rm flav} | \deltat_{\rm true}|} \otimes {\cal {R}}(\delta_{\rm t}; \hat b_i ), \nonumber \\
{\cal B}_{\pm ,i, {\rm peak}}^{\rm flav}&=& \frac{1}{4}\,\Gamma_{i,{\rm peak}}(1 \pm {\cal D}_{i,{\rm peak}}^{\rm flav} \cos \deltam_{i,{\rm peak}} \deltat_{\rm true} )\times \nonumber \\
&& { \rm e}^{-\Gamma_{i,{\rm peak}}^{\rm flav} | \deltat_{\rm true}|} \otimes {\cal {R}}(\delta_{\rm t}; \hat a_i ),
\label{eq:bflavbackground}
\end{eqnarray}
corresponding to prompt, non-prompt, and mixing background components, as well as a peaking contribution.
For the $\etaCP=-1$ sample, the possible background contributions are
\begin{eqnarray}
{\cal B}_{\pm ,i, 1}^{\CP}&=&\frac{1}{2}\,\delta(\Delta t)\otimes {\cal {R}}( \delta_{\rm t} ; \hat b_i ), \nonumber \\
{\cal B}_{\pm ,i, 2}^{\CP}&=& \frac{1}{4}\,\Gamma_{i,2} (1 \pm {\cal D}_{i,2}^{\CP} \sin \deltamd \deltat)\times\nonumber \\
&& { \rm e}^{-\Gamma_{i,2}^{\CP} | \Delta t|} \otimes {\cal {R}}(\delta_{\rm t}; \hat b_i ), \nonumber \\
{\cal B}_{\pm ,i, {\rm peak}}^{\CP}&=& \frac{1}{4}\,\Gamma_{i,{\rm peak}} (1 \pm {\cal D}_{i,{\rm peak}}^{\CP} \sin \deltamd \deltat)\times\nonumber \\
&& { \rm e}^{-\Gamma_{\rm peak}^{\CP} | \Delta t|} \otimes {\cal {R}}(\delta_{\rm t}; \hat a_i ),
\label{eq:cpbackground}
\end{eqnarray}
corresponding to prompt and \CP\ background components, as well as a peaking
contribution.  The background resolution function parameters $\hat b_i$
are common with the background resolution function of the $B_{\rm flav}$
sample.
The likelihood fit includes as free parameters
the fraction of each time component, as well as 
apparent lifetimes, resolutions, mixing frequencies and dilutions that
best describe the events with high weights for being background. 
These parameters are described in Section~\ref{liki:freeparams} below,
along with additional assumptions.

\subsubsection{Background formulation for \boldmath $\Bz\to\jpsi\KL$ }\label{subsubsec:jpsikl-bkgd}

\begin{table}
\caption{Parameters of the probability distribution function for the
non-$\jpsi$ background contribution in the $\Bz \to \jpsi \KL$ channel.}
\begin{center}
\vskip 5 mm
\begin{tabular}{|c|c|} \hline
Parameter                  & Fit result \\ \hline \hline
$F_{\tau=0}$               & $0.16 \pm 0.49$  \\
$\Gamma_{i,2}$ [ps$^{-1}$] & $1.25\pm 0.45$   \\
$S_1$                      & $1.12 \pm 0.26$  \\
$b_1$                      & $-0.11 \pm 0.20$ \\
$S_2$                      & $3.9 \pm 0.8$    \\
$b_2$                      & $-1.2 \pm 1.0$   \\
$f_2$                      & $0.23 \pm 0.14$  \\
$f_3$                      & $0.005$ (fixed)  \\ \hline
\end{tabular}
\end{center}
\label{tab:klnonpsivtx}
\end{table}

The higher background level in the $\Bz \to \jpsi \KL$ channel 
requires a more extensive treatment of its properties.
As discussed in Section~\ref{subsec:sample_KL}, the data are 
used to determine the relative fraction of signal, background
from $B \to \jpsi X$ events, and events with a 
misreconstructed $\jpsi \to \ell \ell$ candidate.  Along with
a Monte Carlo simulation of the channels that contribute
to the $B \to \jpsi X$ background, this information is used
to formulate the PDF model.  In addition, some of the $\jpsi X$ background 
modes, such as $\Bz\to\jpsi K^{*0}$ and $\Bz\to\jpsi\KS$, have a non-zero 
\CP\ asymmetry ($\etaCP$), as given in Table~\ref{tab:psiklComposition}. 
The value of the asymmetry in $\Bz \to \jpsi \Kstarz (\KL \piz)$
is taken from the measurement of $R_\perp=0.160 \pm 0.032 \pm 0.014$ 
in Ref.~\cite{babar0105}. 
The probability density functions ${\cal F}_{\pm}$ 
of Eq.~\ref{eq:Convol} are modified to include contributions for each of the
$B\to\jpsi X$ channels $\alpha$ specified in Table~\ref{tab:psiklComposition} 
and the non-$\jpsi$ background component. The complete PDF is given by
\begin{widetext}
\begin{eqnarray}
{\cal F}_{\pm,i} = f_{i,k,{\rm sig}}(\deltae) {\cal F}_{\pm}
(\deltat;\Gamma,\deltamd,\mistag_i,\stwob, \hat a_i)
&+& \sum_{\alpha={\jpsi X}} f_{i,k,\alpha} (\deltae)\ {\cal F}_{\pm}
(\deltat;\Gamma,\deltamd,\eta_{\CP,\alpha},\mistag_i,\stwob, \hat a_i) \nonumber \\
&+& f_{i,k,{\rm non}-\jpsi} (\deltae) \ {\cal{B}}_{\pm}^{KL}(\deltat;\hat b). 
\end{eqnarray}
\end{widetext}
Each event is classified according to its flavor tagging category ($i$),
flavor tag value ($\pm$), and the $\KL$ reconstruction category ($k$), 
which is either EMC or IFR. The signal fraction $f_{i,k,{\rm sig}}$ and 
background fractions $f_{i,k,\alpha}$ and
$f_{i,k,{\rm non}-\jpsi}$ are determined as a function of
$\deltae$ and are the same for all tagging categories.  The shape of the 
signal and background $\deltae$ functions are determined either
from data (non-$\jpsi$ contribution) or from Monte Carlo samples
(signal and $\jpsi X$ background).  The normalizations 
$\int_{-10\,{\rm MeV}}^{10\,{\rm MeV}} f\,d(\deltae)$
are determined from Tables~\ref{tab:psiklComposition} and~\ref{tab:psikldefit}
so that
\begin{eqnarray}
\int_{-10\,{\rm MeV}}^{10\,{\rm MeV}} &&[\,f_{i,k,{\rm sig}}(\deltae)  + f_{i,k,{\rm non}-\jpsi}(\deltae) + \nonumber \\
&&\sum_{{\jpsi X}} f_{i,k,\alpha}(\deltae)\,]\,d(\deltae) = 1.
\end{eqnarray}
The non-$\jpsi$ background PDF is given by
\begin{equation}
   {\cal{B}}_{\pm}^{KL} = F_{\tau=0}{\cal B}_{\pm, i,1}^{\CP} + (1-F_{\tau=0}) {\cal B}_{\pm, i,2}^{\CP} 
\end{equation}
where the dilutions ${\cal D}^{\CP}_{i,2}=0$ and 
the prompt fraction $F_{\tau=0}$, effective decay width $\Gamma_{i,2}$, 
and \deltat\ resolutions parameters
$\hat{b}$ are fixed to values obtained from an external fit 
to the $m(\ell \ell)$ sideband events
as given in Table~\ref{tab:klnonpsivtx}.
The resolution function ${\cal R}(\Delta t;\hat{b})$
is defined in Eq.~\ref{eq:vtxresolfunct}
with $f_3=0.005$ and with core bias $\delta_1$ equal for all tagging 
categories.

The $\jpsi \KL$ sample has significant background, primarily from 
other $\jpsi$ modes.
The Monte Carlo simulation
is used to check the flavor tagging efficiency of the
inclusive $\jpsi$ background relative to the signal for the $\KL$ mode.
The inclusive $\jpsi$ background fraction in the simulation is consistent
across the flavor tagging categories to within a few percent.
The flavor tagging efficiency for the fake-$\jpsi$ background, determined
from the $\jpsi$ sideband, is also roughly consistent with signal.
The composition of the $\jpsi \KL$ sample is determined from a
fit of the $\Delta E$ spectrum before flavor tagging.
We assume the inclusive $\jpsi$ and fake-$\jpsi$ background fractions are
independent of flavor tag in the nominal fit and adjust the fractions as a
function of tagging category, based on the Monte Carlo simulation and
$\jpsi$ sideband, in order to determine systematic errors.

Some of the decay modes in the inclusive $J/\psi$ background, such as
$\jpsi K^{*0}$ and $\jpsi \KS$, have an expected \CP\ asymmetry.
The mistag fractions for all \CP\ modes in the inclusive $\jpsi$ background
are determined with the Monte Carlo simulation
and found to be consistent with the
values for the signal.
We assume that the signal mistag fractions apply to the \CP\ modes
in the inclusive $\jpsi$ background.

The $\Delta t$ resolution for the $B \rightarrow \jpsi X$ background should be very similar to
the signal resolution.
However, extra tracks associated with
$B^+ \rightarrow \jpsi X^+$ decay, such as the charged $\pi$ from the
$K^{*+}$ decay in $B^+ \rightarrow \jpsi K^{*+}$, could bias the
measurement of $\Delta t$ since they are not
associated with the $B_{CP}$ vertex and therefore can be used
in the $B_{\rm tag}$ vertex.
In the Monte Carlo simulation,
we find that extra tracks in the $B \rightarrow \jpsi X$
decay modes have a negligible effect on the $\Delta t$ resolution.
Therefore, we assume that all $B \rightarrow \jpsi X$ background has the same
resolution as the signal.

The $\Delta t$ resolution of the non-$\jpsi$ background is measured with
the $\jpsi$ sideband sample.
The non-$\jpsi$ $\Delta t$ resolution parameters are varied by their
statistical uncertainties to estimate the systematic uncertainty.

\subsubsection{Background formulation for $\Bz\to\jpsi\Kstarz(\KS \piz)$ }
\label{subsubsec:jpsikstar-bkgd}

Monte Carlo simulation is used to construct the probability density
function for the $\Bz\to\jpsi\Kstarz(\KS \piz)$ channel.  As shown
in Table~\ref{tab:kstarback}, the background for this channel is due
to true $B \to \jpsi X$ decays.  Thus, we assume that the background
has the same resolution function and tagging performance as the 
signal.  
The probability density functions ${\cal F}_{\pm}$ 
of Eq.~\ref{eq:Convol} are modified to include contributions for each of the
$B\to\jpsi X$ channels $\alpha$ specified in Table~\ref{tab:kstarback}. 
The complete PDF is given by
\begin{widetext}
\begin{equation}
{{\cal F}_{\pm,i}} = f_{\rm sig} {\cal F}_{\pm}
(\deltat;\Gamma,\deltamd,\eta_{\CP,{\rm signal}}, \mistag_i,\stwob, \hat a_i)
+ \sum_{\alpha={{\rm bkgd}}} f_{\alpha}\ {\cal F}_{\pm}
(\deltat;\Gamma,\deltamd,\eta_{f,\alpha},\mistag_i,\stwob, \hat a_i), 
\end{equation}
\end{widetext}
where each event is classified according to its flavor tagging category ($i$)
and flavor tag value ($\pm$).  The signal and background fractions as well
as $\eta_{\CP}$ are taken from Table~\ref{tab:kstarback}.

\subsection{Extensions for direct \CP\ search}

While the main likelihood fits are performed with the Standard Model expectation that $|\lambda|=1$, a search for the effects
of direct \CP\ violation is also made. Such a measurement is also particularly sensitive to possible differences in the 
fraction of \Bz\ or \Bzb\ meson that are tagged.
Defining $\epsilon_{\rm tag}$ and $\overline{\epsilon}_{\rm tag}$ as the tagging efficiencies for \Bz\ and \Bzb, and
$\epsilon_{\rm r}$ and $\overline{\epsilon}_{\rm r}$ as the reconstruction efficiencies for \Bz\ and \Bzb\ in the $B_{\rm flav}$ sample,
it is useful to construct
\begin{eqnarray}
  \mu_i = \frac{\epsilon_{{\rm tag},i}-\overline{\epsilon}_{{\rm tag},i}}{\epsilon_{{\rm tag},i}+\overline{\epsilon}_{{\rm tag},i}},
&& \langle \epsilon_{\rm tag}\rangle_i = \frac{\epsilon_{{\rm tag},i}+\overline{\epsilon}_{{\rm tag},i}}{2} 
                                                   \label{eq:mudef}\\
  \nu_i = \frac{\epsilon_{{\rm r},i}-\overline{\epsilon}_{{\rm r},i}}{\epsilon_{{\rm r},i}+\overline{\epsilon}_{{\rm r},i}},
&& \langle \epsilon_{\rm r}\rangle_i = \frac{\epsilon_{{\rm r},i}+\overline{\epsilon}_{{\rm r},i}}{2}.
                                                   \label{eq:nudef}
\end{eqnarray}
For the $B_{\CP}$ sample, the time-dependent decay rate (Eq.~\ref{eq:TimeDep}) becomes
\begin{widetext}
\begin{equation}
f_{\pm,i} (\deltat)  = \frac{\Gamma}{4}\,{\mathrm e}^{-\Gamma|\Delta t|}
        \left[ \frac{(1+|\lambda|^2) ( 1 \pm X_i )}
                    {\displaystyle 1+|\lambda|^2 + \xi_i}
              + ( \mu_i \pm X^\prime_i  )
        \left( 
               \frac{2 Im \lambda}
                    {\displaystyle 1+|\lambda|^2 + \xi_i}\sin\deltamd\deltat
             - \frac{1 - |\lambda|^2}
                    {\displaystyle 1+|\lambda|^2 + \xi_i}\cos\deltamd\deltat
\right) \right],
\label{eq:method-tageff-timeevol}
\end{equation}
\end{widetext}
where
$\xi_i = \mu_i (|\lambda|^2-1)/(1+x_d^2)$,
$X_i = \mu_i\langle{\cal D}\rangle_i + \Delta {\cal D}_i/2$, and
$X^\prime_i = \langle {\cal D}\rangle_i + \mu_i\Delta {\cal D}_i/2$.
Likewise, for the $B_{\rm flav}$ sample the time-dependent decay rate (Eq.~\ref{eq:mixing_rate}) becomes
\begin{eqnarray}
h(\deltat) & = & \frac{\Gamma}{4}\,{\mathrm e}^{-\Gamma|\Delta t|}\frac{1+s_1\nu_i}{\displaystyle\left[1 - \frac{\mu_i \nu_i}{1+x_d^2} \right]}
\left[1+s_2 X_i - \right.\nonumber \\
&&\left.\qquad s_1(\mu_i+s_2 X^\prime_i)\cos \deltamd \deltat\right] 
\end{eqnarray}
where
\begin{eqnarray*}
s_1 & = & 1(-1)\ \mathrm{if\ the\ reconstructed\ }B\ \mathrm{is\ a\ } \Bz (\Bzb) \\
s_2 & = & 1(-1)\ \mathrm{for\ a\ }\Bz (\Bzb)\mathrm{\ tag}.
\end{eqnarray*}

The parameters $\nu_i$, $\langle\epsilon_{\rm tag}\rangle_i$, and $\mu_i$ can be 
extracted from time-integrated numbers of events in the
$B_{\rm flav}$ sample. Defining integrated samples of events by
\begin{eqnarray}
N^{\mathrm{tag}}_i                & = & N(\Bz/\Bzb\mathrm{\ tag\ in\ }i^{th}\mathrm{\ category},B_{\rm flav}=\Bz) \nonumber \\
\overline{N}^{\mathrm{tag}}_i     & = & N(\Bz/\Bzb\mathrm{\ tag\ in\ }i^{th}\mathrm{\ category},B_{\rm flav}=\Bzb) \nonumber \\
N^{\mathrm{no\ tag}}_i            & = & N(\mathrm{no\ tag\ in\ }i^{th}\mathrm{\ category},B_{\rm flav}=\Bz) \nonumber \\
\overline{N}^{\mathrm{no\ tag}}_i & = & N(\mathrm{no\ tag\ in\ }i^{th}\mathrm{\ category},B_{\rm flav}=\Bzb),
\end{eqnarray}
it can be shown that
\begin{eqnarray}
\nu_i & = &
\frac{N^{\mathrm{tag}}_i-\overline{N}^{\mathrm{tag}}_i+N^{\mathrm{no\ tag}}_i-\overline{N}^{\mathrm{no\ tag}}_i}
     {N^{\mathrm{tag}}_i+\overline{N}^{\mathrm{tag}}_i+N^{\mathrm{no\ tag}}_i+\overline{N}^{\mathrm{no\ tag}}_i} \\
\langle\epsilon_{\rm tag}\rangle_i
    & = &
\frac{2N^{\mathrm{tag}}_i\overline{N}^{\mathrm{tag}}_i+\overline{N}^{\mathrm{tag}}_iN^{\mathrm{no\ 
tag}}_i+N^{\mathrm{tag}}_i\overline{N}^{\mathrm{no\ tag}}_i}
     {2(N^{\mathrm{tag}}_i+{N}^{\mathrm{no\ tag}}_i)(\overline N^{\mathrm{tag}}_i+\overline{N}^{\mathrm{no\ tag}}_i)} \nonumber \\
\mu_i & = & 
\frac{(1+x^2_d)(\overline{N}^{\mathrm{tag}}_iN^{\mathrm{no\ tag}}_i-N^{\mathrm{tag}}_i\overline{N}^{\mathrm{no\ tag}}_i)}
{2N^{\mathrm{tag}}_i\overline{N}^{\mathrm{tag}}_i+\overline{N}^{\mathrm{tag}}_iN^{\mathrm{no\
tag}}_i+N^{\mathrm{tag}}_i\overline{N}^{\mathrm{no\ tag}}_i} \nonumber
\end{eqnarray}
under the assumption that nearly all \B\ mesons decay to final states
that can be reached from either \Bz\ or \Bzb, but not both. 
The results for $\langle \epsilon_{\rm tag}\rangle_i$ 
and $\mu_i$ are shown in Table~\ref{tab:dircp-mu}. The value of $\nu_i$, averaged
over all four tagging categories, is $0.004\pm 0.012$. While there is no 
statistically significant difference in the tagging
efficiencies or the reconstruction efficiencies given by $\mu_i$ and $\nu_i$, 
we use the central values obtained
from the $B_{\rm flav}$ sample in performing the fit for $|\lambda|$.

\begin{table}[htb]
\caption{Values of $\langle\epsilon_{\rm tag}\rangle_i$ and $\mu_i$ for the four tagging categories, as determined by counting
 numbers of tagged and untagged events in the $B_{\rm flav}$ sample.}
\begin{center}
\begin{tabular}{|l|c|c|} \hline
Tagging category  &   $\langle\epsilon_{\rm tag}\rangle_i$   & $\mu_i$   \\ \hline\hline
{\tt Lepton}      & $  0.095 \pm 0.002  $  & $ 0.069 \pm 0.032$ \\
{\tt Kaon}        & $  0.358 \pm 0.003  $  & $-0.005 \pm 0.014$ \\
{\tt NT1}         & $  0.080 \pm 0.002  $  & $0.061 \pm 0.035$  \\
{\tt NT2}         & $  0.139 \pm 0.002  $  & $0.017 \pm 0.026$  \\ \hline
\end{tabular}
\end{center}
\label{tab:dircp-mu}
\end{table}

\begin{table}[htb]
\caption{Average mistag fractions $\langle\mistag_i\rangle$ and mistag differences 
$\Delta \mistag_i$ for each tagging category $i$ from a maximum-likelihood
fit to the distribution for the \Bu\ control sample.}
\begin{center}
\begin{tabular}{|l|c|c|} \hline
Tagging category & $\langle\mistag\rangle_i$ [\%] & $\Delta \mistag_i$ [\%] \\ \hline\hline
{\tt Lepton} & $4.6 \pm 0.6$ & $1.1 \pm 1.2$  \\
{\tt Kaon}   & $11.8 \pm 0.5$ &$-0.3 \pm 1.0$ \\
{\tt NT1}    & $21.3 \pm 1.6$ &$-5.9 \pm 3.2$ \\
{\tt NT2}    & $37.2 \pm 1.3$ & $-0.7 \pm 2.7$ \\ \hline
\end{tabular}
\end{center}
\label{tab:bchdilutions}
\end{table}

\subsection{Free parameters for the \stwob\ and \deltamd\ fits}
\label{liki:freeparams}
The unbinned likelihood fit for \stwob\ has a total of 45 free parameters:
\begin{itemize}
\item
{\bf Value of {\boldmath \stwob}}
\item
{\bf Signal resolution function:}
Sixteen parameters $\hat a_i$ to describe the resolution function for the signal.  Due to improvements in the
reconstruction algorithms, the Run 1 and Run 2 resolution functions are found to be different, as described in Section~\ref{sec:run1vsrun2}.  Thus, we allow
for separate resolution function parameters for these data samples, each with eight free parameters,
being a scale factor $S_{1}$ for the 
event-by-event \deltaz\ resolution errors of the core Gaussian
components, individual core bias scale factors $b_{1,i}$ for the four 
tagging categories and a common tail bias $b_2$, and the tail $f_2$ 
and outlier $f_3$
fractions; the scale factor of the tail component is fixed to 3.0 and 
the width of the outlier component is fixed to 8\ps\ with zero bias.
\item
{\bf Signal dilutions:}
Eight parameters to describe the measured average dilutions $\langle {\cal D}\rangle_i$ and dilution
differences $\Delta {\cal D}_i$ in each tagging category.
\item
{\bf Background resolution function:}
Six parameters are used to describe a common resolution function for all non-peaking backgrounds. As
with the signal resolution function, we include separate resolution function parameters for
the Run 1 and Run 2 data samples.  The resolution function  
is taken as a single Gaussian distribution with a scale factor $S_1$ for the event-by-event 
\deltaz\ errors and a common bias scale factor $b_1$, and an outlier fraction $f_3$; the width of the 
outlier component is taken to be a fixed 8\ps\ with zero bias.   
\item
{\bf {\boldmath $B_{\rm flav}$} background composition parameters:} 
A total of 13 parameters describe the $B_{\rm flav}$ background composition.
We make several assumptions to simplify the parameterization 
shown in Eq.~\ref{eq:bflavbackground},
such as removing the mixing background contribution by setting
$f^{\rm flav}_{i,3}=0$, and assign a corresponding systematic
uncertainty. 
The size of the peaking background is determined from Monte Carlo simulation 
to be $\delta_{\rm peak}^{\rm flav}=(1.5 \pm 0.5) \%$ of the signal contribution in each tagging category.  This
contribution is predominately from $\Bu$ events, so $\deltam_{i,{\rm peak}}=0$,
$\Gamma_{i,{\rm peak}}^{\rm flav}=\Gamma_{\Bu}$ and $D^{\rm flav}_{i,{\rm peak}}$
are taken from the $\Bu$ data sample (Table~\ref{tab:bchdilutions}).  
The effective dilutions for the prompt ($D^{\rm flav}_{i,1}$, 4 parameters) 
and non-prompt ($D^{\rm flav}_{i,2}$, 4 parameters) contributions are allowed to vary.
The relative amount of these two contributions is allowed to vary independently 
in each tagging category (4 parameters).
For the non-prompt contribution, $\Gamma_{i,2}^{\rm flav}$ is assumed to be the same for all
tagging categories, giving one free parameter.
 
\item
{\bf {\boldmath $\CP$} background composition parameters:}
One parameter, the fraction of prompt relative to non-prompt background, assumed
to be the same for each tagging category, is allowed to float to describe
the $\CP$ background properties.  The effective dilutions of the non-prompt
and peaking contribution are set to zero ($D^{\CP}_{i,2}=D^{\CP}_{i,{\rm peak}}=0$),
corresponding to no \CP-asymmetry in the background.  The size and parameters of the peaking
background are determined from Monte Carlo simulation.  
The fraction of peaking background is $\delta_{\rm peak}^{\CP}=(1 \pm 1)\%$ of the 
signal contribution, independent
of tagging category.  This contribution is assumed to have
lifetime parameters in common with the signal.
Finally, the lifetime of the non--prompt background is assumed to be $\tau_{\Bz}$
in all tagging categories.
\end{itemize}
The unbinned likelihood fit for \deltamd\ has 44 free parameters, removing \stwob\ and the parameter for
fraction of prompt background in the \CP\ sample and leaving \deltamd\ to float.

\subsection{Blind analysis}

A blind analysis technique was adopted for the extraction 
of \stwob\ and \deltamd\ in order to eliminate possible 
experimenter's bias.  We used a method 
that hides not only the central value for these parameters from the unbinned 
maximum-likelihood fit, but also the visual \CP\ asymmetry
in the \deltat\ distribution.  
The error on both the asymmetry and \deltamd\ is not hidden.

The amplitude of the asymmetry ${\cal A}_{\CP}(\deltat)$ from the fit 
was hidden by a one-time choice of
sign flip and arbitrary offset based on a user-specified key word.
The sign flip hides whether a change in the analysis 
increases or decreases the resulting asymmetry.
However, the magnitude of the change is not hidden. 
The visual \CP\ asymmetry in the  \deltat\ distribution is  
hidden by multiplying \deltat\ by the sign of the tag 
and adding an arbitrary offset.

With these techniques, systematic studies can be performed
while keeping the numerical value of \stwob\  or \deltamd\ hidden.  
In particular, 
we can check that the hidden \deltat\ distributions are 
consistent for \Bz\ and \Bzb\ tagged events. The same is true for all the 
other checks concerning tagging, vertex resolution and the 
correlations between them.   
For instance, fit results in the different tagging 
categories can be compared to each other, since each fit is 
hidden in the same way. The analysis procedure for extracting 
\stwob\ and \deltamd\ were frozen
prior to unblinding.

\renewcommand{\secname}{Mixing}
\section{\boldmath \Bz\ flavor oscillations and mistag rates}
\label{sec:\secname}

\subsection{Likelihood fit results for \boldmath\deltamd}

We extract \deltamd, the dilution factors ${\cal D}_i$, the 
\deltat\ resolution parameters $\hat {a}_i$, and the background 
\deltat\ parameterization
by fitting the \deltat\ distributions of the flavor-eigenstate 
\Bz\ sample with $\mes > 5.2$\gevcc\ with the likelihood function described 
in Section~\ref{sec:Likelihood}. 
The selection of the $B_{\rm flav}$ data sample is described in 
Section~\ref{subsec:sample_hadronicBz}. We also demand a
valid tag and \deltat\ determination for the event,
based on the algorithms described in Sections~\ref{sec:Tagging}
and \ref{sec:vertexing}. The more restrictive requirements
$\vert\deltat\vert<20\ps$ and $\sigma_{\deltat}<1.4\ps$
are applied to the proper time-difference measurement. In addition,
identified kaons in the $B_{\rm tag}$ decay are rejected
in the reconstruction of the tagging vertex. These requirements
are intended to reduce systematic errors on the precision
\deltamd\ measurement. The final sample consists
of 12310 fully-reconstructed and tagged \Bz\ candidates
with $\mes>5.2\gevcc$, of which 7399 are in the signal
region $\mes>5.27\gevcc$.

\begin{figure}[htb]
\begin{center}
    \includegraphics[width=0.85\linewidth]{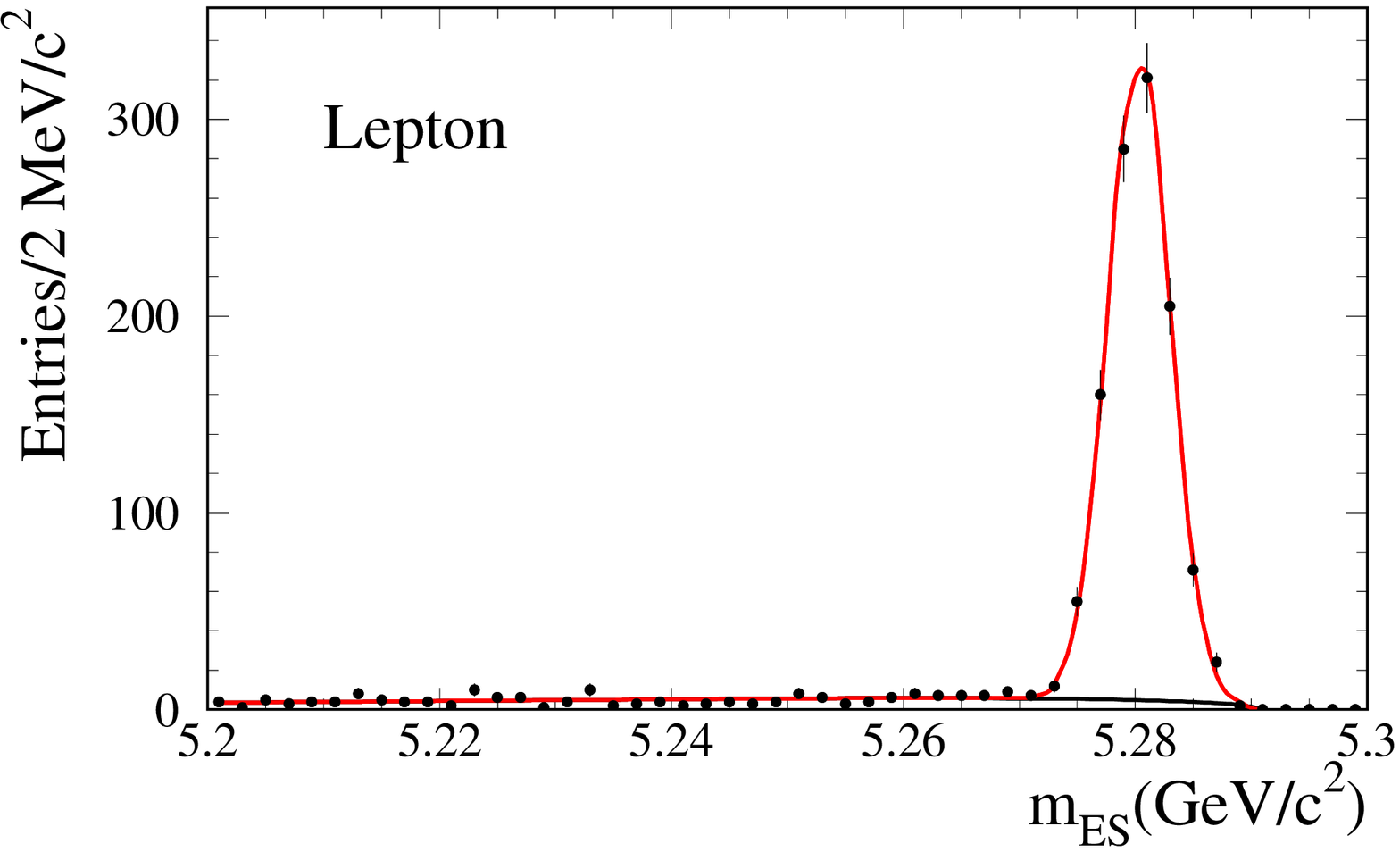}
    \includegraphics[width=0.85\linewidth]{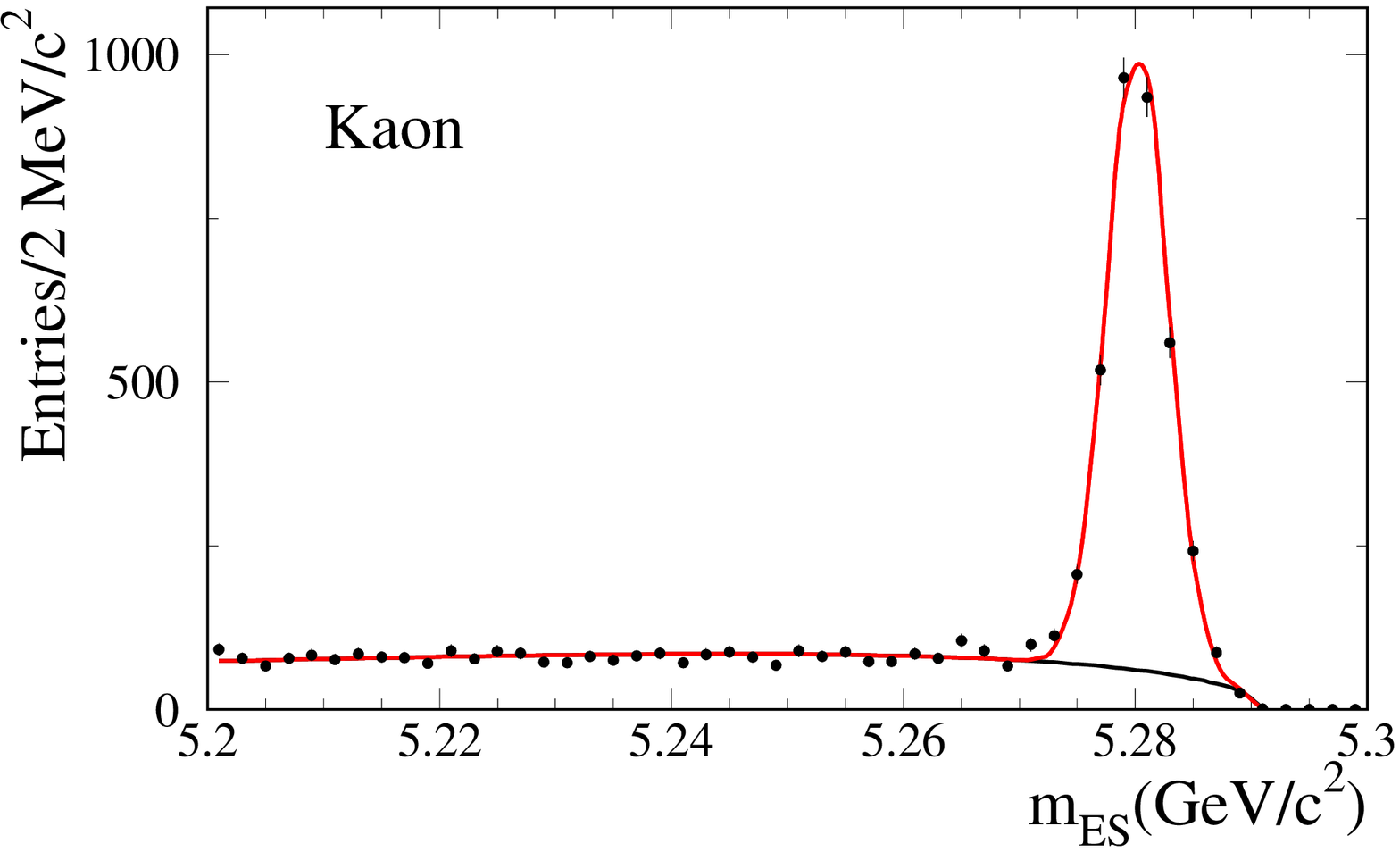}
    \includegraphics[width=0.85\linewidth]{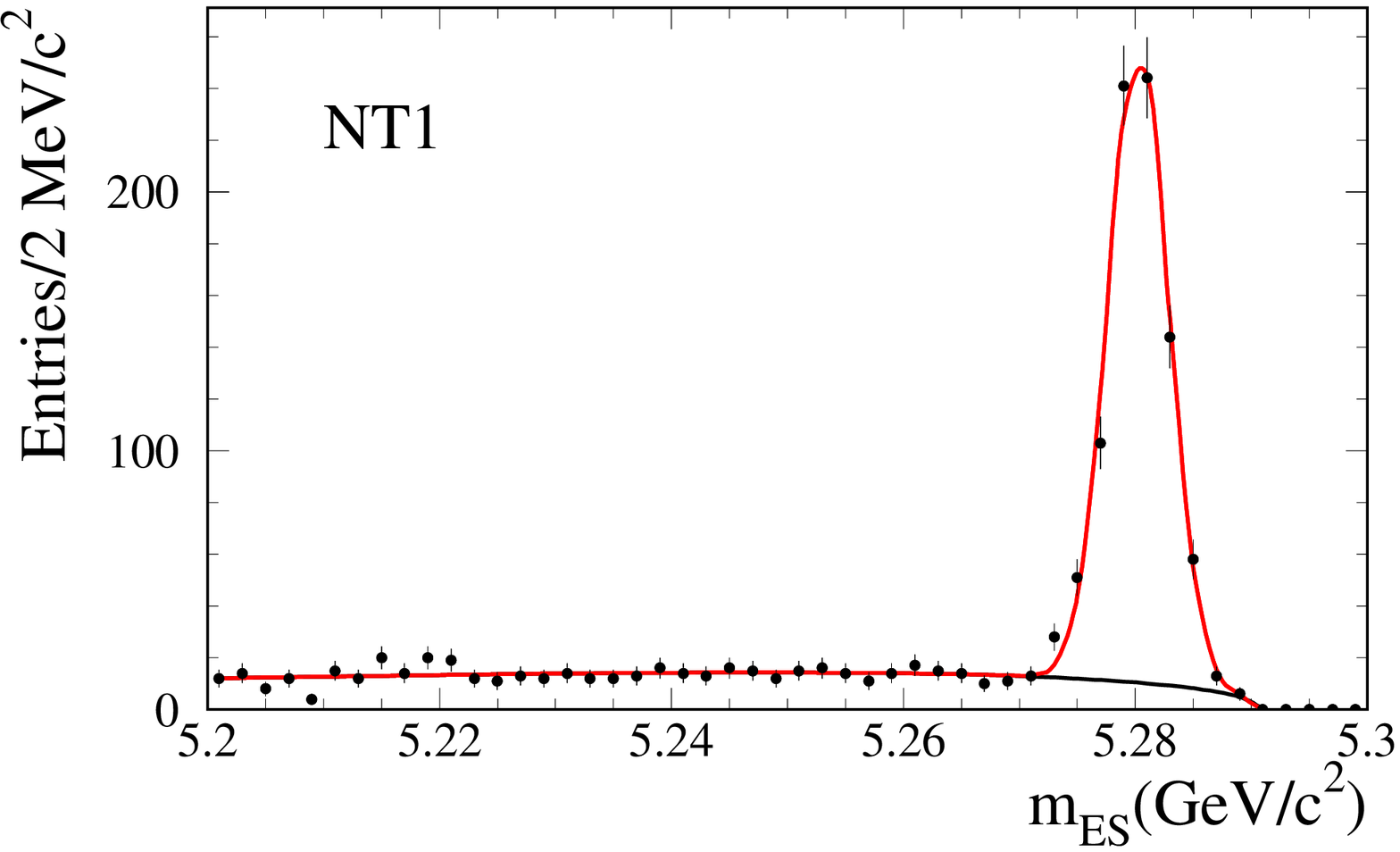}
    \includegraphics[width=0.85\linewidth]{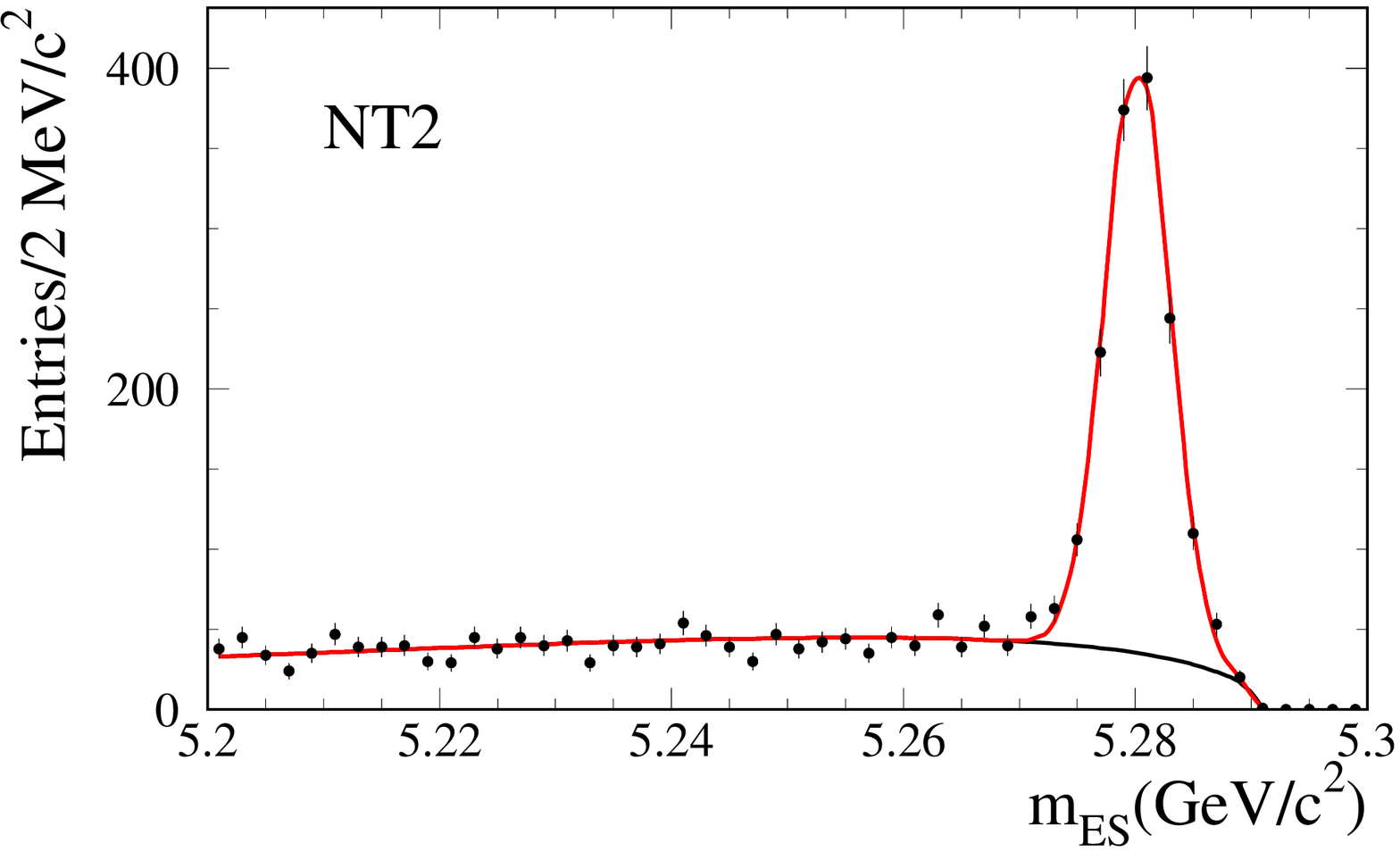}
\end{center}
\caption{Distribution of \mes\ for mixing $B_{\rm flav}$ candidates in  
separate tagging categories ({\tt Lepton}, {\tt Kaon}, {\tt NT1} and {\tt NT2}),
overlaid with the result of a fit with a Gaussian distribution
for the signal and an ARGUS function for the background.
\label{fig:b0mix.excl.mb-cats}}
\end{figure}

The breakdown of this mixing $B_{\rm flav}$ 
sample into individual tagging categories
is shown in Fig.~\ref{fig:b0mix.excl.mb-cats} as a function of \mes.
Superimposed on the observed mass spectra are the results of the
fits with a Gaussian distribution for the
signal and the ARGUS background function for the background.
The tagging efficiency and signal purity for the
individual tagging categories in data are extracted
from fits to the \mes\ distributions and are listed in
Table~\ref{tab:tagging-excl}.
The efficiency for each tagging
category is defined as the ratio of the number of signal events for
each tag over the total number of signal events after imposition of vertex cuts.

\begin{table}[!htb]
  \caption{Tagging efficiencies for hadronic \Bz\ decays and signal purities 
in data, shown separately for the four tagging categories. Signal purities are
estimated for $\mes > 5.27$\gevcc.} \vspace{0.3cm}
\label{tab:tagging-excl} 
\begin{center}
  \begin{tabular}{|l|c|c|c|}\hline
Tagging  & Efficiency & Signal  & Purity  \\
Category & [\%]       &         &  [\%]   \\
    \hline\hline
{\tt Lepton}&   $11.8\pm 0.3$ & $1097\pm 34$   & $96.0\pm 0.7$   \\
{\tt Kaon}  &   $33.9\pm 0.5$ & $3156\pm 63$   & $84.6\pm 0.7$   \\
{\tt NT1}   &   $ 8.6\pm 0.3$ & $ 798\pm 31$   & $88.9\pm 1.2$   \\
{\tt NT2}   &   $13.9\pm 0.4$ & $1293\pm 43$   & $79.4\pm 1.3$   \\
\hline\hline
Full sample &   $68.1\pm 0.4$ & $6347\pm 89$   & $85.8\pm 0.5$   \\ 
    \hline
  \end{tabular}
\end{center}
\end{table}

\begin{figure*}[!htb]
  \begin{center}
    \includegraphics[width=0.49\linewidth]{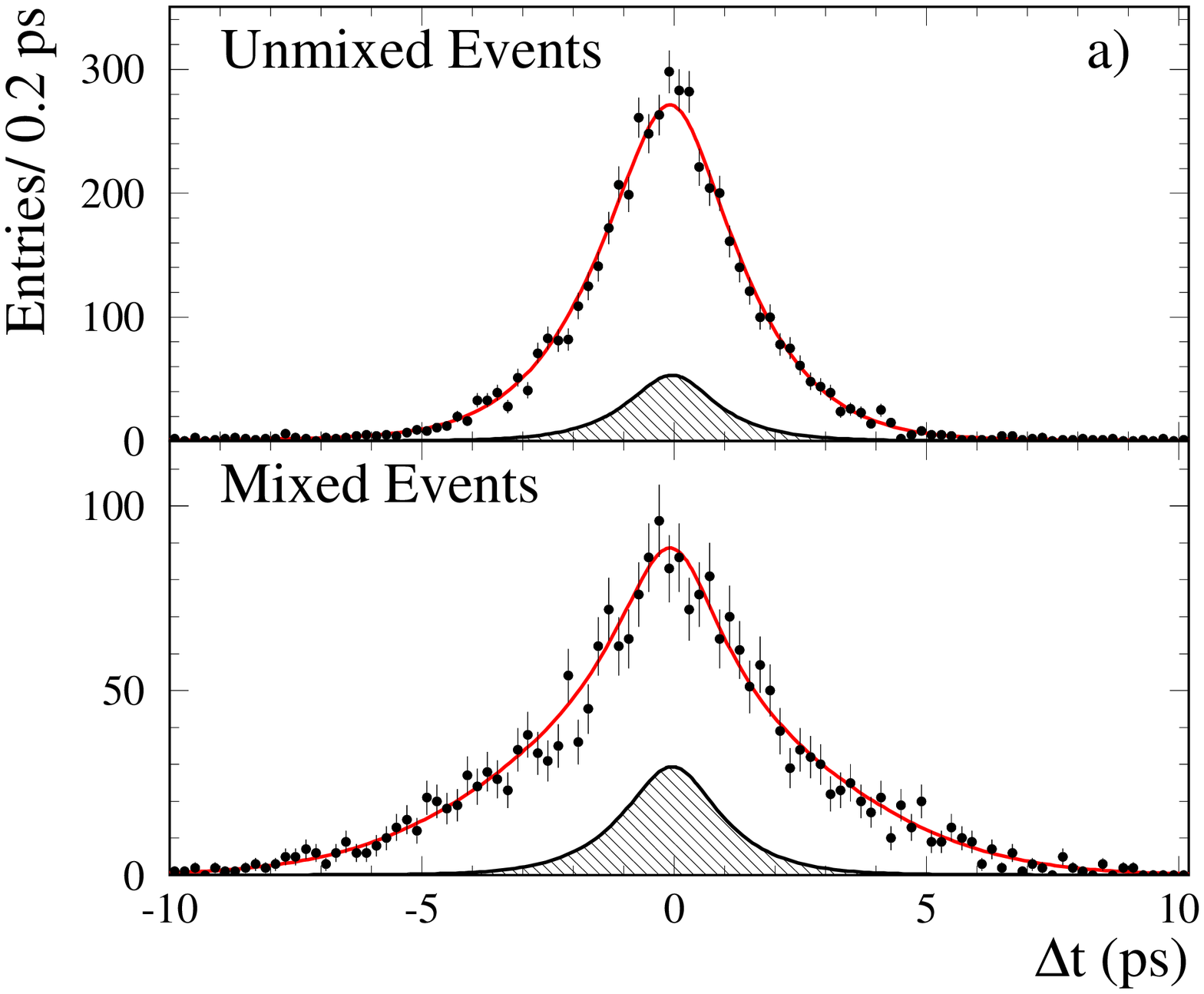}
    \includegraphics[width=0.49\linewidth]{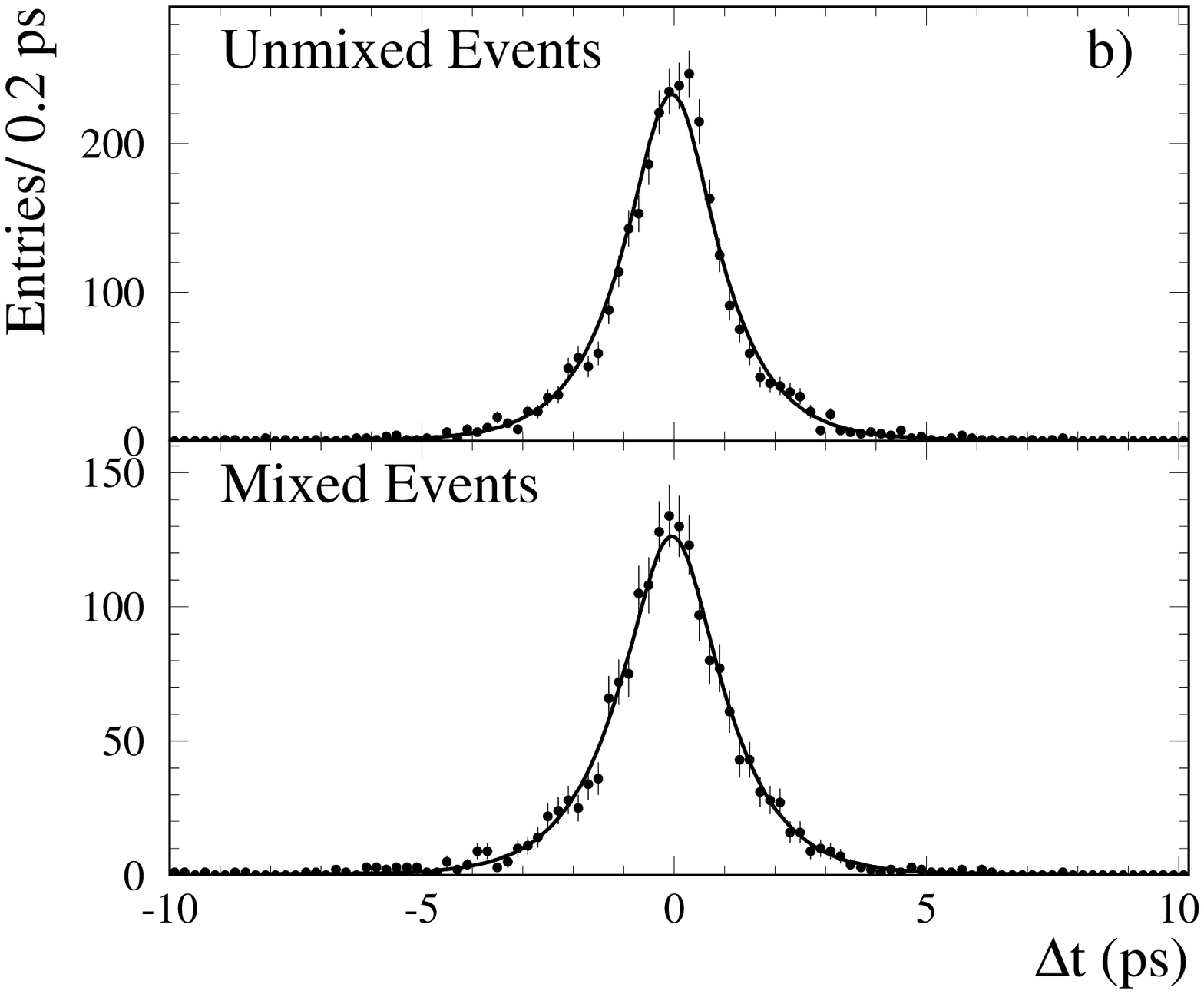}
    \includegraphics[width=0.49\linewidth]{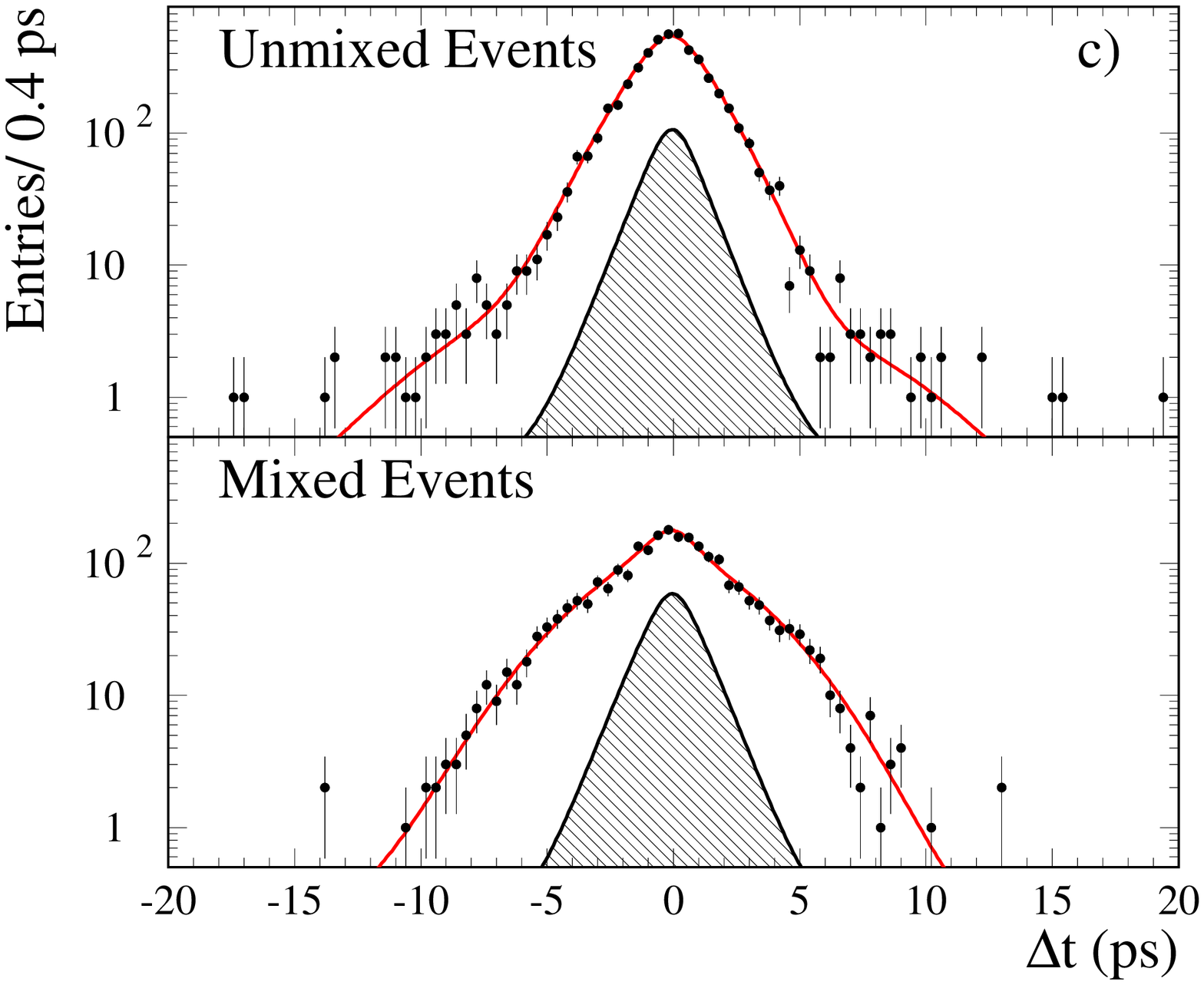}
    \includegraphics[width=0.49\linewidth]{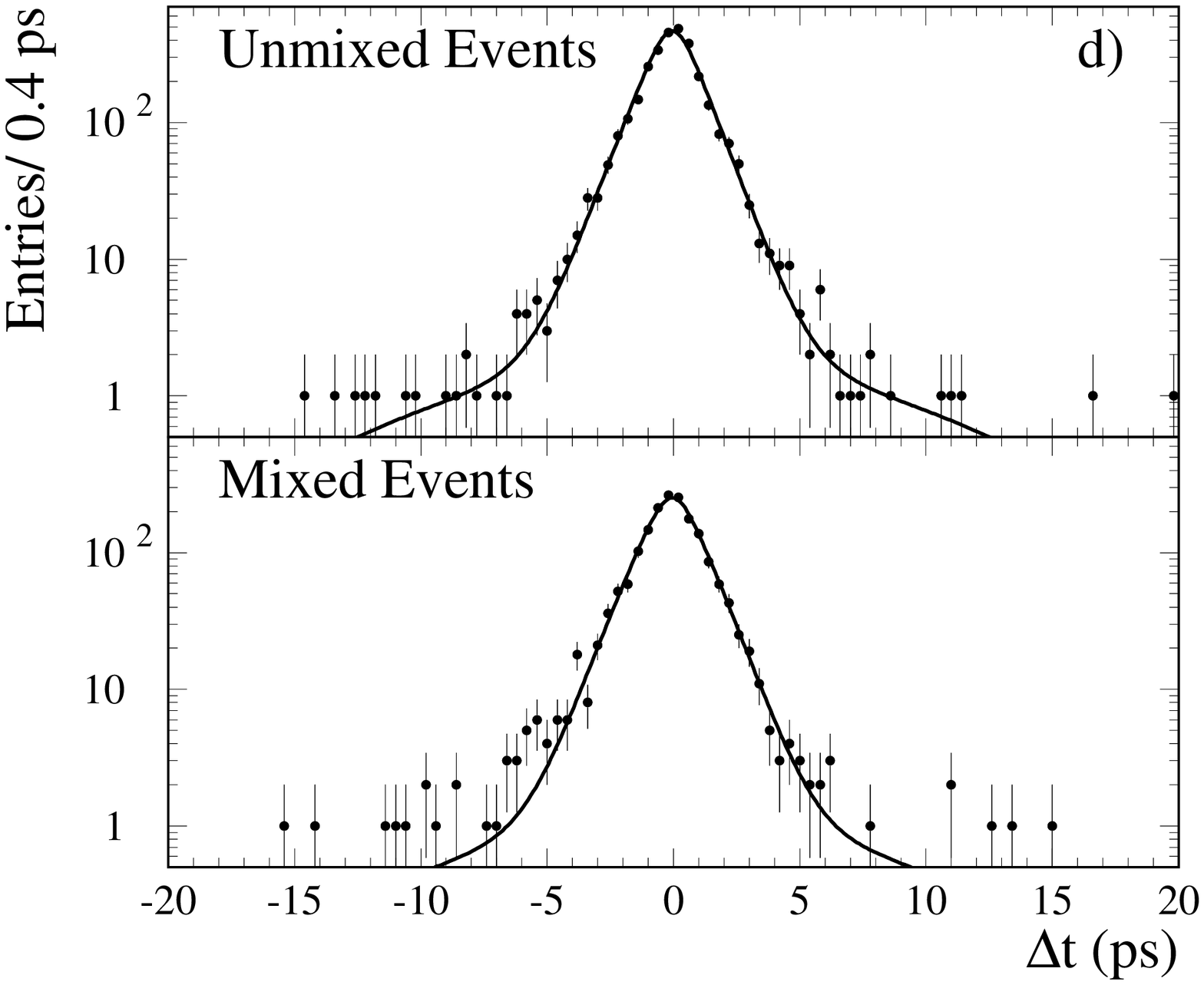}
  \end{center}
\caption{Distributions of \deltat\ for unmixed (upper panel) and
mixed (lower panel) events in the
hadronic $B$ sample, divided into a signal
region $\mes > 5.27$\gevcc\ with a) a linear and 
c) logarithmic scale, and a sideband region $\mes < 5.27$\gevcc\ 
with b) a linear and d) logarithmic scale. 
In all cases, the data points are overlaid with the result from the global
unbinned likelihood fit, projected on the basis of the individual
signal and background probabilities, and event-by-event \deltat\ resolutions,
for candidates in the respective samples. 
In a) and c), the \deltat\ distributions obtained from the likelihood fit 
to the full sample are overlaid, along with the simultaneously determined 
background distribution shown as the curve in b) and d). 
   \label{fig:b0.excl-data.deltat.all}}   
\end{figure*}

\begin{figure*}[!htb]
  \begin{center}
    \includegraphics[width=0.49\linewidth]{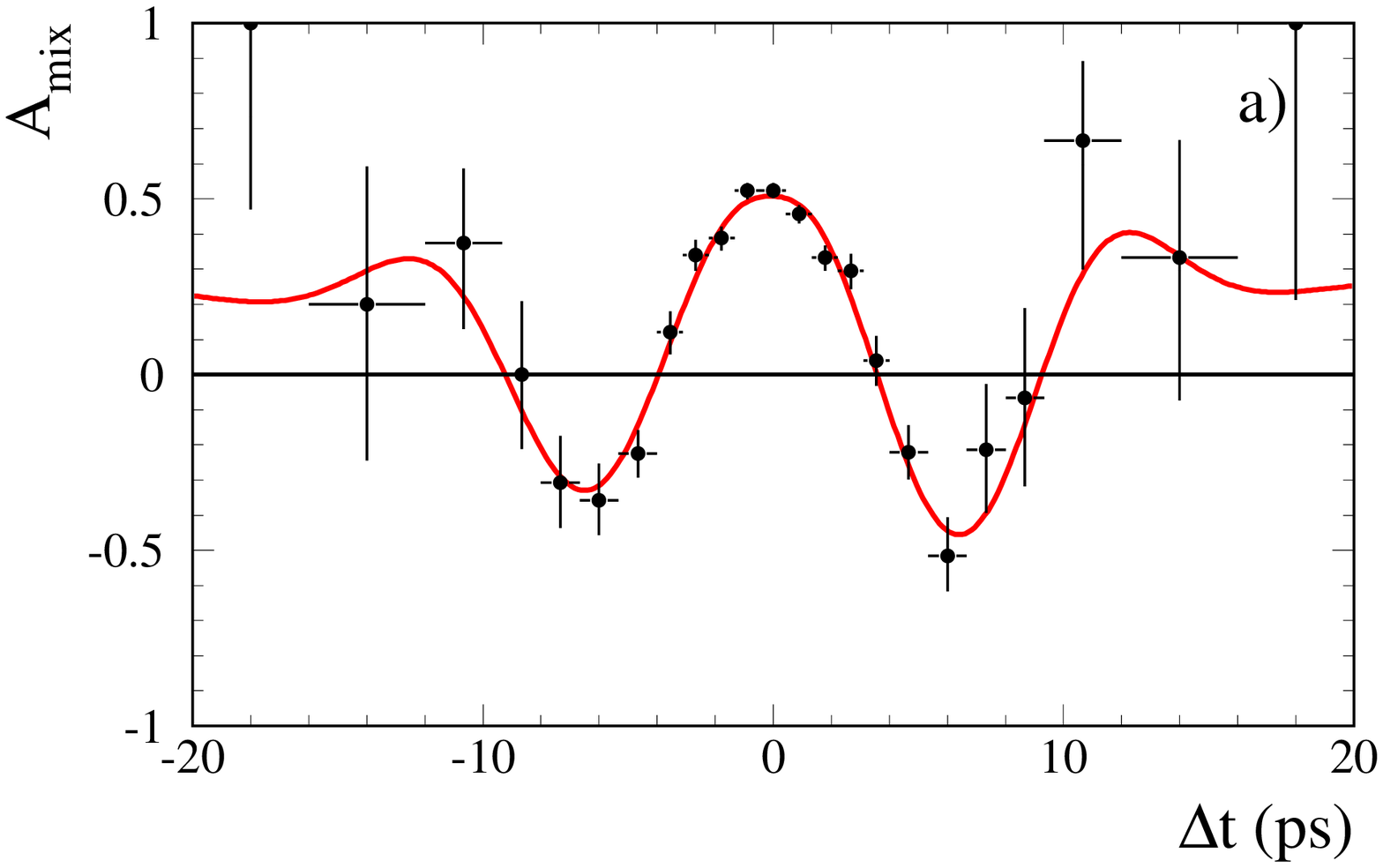}
    \includegraphics[width=0.49\linewidth]{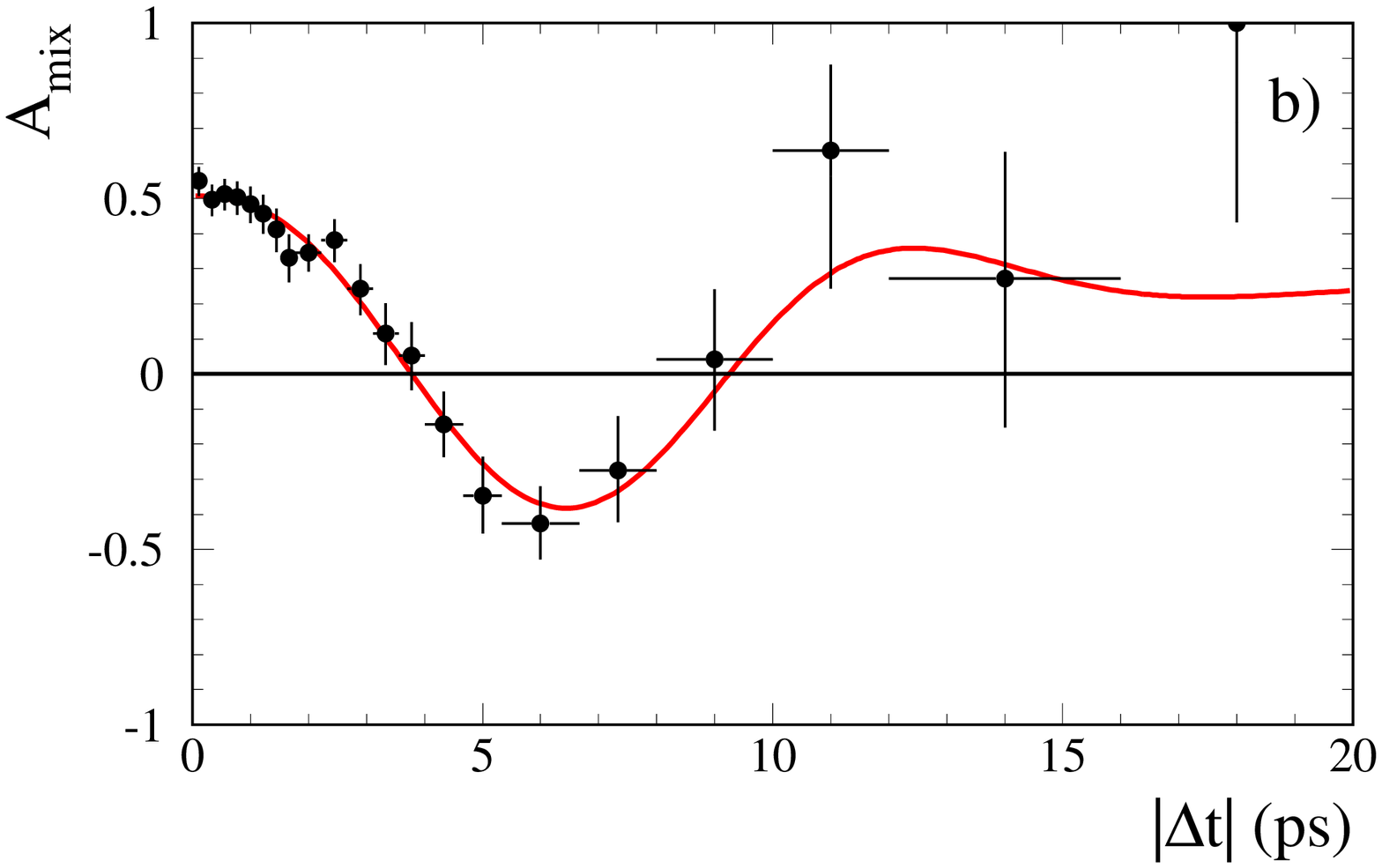}
  \end{center}
  \caption{Time-dependent asymmetry ${\cal A}(\deltat)$ between unmixed and
      mixed events for hadronic $B$ candidates with $\mes > 5.27$\gevcc, a) as
      a function of \deltat; and b) folded as a function of $\vert\deltat\vert$.
      The asymmetry in a) is due to the fitted bias in the \deltat\ resolution function.
      \label{fig:b0.excl-data.cosine}}   
\end{figure*}

The results from the likelihood fit to the mixing sample are summarized
in Table~\ref{tab:mixing-likelihood}. The probability to obtain a
likelihood smaller than the observed value, evaluated with
fast parameterized Monte Carlo simulation of a large number 
of similar experiments, is $(44\pm 1)\%$.
The $\Delta t$ distributions of the signal ($\mes>5.27\gevcc$) and
background ($\mes<5.27\gevcc)$ candidates, overlaid with the projection
of the likelihood fit, are shown in Fig.~\ref{fig:b0.excl-data.deltat.all}.
In Fig.~\ref{fig:b0.excl-data.cosine} the mixing asymmetry 
of Eq.~\ref{eq:asym} is plotted; the time-dependence of the mixing
probability is clearly visible.

The tagging separation $Q=\epsilon_{tag}(1-2\mistag)^2$ is  calculated from 
the efficiencies and the mistag rates quoted in Tables~\ref{tab:tagging-excl} 
and \ref{tab:mixing-likelihood} respectively. Summing over all tagging 
categories, we measure a combined effective tagging efficiency 
$Q\approx 27\%$.

Two small corrections, which
are described in more detail in Sections~\ref{sec:dm_background} and 
\ref{sec:dm_MC_validation} together with their assigned systematic errors,
are applied to the output of the fit.
The value of \deltamd obtained after applying these corrections is
\begin{equation}
\deltamd = 0.516\pm 0.016\pm 0.010\hbarps,\nonumber
\end{equation} 
where the first error is statistical and the second systematic.

We have also examined the fitted value for \deltamd\ with various
subsamples of the full data set, including individual $B$ decay channels,
separate tagging categories, the state of the reconstructed $B_{\rm rec}$
or tagging $B_{\rm tag}$, and different time periods.
As can be seen from Table~\ref{tab:mixing_studies}, the values obtained from the subsample
fits are all consistent with the global result for \deltamd.

\begin{table*}[!htb]
\caption{Results from the likelihood fit to the 
\deltat\ distributions of the hadronic \Bz\ decays. The value for 
\deltamd\ includes small corrections as described in the text.
 The first major column
contains the fit results, while the second major
column contains the correlation coefficients with respect to \deltamd\
for each fit parameter.
} \vspace{0.3cm}
\label{tab:mixing-likelihood}
\begin{center}
\begin{tabular}{|l|c|c|c|c|}
\hline
      Parameter & \multicolumn{2}{c}{Fit Result} & \multicolumn{2}{|c|}{Correlation} \\ \cline{2-5}
                & Run 1 & Run 2                    & Run 1 & Run 2 \\ \hline
      \deltamd [$\hbarps$]                             & \multicolumn{2}{|c|}{$\phanm 0.516 \pm 0.016$} & \multicolumn{2}{c|}{       } \\  \hline
      \multicolumn{5}{|c|}{Signal Resolution Function}\\   \hline
      $S_1$ (core)                     & $\phanm ~1.37  \pm 0.09~$ &$\phanm ~1.18 \pm 0.11$     & $\phanm 0.25 $ & $\phanm 0.16$ \\
      $b_1(\Delta t)$ {\tt lepton} (core)      & $\phanm 0.06   \pm 0.13 $ &$      -0.04   \pm  0.16  $ & $\phanm 0.08 $ & $\phanm 0.00$ \\
      $b_1(\Delta t)$ {\tt kaon} (core)        & $      -0.22   \pm 0.08 $ &$      -0.25   \pm  0.09  $ & $\phanm 0.03 $ & $\phanm 0.00$ \\
      $b_1(\Delta t)$ {\tt NT1} (core)         & $      -0.07   \pm 0.15 $ &$      -0.45   \pm  0.21  $ & $      -0.00 $ & $\phanm 0.00$ \\
      $b_1(\Delta t)$ {\tt NT2} (core)         & $      -0.46   \pm 0.12 $ &$      -0.20   \pm  0.16  $ & $\phanm 0.01 $ & $\phanm 0.03$ \\
      $b_2(\Delta t)$ (tail)             & $      -5.0    \pm 4.2  $ &$      -7.5  \pm 2.4$       & $\phanm 0.04 $ & $\phanm 0.06$ \\
      $f_2$(tail)                             & $\phanm 0.014 \pm 0.020$&$\phanm 0.015 \pm 0.010$ & $\phanm 0.06 $ & $\phanm 0.07$ \\
      $f_3$(outlier)                          & $\phanm 0.008 \pm 0.004$&$\phanm 0.000 \pm 0.014$ & $      -0.09 $ & $\phanm 0.01$ \\    \hline
      \multicolumn{5}{|c|}{Signal dilutions}  \\   \hline
      $\langle D \rangle$, {\tt lepton}           &\multicolumn{2}{c}{$\phanm 0.842  \pm  0.028$}  & \multicolumn{2}{|c|}{$\phanm   0.24$} \\
      $\langle D \rangle$, {\tt kaon}             &\multicolumn{2}{c}{$\phanm 0.669  \pm  0.023$}  & \multicolumn{2}{|c|}{$\phanm   0.30$} \\
      $\langle D \rangle$, {\tt NT1}              &\multicolumn{2}{c}{$\phanm 0.563  \pm  0.044$}  & \multicolumn{2}{|c|}{$\phanm   0.11$} \\
      $\langle D \rangle$, NT2              &\multicolumn{2}{c}{$\phanm 0.313  \pm  0.041$}  & \multicolumn{2}{|c|}{$\phanm   0.11$} \\
      $\Delta D$, {\tt lepton}                    &\multicolumn{2}{c}{$      -0.006  \pm  0.045$}  & \multicolumn{2}{|c|}{$\phanm   0.02$} \\
      $\Delta D$, {\tt kaon}                      &\multicolumn{2}{c}{$\phanm 0.024  \pm  0.033$}  & \multicolumn{2}{|c|}{$\phanm   0.01$} \\
      $\Delta D$, {\tt NT1}                       &\multicolumn{2}{c}{$      -0.086  \pm  0.068$}  & \multicolumn{2}{|c|}{$\phanm   0.00$} \\
      $\Delta D$, {\tt NT2}                       &\multicolumn{2}{c}{$\phanm 0.100  \pm  0.060$}  & \multicolumn{2}{|c|}{$\phanm   0.00$} \\ \hline
      \multicolumn{5}{|c|}{Background properties}\\   \hline
      $\tau$, mixing bkgd [ps]               &\multicolumn{2}{c}{$\phanm 0.853\pm 0.036$}  & \multicolumn{2}{|c|}{$      -0.01$} \\
      $f(\tau=0)$, mixing bkgd, {\tt lepton}       &\multicolumn{2}{c}{$\phanm 0.05 \pm 0.10 $}  & \multicolumn{2}{|c|}{$\phanm 0.01$} \\
      $f(\tau=0)$, mixing bkgd, {\tt kaon}         &\multicolumn{2}{c}{$\phanm 0.42 \pm 0.05 $}  & \multicolumn{2}{|c|}{$\phanm 0.01$} \\
      $f(\tau=0)$, mixing bkgd, {\tt NT1}          &\multicolumn{2}{c}{$\phanm 0.33 \pm 0.08 $}  & \multicolumn{2}{|c|}{$\phanm 0.01$} \\
      $f(\tau=0)$, mixing bkgd, {\tt NT2}          &\multicolumn{2}{c}{$\phanm 0.32 \pm 0.08 $}  & \multicolumn{2}{|c|}{$\phanm 0.01$} \\     \hline
      \multicolumn{5}{|c|}{Background resolution function}\\   \hline
      $S_1$ (core)                          &$\phanm 1.211  \pm  0.043 $&$\phanm   1.131  \pm  0.046$  & $-0.00$ & $\phanm 0.00$ \\
      $b_1(\Delta t)$ (core)                    &$      -0.135  \pm  0.031 $&$        -0.015  \pm  0.038$  & $-0.00$ & $      -0.00$ \\
      $f_3$ (outlier)                          &$\phanm 0.022  \pm  0.004 $&$\phanm   0.036  \pm  0.007$  & $-0.01$ & $\phanm 0.02$ \\        \hline
      \multicolumn{5}{|c|}{Background dilutions}  \\   \hline
      $\langle D \rangle$, {\tt lepton}, $\tau =0$&\multicolumn{2}{c}{$\phanm  0.0   \pm  2.9  $}  & \multicolumn{2}{|c|}{$      -0.02$} \\
      $\langle D \rangle$, {\tt kaon}, $\tau =0$  &\multicolumn{2}{c}{$\phanm  0.52  \pm  0.08  $}  & \multicolumn{2}{|c|}{$      -0.03$} \\
      $\langle D \rangle$, {\tt NT1}, $\tau =0$   &\multicolumn{2}{c}{$\phanm  0.67  \pm  0.27  $}  & \multicolumn{2}{|c|}{$      -0.01$} \\
      $\langle D \rangle$, {\tt NT2}, $\tau =0$   &\multicolumn{2}{c}{$\phanm -0.05  \pm  0.13  $}  & \multicolumn{2}{|c|}{$      -0.00$} \\
      $\langle D \rangle$, {\tt lepton}, $\tau >0$&\multicolumn{2}{c}{$\phanm  0.34  \pm  0.13  $}  & \multicolumn{2}{|c|}{$\phanm 0.02$} \\
      $\langle D \rangle$, {\tt kaon}, $\tau >0$  &\multicolumn{2}{c}{$\phanm  0.26  \pm  0.06  $}  & \multicolumn{2}{|c|}{$\phanm 0.04$} \\
      $\langle D \rangle$, {\tt NT1}, $\tau >0$   &\multicolumn{2}{c}{$\phanm -0.13  \pm  0.11  $}  & \multicolumn{2}{|c|}{$\phanm 0.01$} \\
      $\langle D \rangle$, {\tt NT2}, $\tau >0$   &\multicolumn{2}{c}{$\phanm  0.12  \pm  0.031  $}  & \multicolumn{2}{|c|}{$\phanm 0.01$} \\  \hline
\end{tabular}
\end{center}
\end{table*}

\begin{table}[!htb]
\caption{
Result of fitting for \deltamd\ in the entire $B_{\rm flav}$ sample and in 
various subsamples. The difference in the fitted value of \deltamd\ versus 
the result from the fit to the full $B_{\rm flav}$ sample are reported. 
}
\vspace{0.3cm}
\begin{center}
\begin{tabular}{|l|c|} \hline
 Sample              & $\deltamd -\deltamd(all)$ \\ \hline \hline
\multicolumn{2}{|l|}{Decay mode}         \\ \hline
 \ \ $\Dstarm\pip$        & $-0.029\pm 0.030$ \\
 \ \ $\Dstarm\rho^+$      & $+0.017\pm 0.039$ \\
 \ \ $\Dstarm a_1^+$      & $+0.066\pm 0.063$ \\
 \ \ $\Dm\pip$            & $+0.022\pm 0.030$ \\
 \ \ $\Dm\rho^+$          & $-0.031\pm 0.038$ \\
 \ \ $\Dm a_1^+$          & $-0.033\pm 0.041$ \\
 \ \ $\Dstarm X$          & $+0.000\pm 0.025$ \\
 \ \ $\Dm X$              & $-0.005\pm 0.023$ \\
\hline\hline
\multicolumn{2}{|l|}{Tagging category}   \\ \hline
 \ \ {\tt Lepton}         & $+0.005\pm 0.026$\\ 
 \ \ {\tt Kaon}           & $+0.002\pm 0.023$\\
 \ \ {\tt NT1}            & $-0.032\pm 0.044$\\
 \ \ {\tt NT2}            & $+0.12 \pm 0.10$\\ \hline
\multicolumn{2}{|l|}{$B_{\rm rec}$ state} \\ \hline
 \ \ $B_{\rm rec} = \Bzb$ & $+0.015\pm 0.023$\\
 \ \ $B_{\rm rec} = \Bz$  & $-0.003\pm 0.023$\\ \hline
\multicolumn{2}{|l|}{$B_{\rm tag}$ state} \\ \hline
 \ \ $B_{\rm tag} = \Bzb$ & $+0.019\pm 0.023$ \\
 \ \ $B_{\rm tag} = \Bz$  & $-0.007\pm 0.022$\\ \hline
\multicolumn{2}{|l|}{Data sample} \\ \hline
 \ \ Run 1                & $ -0.012\pm 0.022$ \\
 \ \ Run 2                & $ +0.019\pm 0.025$ \\ \hline
\end{tabular}
\end{center}
\label{tab:mixing_studies}
\end{table}

\subsection{Systematic error estimation} 

Systematic errors can be grouped into four categories:
signal properties and description, background properties and description, 
fixed external parameters
and statistical limitations of Monte Carlo validation tests of the 
fitting procedure. A summary
of these sources for the hadronic \Bz\ sample
is shown in Table~\ref{tab:syst-had}. In the following,
the individual contributions are referenced by the lettered lines in this table.

\begin{table}[htb]
\begin{center} 
 \caption{Systematic uncertainties and contributions to statistical errors for
    \deltamd\ obtained with the likelihood fit to the hadronic \Bz\ sample.
    \label{tab:syst-had}}  \vspace{0.3cm}
  \begin{tabular}{|l|c|}\hline
    Source  & $\sigma(\deltamd)$  \\
            & $[{\rm ps}^{-1}]$         \\
    \hline\hline
\multicolumn{2}{|c|}{Signal properties} \\ \hline
(a) SVT alignment                  &  0.004 \\  
(b) \deltat\ outlier description   &  0.002 \\  
(c) Beamspot position/size         &  0.001 \\  
(d) $\sigma_{\deltat}$ requirement &  0.003 \\ \hline
\multicolumn{2}{|c|}{Background properties} \\ \hline
(e) Background fraction            &  0.002 \\ 
(f) Background \deltat\ structure  &  0.001 \\ 
(g) Background \deltat\ resolution &  0.001 \\ 
(h) Sideband extrapolation         &  0.002 \\ 
(i) Peaking \Bu\ background        &  0.002 \\ \hline 
\multicolumn{2}{|c|}{External parameters} \\ \hline
(j) $z$ scale                      &$<$0.002\\
(k) $z$ boost (parameters)         &  0.001\\
(l) $z$ boost (method)             &  0.001\\ 
(m) \Bz\ lifetime                  &  $0.006$ \\ \hline
\multicolumn{2}{|c|}{Monte Carlo studies} \\ \hline
(n) Signal MC statistics           &  0.003 \\ 
(o) Tag-side $D$ composition \& lifetime & 0.001   \\ 
(p) Right/wrong tag resolution differences& 0.001 \\
    \hline\hline
    Total systematic error & 0.010 \\
    \hline
    Statistical error      & 0.016 \\
    $\;\;\;$Contribution due to resolution function   & 0.005 \\
    $\;\;\;$Contribution due to mistag rate           & 0.005 \\
    \hline\hline
    Total error            & 0.019 \\
        \hline  
   \end{tabular}
\end{center}
\end{table}

\subsubsection{Signal properties and description}
\label{sec:sigpropsyst}

For the signal events, the use of a double Gaussian plus 
outlier model for re-scaling the event-by-event
$\Delta t$ errors as part of the likelihood fit
means that uncertainties in the vertex resolution are incorporated 
into the statistical error on \deltamd, including proper treatment 
of all correlations. Assuming that this model is sufficiently
flexible to accommodate the observed distribution in data, no 
additional systematic error need be assigned. The contribution 
to the total statistical error due to the vertex resolution can 
be extracted by fitting the data twice: once holding all parameters 
except \deltamd\ fixed, and once allowing the
resolution function parameters to vary in addition to \deltamd. 
Subtracting in quadrature the
respective errors on \deltamd\ from the two fits shows that 
$\pm 0.005\hbarps$ of the statistical
error can be attributed to the resolution parameters.

To determine the systematic error due to the assumed parameterization 
of the resolution model, we apply a number of possible misalignment 
scenarios to a sample of simulated events.
By comparing the value of \deltamd\ derived from these misaligned 
samples to the case of perfect alignment, we derive a systematic 
uncertainty of $\pm 0.004\hbarps$ (a).

An additional systematic error is attributed to uncertainties in 
the treatment of the small fraction of \deltat\ outliers that are 
the result of misreconstructed vertices. The stability of the
\deltamd\ result is examined under variation of
the width of the third Gaussian component in the resolution 
function between 6 and 18\ps, and through replacement of
the third Gaussian with a uniform distribution and varying the 
width between 8 and 40\ps. On this basis, we attribute
a systematic uncertainty of $\pm 0.002\hbarps$ to the outlier treatment (b).

As described in detail in Section~\ref{subsec:dz_reconstruction}, 
the beamspot position is an integral part of
the determination of \deltat. Increasing its vertical size by up 
to $80\mum$, and systematically biasing its vertical position by 
up to $80\mum$, results in a corresponding variation of \deltamd\ by 
less than $0.001\hbarps$ (c).

The requirement on the maximum allowed value of $\sigma_{\deltat}$ 
is varied between 1 and 2.4\ps, and the observed variation of 
$0.003\hbarps$ in \deltamd is assigned as a systematic uncertainty (d). 
The observed dependence is mainly due to correlations
between tagging and vertexing, as described in Sec.~\ref{sec:dm_MC_validation}.

\subsubsection{Background properties}
\label{sec:dm_background}

A systematic uncertainty in \deltamd\ arises from our ability
to separate signal from background as a function of $m_{ES}$.
We estimate this uncertainty by varying the width and height of the 
fitted Gaussian peak in \mes, the slope parameter of the ARGUS background 
shape, and the normalizations of the signal and backgrounds
by one standard deviation around their central values, resulting in
an uncertainty of $\pm 0.002\hbarps$ in \deltamd (e).

As discussed in Sec.~\ref{cpm1background}, 
the \deltat\ distribution of the background is described
by the combination of a prompt component and a lifetime component. To estimate
the systematic uncertainty due to this choice, we add an additional component, 
with its own separate lifetime, that is allowed to mix;  the
observed value of \deltamd\ changes by $0.001\hbarps$ (f). Similarly, 
adding an additional Gaussian distribution to the \deltat\ background
resolution function changes \deltamd\ by no more than $0.001\hbarps$ (g).

Finally, the composition of the background changes slightly as 
a function of \mes, since the fraction of background due to continuum 
production slowly decreases towards the \B\ mass. As a result,
the \deltat\ structure of the background could change as well. To
study this dependence, we split the \mes\ sideband region into 
seven mutually exclusive, 10\mevcc-wide intervals, and repeat the \deltamd\ fit
with each of these slices in turn. The variation of \deltamd\ is then 
extrapolated as a function of the position of the sideband slice relative to
the $B$ mass. We correct the value of \deltamd\ by $-0.002\hbarps$
obtained from this extrapolation, and assign the statistical uncertainty 
of $0.002\hbarps$ of this procedure as a systematic error on \deltamd (h).

A small fraction (about 1.5\%) of the events attributed to the \Bz\ 
signal by the fit to the \mes\ distribution consists of \Bu\ events, 
mainly due to the swapping of a soft \piz\ with a charged pion
as described in Section~\ref{subsec:sample_hadronicBz}. 
The uncertainty on this peaking fraction is propagated to \deltamd, 
yielding a systematic error of $0.002\hbarps$ (i).

\begin{figure}[htb]
\begin{center}
    \includegraphics[width=\linewidth]{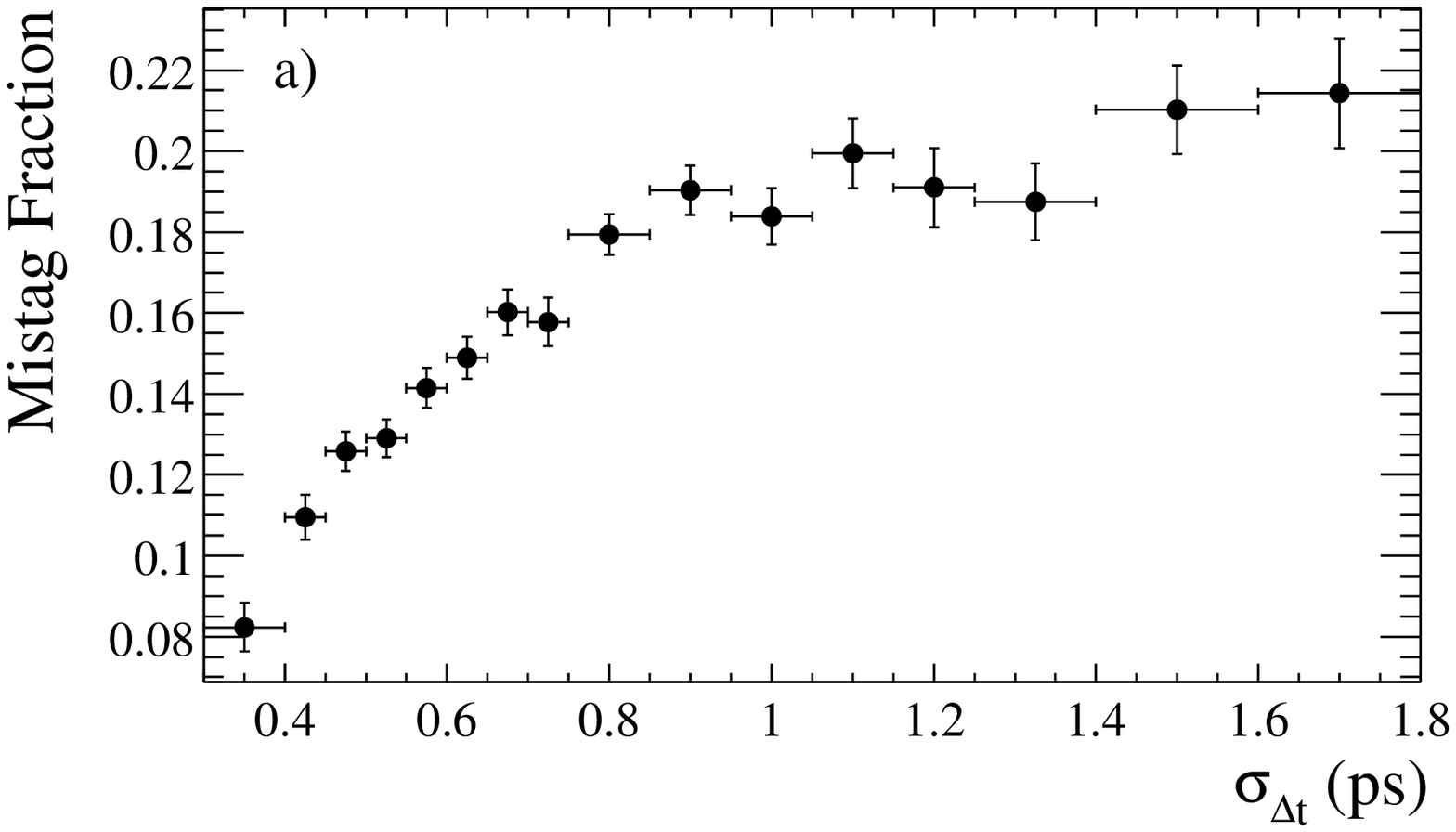}
    \includegraphics[width=\linewidth]{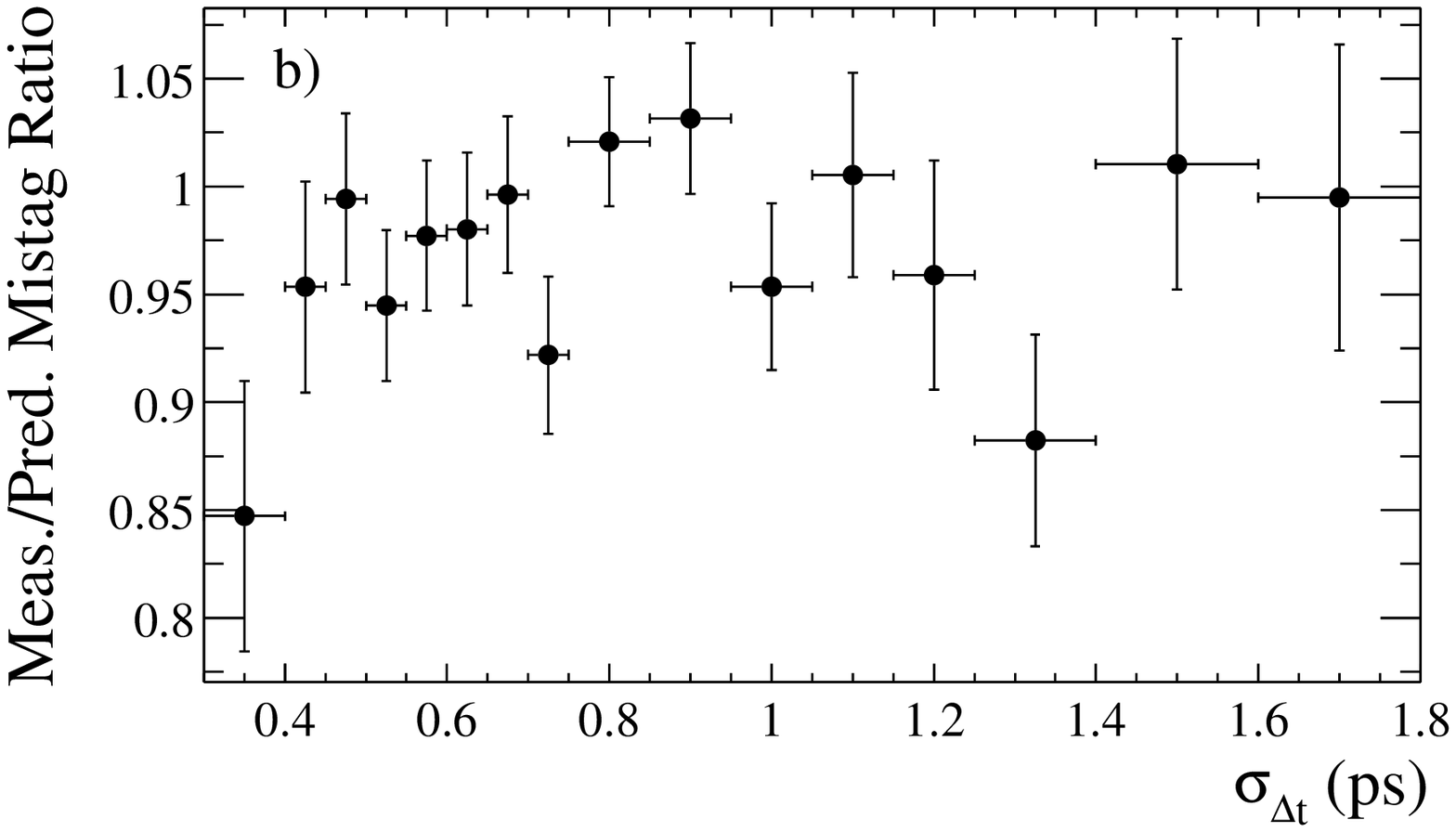}
\end{center}
\caption{a) Correlation between the event-by-event error on 
\deltat ($\sigma_{\deltat}$) and the mistag rate in the {\tt Kaon}
category from Monte Carlo simulation; b) Dependence of mistag
rate on $\sigma_{\deltat}$ after scaling the mistag rate by $\sqrt{\sum p^2_t}$.
\label{fig:mistag_sigt_correlations}}
\end{figure}

\subsubsection{External parameters}

An error in the boost of the \FourS\ system (0.1\%)
or in the knowledge of the $z$ scale of the
detector, as described in Section~\ref{subsec:dz_reconstruction},
could bias the \deltamd\ measurement because these parameters are used to
reconstruct the decay length difference $\Delta z$ and to convert it 
to the decay time difference \deltat. The uncertainties on these quantities 
are propagated to \deltamd\ and lead to systematic uncertainties of 
$0.001\hbarps$ (l) and less than $0.002\hbarps$ (j), respectively.
In addition to these, we assign the difference of $0.001\hbarps$ (k) in 
the value of \deltamd\ obtained by using the \deltaz\ to \deltat\ conversion
described in Eq.~\ref{eq:dt_boost_approx} instead of 
Eq.~\ref{eq:dt_taub_approx} as a systematic error.
Finally, in the likelihood fit, we fix the \Bz\ lifetime to the PDG
value~\cite{PDG2000}. The present uncertainty on this value of
$\pm 0.032\ps$ leads to a systematic error of $\mp 0.006\hbarps$ (m).

\subsubsection{Monte Carlo validation of measurement technique}
\label{sec:dm_MC_validation}

Candidate selection criteria, or the analysis and fitting
procedure, could potentially cause systematic biases
in the measurement of \deltamd. These potential biases are
estimated by repeating the analysis with a large sample of 
Monte Carlo events, which are generated with the full GEANT3~\cite{geant}
detector simulation.  In the Monte Carlo sample, the fitted result for 
\deltamd\ is shifted by $+0.007\pm 0.003 \hbarps$ from the input value. 
A corresponding correction with this central value is applied to the 
fitted result with data, and the uncertainty is assigned as a systematic 
error (n).

The main cause of this bias is a small
correlation between the mistag rate and the \deltat\ resolution 
that is not modeled in the likelihood function. This correlation is 
seen most readily in data for {\tt Kaon} tags and is 
shown for simulation in Fig.~\ref{fig:mistag_sigt_correlations}a.
We find that both the mistag rate for kaon tags and the event-by-event error
$\sigma_{\deltat}$ depend inversely on $\sqrt{\sum p^2_t}$, where $p_t$
is the transverse momentum with respect to the $z$ axis of tracks from
the $B_{\rm tag}$ decay. Correcting for this dependence of the mistag
rate removes most of the correlation between the mistag rate and
$\sigma_{\deltat}$, as can be seen in Fig.~\ref{fig:mistag_sigt_correlations}b.
The mistag rate dependence originates from the kinematics of
the physics sources for wrong-charge kaons. The
three major sources of mistags are wrong-sign $D^0$ mesons from 
\B\ decays to double charm, wrong-sign kaons from $D^+$ decays,
and kaons produced directly in \B\ decays. All these sources produce a spectrum
of charged tracks that have smaller $\sqrt{\sum p^2_t}$ than \B\ decays that
produce a correct tag. The \deltat\ resolution dependence originates from
the $1/p_t^2$ dependence of $\sigma_z$ for the individual contributing
tracks.

Since the effect is small and well described by the Monte Carlo simulation,
we have chosen to treat the impact of this correlation as a correction, rather
than building the effect into the likelihood function.
We include additional systematic errors related to the tag-side properties
that could affect the accuracy of the description of this correlation in
the simulation. In particular, the \Dz, \Dp, and \Ds meson branching 
fractions, the $D$ meson lifetimes, and the wrong-sign kaon production rates
in $B$ meson decays are all varied. These studies lead to an
assigned systematic error of $\pm 0.001\hbarps$ (o). 

In addition, we consider
the possibility that correctly and incorrectly tagged events could have
different resolution functions. Based on Monte Carlo studies of the variation
in the fitted value for \deltamd\ with and without allowing for
independent resolution functions for correctly and incorrectly tagged
events, an uncertainty of $\pm 0.001\hbarps$ is assigned to this source (p).

\subsection{Validation studies and cross checks}

\subsubsection{Monte Carlo studies}

A high-precision test of the fitting procedure was performed with
fast parameterized Monte Carlo simulations, where 2000 experiments
were generated with sample size and composition corresponding to 
that obtained from the actual data. The mistag rates and 
\deltat\ distributions were generated according to the model used in the 
likelihood fit. The full fit was then performed on each of these experiments.
The resulting distribution of pulls (defined as the difference between 
the fitted and generated value of a parameter divided by the statistical 
error as obtained from the likelihood fit) has a mean $-0.038\pm 0.022$ 
and standard deviation $1.012\pm 0.023$, consistent
with no measurement bias in either the value of \deltamd\ or its
estimated error.

\subsubsection{Simple counting experiment}

If the mistag rate is known, the time-integrated
fraction $\chi_d=\frac{1}{2}x_d/(1+x_d^2)$ of mixed events 
can be determined from the $B_{\rm flav}$ sample by counting mixed
and unmixed events. 
The value for $\chi_d$ obtained by this means, after correcting
for the mistag rates obtained from the full time-dependent fit and assuming the PDG value
for the \Bz\ lifetime, leads to a value of $\deltamd=x_d/\tau_{\Bz}$ that differs from the
likelihood-fit result by $-0.003\pm 0.013\hbarps$, where the quoted error is the 
difference in quadrature of the statistical errors of both measurements.

Due to the choice for normalization of the likelihood ${\cal L}_{\rm mix}$, 
the time-integrated ratio of the number of mixed to unmixed events
contributes to our measurement of \deltamd. Alternatively, 
it is possible to normalize the
likelihoods of mixed and unmixed events individually, in which case
\deltamd\ is determined
solely from the shape of \deltat\ distributions.  
The value of \deltamd\ determined by
the \deltat\ distributions alone
differs from the full measurement by $0.003\pm 0.015 \hbarps$,
where the quoted error is given by the difference in quadrature of the
statistical errors of the two measurements.

\subsubsection{Cross check with $\tau_{\Bz}$}

If we allow the value of $\tau_{\Bz}$ to float in the \deltamd fit
the value of \deltamd increases by $0.008\pm 0.007\hbarps$, and
the lifetime is found to be $1.51 \pm 0.03\ps$, consistent with
our recent measurement~\cite{babar0106}.  We have also
performed a series of fits with fixed values for
$\deltamd$ and $\tau_{\Bz}$
in order to determine the dependence of \deltamd\ on
$\tau_{\Bz}$, which is found to be 
\begin{equation}
\deltamd = \left[0.516 - 0.279\left(\frac{\tau_{\Bz}}{1.548\ps}-1\right)\right]\hbarps.
\end{equation}

\renewcommand{\secname}{Sin2beta}
\section{\boldmath \CP\ violation in neutral \boldmath \B\ decays}
\label{sec:\secname}
\subsection{Likelihood fit results for \boldmath \stwob}
\label{sec:fitresults}

The value of \stwob, 
the dilution factors ${\cal D}_i$, the \deltat\
resolution parameters $\hat {a}_i$, and the background fractions and
time distribution parameters are extracted with an unbinned maximum-likelihood fit
to the flavor-eigenstate $B_{\rm flav}$ and $B_{\CP}$ samples
as described in Section~\ref{sec:Likelihood}.
We also demand a
valid tag and \deltat\ determination for the event,
based on the algorithms described in Sections~\ref{sec:Tagging}
and \ref{sec:vertexing}. The looser requirements
$\vert\deltat\vert<20\ps$ and $\sigma_{\deltat}<2.4\ps$
are applied to the proper time difference measurement. 
The fit results are summarized in Table~\ref{tab:summary-sin2beta} 
together with the correlation of the parameters with \stwob.
The mistag fractions and
vertex parameters are predominantly determined by the $B_{\rm flav}$
sample. The \CP\ asymmetry and parameters describing the background 
for the \CP\ events
are determined by the \CP\ sample. The value of \stwob\ obtained from the
combined $\etaCP=-1$, $\etaCP=+1$, and $\jpsi\Kstarz$ \CP\ samples is
\begin{equation}
\stwob=0.59\pm 0.14\pm 0.05, \nonumber
\end{equation} 
where the first error is statistical and the second systematic.

\begin{figure}[tbh]
\begin{center}
    \includegraphics[width=0.85\linewidth]{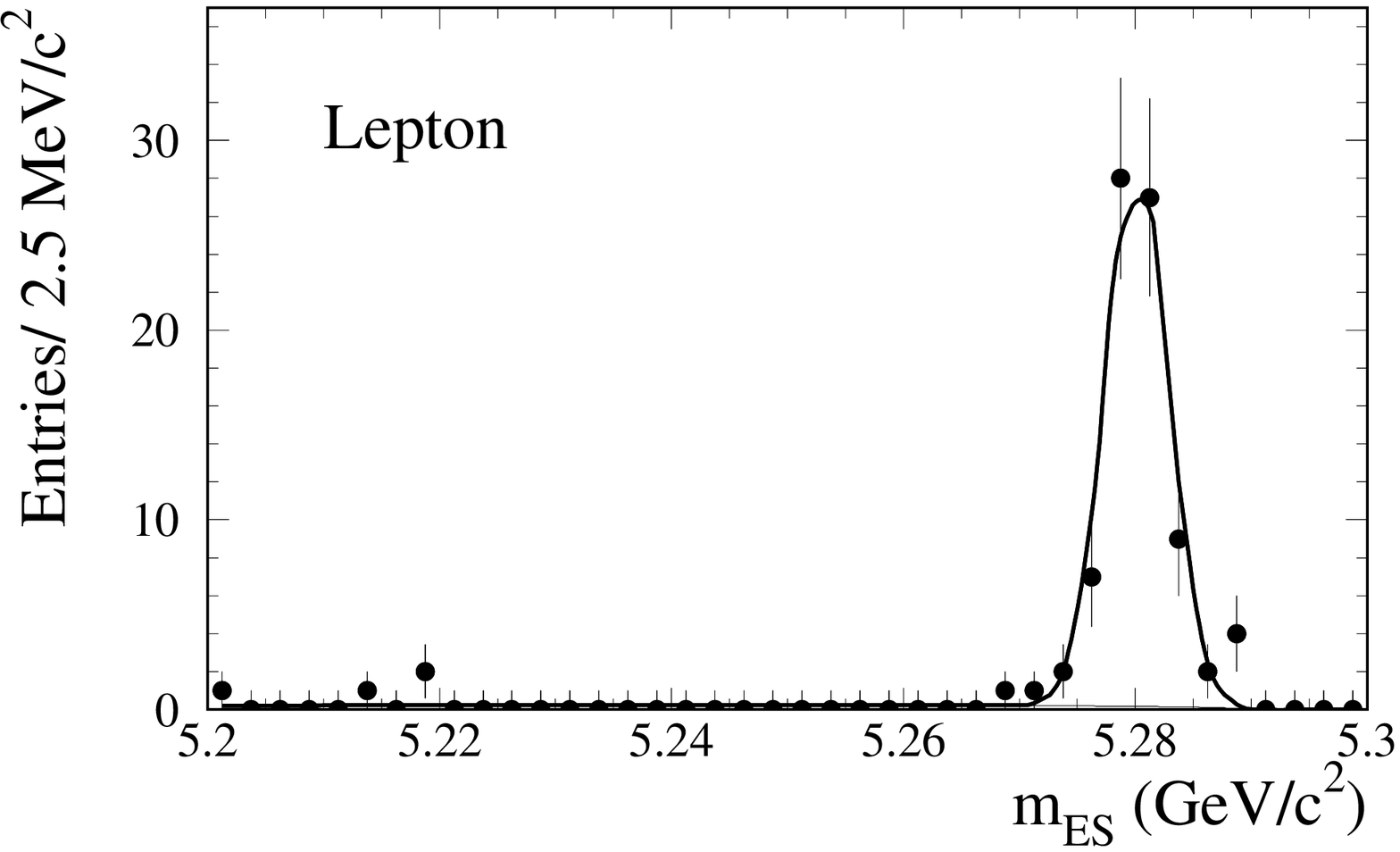}
    \includegraphics[width=0.85\linewidth]{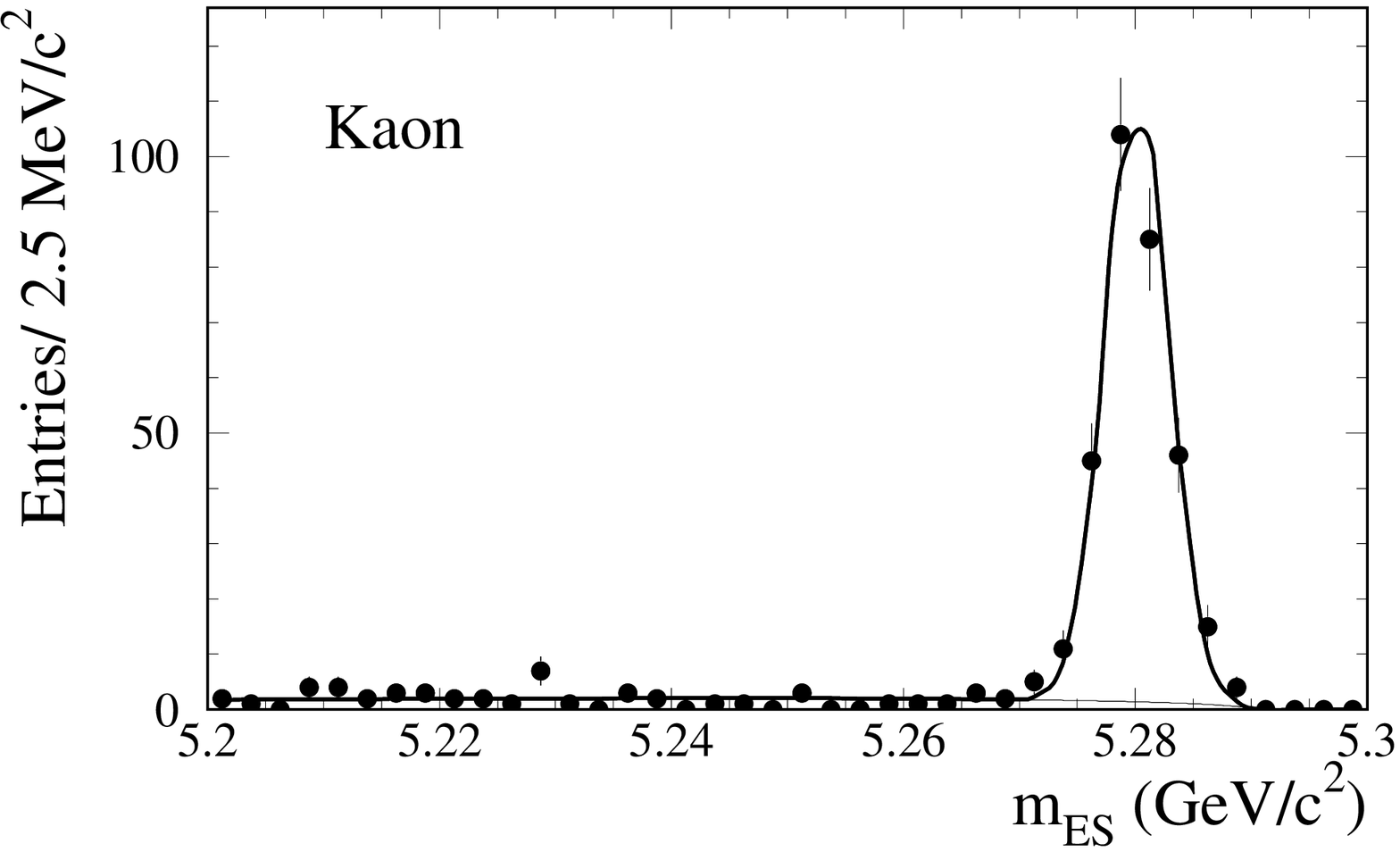}
    \includegraphics[width=0.85\linewidth]{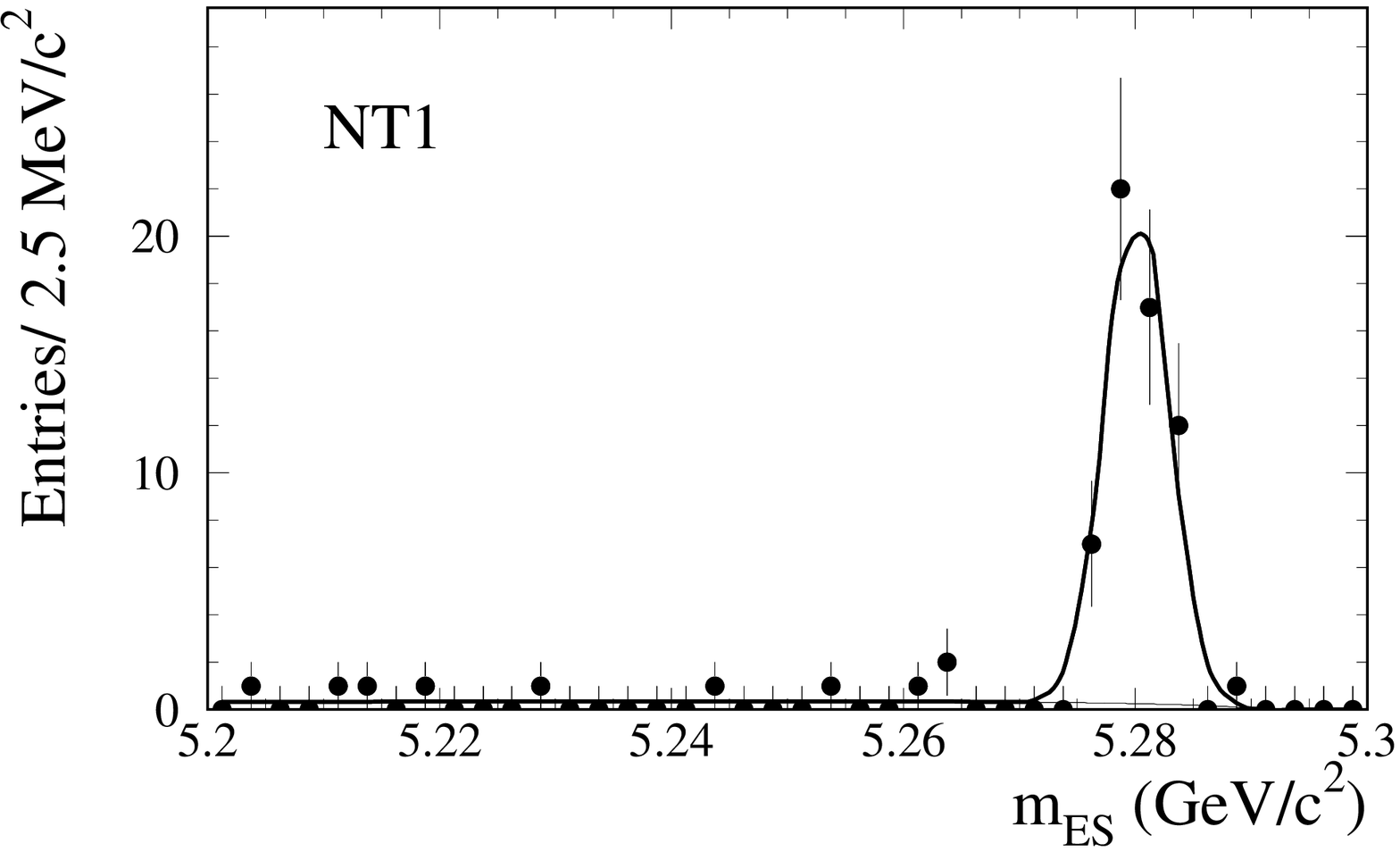}
    \includegraphics[width=0.85\linewidth]{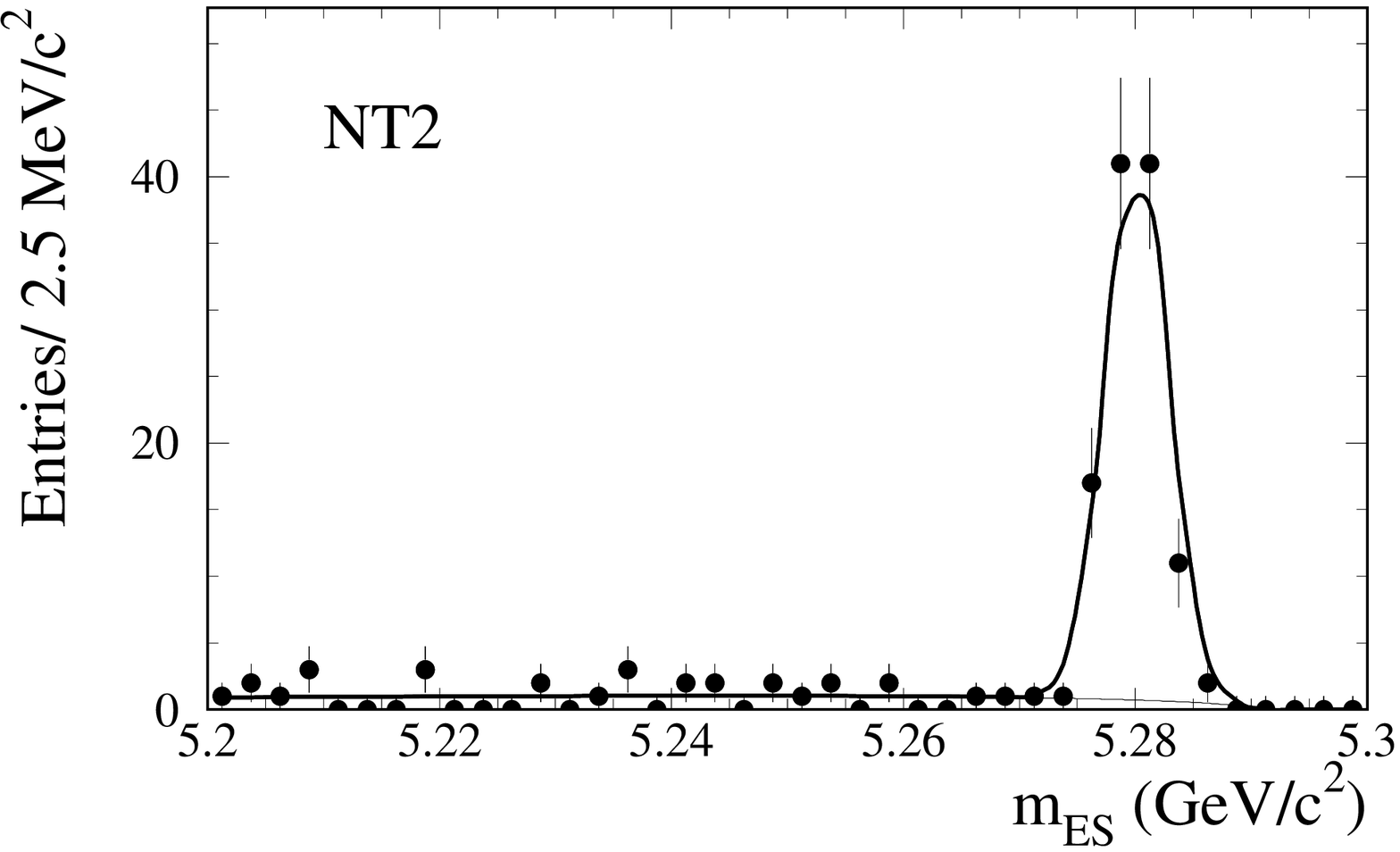}
\end{center}
\caption{Distribution of \mes\ for $\etaCP=-1$ candidates in
separate tagging categories ({\tt Lepton}, {\tt Kaon}, {\tt NT1} and {\tt NT2}),
overlaid with the result of a fit with a Gaussian distribution
for the signal and an ARGUS function for the background.
\label{fig:CPminus.mb-cats}}
\end{figure}

The \mes\ distribution for events in $\etaCP=-1$ modes, 
separated into tagging categories, is shown in
Fig.~\ref{fig:CPminus.mb-cats}. 
The signal probability $f^{\CP}_{i,{\rm sig}}(\mes)$ 
for the $\etaCP=-1$ sample
is determined from these fits as described 
in Section~\ref{cpm1background}.

\begin{figure*}[!htb]
  \begin{center}
  \includegraphics[width=0.49\linewidth]{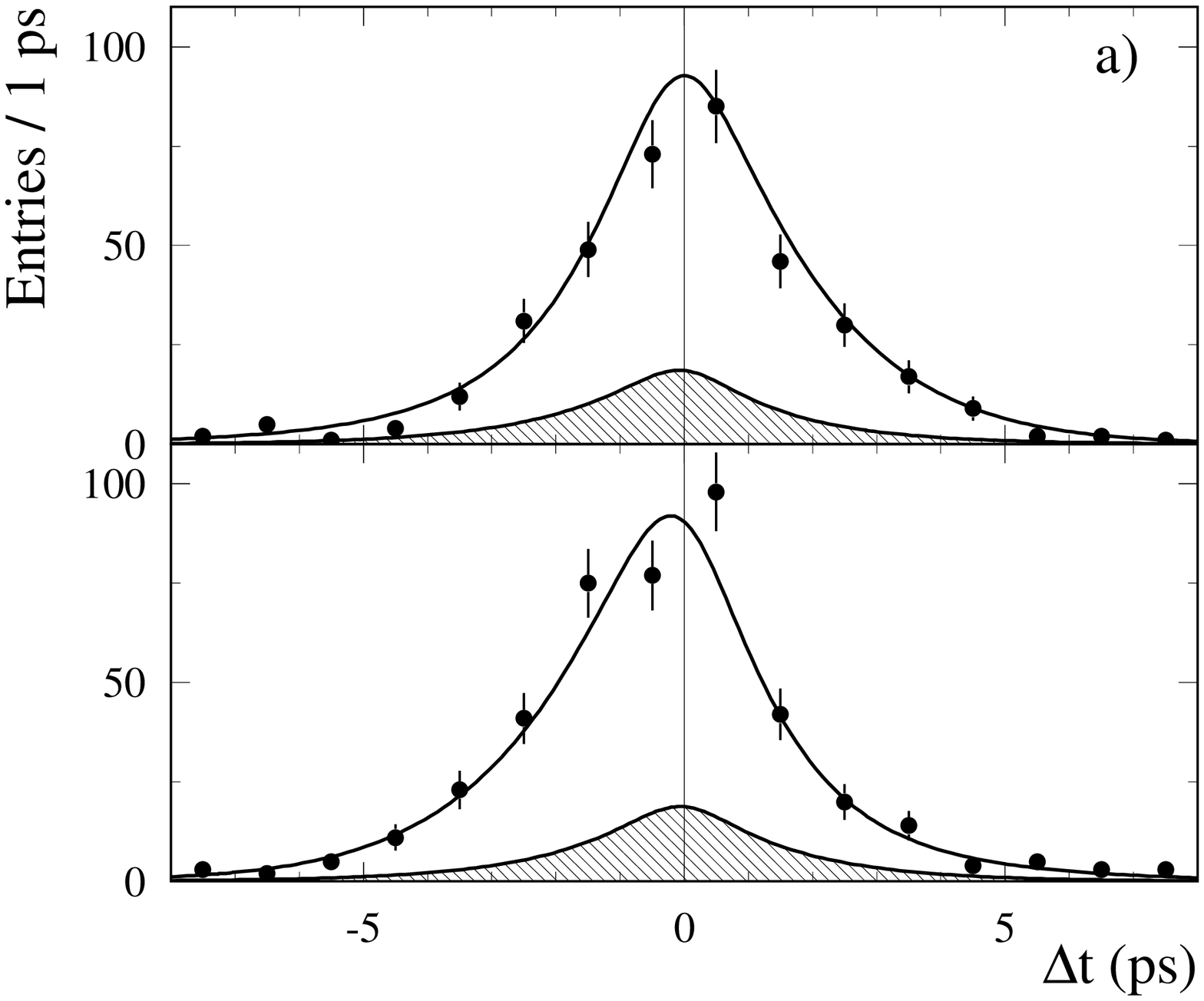}
  \includegraphics[width=0.49\linewidth]{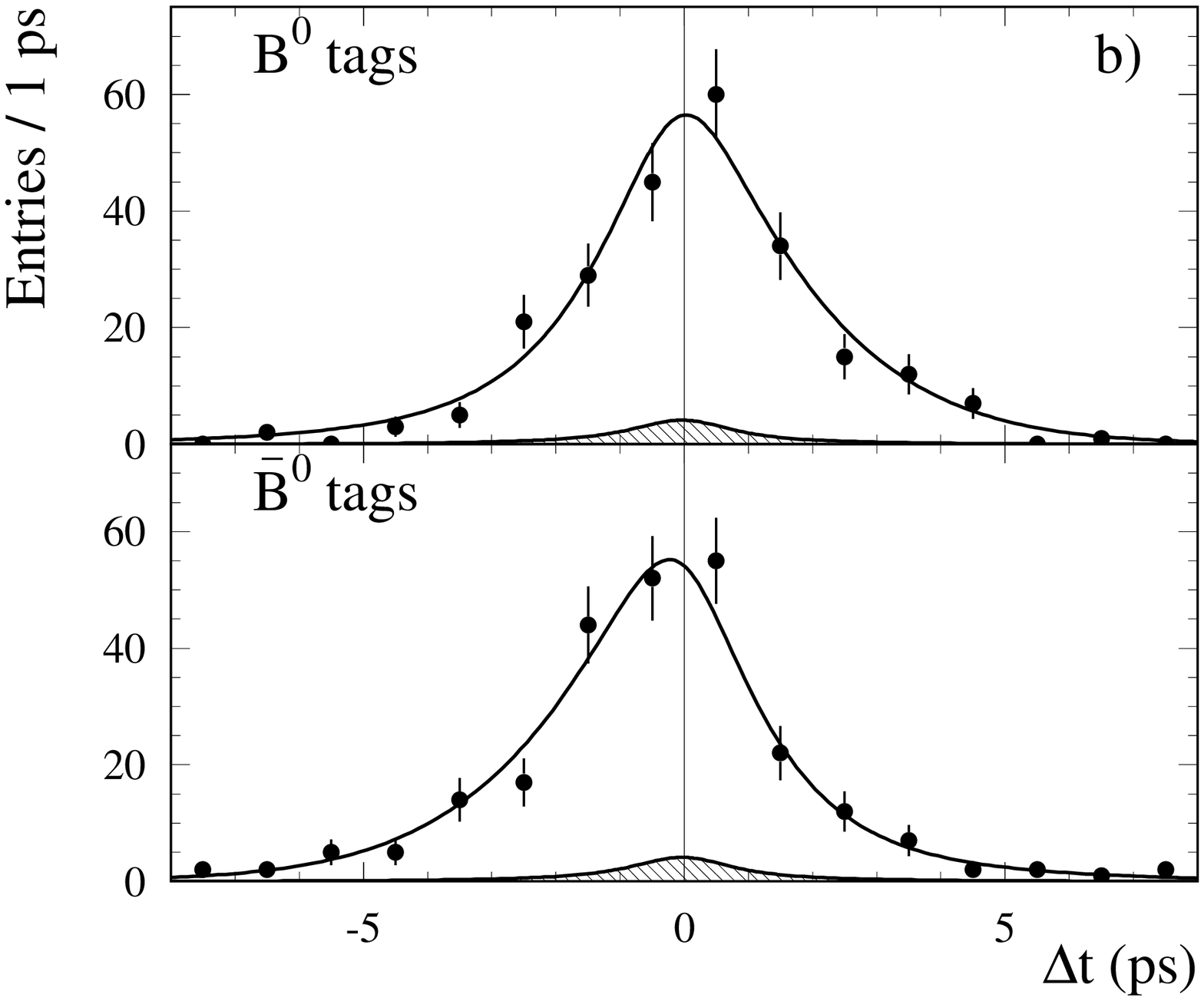}
  \includegraphics[width=0.49\linewidth]{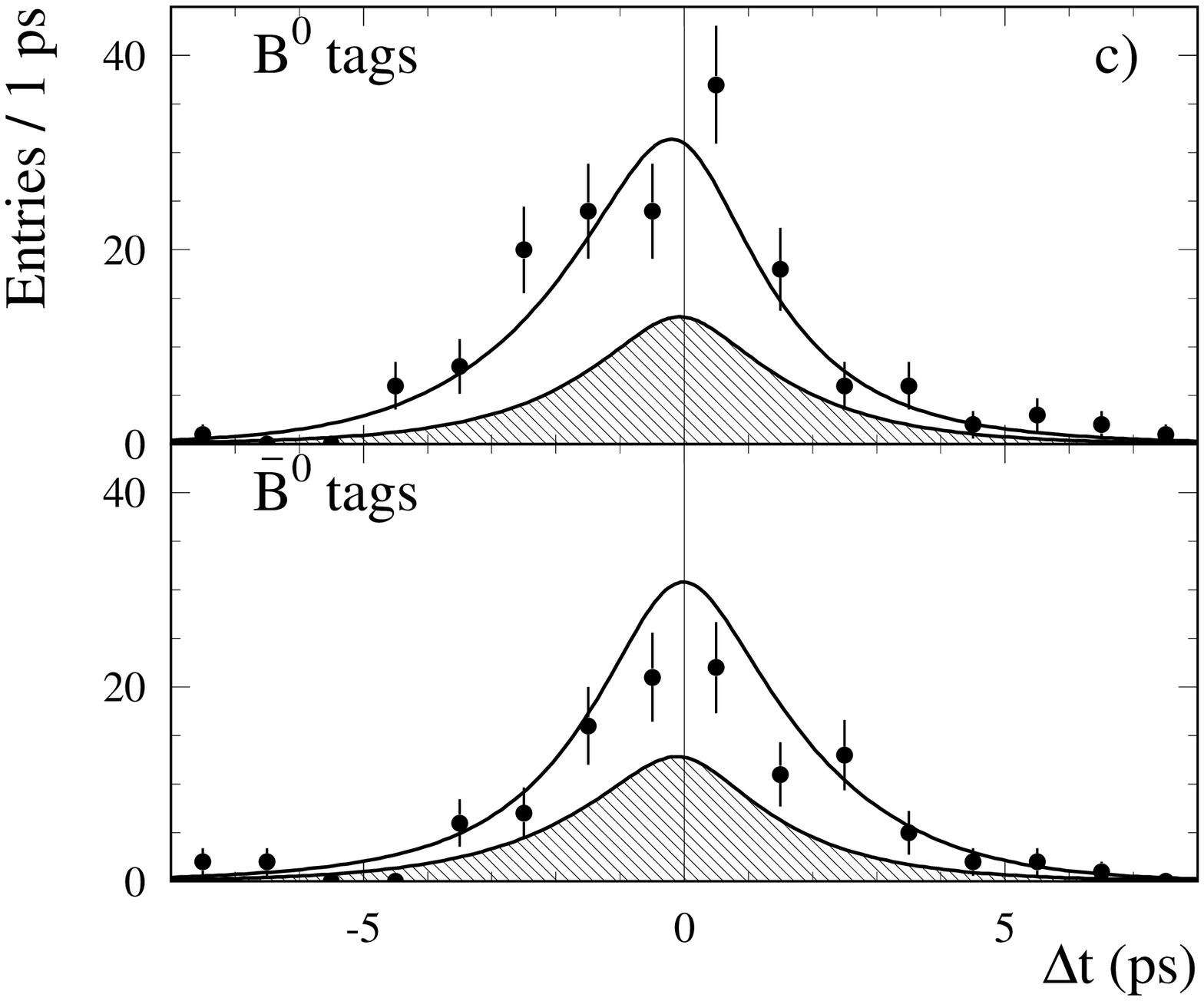}
  \includegraphics[width=0.49\linewidth]{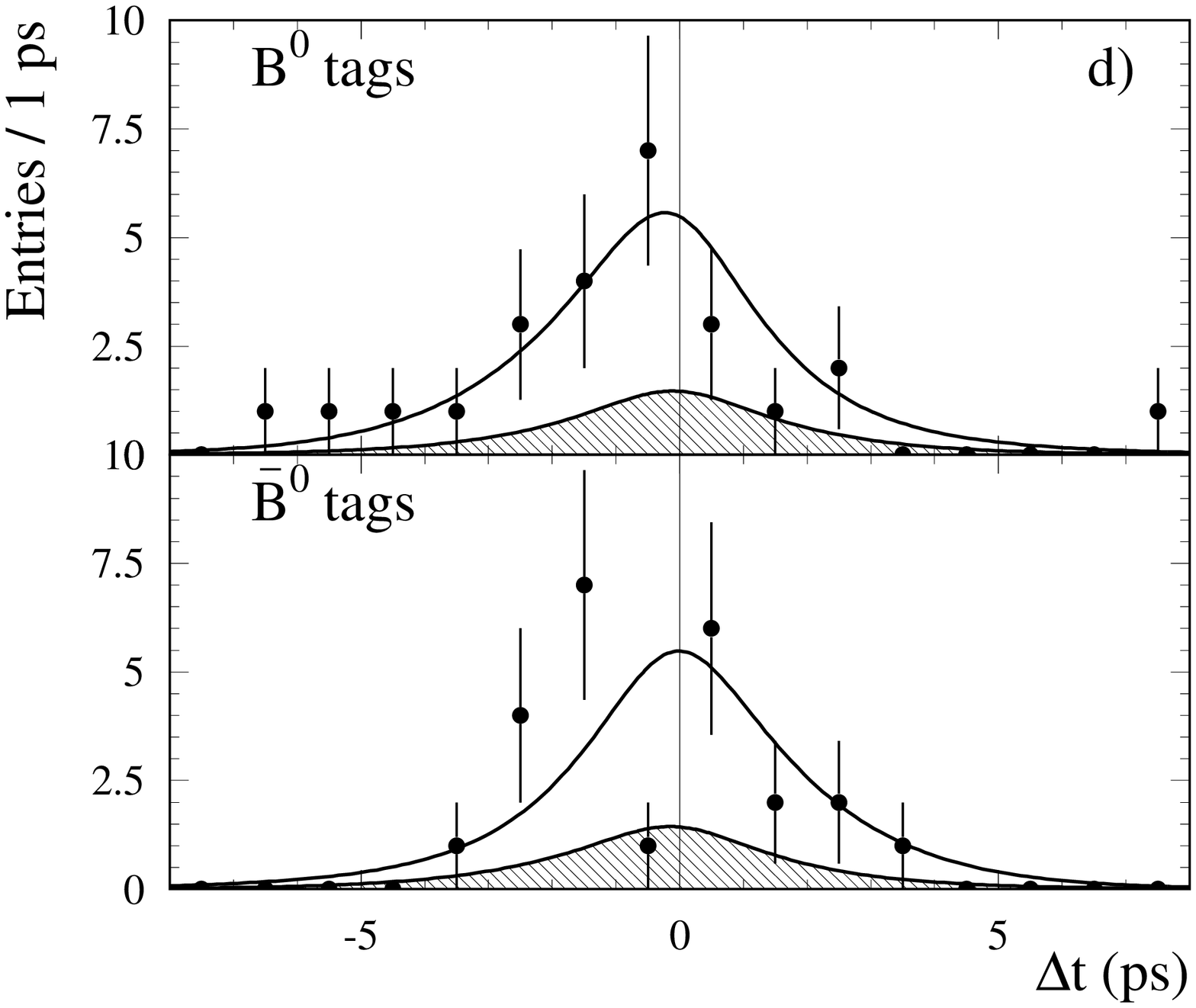}
    \caption{
a) Distribution of \deltat\ for tagged events in the 
full \CP\ sample. The upper (lower) panel is the sum of \Bz (\Bzb)-tagged
events in the $\etaCP=-1$ and $\jpsi\Kstarz$ samples, combined with
the \Bzb (\Bz)-tagged events in the $\jpsi\KL$ sample.
Corresponding \Bz- (lower panel) and \Bzb-tagged (upper panel)
distributions for the b) $\etaCP=-1$, c) $\jpsi\KL$, 
and d) $\jpsi\Kstarz$ samples are also shown.
In all cases, the data points are overlaid with the result from the global
unbinned likelihood fit, projected on the basis of the individual
signal and background probabilities, and event-by-event \deltat\ resolutions,
for candidates in the respective samples. Therefore,
the curves correspond to $\stwob=0.59$, rather than the fitted value 
obtained with the individual subsample.
The probability-weighted \deltat\ spectra of the background
candidates obtained from the fit are indicated by the shaded areas.
      \label{fig:deltatfitCPall}}
  \end{center}
\end{figure*}

Table~\ref{tab:result}  summarizes the event yields and \stwob\ values determined for
the full \CP\ sample and various subsamples.
Results are provided by \CP\ channel, tagging category, $\Bz$ versus $\Bzb$
tag, $\jpsi$ decay mode and data-taking period.
The consistency between the six \CP\ modes is satisfactory, the 
probability of finding a worse agreement being 8\%. 
The large observed asymmetry in $\Bz \to\chic1 \KS$ causes the likelihood
for this channel
to become negative in certain regions of \deltat. The likelihood of
each of the selected candidates is of course positive. 
Fast parameterized Monte Carlo
studies show that \stwob\ is unbiased if the likelihood is not required
to be positive for all values of \deltat\ and that the probability to
measure such a large asymmetry is about 1\%. 
The observed asymmetry in the number of \Bz (160) 
and \Bzb (113) tags in the $\jpsi\KL$ sample 
has no impact on the \stwob measurement.
The results obtained with the full $B_{\CP}$
samples for Run 1 and Run 2 are consistent at the 1.8 sigma level.
The yields and fitted values for \stwob\ are also listed in
Table~\ref{tab:result} for the high purity $\etaCP=-1$ sample alone, along
with a similar breakdown into subsamples; again, no significant variation
is seen.

\begin{figure}[!htb]
  \begin{center}
    \includegraphics[width=\linewidth]{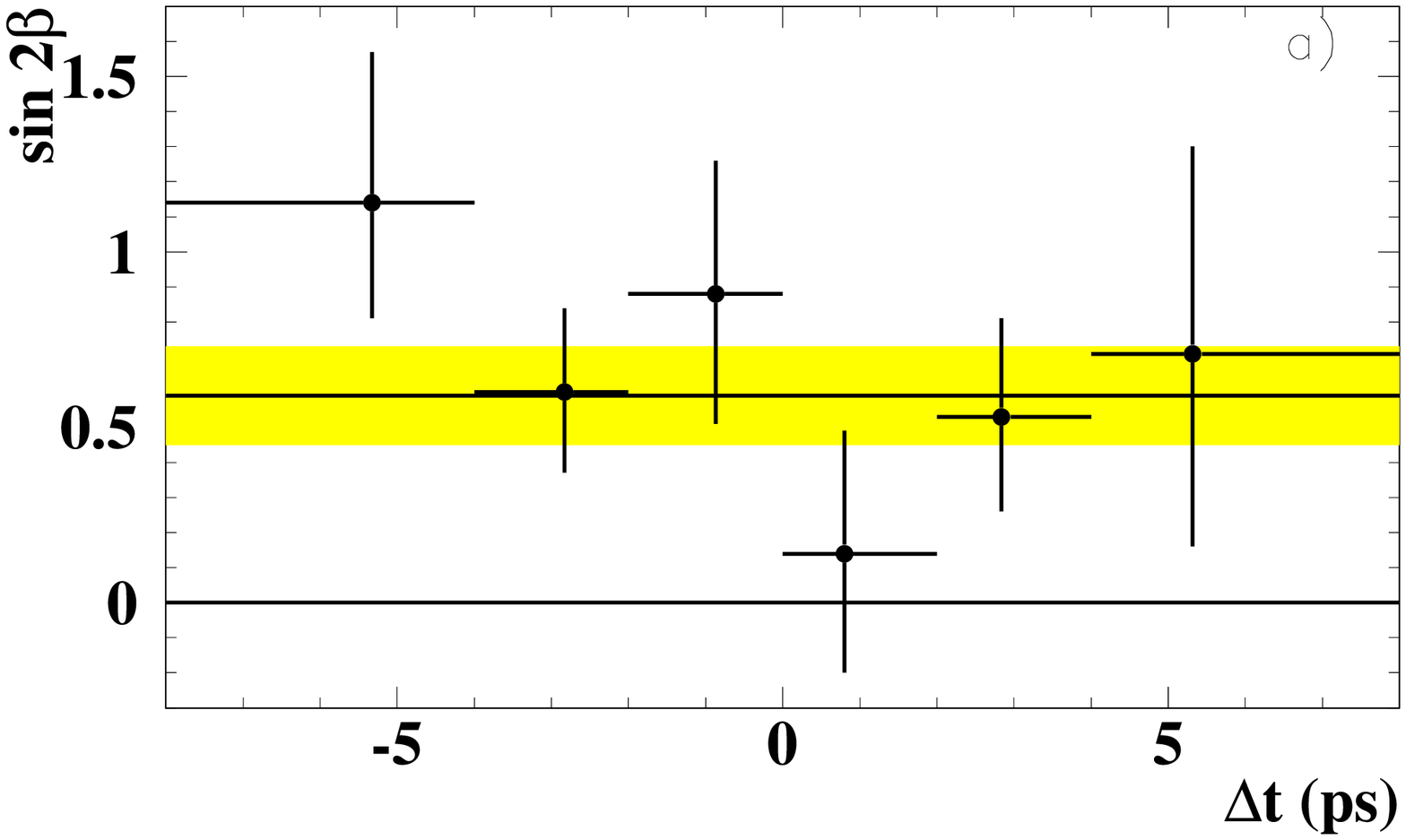}
    \includegraphics[width=\linewidth]{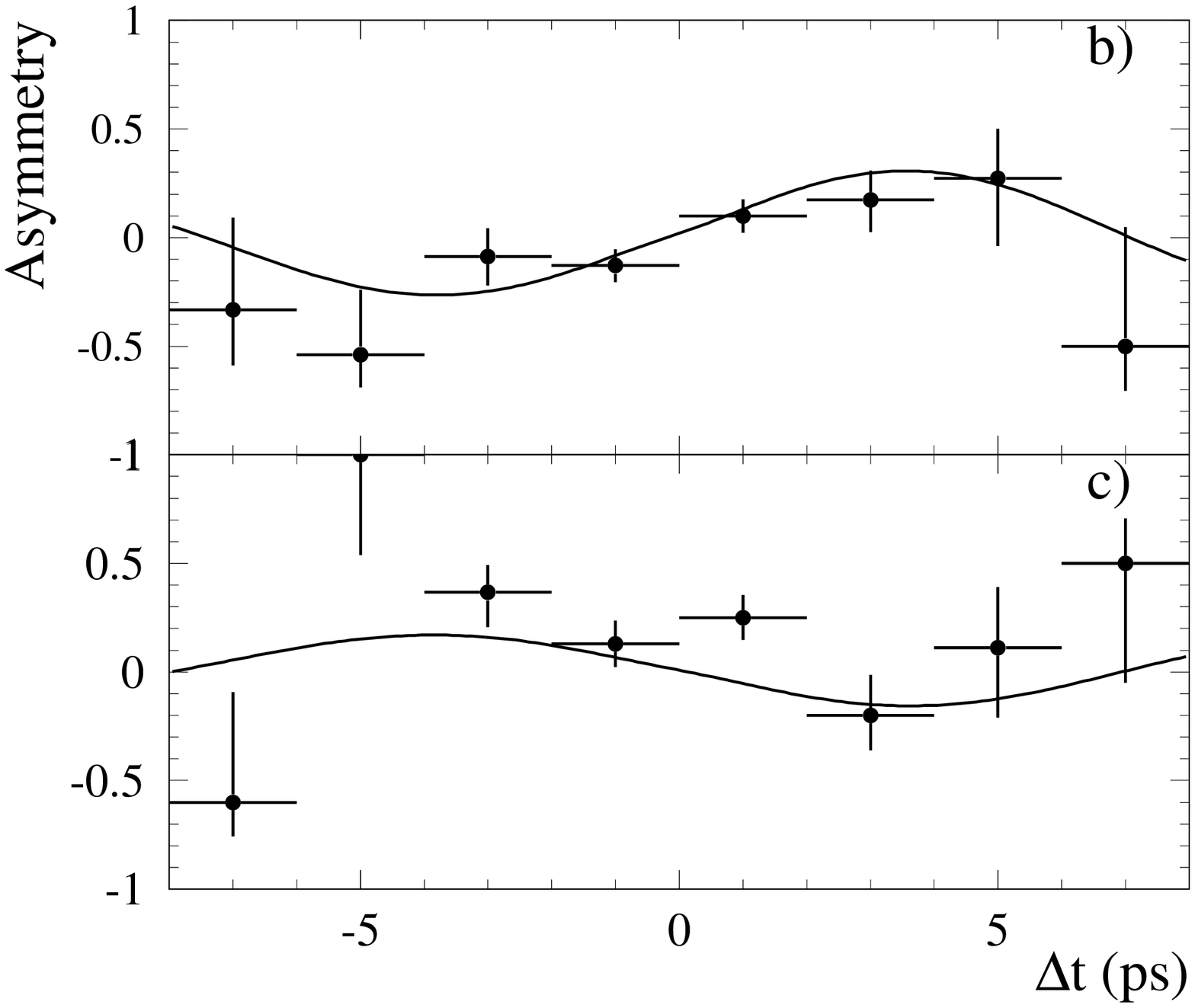}
    \caption{
      a) Fitted value of \stwob\ obtained in bins of \deltat\ by fixing
      all other parameters to the values obtained with the full \CP\ sample;
      Raw asymmetry in the number of \Bz\ and \Bzb\ tags in the signal region,
      $(N_{\Bz}-N_{\Bzb})/(N_{\Bz}+N_{\Bzb})$, with asymmetric binomial errors,
      as a function of \deltat\ for b) $\etaCP=-1$ and c) $\jpsi\KL$ samples. 
      The data points are overlaid with the separate fit results for the two samples.
      \label{fig:asymCPall}}
  \end{center}
\end{figure}

The distribution of events as a function of \deltat\ for \Bz\ and \Bzb\ tags
is shown in Fig.~\ref{fig:deltatfitCPall}a for the full \CP\ sample. 
For this purpose, only those events with $\mes > 5.27\gevcc$ in the
$\etaCP=-1$ and $\jpsi\Kstarz$ samples or $\Delta E<10\mev$ in the $\etaCP=+1$ sample are 
included. Overlaid
on the data are the projections of the signal and background
$\deltat$ distributions obtained from the fit, where the latter is
normalized to the projected background level.
Figure~\ref{fig:deltatfitCPall}b-d 
shows the corresponding \deltat\ distributions for the $\etaCP=-1$,
$\etaCP=+1$ samples and $\Bz\to\jpsi \Kstarz$ ($\Kstarz \to \KS \piz$, $\KS \to \pi^+ \pi^-$). The
superimposed likelihood curves show the quality of the fit for each subsample. 
The value of \stwob\ obtained by fixing all other parameters
to results obtained with the full \CP\ sample and then fitting
for \stwob\  in bins of \deltat\ is  shown in Fig.~\ref{fig:asymCPall}a. 
The values obtained for \stwob\ are all consistent, demonstrating that the 
oscillation as a function of \deltat\ has the expected behavior.
The observed asymmetry ${\cal A}(\deltat)$ is 
shown in 
Fig.~\ref{fig:asymCPall}b and c for the $\etaCP=-1$ and
$\etaCP=+1$ samples respectively, along with the projections from the fit results.

\begin{table*}[htb]
\caption{Parameters for the combined likelihood fit to the $B_{\CP}$ and
$B_{\rm flav}$ samples. The first major column contains the fit results,
while the second major column contains the correlation
coefficients with respect to \stwob\ for each fit parameter.}
  \vspace{0.3cm}  
  \label{tab:summary-sin2beta}
\begin{center}
\begin{tabular}{|l|c|c|c|c|}
\hline
      Parameter & \multicolumn{2}{c}{Fit Result} & \multicolumn{2}{|c|}{Correlation} \\  \cline{2-5}
                & Run 1 & Run 2                    & Run 1 & Run 2 \\ \hline
      \stwob                                & \multicolumn{2}{c}{$\phanm 0.59 \pm 0.14$} & \multicolumn{2}{|c|}{ } \\  \hline
      \multicolumn{5}{|c|}{Signal Resolution Function}\\   \hline      
      $S_1$ (core)                          & $\phanm ~1.2 \pm 0.1~$  &$\phanm ~1.1 \pm 0.1$    & $\phanm 0.018$ & $\phanm 0.020$\\         
      $b_1(\Delta t$) {\tt lepton} (core)      & $\phanm 0.07 \pm 0.12$  & $\phanm 0.04 \pm 0.16$  & $\phanm 0.008$ & $\phanm 0.045$ \\     
      $b_1(\Delta t$) {\tt kaon} (core)        & $-0.26 \pm 0.08$        &$-0.18 \pm 0.09$         & $\phanm 0.002$ & $\phanm 0.021$\\     
      $b_1(\Delta t$) {\tt NT1} (core)         & $-0.21 \pm 0.15$        &$-0.33 \pm 0.21$         & $\phanm 0.004$ & $\phanm 0.001$ \\     
      $b_1(\Delta t$) {\tt NT2} (core)         & $-0.31 \pm 0.11$        &$-0.17 \pm 0.15$         & $-0.001$       & $-0.002$ \\     
      $b_2(\Delta t$) (tail)             & $-1.7 \pm 1.5$          & $-3.3 \pm 2.8$          & $\phanm 0.001$ & $\phanm 0.006$ \\     
      $f_2$(tail)                             & $\phanm 0.08 \pm 0.06$  &$\phanm 0.04 \pm 0.04$   & $\phanm 0.009$ & $\phanm 0.005$ \\     
      $f_3$(outlier)                          & $\phanm 0.005 \pm 0.003$&$\phanm 0.000 \pm 0.001$ & $-0.001$       & $\phanm 0.000$ \\    \hline 
      \multicolumn{5}{|c|}{Signal dilutions}  \\   \hline                               
      $\langle D \rangle$, {\tt lepton}           &\multicolumn{2}{c}{$\phanm 0.82 \pm 0.03$}  & \multicolumn{2}{|c|}{$-0.042$} \\   
      $\langle D \rangle$, {\tt kaon}             &\multicolumn{2}{c}{$\phanm 0.65 \pm 0.02$}  & \multicolumn{2}{|c|}{$-0.083$} \\      
      $\langle D \rangle$, {\tt NT1}              &\multicolumn{2}{c}{$\phanm 0.56 \pm 0.04$}  & \multicolumn{2}{|c|}{$-0.015$} \\      
      $\langle D \rangle$, {\tt NT2}              &\multicolumn{2}{c}{$\phanm 0.30 \pm 0.04$}  & \multicolumn{2}{|c|}{$-0.032$} \\      
      $\Delta D$, {\tt lepton}                    &\multicolumn{2}{c}{$-0.02 \pm 0.04$}        & \multicolumn{2}{|c|}{$\phanm 0.010$} \\     
      $\Delta D$, {\tt kaon}                      &\multicolumn{2}{c}{$\phanm 0.04 \pm 0.03$}  & \multicolumn{2}{|c|}{$\phanm 0.005$} \\      
      $\Delta D$, {\tt NT1}                       &\multicolumn{2}{c}{$-0.11 \pm 0.06$}        & \multicolumn{2}{|c|}{$\phanm 0.014$} \\     
      $\Delta D$, {\tt NT2}                       &\multicolumn{2}{c}{$\phanm 0.12 \pm 0.05$}  & \multicolumn{2}{|c|}{$-0.008$} \\ \hline  
      \multicolumn{5}{|c|}{Background properties}\\   \hline                                 
      $\tau$, mixing bkgd [ps]               &\multicolumn{2}{c}{$\phanm ~1.3 \pm 0.1~$}  & \multicolumn{2}{|c|}{$-0.001$} \\ 
      $f(\tau=0)$, \CP\ bkgd                 &\multicolumn{2}{c}{$\phanm 0.60 \pm 0.12$}  & \multicolumn{2}{|c|}{$-0.011$} \\   
      $f(\tau=0)$, mixing bkgd, {\tt lepton}       &\multicolumn{2}{c}{$\phanm 0.31 \pm 0.10$}  & \multicolumn{2}{|c|}{$-0.001$} \\    
      $f(\tau=0)$, mixing bkgd, {\tt kaon}         &\multicolumn{2}{c}{$\phanm 0.65 \pm 0.04$}  & \multicolumn{2}{|c|}{$-0.001$} \\   
      $f(\tau=0)$, mixing bkgd, {\tt NT1}          &\multicolumn{2}{c}{$\phanm 0.62 \pm 0.06$}  & \multicolumn{2}{|c|}{$-0.001$} \\   
      $f(\tau=0)$, mixing bkgd, {\tt NT2}          &\multicolumn{2}{c}{$\phanm 0.64 \pm 0.04$}  & \multicolumn{2}{|c|}{$-0.001$} \\     \hline
      \multicolumn{5}{|c|}{Background resolution function}\\   \hline                                 
      $S_1$ (core)                          &$\phanm ~1.5 \pm 0.1~$&$\phanm ~1.3 \pm 0.1~$  & $\phanm 0.004$ & $-0.003$\\      
      $b_1(\Delta t$) core [ps]          &$      -0.16 \pm 0.03$&$\phanm 0.02 \pm 0.04$  & $\phanm 0.000$ & $-0.001$ \\     
      $f_2$(outlier)                          &$\phanm 0.016 \pm 0.004$&$\phanm 0.017 \pm 0.005$  & $-0.001$ & $0.000$ \\        \hline
      \multicolumn{5}{|c|}{Background dilutions}  \\   \hline                               
      $\langle D \rangle$, {\tt lepton}, $\tau =0$&\multicolumn{2}{c}{$\phanm 0.33 \pm 0.27$}  & \multicolumn{2}{|c|}{$\phanm 0.003$} \\    
      $\langle D \rangle$, {\tt kaon}, $\tau =0$  &\multicolumn{2}{c}{$\phanm 0.45 \pm 0.03$}  & \multicolumn{2}{|c|}{$\phanm 0.008$} \\
      $\langle D \rangle$, {\tt NT1}, $\tau =0$   &\multicolumn{2}{c}{$\phanm 0.25 \pm 0.10$}  & \multicolumn{2}{|c|}{$\phanm 0.002$} \\
      $\langle D \rangle$, {\tt NT2}, $\tau =0$   &\multicolumn{2}{c}{$\phanm 0.11 \pm 0.06$}  & \multicolumn{2}{|c|}{$\phanm 0.003$} \\
      $\langle D \rangle$, {\tt lepton}, $\tau >0$&\multicolumn{2}{c}{$\phanm 0.33 \pm 0.14$}  & \multicolumn{2}{|c|}{$\phanm 0.000$} \\   
      $\langle D \rangle$, {\tt kaon}, $\tau >0$  &\multicolumn{2}{c}{$\phanm 0.24 \pm 0.06$}  & \multicolumn{2}{|c|}{$\phanm 0.000$} \\ 
      $\langle D \rangle$, {\tt NT1}, $\tau >0$   &\multicolumn{2}{c}{$\phanm 0.05 \pm 0.14$}  & \multicolumn{2}{|c|}{$-0.001$} \\
      $\langle D \rangle$, {\tt NT2}, $\tau >0$   &\multicolumn{2}{c}{$\phanm 0.09 \pm 0.09$}  & \multicolumn{2}{|c|}{$\phanm 0.000$} \\  \hline
\end{tabular}
\end{center}
\end{table*}

The average dilutions and dilution differences for \Bz\
and \Bzb\ tags obtained from the fit to the \Bz\ flavor eigenstate
and full \CP\  sample, and the corresponding tagging efficiencies, are
summarized  in Table~\ref{tab:mistagCPall}. We find a total tagging 
efficiency of  $(68.4\pm0.7)\%$ (statistical error only). The lepton
categories have the lowest mistag fractions, but also have low efficiency.  
The {\tt Kaon} category, despite having a larger mistag fraction
(17.6\%),  has a higher effective tagging efficiency; one-third of
events are assigned to this category. Altogether, lepton and kaon
categories have an effective tagging efficiency $Q \approx 22.4\%$.  The
neural network categories increase the effective tagging efficiency by
$\sim 4\%$ to an overall $Q = (26.1 \pm 1.2) \%$ (statistical error only).
These mistag fractions are very similar to the mistag fraction that are
obtained from the $B_{\rm flav}$ sample alone (see
Table~\ref{tab:mixing-likelihood}). The small differences are due to
the correlation between the mistag fractions and the $\Delta t$
resolution function parameters. 

\begin{table*}[!htb]
\caption{
Result of fitting for \CP\ asymmetries in the entire \CP\ sample and in 
various subsamples. The yields are obtained with likelihood fits and are
therefore background subtracted.} 
\vspace{0.3cm}
\begin{center}
\begin{tabular}{|l|c|c|c|c|c|} \hline
 Sample       &  $N_{\rm tag}$ &  Purity (\%)  &  \stwob                    & \etaimlambdaoverabslambda & \abslambda  \\ \hline \hline
 \CP\ sample                                     
              & 803            &  80           &  {\bf 0.59}$\pm${\bf 0.14} & &  \\  
\hline
 \ \ $\jpsi \KS$ ($\KS \to \pi^+ \pi^-$)
              & 316            & 98            &  $0.45 \pm 0.18$           & $0.45 \pm 0.18$ & $0.91 \pm 0.11$  \\   
 \ \ $\jpsi \KS$ ($\KS \to \pi^0 \pi^0$)
              &  64            & 94            &  $0.70 \pm 0.50$           & $0.71 \pm 0.50$ & $0.95 \pm 0.27$   \\   
 \ \ $\psitwos\KS$ ($\KS \to \pi^+ \pi^-$)
              &  67            & 98            &  $0.47 \pm 0.42$           &$0.48 \pm 0.45$ & $1.22 \pm 0.33$ \\   
 \ \ $\chic1 \KS$ ($\KS \to \pi^+ \pi^-$) 
              &  33            & 97            &  $2.59 \pm^{0.55}_{0.67}$  &$2.67 \pm 0.59$ & $0.71 \pm 0.23$\\
 \ \ $\jpsi\KL$                                  
              & 273            & 51            &  $0.70 \pm 0.34$           && \\  
 \ \ $\jpsi \Kstarz$ ($\Kstarz \to \KS \piz$, $\KS \to \pi^+ \pi^-$) 
              & 50             & 74            &  $0.82 \pm 1.00$           && \\
\hline
 \ \ {\tt Lepton}                                
              & 130            & 82            &  $0.54 \pm 0.26$ &   & \\ 
 \ \ {\tt Kaon}                                                        
              & 438            & 79            &  $0.58 \pm 0.18$ &   & \\
 \ \ {\tt NT1}                                                         
              &  79            & 74            &  $0.89 \pm 0.30$ &   & \\
 \ \ {\tt NT2}                                                         
              & 156            & 80            &  $0.40 \pm 0.65$ &   & \\
\hline
 \ \ {\tt $\Bz$}                                 
              & 420            & 79            &  $ 0.54 \pm 0.19$ && \\
 \ \ {\tt $\Bzb$}                                
              & 383            & 78            &  $0.64 \pm 0.20 $ && \\
\hline
 \ \ $\jpsi\to\epem$                             
              & 385            & 78            &  $0.49 \pm 0.20$ &   &  \\
 \ \ $\jpsi \to \mu^+ \mu^-$                                            
              & 418            & 84            &  $0.70 \pm 0.18$ &   &  \\
\hline
 \ \ Run 1                              
              & 533            & 80            &  $0.49 \pm 0.20$ && \\
 \ \ Run 2                     
              & 270            & 84            &  $0.82 \pm 0.22$ && \\
\hline\hline
 $\etaCP=-1$ sample                                 
              & 480          & 96              &  $0.56 \pm 0.15$           & $0.56 \pm 0.15$ & $0.93 \pm 0.09$  \\  \hline
 \ \ {\tt Lepton}                                 
              & 74           & 100             &  $0.54 \pm 0.29$           & $0.57 \pm 0.29$  & $0.77 \pm 0.14$ \\ 
 \ \ {\tt Kaon}                                
              & 271          & 98              &  $0.59 \pm 0.20$           & $0.59 \pm 0.20$  & $0.98 \pm 0.12$ \\
 \ \ {\tt NT1}                                   
              & 46           & 97              &  $0.67 \pm 0.45$           & $0.57 \pm 0.46$  & $0.73 \pm 0.29$ \\
 \ \ {\tt NT2}                                  
              & 89           & 95              &  $0.10 \pm 0.74$           & $0.28 \pm 1.29$  & $2.95 \pm 3.83$ \\
\hline
 \ \ {\tt $\Bz$}                                 
              & 234          & 98              &  $0.50 \pm 0.22$           &&\\
 \ \ {\tt $\bar\Bz$}                             
              & 246          & 97              &  $0.61 \pm 0.22$           &&\\
\hline 
 \ \ $\jpsi\to\epem$                             
              & 219          & 94              &  $0.54 \pm 0.22$           & $0.52 \pm 0.22$  & $1.00 \pm 0.15$ \\
 \ \ $\jpsi \to \mu^+ \mu^-$                     
              & 261          & 98              &  $0.60 \pm 0.21$           & $0.63 \pm 0.21$  & $0.87 \pm 0.11$ \\
\hline
 \ \ Run 1                              
              & 310          & 95              &  $0.37 \pm 0.20$           & $ 0.37 \pm 0.20 $ & $1.16 \pm 0.15   $\\
 \ \ Run 2                     
              & 170          & 98              &  $0.86 \pm 0.24$           & $0.96 \pm 0.26  $ & $ 0.66 ^{-0.11}_{+0.12}   $ \\
\hline\hline
\multicolumn{6}{|l|}{Control samples} \\ \hline
\ \ $\Bz \to D^{(*)-} \pi^+/\rho^+/a_1^+$                                  
              & 7579         & 84              &  $0.00 \pm 0.04$            &&\\
\ \ $\Bu \to \Dbar^{(*)0} \pi^+$                                  
              & 6800         & 86              &  $-0.02 \pm 0.04 $          &&\\
\ \ $\Bz\to\jpsi\Kstarz$ ($\Kstarz\to\Kp\pim$)       
              & 705          & 95              &  $0.12 \pm 0.12$            &&\\
\ \ $\Bu\to\jpsi K^{(*)+}$, $\psitwos\Kp$            
              & 2031         & 94              &  $0.07 \pm 0.07$            &&\\
\hline 
\end{tabular}
\end{center}
\label{tab:result}
\end{table*}

\begin{table*}[!htb] 
\caption
{ Average mistag fractions $\mistag_i$ and mistag differences
  $\Delta\mistag_i=\mistag_i(\Bz)-\mistag_i(\Bzb)$ extracted for each
  tagging category $i$ from the maximum-likelihood fit to the time
  distribution for the fully-reconstructed \Bz\ sample ($B_{\rm
  flav}$+$B_{\CP}$). The figure of merit for tagging is the effective
  tagging efficiency $Q_i = \eps_i (1-2\mistag_i)^2$, where $\eps_i$ is
  the fraction of events with a reconstructed tag vertex that are
  assigned to the $i^{th}$ category. $\eps_i$ is computed for the $\etaCP=\pm1$
 samples as well as the combined $B_{\CP}$ and $B_{{\rm flav}}$ samples.
Uncertainties are statistical
  only. The statistical error on \stwob is proportional to $1/\sqrt{Q}$,
  where $Q=\sum Q_i$. \label{tab:mistagCPall}}
\vspace{0.3cm}
\begin{center}
  \begin{tabular}{|l|c|c|c|c|c|c|} \hline
    Category & $\etaCP=-1$ & $\etaCP=+1$ & \multicolumn{4}{|c|}{$B_{\rm flav} + B_{\CP}$} \\ \cline{2-7}
             & $\varepsilon$ [\%] & $\varepsilon$ [\%] & $\varepsilon$ [\%] & $\mistag$ [\%] & $\Delta \mistag$ [\%] & $Q$ [\%]      \\ \hline \hline
    {\tt Lepton}     & $11.0 \pm 1.2$                 & $10.4 \pm 3.0$                 & $10.9 \pm 0.3$     & $~9.0 \pm 1.4$ & $\phanm 0.9 \pm 2.2$  &  $~7.4 \pm 0.5$  \\
    {\tt Kaon}       & $38.9 \pm 1.9$                 & $28.3 \pm 4.5$                 & $35.8 \pm 1.0$     & $17.6 \pm 1.0$ & $-1.9 \pm 1.5$        &  $15.0 \pm 0.9$ \\
    {\tt NT1}        & $~6.9 \pm 0.9$                 & $~4.8 \pm 2.3$                 & $~7.8 \pm 0.3$     & $22.0 \pm 2.1$ & $\phanm 5.6 \pm 3.2$  &  $~2.5 \pm 0.4$  \\
    {\tt NT2}        & $13.0 \pm 0.4$                 & $13.9 \pm 3.3$                 & $13.8 \pm 0.3$     & $35.1 \pm 1.9$ & $-5.9 \pm 2.7$        &  $~1.2 \pm 0.3$  \\  \hline \hline
    All              & $69.8 \pm 2.7$                 & $57.4 \pm 6.7$                 & $68.4 \pm 0.7$     &                &                       &  $26.1 \pm 1.2$ \\ 
    \hline
  \end{tabular}
\end{center}
\end{table*}

Based on a large number of fast parameterized Monte Carlo experiments with the same
number of events as our full $B_{\CP}$ and $B_{\rm flav}$ data samples,
we estimate a probability of 27\% for finding a value of the maximum
likelihood lower than that observed. 
These same studies, based on samples 
with the size and composition of the data,
show that the expected statistical error is 0.132 with a spread of 0.005, 
in very good agreement with the observed error of 0.137.

\subsection{Systematic error estimation}
\label{subsec:stwob-systematics}

Just as for the \deltamd\ measurement, systematic errors can usefully be grouped into
signal description, including detector reconstruction effects,
background description, fixed external parameters, 
and statistical limitations of Monte Carlo validation tests for the fitting procedure
(discussed in Section~\ref{sect:mcvalid}). A summary of these sources of systematic error
is shown in Table~\ref{tab:systematics} for the various \CP\ samples. In the following,
the individual contributions are referenced by the lettered lines in this table.

\begin{table*}[!htb]
\caption{Summary of contributions to the systematic error on \stwob, \etaimlambdaoverabslambda\ and \abslambda. 
Note that the last two measurements use only the $\etaCP=-1$ sample.} 
\begin{center}
\begin{tabular}{|l|c|c|c|c|c|c|} \hline
 Source    &  \multicolumn{6}{c|}{\CP\ Sample} \\ \cline{2-7}
           & $\etaCP=-1$ & $\jpsi\KL$ & $\jpsi \Kstarz$ & Full &  \etaimlambdaoverabslambda & \abslambda \\ \hline  \hline 
           \multicolumn{7}{|c|}{Signal parameters} \\ \hline
(a) \deltat\ signal resolution model             & $\pm 0.009$& $\pm 0.01$&$\pm 0.07$ & $\pm 0.009$& $\pm 0.003$ & $\pm 0.003$ \\ \hline
(b) SVT alignment                                & \multicolumn{5}{c|}{$\pm 0.027$} & $\pm 0.012$ \\ \hline 
(c) \deltat\ for right/wrong tagged events       & \multicolumn{4}{c|}{$\pm 0.012$}&$\pm 0.011$ &$\pm 0.003$ \\ \hline
(d) \deltat\ signal resolution outliers          & $\pm 0.002$& $\pm 0.018$& $\pm 0.03$ & $\pm 0.002$ &$\pm 0.003$ & $\pm 0.002$\\ \hline
(e) \deltat\ signal resolution                   & \multicolumn{5}{c|}{$\pm 0.003$}  & $\pm 0.009$\\ \hline
(f) Signal dilutions for \CP\ vs.~$B_{\rm flav}$ & \multicolumn{5}{c|}{$\pm 0.027$}  & $\pm 0.011$\\ \hline
(g) Tagging Efficiencies                         & \multicolumn{4}{c|}{$\pm 0.003$} &$\pm 0.004$&$\pm 0.012$  \\ \hline
           \multicolumn{7}{|c|}{Background properties: $\etaCP=-1$} \\ \hline
(h) Background fraction               &$\pm 0.006$ & --- & ---  & $\pm 0.005$ &$\pm 0.006$ &$\pm 0.004$ \\ \hline
(i) \CP\ bkgd peaking component       &$\pm 0.004$ & --- & --- & $\pm 0.003$ & $\pm 0.005$& $\pm 0.001$ \\ \hline
(j) \CP\ bkgd \CP\ content (ARGUS)    &$\pm 0.015$ & --- & --- & $\pm 0.015$ & $\pm 0.015$& $\pm 0.001$ \\ \hline
(k) \CP\ bkgd \CP\ content (Peak)     &$\pm 0.004$ & --- & --- & $\pm 0.004$ & $\pm 0.004$& $\pm 0.001$ \\ \hline
(l) \CP\ bkgd effective lifetime      & 0 & --- & --- & 0  &  0 & 0 \\ \hline
(m) \CP\ bkgd resolution              & $\pm 0.002$ & --- & --- & $\pm 0.002$  &$\pm 0.002$ &$\pm 0.001$ \\ \hline
           \multicolumn{7}{|c|}{Background properties: $\jpsi \KL$} \\ \hline
(n) Background fraction                     & --- & $\pm 0.075$ & --- & $\pm 0.01$  & --- & --- \\ \hline
(o) $\Delta E$ distribution                 & --- & $\pm 0.04$  & --- & $\pm 0.007$ & --- & --- \\ \hline
(p) Effective \CP\ of backgrounds           & --- & $\pm 0.020$ & --- & $\pm 0.001$ & --- & --- \\ \hline
(q) Background composition                  & --- & $\pm 0.014$ & --- & $\pm 0.002$ & --- & --- \\ \hline
(r) Background $\Delta t$ and dilution      & --- & $\pm 0.023$ & --- & $\pm 0.003$ & --- & --- \\ \hline
           \multicolumn{7}{|c|}{Background properties: $\jpsi\Kstarz$} \\ \hline
(s) Sample composition                       & --- & --- & $\pm 0.08$ & $\pm 0.001$ & --- & --- \\ \hline
(t) $R_{\perp}$                              & --- & --- & $\pm 0.08$ & $\pm 0.001$ & --- & --- \\ \hline
           \multicolumn{7}{|c|}{Background properties: $B_{\rm flav}$} \\ \hline
(u) Background fraction                      & $\pm 0.001$ & $\pm 0.008$ & $\pm 0.003$ & $\pm 0.002$ & $\pm 0.002$ & $\pm 0.001$ \\ \hline
(v) $B_{{\rm flav}}$ bkgd mixing contrib.    & $\pm 0.001$ & $\pm 0.002$ & $\pm 0.001$ & $\pm 0.002$ & $\pm 0.001$ & 0 \\ \hline 
(w) $B_{{\rm flav}}$ bkgd peaking component  & 0           & $\pm 0.001$ & $\pm 0.001$ & 0           & 0           & 0 \\ \hline
\multicolumn{7}{|c|}{External parameters} \\ \hline
(x) $z$ scale and boost                           &\multicolumn{5}{c|}{$\pm 0.003$}  &$\pm 0.001$  \\ \hline
(y) Beam spot                                     &\multicolumn{5}{c|}{$\pm 0.002$}  &$\pm 0.006$  \\ \hline
(z) \Bz\ lifetime               &$\pm 0.008$ &$\pm 0.011$ & $\pm 0.022$  &$\pm 0.009$ & $\pm 0.009$ & $\pm 0.012$\\ \hline
(aa) \deltamd\                   &$\pm 0.015$ &$\pm 0.012$ & $\pm 0.082$  &$\pm 0.013$ &$\pm 0.015$ & $\pm 0.001$\\ \hline 
\multicolumn{7}{|c|}{Monte Carlo studies} \\ \hline
(bb) Monte Carlo statistics      &\multicolumn{5}{c|}{$\pm 0.012$} &$\pm 0.007$ \\ \hline \hline
Total systematic error      &$\pm 0.05$ & $\pm 0.10$ & $\pm 0.16$ & $\pm 0.05$ & $\pm 0.05$&$\pm 0.02$ \\ \hline
Statistical error           &$\pm 0.15$ &$\pm 0.34$ & $\pm 1.01$  & $\pm 0.14$ & $\pm 0.15$ & $\pm 0.09$\\ \hline
\end{tabular}
\end{center}
\label{tab:systematics}
\end{table*}

\subsubsection{Signal properties and description}

The parameters of the \deltat\ resolution function, the dilutions and 
dilution differences are determined from the data sample itself with
the likelihood fit. Thus, they do not contribute to the systematic error, but rather
are incorporated into the statistical uncertainty at a level determined by
the size of the data sample itself. Their overall contribution to the total error on 
\stwob\ is 0.02, as determined from the difference in quadrature between the 
statistical error on \stwob\ from the full likelihood fit and 
from a fit with only \stwob\ allowed to vary.

While the bulk of the uncertainties from these sources is thus
incorporated into the statistical error, we assign additional
systematic uncertainties due to
the fixed form of the parameterization for the \deltat\ resolution function.
This form may not be flexible enough to account for all possible 
effects. In addition, tests of the
assumption that the resolution function and dilution parameters
are the same for the $B_{\rm flav}$ and $B_{\CP}$ samples
are limited in precision by the size of the available Monte Carlo
samples.

The resolution function, described in Section~\ref{sec:vtxresolfunct},
is one of several possible functional forms. In order to test possible biases induced
by this particular choice, an alternative model has been considered where a 
Gaussian distribution is convolved with an exponential, with the effective lifetime in the 
exponential depending on the tagging category.
No difference between the fit results with the two models 
is observed in Monte Carlo simulation. We assign as a systematic uncertainty 
the difference in the fit results observed
in the data (Table~\ref{tab:systematics}, line a).
The largest systematic uncertainties from the \deltat\ behavior
arises from possible effects that our model of the resolution
function cannot accommodate or completely parameterize.
These include residual 
uncertainties in the SVT alignment (b) and
possible differences in the \deltat\ determination for
correctly and incorrectly tagged events (c).
An additional uncertainty 
is assigned due to the treatment of the \deltat\ outliers (d).    
Fits with Monte Carlo samples of $B_{\rm flav}$ and $B_{\CP}$ signal events show
no significant difference between resolution function parameters for the two samples.  
We assign a systematic uncertainty of $\pm 0.003$ 
due to the residual shift in \stwob\ between the two
sets of fitted \deltat\ resolution parameters (e).

An underlying assumption of the global fit is that dilutions and 
dilution differences are the same for the  
$B_{\rm flav}$ and $B_{\CP}$ samples. We assign the full difference as
seen in Monte Carlo simulation as
systematic error,  $\pm 0.027$ (f).
In addition, the $B^{\pm}$ data sample was used to study any possible dependence
of the dilutions on \deltat. No significant effect was observed. However, a
dependence of the dilutions on $\sigma_{\deltat}$ has been seen, both
in data and the Monte Carlo simulation (see Section~\ref{sec:dm_MC_validation}). 
Finally, it is possible that tagging efficiencies could be different for \Bz\ and \Bzb\ mesons. 
A separate study of the relative tagging efficiencies is described in Section~\ref{sub:dCP},
since the relative efficiencies form an important part of the direct \CP\ violation search.
The systematic error on \stwob\ due to this effect is estimated to be $\pm 0.003$ (g).

\subsubsection{Background properties}

The fraction of background events in the $\etaCP=-1$ sample
is estimated from fits to the \mes\ distribution.  
Varying this fraction within the stated errors and changing the signal probability
as a function of \mes\ results in a systematic error 
of $\pm 0.005$ on \stwob\ (h). The uncertainty on the fraction
of peaking background contributes a systematic error of $\pm 0.003$ (i).
Varying the effective \stwob\ assumed for the ARGUS 
(${\cal A}$ in Section~\ref{cpm1background}) and
peaking ($\delta_{\rm peak}$ in Section~\ref{cpm1background}) backgrounds
in the \CP\ sample from $-1$ to $+1$ contributes a systematic error of $\pm 0.015$ (j) and $\pm 0.004$ (k),
respectively.
In addition, the contributions due to the
uncertainty of the \deltat\ resolution model ($\pm 0.002$),
and the effective lifetime (negligible)
of the \CP\ background, have been evaluated (l-m).

For the $\Bz\to\jpsi\KL$ channel, the signal and non-$\jpsi$ background fractions
are varied within their statistical
uncertainties ($\pm 1 \sigma$) as obtained with the fit 
to the $\Delta E$ distribution of the sample.
This contributes a systematic error of $\pm 0.075$ to the $\Bz\to\jpsi\KL$ \stwob\ result 
and $\pm 0.01$ to the final result (n).
We also vary background parameters for the $\Bz\to\jpsi\KL$ sample, including the
$\jpsi X$ branching fractions according to Table~\ref{tab:psiklComposition}, 
the assumed \etaCP, the mistag rates and efficiencies,
the \deltat\ resolution function, and $\Delta E$ shape (o-r).
The total $\Bz\to\jpsi\KL$ background systematic error, 
summing these contributions in quadrature,
is $\pm 0.09$ for the \stwob\ fit to the $\Bz\to\jpsi\KL$ sample alone and 
$\pm 0.013$ for the full sample.

For the $\Bz \to \jpsi \Kstarz$ ($\KS \piz$) sample, the value of $R_{\perp}$ 
as well as the
sample composition are varied (s-t) according to Table~\ref{tab:kstarback}.

The effect of the uncertainty on background component in the $B_{\rm flav}$ sample on
\stwob\ has also been evaluated.  The only significant sources of uncertainty are
the fraction of background that mixes (v) and
the signal probability distribution as a function of \mes\ (u,w).

\subsubsection{External parameters}

The residual uncertainty on the physical $z$ scale (x) and the boost parameters
of the \FourS\ center of mass (y) contribute systematic uncertainties.
We fix the \Bz\ lifetime to the current world average values $\tau_{\Bz} = 1.548$\ps\ 
and ${{\rm \Delta} m_d } = 0.472\hbarps$~\cite{PDG2000}.
The errors on \stwob\ due to uncertainties in  
$\tau_{\Bz}$ and \deltamd\ are $\pm 0.009$ 
and $\pm 0.013$, respectively (z-aa). 

\subsubsection{Monte Carlo validation of measurement technique}

The analysis method has been studied with a high-statistics
Monte Carlo sample.  A fit result that is consistent with the generated value for \stwob\ was found.
We assign a $\pm 0.012$ systematic error
due to the statistical limitation of the Monte Carlo sample size (bb).  
Section~\ref{sect:mcvalid} describes this
study in more detail.

\subsection{Validation studies and cross checks}
\label{sub:validation}
We have used data and Monte Carlo samples to perform validation studies of the
analysis technique.  These tests include studies with
parameterized Monte Carlo samples, full GEANT3~\cite{geant} simulation samples, as well
as data samples where no \CP\ asymmetry is expected.

\subsubsection{Monte Carlo studies}
\label{sect:mcvalid}

The highest precision test of the fitting procedure was
performed with fast parameterized Monte Carlo simulation, where 1000 experiments
were generated with sample sizes corresponding to the observed
$B_{\rm flav}$ and $B_{\CP}$ events in data, including mistag rates, 
\deltat\ resolutions, and background fractions and time dependence.
The full fit is performed on each of these experiments.  
The resulting pull distribution (defined as the difference between 
the fitted and generated value of a parameter divided by the statistical 
error as obtained from the likelihood fit)
has a mean $-0.029 \pm 0.032$ and standard deviation $1.007 \pm 0.022$,
consistent with no measurement bias in either the value of \stwob\ or
its estimated error.

In addition, large samples of signal and background Monte Carlo events 
generated with a GEANT3~\cite{geant} detector simulation are used to 
validate the measurement.  For these tests, we obtained the
resolution function parameters as well as the dilutions from
a Monte Carlo sample of $B_{\rm flav}$ events.  Using these
parameters, we fit for \stwob\ in Monte Carlo samples of \CP\ signal
events that correspond in number to the reconstructed
data sample.  These Monte Carlo events are generated with
various values of \stwob\ (0.1 to 0.9) and different \CP-eigenstate modes, 
corresponding to those used in the
measurement with data.  The mean and spread of the pull distribution for these 
Monte Carlo samples can be used to check for any measurement bias and to confirm
the validity of the reported error.  We find that the mean pull
is consistent with zero and the spread is consistent 
with the reported error.  A systematic error
of $\pm 0.012$ is assigned to \stwob\ due to the 
limited Monte Carlo statistics for this test.

The effect of background has been evaluated by
adding an appropriate fraction of background events to our signal
Monte Carlo sample and performing the likelihood fit. The
background samples are obtained either from simulated
$B \to \jpsi X$ events or $\Delta E$ sidebands in data
($|\Delta E|< 120\mev$ but outside the signal region). 
We find no significant bias for \stwob\ with the addition of either
source of background.

\subsubsection{Cross checks with $\tau_{\Bz}$ and \deltamd}

Table~\ref{tab:sys_floattaumd} shows results for \stwob\ if
$\deltamd$ and $\tau_{\Bz}$ are allowed to float in the combined fit
to the $\CP$ and $B_{{\rm flav}}$ samples.  
The fitted value of \deltamd\ is somewhat larger than that
reported in Section~\ref{sec:Mixing}. However, with no kaon veto applied
to the tagging vertex, the correction for the bias
introduced by known correlations between
mistag rates and the \deltat\ resolution is also larger.
Taking this into account, the two results are consistent
within the independent statistical errors. 
Likewise the
lifetime is found to be consistent with our recent 
measurement~\cite{babar0106}.
We have also performed
fits with $\deltamd$ and $\tau_{\Bz}$ fixed to a series of 
values around the world average
in order to determine the dependence of \stwob\ on
these two parameters, thereby finding that 
\begin{eqnarray}
\stwob = \left[\,0.59\right.&-&0.35\left(\frac{\deltamd}{0.472\hbarps}-1\right)\nonumber\\
                   &-&\left.0.45\left(\frac{\tau_{\Bz}}{1.548\ps}-1\right)\right].
\end{eqnarray}

\begin{table}
\caption{Results when \deltamd\ and (or) $\tau_{\Bz}$ are floated in
the \stwob\ fit to the full \CP\ sample and the $\etaCP=-1$ subsample alone.}
\begin{center}
\begin{tabular}{|l|c|c|c|} \hline
Fit                            & \stwob           & \deltamd (ps$^{-1}$)& $\tau_{\Bz}$ (ps)\\ \hline\hline
\multicolumn{4}{|c|}{All \CP\ modes} \\ \hline
Nominal fit                    & $0.59 \pm 0.14$ & $0.472$            &$1.548$\\
Float \deltamd                 & $0.55 \pm 0.13$ & $0.533 \pm 0.015$ & $1.548$    \\
Float $\tau_{\Bz}$             & $0.60 \pm 0.14$ & $0.472$           & $1.53 \pm 0.03$\\
Float \deltamd and $\tau_{\Bz}$& $0.56 \pm 0.13$ & $0.542 \pm 0.016$ & $1.50 \pm 0.03$ \\ \hline
\multicolumn{4}{|c|}{$\etaCP=-1$ modes} \\ \hline
Nominal fit                    & $0.56 \pm 0.15$ & $0.472$           & $1.548$\\
Float \deltamd                 & $0.51 \pm 0.15$ & $0.531 \pm 0.015$ & $1.548$\\
Float $\tau_{\Bz}$             & $0.57 \pm 0.15$ & $0.472$           & $1.53 \pm 0.03$ \\
Float \deltamd and $\tau_{\Bz}$& $0.52 \pm 0.15$ & $0.540 \pm 0.016$ & $1.50 \pm 0.03$ \\ \hline
\end{tabular}
\end{center}
\label{tab:sys_floattaumd}
\end{table}

\subsubsection{Asymmetries in data control samples}

\begin{figure}[tbph]
\begin{center}
    \includegraphics[width=\linewidth]{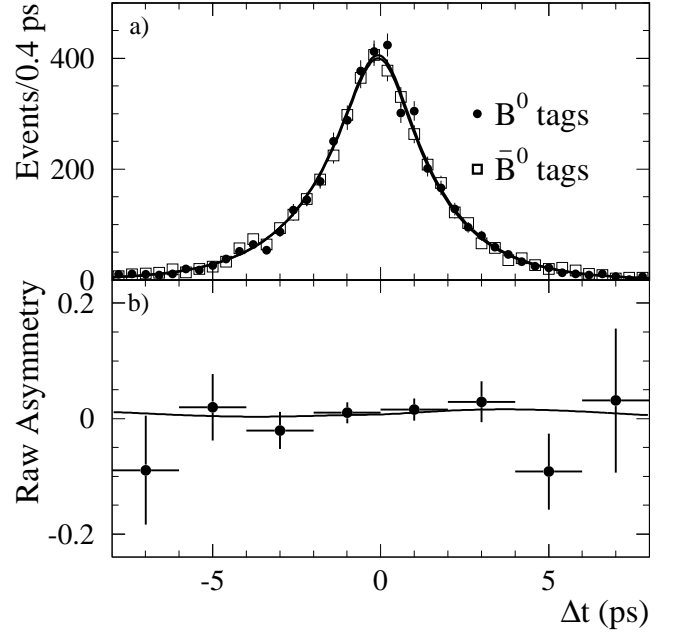}
\end{center}
\caption{a) Distribution in \deltat\ for \Bz- and \Bzb-tagged and b) observed asymmetry for
events in the flavor-eigenstate \Bz\ sample. The projections of the likelihood fit for
the \Bz- and \Bzb-tagged samples are shown in a) as the overlapping solid lines.
\label{fig:dt_allb0}}
\end{figure}

Control samples in data
where the reconstructed \Bz\ and \Bu\ meson decays 
to a flavor-eigenstate mode with a $D^{(*)}$ or charmonium meson in the final state
can be used to validate the \stwob\ measurement, since
the asymmetry is expected to be zero in this case.
For these samples, the \deltat\ resolution function parameters
and the dilutions are fixed to the values obtained with the $B_{\rm flav}$
sample.  The \CP\ asymmetry and the fraction of prompt background (identical
for each tagging category, as is the case for the fit to the \CP\ data sample) are
allowed to float.  The measured asymmetries are 
all consistent with zero, as shown in Table~\ref{tab:result}.
The observed \deltat\ distributions for the
\Bz- and \Bzb-tagged events in the $B_{\rm flav}$ sample 
is shown in Fig.~\ref{fig:dt_allb0}a, where
good agreement is clearly visible. 
Figure~\ref{fig:dt_allb0}b demonstrates that
there is no visible asymmetry
as a function of \deltat.

\begin{figure}[!htb]
\begin{center}
 \includegraphics[width=\linewidth]{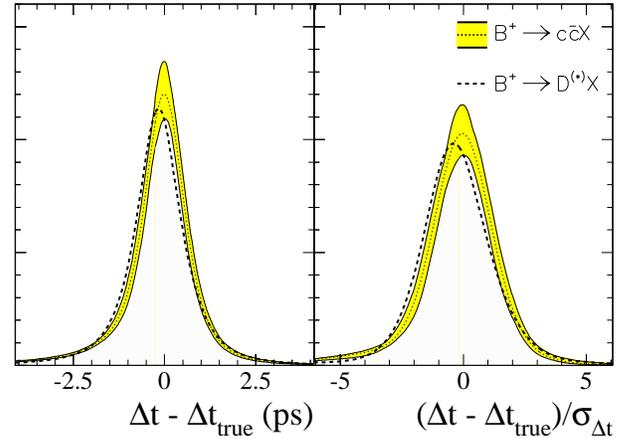}
\caption{
Comparison of the fitted $\Delta t$ resolution obtained
with the data control samples
$B^+\to D^{(*)}X$ and $B^+\rightarrow c\overline{c}X$,
showing the fitted distribution 
for a) $\delta_{\rm t} = \deltat-\deltat_{\rm true}$ and b) the normalized
difference $\delta_{\rm t}/\sigma_{\deltat}$.
The one sigma error envelope from the
fit to the $B^+\to c\overline{c}X$ sample (shaded region), 
overlaps the central value for the five-times larger
$B^+\to D^{(*)}X$ sample (dashed line). 
}
\label{vali:brec-cc-resolution-comp}
\end{center}
\end{figure}

Control samples are also used to check the assumption that the $\Delta t$
resolution function, which is primarily determined by the $B_{\rm flav}$
sample, can be applied to the charmonium decay modes in the \CP\ sample.
Figure~\ref{vali:brec-cc-resolution-comp} graphically compares the fitted
\deltat\ resolution function for the $B^+\rightarrow D^{(*)}X$ control sample
with that of the $B^+\rightarrow c\overline{c}X$ control sample.
A 1$\sigma$ error envelope encompasses 
the fit to the $B^+\rightarrow c\overline{c}X$ 
sample, which has five times fewer events.
The level of agreement is acceptable.
The same comparison between the $B^0\rightarrow D^{(*)}X$  and
$B^0\rightarrow c\overline{c}X$ samples was inconclusive due to the low
statistics of the $B^0\rightarrow c\overline{c}X$ sample.

\subsubsection{Time-integrated measurement of mistag rates}
\label{sec:onebin}
As described in Section~\ref{sec:Introduction}, a
time-integrated technique can also be used to measure the mistag fractions in data, thereby
providing a simple check of the likelihood fit method. The statistical precision
of the time-integrated measurement is enhanced by restricting the sample to
events in a single optimized \deltat\ interval.
Taking into account detector vertex resolution, the
optimal interval is found to be  $|\deltat| < 2.5$\ps.  
Events with $|\deltat|> 2.5$\ps\
have, on average, equal numbers of mixed and unmixed events due to
flavor oscillations, and therefore
contribute nothing to the determination of the mistag rate.
We refer to this time-integrated
technique using a single optimized \deltat\ interval as the ``single-bin''
method and apply it to both the $B_{\rm flav}$ sample described in Section~\ref{subsec:sample_hadronicBz}
and the semileptonc \Bz\ sample described in Section~\ref{subsec:sample_semileptonic}.

To correct for the presence of backgrounds, a term is added to
Eq.~\ref{eq:TagMix:Integrated} to account for the
contribution of
each background source to the fraction of mixed events in the sample:
\begin{equation}
\label{eq:onebin_bkgd}
\chi_{\rm obs} = f_{\rm sig} (\chi_d  + (1 - 2 \chi_d )\, \mistag) + 
\sum_{\beta} f_{\beta} \chi_{\beta}, 
\end{equation}
where $f_{\rm sig}$ and $f_{\beta}$ are the fraction of signal and background
source $\beta$, respectively, $\chi_{\beta}$ is the fraction of
mixed events in each background source, 
and $\chi_{\rm obs}$ is the
observed fraction of mixed events.  
In this expression,
$\chi_d$ must also be modified to represent the integrated mixing
probability for $|\deltat| < 2.5$\ps.  Using the world-average values
for \deltamd\ and $\tau_{\Bz}$~\cite{PDG2000}, and
taking into account the 
\deltat\ resolution function ${\cal{R}}(\deltat)$, 
we find
\begin{eqnarray}
\chi_d^{\prime} & =& \frac{1}{2} \lbrack 1 - 
\int e^{-|\deltat|/\tau}\cos(\deltamd \deltat) \otimes {\cal{R}}(\deltat)\,d(\deltat)/ \nonumber \\
&&\int e^{-|\deltat|/\tau} \otimes {\cal{R}}(\deltat)\,d(\deltat) \rbrack
                 = 0.079,
\end{eqnarray}
where the integral is performed over the range $\vert\deltat\vert<2.5\ps$ and
${\cal{R}}(\deltat)$ is modeled by a double-Gaussian distribution with five
parameters (one fraction, two biases and two widths) 
determined directly from data using the hadronic sample.
Solving Eq.~\ref{eq:onebin_bkgd} for $\mistag$, and using the calculated
value for $\chi_d^{\prime}$, the mistag rates are obtained:
\begin{equation}
\label{eq:onebin_w}
\mistag = {{\chi_{\rm obs} - f_{\rm sig} \chi_d^{\prime}  - \sum_{\beta} f_{\beta} \chi_{\beta}}
\over {f_{\rm sig} (1 - 2 \chi_d^{\prime} )} } .
\end{equation}

\begin{table}[htb]
\begin{center}
  \caption{Yields, efficiencies, mistag rates $\mistag$, and
tagging separation $Q=\epsilon_{tag}(1-2\mistag)^2$
as measured by the single-bin method in the hadronic
$B_{\rm flav}$ event sample. 
A comparison of the mistag rates
measured in the same sample with the single-bin method, $w_{\rm sb}$,
and the likelihood fit, $w_{\rm like}$ (Table~\ref{tab:mistagCPall}),
are reported as the differences $\Delta_{\rm like}=w_{\rm sb}-w_{\rm like}$
between the two extraction techniques, 
normalized to the uncorrelated statistical and systematic errors. 
\label{tab:onebin_hadronic}}   \vspace{0.3cm}
  \begin{tabular}{|l|c|c|c|c|c|}\hline
Category & Yield & Efficiency & Mistag rate \mistag & $Q$ & $\Delta_{\rm like}$ \\
& & [\%] & [\%] & [\%] & [$\sigma$] \\
    \hline\hline
{\tt Lepton}    & 1128 & $11.0\pm 0.3$ &  $9.5\pm 1.5 \pm 0.6$   &  7.2 & $-0.8$ \\
{\tt Kaon}      & 3687 & $35.8\pm 0.5$ & $17.8\pm 1.0 \pm 0.7$   & 14.8 & $-0.4$ \\
{\tt NT1}       &  819 &  $7.9\pm 0.3$ & $22.0\pm 2.2 \pm 0.9$   &  2.5 & $+0.0$ \\ 
{\tt NT2}       & 1428 & $13.9\pm 0.3$ & $34.3\pm 1.9 \pm 1.1$   &  1.4 & $+0.8$ \\
    \hline
  \end{tabular}
\end{center}
\end{table}

All tagged events in the $B_{\rm flav}$ sample with  $|\Delta t| < 2.5$\ps\ are used for a single-bin study.
The combinatorial background fraction
in the signal sample is determined from a fit to the \mes\ distribution
as described in Section~\ref{sec:Mixing}.  
The signal
region is defined as events with $\mes > 5.27 $\gevcc.
The \Bu\ peaking background in this signal is estimated to be $(1.3 \pm 0.8)\%$.
The fraction of
mixed events in the combinatorial background is determined by tagging category
with the sideband
control sample, $5.20< \mes < 5.27$\gevcc, and 
the mistag fraction associated with the \Bu\ peaking background has 
been measured directly in data. 
The number of
tagged events in each category is summarized in 
Table~\ref{tab:onebin_hadronic}.

\begin{table}[htb]
\begin{center}
  \caption{Yields, efficiencies, mistag rate $\mistag$, and
tagging separation $Q=\epsilon_{tag}(1-2\mistag)^2$ 
as measured by the single-bin method in the
semileptonic \Bz\  event sample. 
A comparison of the mistag rates
measured with the single-bin method  
are reported as the differences $\Delta_{\rm sample}=w_{\rm flav}-w_{\rm sl}$
between the mistag rates in
the $B_{\rm flav}$ sample, $w_{\rm flav}$ (Table~\ref{tab:onebin_hadronic}),
and semileptonic \Bz\ samples, $w_{\rm sl}$,
normalized to the quadratic sum of statistical and uncorrelated systematic errors. 
\label{tab:onebin_semileptonic}}   \vspace{0.3cm}
  \begin{tabular}{|l|c|c|c|c|c|}\hline
Category & Yield & Efficiency & Mistag rate \mistag & $Q$ & $\Delta_{\rm sample}$ \\
& & [\%] & [\%] & [\%] & [$\sigma$] \\
    \hline\hline
{\tt Lepton}    &  3046 & $11.9\pm 0.4$ &  $8.7\pm 0.9 \pm 1.4$ &  8.1 & $+0.4$ \\
{\tt Kaon}      & 10270 & $36.2\pm 1.9$ & $19.5\pm 0.7 \pm 1.2$ & 13.5 & $-1.1$ \\
{\tt NT1}       &  2127 &  $8.1\pm 0.4$ & $22.3\pm 1.4 \pm 1.2$ &  2.5 & $-0.1$ \\
{\tt NT2}       &  3967 & $13.5\pm 0.9$ & $36.0\pm 1.2 \pm 1.3$ &  1.1 & $-0.7$ \\
    \hline
  \end{tabular}
\end{center}
\end{table}

A separate single-bin analysis is also performed
with the sample of \bzdstlnu\ events described in 
Section~\ref{subsec:sample_semileptonic}.
We use tagged events with $|\Delta t| < 2.5$\ps\ and evaluate
the backgrounds for events in this time interval. 
The backgrounds and mixed-event fractions
are evaluated separately for each tagging category.
Backgrounds are larger for the semileptonic modes than for the
hadronic modes and originate from a variety of sources.
In the case of the combinatorial background, the estimate is obtained from 
the $m(\Dzb\pim)-m(\Dzb)$ sideband. For the continuum background, 
off-resonance data is used after correction for
the combinatorial component. The  mixed-event fraction for 
\BB\ background is estimated with generic \BB\ Monte Carlo simulation. 
The mistag fraction of the last background component, the decay 
\bchdstxlnu, has been determined with data.  
The estimates of the contributions of
the various backgrounds are described in 
Section~\ref{subsec:sample_semileptonic}.
The number of
tagged events in each category are summarized in 
Table~\ref{tab:onebin_semileptonic}.

We use Eq.~\ref{eq:onebin_w} to obtain the mistag rates in each tagging
category shown in Table~\ref{tab:onebin_hadronic} 
for the $B_{\rm flav}$ sample and Table~\ref{tab:onebin_semileptonic}
for the \Bz\ semileptonic sample.
The sources of systematic error on these results
are summarized in  Tables~\ref{tab:syst_onebin_had} and 
\ref{tab:syst_onebin_semil} respectively.

\begin{table}[!htb]
\begin{center}
\caption {Sources of systematic error for the mistag measurement
on the $B_{\rm flav}$ sample in the single-bin method.
\label{tab:syst_onebin_had}} \vspace{0.3cm}
  \begin{tabular}{|l|c|c|c|c|c|}
\hline
Type                    & Variation     &{\tt Lepton}   &{\tt Kaon}& {\tt NT1} & {\tt NT2}  \\
\hline\hline
$\tau(\Bz)$, \deltamd\  &$\pm1\sigma$   & 0.005         &0.004          &0.003      &0.002  \\
Resolution              & see text      & 0.002         &0.002          &0.001      &0.001  \\
Wrong-tag resolution    & see text      & 0.003         &0.006          &0.007      &0.009  \\
Combinatorial bkgd      &$\pm1\sigma$   & 0.002         &0.002          &0.005      &0.004  \\
$B^{\pm}$ peaking bkgd  &$\pm1\sigma$   & 0.001         &0.001          &0.000      &0.000  \\
\hline
Total                   &               & 0.006         &0.007          &0.009      &0.011  \\
\hline

\end{tabular}
\end{center}
\end{table}

\begin{table}[!htb]
\begin{center}
\caption {Sources of systematic error for the mistag measurement
from the semileptonic \Bz\ sample in the single-bin method.
\label{tab:syst_onebin_semil}} \vspace{0.3cm}
  \begin{tabular}{|l|c|c|c|c|c|}
\hline
Type                    & Variation & {\tt Lepton} & {\tt Kaon} & {\tt NT1} & {\tt NT2}  \\
\hline\hline
$\tau(\Bz)$, \deltamd\  & $\pm1\sigma$     & 0.006 &0.004  &0.004 &0.002   \\
Resolution              & see text         & 0.001 &0.001  &0.001 &0.001 \\  
Wrong-tag resolution    & see text         & 0.003 &0.006  &0.007 &0.009  \\
Combinatorial bkgd      & $\pm1\sigma$     & 0.001 &0.006  &0.004 &0.004 \\
Continuum bkgd          & $\pm1\sigma$     & 0.001 &0.003  &0.006 &0.007 \\
\BB\ bkgd               & $\pm1\sigma$     & 0.011 &0.005  &0.005 &0.004 \\
\bchdstxlnu\ bkgd       & $\pm1\sigma$     & 0.003 &0.004  &0.002 &0.001   \\   
\hline
Total                   &                  & 0.014 &0.012  &0.012 &0.013 \\
\hline
\end{tabular}
\end{center}
\end{table}

Three sources of systematic uncertainties are common to both the hadronic and
semileptonic samples. The first is the uncertainty due to the errors on the
world-average values for the \Bz\ lifetime and  \deltamd\ values. The second is 
due to the \deltat\ resolution function, whose fit parameters in data are varied
within errors. 
The third common uncertainty is related to 
the possibility that wrong tags have worse
\deltat\ resolution than correct tags.  This effect has
been studied with Monte Carlo simulation, where we observe a slightly larger RMS 
width for events with wrong-sign tags. From this study,
scale factors comparing the right and wrong-tag resolution functions have been 
extracted and then applied 
to the resolution function for wrong tags. 

The systematic uncertainties unique to each sample are
due to the background components. These are estimated by varying both the background 
fractions $f_{\beta}$ and the fraction of mixed events associated with each background source,
$\chi_{\beta}$, by one standard deviation in their uncertainty.

For the semileptonic sample, the systematic error due to backgrounds
is the dominant source. The characterization 
of these various backgrounds is described in Section~\ref{subsec:sample_semileptonic}. 
For the combinatorial 
background fraction, a relative systematic uncertainty of 20\% 
is added in quadrature to the statistical error
to cover the range of results obtained with various $m(\Dzb\pim)-m(\Dzb)$ 
fitting functions. 

The systematic error due to the continuum background is 
determined by varying both the background level and the mixed fractions. 
The \BB\ background fraction uncertainty is obtained by combining the 
statistical uncertainty 
and the systematic error given in Section~\ref{subsec:sample_semileptonic}. 
The systematic errors introduced by uncertainties 
on the background from the decay \bchdstxlnu\ are obtained by varying 
the fraction described in Section~\ref{subsec:sample_semileptonic}
as well as the mistag fraction of \Bu\ mesons measured on data. 
Studies with Monte Carlo simulation have been 
performed to verify that the mistag fractions are not affected by the 
presence of the extra pions 
in the decay \bchdstxlnu. An additional uncertainty due to the
statistical precision of the 
Monte Carlo study has been added to the charged $B$ mistag fractions measured with data.  

Table~\ref{tab:onebin_hadronic} shows the difference 
$\Delta_{\rm like}=w_{\rm sb}-w_{\rm like}$
between the mistag rates 
measured with the single-bin method in the $B_{\rm flav}$ sample, $w_{\rm sb}$, and
the likelihood fit result, $w_{\rm like}$ (Table~\ref{tab:mistagCPall}). 
The difference is reported in terms of the
uncorrelated statistical and systematic errors for the two methods, when 
applied to the same data sample. The component of
uncorrelated statistical error is estimated with a fast parameterized 
Monte Carlo simulation. It varies with
category due to different event yields.
The differences $\Delta_{\rm sample}=w_{\rm flav}-w_{\rm sl}$ 
in the mistag rates measured with the 
single-bin method in the $B_{\rm flav}$ sample, $w_{\rm flav}$, and in the 
semileptonic \Bz\ sample, $w_{\rm sl}$, 
are reported in Table~\ref{tab:onebin_semileptonic}. The 
quadratic sum of the statistical and uncorrelated systematic errors is used to 
estimate the consistency of the measurements.

\subsubsection{Vertexing algorithm checks}
\label{sec:algchecks}

In order to verify that the results are stable under variation of the vertexing algorithm that is
used for the measurement of \deltat, several less powerful alternatives to the default method
have been considered:
\begin{itemize}

\item {\bf{Charmonium mass constraint for vertex fit:}} The mass constraint on the charmonium 
daughter, used in the 
selection of the events, is also applied in the determination of the vertex.
\item {\bf{No \mbox{\boldmath $\KS$} mass constraint:}} The mass constraint on the \KS\ candidate 
is not applied during the vertex reconstruction.
\item {\bf{No Bremsstrahlung recovery:}} Only events without an associated Bremsstrahlung photon
for the \jpsi\ daughter electrons are considered in the likelihood fit. 
\item {\bf{Charmonium daughters only:}} The vertex of the fully reconstructed \B\ meson is reconstructed only with
the tracks from the charmonium daughter.

\item {\bf{No converted photon veto:}} Pairs of tracks from gamma conversions 
are retained in the vertex fit.
\item {\bf{\mbox{\boldmath $\sigma_{\deltat}$} requirement:}} Only events with  
$\sigma_{\deltat}<1.4\ps$ are retained, as is required
in the mixing analysis.
\item {\bf{Boost approximation:}} The boost approximation (Eq.~\ref{eq:dt_boost_approx}) is used 
to convert the $\Delta z$ measurement into \deltat.
\item {\bf{Kaon veto:}} The more restrictive requirement from 
the mixing analysis that no kaons participate in the tagging vertex 
is applied.
\item {\bf{No \mbox{\boldmath $\Upsilon(4S)$} constraint:}} The algorithm described in section \ref{subsec:dz_reconstruction}
is simplified by dropping the $\Upsilon(4S)$ momentum constraint.

\item {\bf{Dilution dependence on \mbox{\boldmath $\sigma_{\deltat}$}:}} Dilutions for the kaon category are parameterized as a function of the error on \deltat.
\end{itemize}

\begin{figure}[!htb]
\begin{center}
 \includegraphics[width=\linewidth]{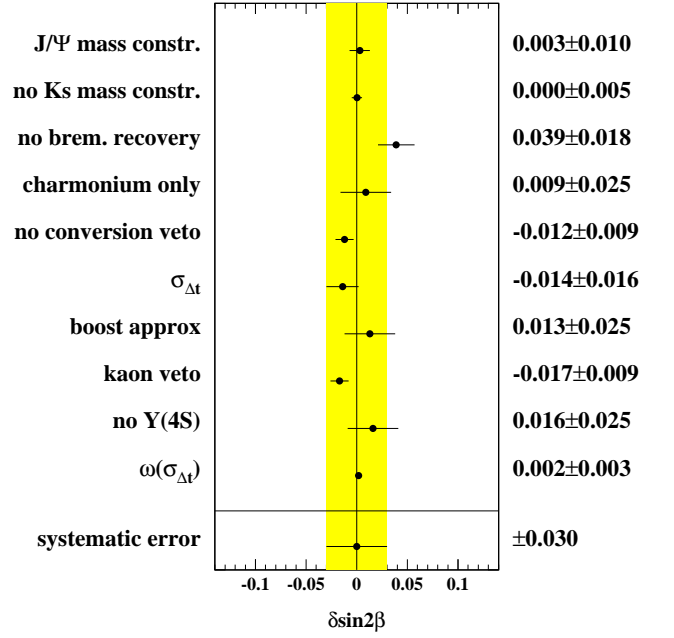}
\caption{
Results obtained with several alternative variations on the vertexing algorithm 
that impact the \deltat\ measurement. The shaded band is 
the systematic error assigned to the parameterization of the resolution function. 
The full range corresponds to one statistical standard deviation.}
\label{vali:altfits}
\end{center}
\end{figure}

A summary of the results obtained with these different
configurations for the \deltat\ determination is provided
in Fig~\ref{vali:altfits}. In all cases the variation of the measured asymmetry 
is consistent with the error assigned to the parameterization of the resolution 
function.

\subsubsection{$\jpsi\KL$ background cross checks}

As a cross check, a likelihood fit was performed to $\jpsi\KL$ candidates
in a $\Delta E$ sideband region 
($20<\Delta E<80$\mev) treated entirely as signal events.
This sample is actually a mixture of 
\B\ decay modes with an expected average $\eta_{CP}$ of $+0.04$.
The true value for \stwob\ in the Monte Carlo simulation is 0.7 and
consequently the
expected result from the likelihood fit to the \deltat\ distribution
of the control sample is 0.03.
The actual fits to sideband regions in data and Monte Carlo simulation find
$\stwob=0.16\pm0.18$ and $-0.03\pm0.10$ respectively, 
both of which are consistent with expectations.

As another cross check,  a sample of $\jpsi\KS$ events was selected
in the data, where only the \KS\ direction information was used, 
thereby emulating the \KL\ selection.
The purity and background composition of this control sample is very similar
to that of the $\jpsi\KL$ sample.
However, in this case, 
the subset of true $\jpsi\KS$ events can be identified with
the normal $\jpsi\KS$ selection criteria.
A fit to the $\Delta E$ distribution
of the full control sample finds $(49\pm 3)$\% signal, which is
in good agreement with the fraction, 47\%, observed for
the cleanly identified $\jpsi\KS$ subset.
Likewise, a likelihood fit to the \deltat\ distribution
of the full control sample agrees well with the value of
\stwob\ obtained with the true $\jpsi\KS$ subsample.

\subsubsection{Graphical display of the asymmetry}

An elegant display of the \CP\ asymmetry in the data can be 
obtained with the use of the so-called {\em Kin}
variable, hereafter denoted as $\kin$. It is also possible
to verify directly the fitted value for \stwob\ from the ratio
of appropriate weighted averages for \kin.  In particular,
\kin\ has a PDF with an asymmetry known to be linearly dependent
with a slope given by \stwob\ regardless of the details of
the analysis.

Writing the PDF ${\cal F}_+ ({\cal F}_-)$
for events with a \Bz\ (\Bzb) tag
in terms of the general functions $F_1(\deltat)$ and
$F_2(\deltat)$
\begin{equation}
{\cal F}_\pm(\deltat) = F_1(\deltat)\pm \stwob\, F_2(\deltat) 
\end{equation}
allows us to introduce
\begin{equation}
\kin(\deltat)=\pm F_2(\deltat)/F_1(\deltat),
\end{equation}
where $+$ applies to events with a \Bz\ tag and $-$
with a \Bzb\ tag. Ignoring resolution
effects, dilutions and background, the Standard Model expectation
for the \deltat\ distribution of tagged \Bz\ decays
into \CP\ modes (Eq.~\ref{eq:TimeDep}) gives
$\kin(\deltat)=-\etaCP\sin \deltamd \deltat$.  
When these effects are included, we can still
write
\begin{equation}
{\cal F}_\pm(\deltat) = F_1(\deltat)(1+\kin(\deltat)\stwob),
\end{equation}
although $\kin$ will be a more complicated function of $\deltat$
and could depend on kinematic variables as well.

\begin{figure}[!htb]
\begin{center}
 \includegraphics[width=\linewidth]{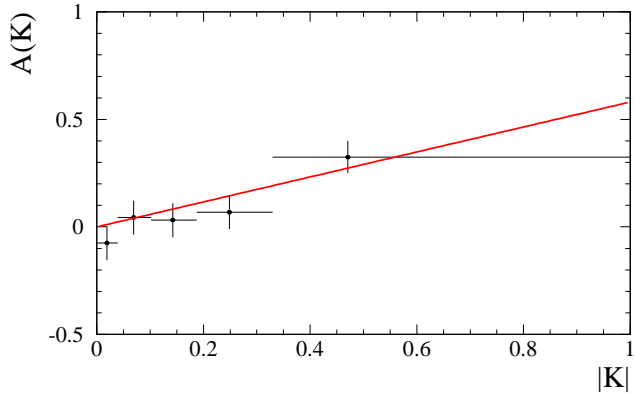}
\caption{Observed asymmetry ${\cal A}(\kin)$ as a function
of $|{\cal K}|$, with an overlay of the expected linear
dependence superimposed. }\label{fig:KinCheck}
\end{center}
\end{figure}

The distribution of events as a function of $\kin$ is
\begin{eqnarray}
\frac{dN}{d\kin}&=&\int d\deltat\,
\left[{\cal F}_+ \delta\left(\kin-\frac{F_2}{F_1}\right)+{\cal F}_-\,\delta\left(\kin+\frac{F_2}{F_1}\right)\right]\nonumber \\
&=&(1+\kin\stwob)\int d\deltat\times \nonumber \\
& & \qquad F_1\left[\delta\left(\kin-\frac{F_2}{F_1}\right)+\delta\left(\kin+\frac{F_2}{F_1}\right)\right]\nonumber \\
&=&(1+\kin\stwob)\Psi(\kin),\label{eq:kin}
\end{eqnarray}
where $\Psi(\kin)$ is an even function of \kin.  It follows that the ratio of
the odd to the even part of the distribution for \kin\ is a linear
function of \kin\ with coefficient \stwob.  Thus, the distribution of
\kin\ can be used to test for the effect of an \CP\ violation simply by
examining the dependence of the asymmetry
\begin{equation}
{\cal A}(\kin) = 
\frac{dN_{{\cal K}>0}/d\kin-dN_{{\cal K}<0}/d\kin}
     {dN_{{\cal K}>0}/d\kin+dN_{{\cal K}<0}/d\kin}.
\end{equation}  
The observed asymmetry for the \CP\ sample is shown 
in Fig.~\ref{fig:KinCheck} as a function of \kin, 
along with an overlay of the expected linear 
dependence. The data agree with this hypothesis at the 55\% C.L.

From the expression in Eq.~\ref{eq:kin}  we find
\begin{equation}
\stwob = {\sum_i\kin_i\over \sum_i\kin_i^2}\pm
           {1\over \sqrt{\sum_i\kin_i^2}}\sqrt{1-(\stwob)^2
                 {\sum_i\kin_i^4\over \sum_i\kin_i^2}}
\end{equation}
In averaging \kin\ the even component of the $\Psi(\kin)$ cancels out, while 
the odd component cancels in averaging $\kin^2$. This offers a 
method of measuring \stwob\ that is mathematically equivalent 
to the result with the global likelihood fit. However, it can only be 
applied when \stwob\ is the one remaining free parameter. The
moments of \kin\ for the full \CP\ sample 
give results that are numerically identical to the likelihood fit,
thereby confirming the minimization procedure used for the fit.

\begin{figure}[!htb]
\begin{center}
 \includegraphics[width=\linewidth]{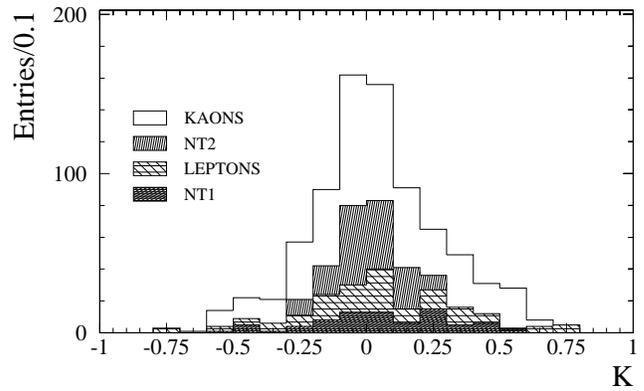}
\caption{Distribution of the observable ${\cal K}$ for the 
individual tagging categories.}\label{fig:KinCP}
\end{center}
\end{figure}

The fact that the mean value of \kin\ is proportional to \stwob\ also 
allows a visual representation of the \CP\ asymmetry. 
Fig.~\ref{fig:KinCP} shows the distribution of \kin\ in data, with 
events in the individual tagging categories indicated as well. 
The larger the value of \kin\ for a given event, the larger
the weight that this event carries in the measurement of \stwob.
Again, the \CP\ asymmetry in the data is clearly evident in the
distribution of \kin.

\subsection{Fits results without assuming $\abslambda\ = 1$}
\label{sub:dCP}

A more general description of the time evolution of neutral $B$ decays
to \CP\ eigenstates contains a term proportional to $\cos \deltamd
\deltat$ (Eq.~\ref{eq:direct}). The coefficient of the cosine term is
expected to be negligible in the Standard Model, where
$\abslambda = 1$. In order to search for a non-Standard Model effect, we
fit the $\etaCP = -1$ sample for \abslambda\ and \etaimlambdaoverabslambda. 
The latter is equal to \stwob\ if $\abslambda = 1$. The $\etaCP = -1$
sample has the advantage of having very little background, while the other \CP\ modes have
backgrounds that are both significantly larger and dominated by
other \B\ decay modes with possible direct \CP\ contributions.

The fitted values for \abslambda\ and \etaimlambdaoverabslambda\ with the
$\CP =-1$ sample and various subsamples are listed in
Table~\ref{tab:result}.  
The two \CP parameters are almost uncorrelated, with the
coefficient between \etaimlambdaoverabslambda\ and \abslambda\ of $-1.7\%$.
The same systematic error studies as described in
detail in Section~\ref{subsec:stwob-systematics} were repeated for the fit to the $\etaCP
= -1$ sample for \etaimlambdaoverabslambda\ and \abslambda. The
estimated uncertainties from these sources are listed in
Table~\ref{tab:systematics}.

We have also performed detailed cross checks, similar to those described 
in Section~\ref{sub:validation}. 
In particular, large samples of parameterized simulation, as well as full Monte Carlo samples, 
have been used to verify the fitting 
procedure. The $B_{\rm  flav}$ sample has also been used to demonstrate that 
no bias is introduced in the measurement. 
The relative normalization of the tagged events in the two flavors is in 
fact sensitive to the  coefficient of 
the cosine term in Eq.~\ref{eq:direct}, and therefore \abslambda.  The systematic 
error introduced by the uncertainty on
the parameters $\langle\epsilon_{\rm tag}\rangle_i$ and $\mu_i$ listed in 
Table~\ref{tab:dircp-mu} are uncorrelated between tagging categories. Therefore,
they are added in quadrature to obtain
the systematic error contribution listed in Table~\ref{tab:systematics}(g).

The final result of the fit with the $\etaCP=-1$ sample is:
\begin{eqnarray}
\abslambda = 0.93\pm 0.09\pm 0.02\ \mbox{and} \nonumber \\
\etaimlambdaoverabslambda = 0.56\pm 0.15\pm 0.05.
\end{eqnarray} 
Thus, we find no evidence for direct \CP\ violation in the $\etaCP=-1$ sample and 
the value of  \etaimlambdaoverabslambda\ is
consistent with the result from the nominal \CP\ fit with $\abslambda=1$.

\renewcommand{\secname}{Conclusions}
\section{Conclusions and prospects}
\label{sec:\secname}

In 29.7\invfb\ of \epem\ annihilation data collected
near the \FourS\ resonance, we have obtained a new measurement of the 
time-dependent \Bz-\Bzb\ oscillation frequency with a sample of 6350 
tagged flavor-eigenstate
\Bz\ meson decays that are fully-reconstructed in hadronic final states:
\begin{eqnarray*}
\dmresult.
\end{eqnarray*}
This result is at a level of precision comparable to the most recent
world average for \deltamd\ and lies about $1.7\sigma$ above the 
combined value of $0.472\pm 0.017\hbarps$~\cite{PDG2000}. It is also
quite compatible with our own recent measurement~\cite{babar0123}
with a dilepton sample.  The \deltamd\ study reported here
confirms our understanding of $B$ reconstruction, flavor tagging,
and \deltat\ resolution in our data sample. Our measurement contributes
significantly to the precision of the determined value for \deltamd,
one of the fundamental parameters constraining our knowledge of the CKM matrix,
and remains dominated by statistical errors that will improve with more data. 

We have presented a measurement of the \CP-violating 
asymmetry parameter \stwob\ in the neutral $B$ meson system:
\begin{equation}
\result,
\end{equation}  
which establishes \CP\ violation in the \Bz\ system at the $4.1\sigma$ level.
This significance is computed from the sum in quadrature of 
the statistical and additive systematic errors.
The probability of obtaining the observed value or higher in the 
absence of \CP violation is less than 
$3 \times 10^{-5}$. The corresponding probability for the 
$\etaCP=-1$ sample alone is $2 \times 10^{-4}$.
Our measurement is consistent at the $1.9\sigma$ level
with the recently reported result from Belle of 
$\stwob = 0.99\pm 0.14\, {\rm (stat)}\pm 0.06\, {\rm (syst)}$~\cite{Belle}, 
and with previous measurements~\cite{s2bave}.
The observed value for \stwob\ is currently limited by the size 
of the \CP\ sample, allowing for substantial improvement as more data
is recorded in the next few years.

We have also used the $\etaCP=-1$ sample to search for possible
direct \CP\ violation through interference of decay amplitudes. 
The direct \CP\ parameter $\lambda$ is found to be:
\begin{equation*}
\abslambda = 0.93\pm 0.09\,{\rm (stat)}\pm 0.02\,{\rm (syst)}.
\end{equation*} 
This result is consistent with the Standard Model expectation, 
where $|\lambda|=1$
and no significant direct \CP\ violation should exist in charmonium decays.

As already noted in Section~\ref{sec:Introduction}, measurements of 
\CP\ asymmetries in $B$ decays to 
charmonium can be used to constrain, with little theoretical ambiguity, 
the parameters of the CKM matrix.
In the Standard Model with three families, the CKM
matrix $V$~\cite{CKM} incorporates three real parameters and 
one phase $\delta$ generating \CP\ violation if 
$\delta \neq 0$ or $\pi$. The Wolfenstein parameterization~\cite{wolfenstein} 
of $V$ takes advantage of the observed hierarchy in the matrix elements 
in terms of the expansion parameter $\lambda_{CKM} = |V_{us}|$. 
The remaining parameters in this representation are 
denoted $A$, $\rho$, and $\eta$, where \CP\ violation requires $\eta \neq 0$.

The parameter $\lambda_{\rm CKM}$ is determined from 
semileptonic kaon decays and nuclear 
$\beta$ decays. Semileptonic $B$ meson decays to charm
are used to determine the parameter $A$. 
Constraints on $\rho$ and $\eta$ are obtained from 
\CP\ violation in mixing in the kaon sector $|\epsilon_{K}|$, 
the ratio $|V_{ub}/V_{cb}|$, and the oscillation frequency 
\deltamd\ for \Bz-\Bzb\ mixing. The oscillation frequency 
$\Delta m_{s}$ has not been measured, since $\Bz_{s}$-$\Bzb_{s}$ mixing 
has not been observed yet. However, the observed amplitude 
spectrum ${\cal{A}}(\Delta m_{s})$ improves the constraints 
on $\rho$ and $\eta$. Together, these measurements provide 
indirect constraints on \stwob. 

Our overall knowledge of the CKM parameters is limited 
by the relatively large 
uncertainties in some of the theoretical quantities, mainly due 
to non-perturbative QCD effects. 
In particular, the constraints on $\rho$ and $\eta$ suffer from theoretical 
and systematic uncertainties in the determination of 
$|V_{ub}/V_{cb}|$ and from theoretical 
uncertainties in QCD parameters entering the prediction of 
$|\epsilon_{K}|$, \deltamd, and $\Delta m_{s}$. Recent analyses 
constraining the CKM matrix have been performed with 
different statistical approaches~\cite{hlll,PlaszczynskiSchune,ciuchinietal,Bargiottietal,Faccioli,AtwoodSoni,AliLondon}.
They mainly differ in the treatment of theoretical 
uncertainties and also in the choice 
of the input values and their errors.

\begin{figure}[!htb]
\begin{center}
 \includegraphics[width=\linewidth]{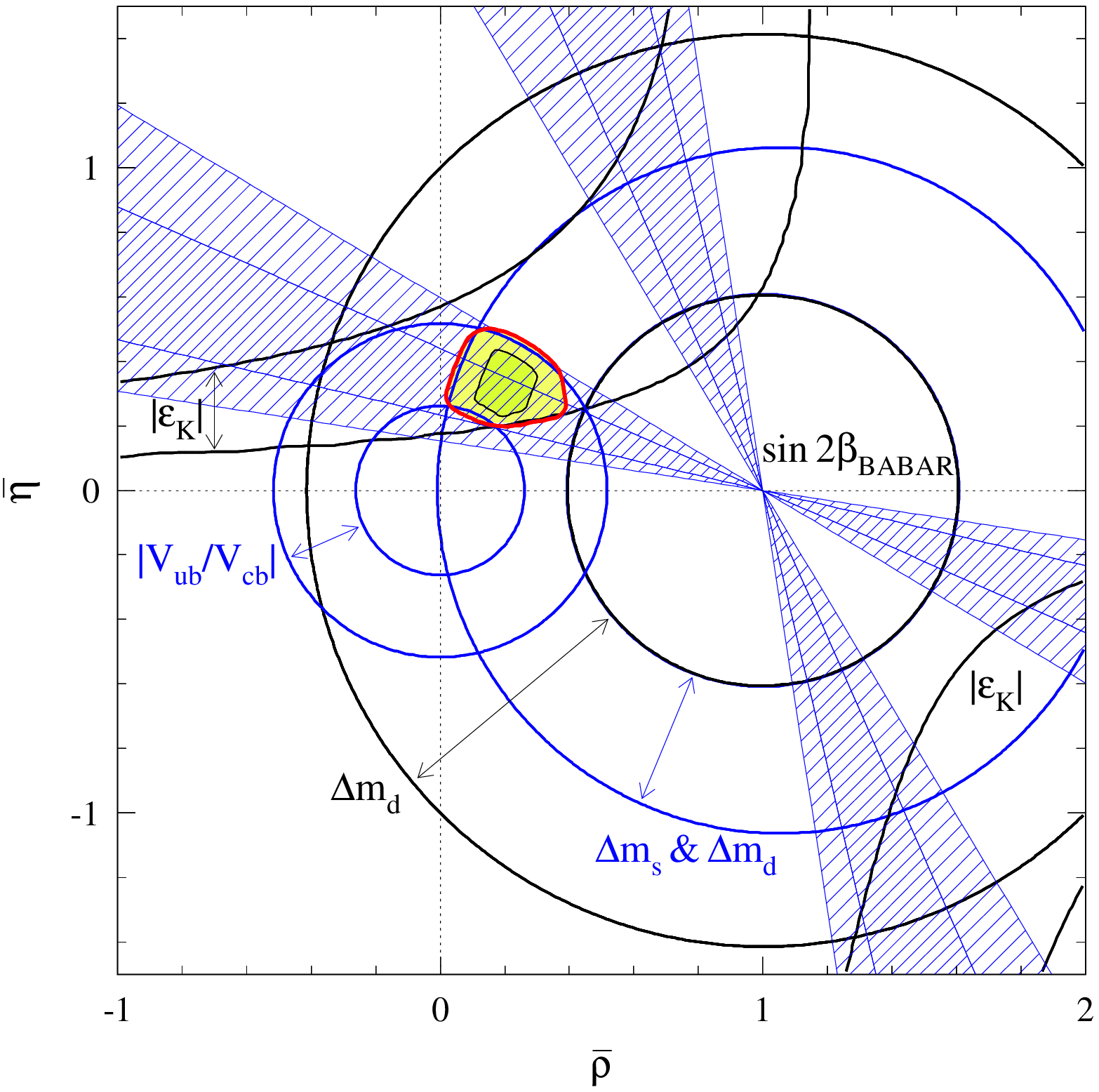}
\caption{
Present indirect constraints on the position of the apex of the 
Unitarity Triangle in the 
($\overline{\rho},\overline{\eta})$ plane,
not including our measurement of \stwob.  
The fitting procedure is described in 
Ref~\cite{hlll}. 
Our result \result\ is represented by diagonally hatched regions,
corresponding to 
one and two statistical standard deviations. The individual indirect constraints
lie between the pairs of solid lines that are connected by
the double-ended arrows with labels.
}
\label{fig:UnitarityTriangle}
\end{center}
\end{figure}

Due to the four-fold ambiguity in the value of 
$\beta$ obtained from the $\stwob$ 
measurement, there are four allowed regions in the 
$\rho$-$\eta$ plane. One of these 
regions is found to be in agreement with the 
allowed $\rho$-$\eta$ region obtained from 
CKM fits within the Standard Model. Figure~\ref{fig:UnitarityTriangle}, taken
from Ref.~\cite{hlll}, shows our direct measurement
and the indirect constraints
in the $\overline{\rho}-\overline{\eta}$ plane in terms of the 
renormalized parameters $\overline{\rho}=\rho(1-\lambda_{CKM}^2/2)$ 
and $\overline{\eta}=\eta(1-\lambda_{CKM}^2/2)$. 
The contributions of the individual measurements 
$|\epsilon_{K}|$, $|V_{ub}/V_{cb}|$, 
\deltamd, and $\Delta m_{s}$~\cite{hlll} are indicated, 
as well as the allowed region 
if all the constraints are considered simultaneously. Overlaid 
as the diagonally-hatched area are the regions corresponding
to one and two times the one-standard-deviation experimental uncertainty
on our \stwob\ measurement. 

It should be emphasized that, beyond being a direct constraint
on $\beta$, the measurement of \stwob\ differs 
qualitatively in its interpretation from the indirect 
constraints on $\beta$ obtained from $|\epsilon_{K}|$, 
$|V_{ub}/V_{cb}|$, \deltamd, 
and eventually $\Delta m_{s}$. 
For \stwob, the size of the allowed 
domain is determined by well-defined experimental 
uncertainties that are predominantly statistical in origin, 
while in contrast the region allowed by the 
indirect measurements is mostly defined by theoretical uncertainties,
which makes a statistical interpretation difficult.

The current experimental uncertainty on \stwob\ has now 
reached a level of precision that offers
significant constraint on the Standard Model. Over the 
next few years there will continue to 
be substantial improvements in precision of the 
\stwob\ determination, including measurements 
for other final states in which \CP-violating asymmetries are proportional to 
\stwob. Beyond this, studies of time-dependent asymmetries in modes involving
$b\to u$ transitions have already begun~\cite{babar0121} 
and may provide additional constraints, although here the
interpretation in terms of \stwoa\ from the Unitarity Triangle is likely to
be made difficult due to significant penguin contributions. 
Nevertheless, these measurements will be able to directly test the validity 
of the CKM picture as the origin for the
observed \CP\ violation in neutral \B\ decays.

\section{Acknowledgments}
\label{sec:Acknowledgments}


We are grateful for the 
extraordinary contributions of our \pep2\ colleagues in
achieving the excellent luminosity and machine conditions
that have made this work possible.
The collaborating institutions wish to thank 
SLAC for its support and the kind hospitality extended to them. 
This work is supported by the
US Department of Energy
and National Science Foundation, the
Natural Sciences and Engineering Research Council (Canada),
Institute of High Energy Physics (China), the
Commissariat \`a l'Energie Atomique and
Institut National de Physique Nucl\'eaire et de Physique des Particules
(France), the
Bundesministerium f\"ur Bildung und Forschung
(Germany), the
Istituto Nazionale di Fisica Nucleare (Italy),
the Research Council of Norway, the
Ministry of Science and Technology of the Russian Federation, and the
Particle Physics and Astronomy Research Council (United Kingdom). 
Individuals have received support from the Swiss 
National Science Foundation, the A. P. Sloan Foundation, 
the Research Corporation,
and the Alexander von Humboldt Foundation.


\end{document}